# Investigations on Lorentzian Spin-foams and Semiclassical Space-times

## Dissertation

zur Erlangung des akademischen Grades
Doctor Rerum Naturalium.
Vorgelegt dem Rat der Physikalisch-Astronomischen
Fakultät der Friedrich-Schiller-Universität Jena.

von

José Diogo de Figueiredo e Simão

geboren am 13.02.1997 in Coimbra, Portugal

Dissertation submitted in accordance with the requirements for the degree of Doctor Rerum Naturalium,

presented to the Council of the Faculty of Physics and Astronomy of the Friedrich Schiller University of Jena.

Gutachter:

1. Dr. Sebastian Steinhaus, *Friedrich-Schiller Univ. Jena*
2. Prof. Dr. Simone Speziale, *Aix-Marseille Univ.*
3. Prof. Dr. Bianca Dittrich, *Perimeter Institute*

seit 1558

Tag der Disputation: 30. April 2024

Para o meu avô,
a quem tanto devo;

e às minhas manas,
a quem tanto quero.

*Os navegadores, dado que com o fervor da obra*
*e alvoroço daquela empresa embarcaram contentes,*
*também passado o termo do desferir das velas,*
*vendo ficar em terra seus parentes e amigos*
*e lembrando-lhes que sua viagem estava posta em esperança*
*e não em tempo certo nem lugar sabido,*
*assi os acompanharam em lágrimas (aos que ficaram)*
*como em o pensamento das cousas*
*que em tão novos casos se representam na memória dos homens.*

— João de Barros, Décadas 1, Livro IV (1552)

# Preface

The following dissertation is the cumulative result of about three and a half years of research carried out at the Friedrich-Schiller University of Jena. Formally it serves as a necessary partial fulfillment of the requirements for a doctoral degree. Effectively it is an entire fulfillment of my own notion of personal satisfaction. I have been lucky enough to have had the opportunity to not only conduct the research I deemed meaningful; but more importantly to have experienced all that grounds it in substance - the people I have met, the places I have visited, and the lessons I have learned. As I condense this chapter of my life in a handful of pages I am reminded of the words that line the door to the office I have visited countless times: *We have not succeeded in answering all our problems; indeed we sometimes feel we have not completely answered any of them. The answers we have found have only served to raise a whole set of new questions. In some ways we feel that we are as confused as ever, but we think we are confused on a higher level and about more important things.* Though I have learned much I remain helplessly confused, and I cherish these past years for having kept me so.

## *Abstract*

This work is developed in the context of the spin-foam approach to quantum gravity; all results are concerned with the Lorentzian theory and with semiclassical methods.

A correspondence is given between Majorana 2-spinors and time-like hypersurfaces in Minkowski 3-space based on complexified quaternions. It is shown that the former suggest a symplectic structure on the spinor phase space which, together with an area-matching constraint, yields a symplectomorphism to $T^*\mathrm{SU}(1,1)$. A complete 3-dimensional Lorentzian spin-foam amplitude for both space- and time-like triangles is proposed. It is shown to asymptote to Regge theory in the semiclassical regime.

The asymptotic limit of the 4-dimensional Conrady-Hnybida model for general polytopes is scrutinized. Minkowski's theorem on convex polyhedra is generalized to Lorentzian signature, and new boundary states for time-like polygons are introduced. It is found that the semiclassical amplitude for such polygons is insufficiently constrained.

A method for the asymptotic evaluation of integrals subject to external parameters is discussed. The method is developed in detail for the special problem of spin-foam gluing constraints away from their dominant critical points. A relation to the gluing constraints of effective spin-foams is suggested.

## Organization of the text

The body of this monograph is separated into four main chapters, a final overview, and a number of appendices. Many of the appendices are strictly necessary for the arguments made throughout.

The first chapter serves as an introduction to the theoretical framework which underpins the work. It contains a general discussion of classical general relativity as well as a review of the spin-foam quantization programme. Special attention is payed to the underlying assumptions of both theories. The Barrett-Crane, Engle-Livine-Pereira-Rovelli, Conrady-Hnybida and coherent-state spin-foam models are discussed. I make a number of run-through remarks on spin-foam quantum gravity.

The second chapter is concerned with the spinorial description of geometrical objects in the spin-foam context. A correspondence between different types of spinors and the surfaces of transitivity of the Lorentz group is established by means of complexified quaternions; the representation theory of the spin Lorentz group in two dimensions is developed in these terms. It is shown that there exists a symplectomorphism between the phase space on the group and $\mathbb{C}^4$ equipped with an appropriate symplectic structure and constraints; such constraints relate to the area of time-like hypersurfaces. A new coherent-state vertex amplitude is proposed for Lorentzian 3-dimensional quantum gravity. Both space- and time-like triangles are considered. It is shown that the semiclassical limit of the amplitude matches the Regge action.

The third chapter is dedicated to an exploration of the semiclassical limit of the Conrady-Hnybida extension. New boundary states for time-like polygons are introduced, and general methods for critical point analysis are developed. A Lorentzian-signature Minkowski theorem for convex 3d polyhedra is proven.

The fourth and last main chapter introduces the notion of gluing constraints to the spin-foam amplitude. A method is developed - based on the work of Hörmander - to explore the semiclassical regime of the gluing constraints for arbitrary boundary data, i.e. away from their critical points. The constraints are shown to qualitatively match their homonym proposal in effective spin-foams.

## Acknowledgments

*Querias agora, portanto,*
*insistir no avanço depois da suspensão...*
*tardia e obstinadamente,*
*desimpedir os planos*
*ao secreto enlevo de uma primeira canção.*

— Paulo Tavares, Órbitas (2023)

The greatest privilege of submitting a doctoral dissertation is having the opportunity to look back and give recognition to all the people that made it possible.

I am first of all thankful to my academic references, to whom I look up to and whose teachings shaped my scientific beliefs. I am indebted to my supervisor, Sebastian Steinhaus, for having guided me with patience and care throughout this journey, not only on technical aspects. I must also thank Martin Ammon, who I think of as a mentor, for both fruitful scientific discussions and for his support on a number of academic matters, present and future. I am grateful to Fernando Nogueira for having set me on the path of theoretical physics; to Stefan Hofmann for rekindling that flame; and to Daniele Oriti for introducing me to the beautiful subject which was to be my work. I thank Katrin Kanter and Lisann Schmidt for their attention and help in all matters bureaucratic.

Then come the people I have been surrounded by over the past three years. A special word of acknowledgement must be given to my friend, flatmate and colleague Ivan Soler, who I stood with in braving the insurmountable jungles of research and social life in Jena. Of Seán Gray and Michael Mandl I keep fond memories of a deep and meaningful friendship, even though practical matters of life may have taken us on different paths. I cherish all the time I spent with Richard Schmieden, whose kindness is perhaps only surpassed by his baking skills. To Kemal Döner and Leonhard Klar I am grateful for everything that we shared, from cinema to philosophy. I am indebted to Alexander Jercher for his friendship, and for that special kind of intimacy which comes from working on the same kind of abstract nonsense. I am thankful to Hrólfur Ásmundsson, Markus Schröfl, Marta Picciau, Dimitrios Gkiatas, Johannes Schmechel, Christiane Klein, Jan Mandrysch and Seth Asante for their comradery, and for all the discussions on matters of physics and life. I benefited from fruitful discussions on conformal field theories with Jacob Hollweck. In Patricia Peinado I have found a kindred soul, and in Ana Bebeti incomparable generosity.

My greatest gratitude is to my family, who unwaveringly stood behind me all my life: to my parents, to whom I owe myself; to my sisters who I dearly miss; to my cousin, godfather and friend, who taught me so much; to my grandfather, who - even in his confusion brought about by old age - not once forgot that it was physics that kept me away in a distant land. And to my dear friends (extended family really) that I so admire - João Diogo, José Bernardo, Maximilian Ruep, Aleksander Strzelczyk and João Oliveira - I simply refer you to those times I have told you how much our friendship means. I hope to have you all by my side for many years to come.

It has been a pleasure. The future is uncertain.

*Mas em boa verdade vos digo: daqui para a frente só para trás.*

I thank Richard Schmieden, Seth Asante, Alexander Jercher and João Oliveira for proof-reading this thesis. The work was partially funded by Deutsche Forschungsgemeinschaft (DFG) Project No. 422809950 and DFG Grant No. 406116891 within the Research Training Group RTG 2522/1.

# Contents

# I. Introduction: Through It All Things Fall

*Und als ich meinen Teufel sah, da fand ich ihn ernst, gründlich, tief, feierlich:*
*es war der Geist der Schwere - durch ihn fallen alle Dinge.*

— Friedrich Nietzsche, Also sprach Zarathustra (1883)

We are to concern ourselves throughout this thesis with the problem of understanding gravity. As one of the fundamental forces, the gravitational interaction is certainly the most immediate to human experience: it manifests its influence not by some abstract phenomena at the subatomic level, like the strong and weak interactions; not by visual sensations or special metals, as electromagnetism; but rather by a very direct - and very unavoidable - actual force. One constantly *feels* gravity. It hovers over humanity like a formless specter, perpetually weighting on its shoulders. Such a constant influence ascribes to gravity a certain prosaic quality, one which justifies it being the first of the forces to be reasonably understood (via Newton's theory of universal gravitation). And herein lies one of the most curious circumstances of theoretical physics: that, in searching for an increasingly accurate description of such a seeming triviality as why things fall, it has turned out that one ought to understand instead why space and time themselves bend as they do. This is the domain of general relativity (henceforth GR), currently our best attempt at describing gravitational phenomena.

Although GR is an exceedingly well established theory (yielding to this day impressive experimental results, as the recently obtained image of the M87 black hole attests [1]), its scope of validity is not expected to hold beyond that which one would consider to be the "classical regime". All remaining basic interactions admit a more fundamental (and extremely well-tested) quantum-field-theoretic description, together with the ontological and epistemological baggage quantum theory demands. If one is to retain a mechanistic understanding of the world, the ordering of its parts better be coherent; as it stands, GR is a bona-fide classical theory. It is realistic, its observables have a deterministic dynamics, and it accounts for no quantum phenomena. Such a blatant lack of consistency would be enough reason for the theoretician to consider a possible adaptation of the theory to the quantum regime [2, 3], but there exist also natural phenomena expected to be understood only in that context. Black hole thermodynamics [4,5], space-time singularities [6] and early-universe cosmology [7] are three examples that cannot be described strictly from within the confines of GR. One expects on general grounds for quantum gravitational effects to be definitely non-negligible at the Planck length scale $l_P \sim 10^{-35}$m, comfortably away from contemporary experimental technology (making quantum gravity a truly strictly theoretical problem for now, notwithstanding the undergoing efforts to establish a phenomenology of QG, see e.g. [8]). Still, to the best of my knowledge, there is currently no definite proof that gravity should be quantized, and this remains a point of contention in theoretical physics [9–12]. The assumption that gravity ought to admit a more fundamental quantum description is central to this monograph.

***

With the goal of laying the foundations for the following chapters, this introduction is dedicated to a general overview of both the classical theory of gravity and the state-of-the-art of the spin-foam framework. I shall start with the former.

## I.1 Elements of general relativity

GR is an ab-initio theory of principles, meaning it is the synthesis of a set of expectations we hold to be true about the world, rather than a model retroactively adapted to predict particular phenomena. These expectations are born from a long history of successful and unsuccessful attempts at describing nature, and they constitute the general world-view of the physicist's orthodoxy. Following Einstein [13], we may identify two foundational ideas:

- General Principle of Relativity: *the content of physical laws should be independent of the reference frame used to describe them;*

- Principle of Equivalence: *an inertial reference frame subject to gravity is indistinguishable from an accelerated one in the absence of gravity.*

It is remarkable that one can make substantial progress in the direction of formulating the theory simply from these two basic principles (together with a healthy dose of guessing guided by practice).

**The general principle of relativity**

The requirement that a physical theory should be independent from a particular choice of reference frame immediately suggests making use of the machinery of (smooth; *natura non saltum facit*) geometry: the theory is to be constructed as an action functional $S$ of a Lagrangian 4-form $\mathcal{L}$ over a 4-dimensional smooth manifold $M$,

$$S[\phi] = \int_M \mathcal{L}\,(\phi) \ ,$$

depending on fields $\phi$ associated to the phenomena of interest in a manner which is yet to be determined. By construction, the above functional is invariant under diffeomorphisms of the underlying manifold and relevant induced pull-backs and push-forwards of additional structures on which the action depends, and it is thus also invariant under chart (i.e. coordinate) transformations. Consequently, any dynamics determined from an extremal action will also be invariant under coordinate changes, and hence so will physical predictions. One devises in this manner a diffeomorphism-invariant theory, one of the hallmarks of GR[1].

**The principle of equivalence**

Implementing the second principle requires a bit more effort. It being a consequence of the equivalence of the inertial and gravitational masses, it may be understood as implying that a theory of gravity is in essence a theory of accelerated frames. A moment of thought suffices to show that, for a sufficiently small region, any accelerating frame can be taken as inertial (as in the classic example of the experimentalist on a free-falling elevator). In the limit of abstraction, then, an accelerated frame is point-wise described by an inertial one; this we take as structural to the theory, and we are interested in local isomorphisms (vector space isomorphisms at each fiber together with local diffeomorphisms between the bases)

$$\begin{aligned} e : U \subset E &\to TM \\ v_I &\mapsto e_I = e_I^\mu \partial_\mu \,, \end{aligned}$$

mapping from a 4-dimensional vector bundle $E$ to the tangent space of the base manifold $M$, where $v_I$ are a choice of basis for $E$. The mapping $e$ assigns to such a basis a set of four tangent vectors to the base manifold, which are interpreted as an inertial frame at every point. We expect the dynamics of the theory to determine the particular form of $e$, and thus too the behavior of all inertial frames. Realizing that Lorentz transformations of inertial frames must be redundant, we further prescribe a gauge-theory structure by introducing a $G$-principal bundle $P$, with $G = \mathrm{SO}(1,3)$ the Lorentz group:

$$\mathrm{SO}(1,3) \hookrightarrow P \xrightarrow{\pi} M \,.$$

[1]Note that in usual quantum field theories (QFT) the Lagrangian form depends on structures beyond the domain of the action functional, namely the Minkowski metric. Thus if one considers a diffeomorphism of only the configuration fields the action will not remain invariant. This is why such QFTs are said to *not* be diffeomorphism-invariant.

We proceed by upgrading $E$ to an associated vector bundle constructed on Minkowski space $\mathbb{R}^4$, together with an action of $g \in G$ via the fundamental representation $\rho(g)$:

$$E \simeq P \times \mathbb{R}^4/G^{\sim} \,, \quad (u \cdot g, v) \sim (u, \rho(g)v) \ .$$

Note that a choice of connection $\omega \in \Omega\,(P, \mathfrak{g})$ for $P$ induces a covariant derivative on $E$ [14, Sect. 10.4], and consequently a covariant derivative $\nabla$ on $TM$ via the $e$ isomorphism. This construction allows one to recover a Lorentzian metric $g$ locally by setting

$$g = \eta_{IJ}\theta^I \otimes \theta^J \ ,$$

where $\theta$ is the dual form to $e$, and in this manner $g(e_I, e_J) = \eta_{IJ}$, as intended for an inertial frame. One may moreover show that the covariant derivative $\nabla$ is metric-compatible with respect to $g$. The interested reader is directed to [14, 15] for further mathematical details on the gauge theory framework.

Having laid down the objects of the theory, it remains to propose a Lagrangian. This requires some guess-work. Since we have a gauge theory, we expect the local connection $A$ (a pull-back of $\omega$ under a section of $P$) to be a configuration variable, and for the curvature $F[A] = \mathrm{d}A + A \wedge A$ to figure in the Lagrangian. This still needs to be composed with a 2-form in a scalar manner, and our only remaining object is the tetrad one-form $\theta$; a particularly simple action[2] is given by

$$S\,[\theta, A] = \int_M \star\,(\theta_I \wedge \theta_J) \wedge F^{IJ}[A]\,, \tag{I.1.1}$$

which is indeed the well-known tetrad action for general relativity, a first-order formulation of the more conventional Einstein-Hilbert action. In the absence of matter, variation of $S[e, A]$ with respect to the connection implies vanishing torsion, from where one recovers dynamically the usual Levi-Civita connection. By extremizing the action with respect to the tetrad field $\theta$ one recovers the Einstein field equations in the absence of matter, i.e. $\epsilon_{IJKL}\theta^J \wedge F^{KL} = 0$. The reader is referred to [16] for a more extensive discussion of the tetrad framework in gravity.

[2] The inclusion of the Hodge star is necessary, as otherwise the Lagrangian is identically zero when the torsion vanishes. I will return to this point in section I.2.2.

**Interpretational consequences**

As mentioned above, GR is inherently a theory of inertial frames; gravity dictates at every point which frames are inertial, up to a Lorentz group redundancy. It is furthermore, by virtue of the absence of additional a-priori (or external) structures in (I.1.1), a fully diffeomorphism-invariant theory. This has a rather profound ontological consequence: that the theory requires no objects beyond the tetrad field and the connection, and in particular that the underlying manifold $M$ serves uniquely a parametrization purpose. Indeed, invariance under coordinate transformations removes any operational significance from charts $\psi : U \subset M \to \mathbb{R}^4$, which would otherwise (as in the case e.g. of classical electromagnetism) be interpreted as a laboratory frame. This description of gravity, then, up to the domain where we know it to be empirically valid, deals away with the notion of an absolute background space-time, a staple of pre-relativistic physics dating back to Newton's *Principia*. All measurements of space distances or time durations are to be determined via diffeomorphism-invariant constructions from the tetrad - or equivalently the metric -, and thus time and space acquire in GR the character of bona-fide physical (i.e. dynamical, subject to influences and influencing) objects. Moreover, although abstract charts are operationally meaningless, one can still introduce other physical fields (e.g. matter), with respect to which gravitational observables may be constructed. This is the foundation of relational observables [17]; mathematical charts are replaced by dynamical fields such as clocks and rods, themselves subject to interactions, and thus representing a more faithful description of physical reality. The world-view afforded by the theory is hence one where there is no substantial underlying space-time where systems reside and move about; rather, locally[3] there exist *only* systems (of which space-time is but one) and their relative motions. I believe these conceptual lessons to be meaningful for the construction of the quantum theory.

## I.2 The spin-foam approach to quantum gravity

Obtaining a full-fledged quantum theory of gravity remains an open problem. The conservative approach of adapting gravity to the QFT framework - where the metric field is treated as a perturbation of some fixed background, usually a flat Minkowski one - is well-known to fail as a fundamental theory (although still useful effectively [18]), insofar as it is perturbatively non-renormalizable; this circumstance could be argued to be a manifestation of GR's inherent lack of a-priori objects. There exist however a large number of candidate theories (or rather putative theories) employing methods and ideas beyond QFT, each with their own points of emphasis, underlying assumptions, scope and mathematical structure. An incomplete list would include string-theory [19], loop quantum gravity (LQG) [20], group field theory [21], causal dynamical triangulations [22] and asymptotic safety [23], to name but a few; see also [24]. What follows is a non-exhaustive general review of the spin-foam framework.

### I.2.1 Guiding principles

Deriving a quantum theory from a classical one is a very complicated matter, because the order of deduction is inverted: a quantum theory ought to be more fundamental than its classical counterpart. It must give rise to the classical theory, and not the other way around. Alas, human beings - and physicists in particular - are irremediably classical beings, for which most physical intuitions and expectations follow from apprehending a classical world. Even the experimentalist who observes quantum phenomena does so through the lens of classical measuring devices, like a pointer in a gauge. Fortunately, physicists have derived certain heuristics which help discerning what a quantum theory should look like.

[3]A geometrical formulation of GR requires a choice of underlying manifold. This amounts to imposing certain topological restrictions on how local solutions to the theory may be patched together, and this choice is thus empirically substantial at a more global level. It is an interesting interpretational question whether this manifold can be argued to have ontological significance, or whether such restrictions can be thought of as measurable "boundary data".

One of the most powerful of such heuristics, due to Feynman [25], is the path-integral (or sum-over-histories) construction: the notion that systems behave probabilistically in a manner which is weighted by their behavior in an otherwise classical regime. Not only has this method been extraordinarily successful in quantum field theory, e.g. as the foundation to functional methods applied to the Standard Model, it also prescribes a rather straightforward mold onto which one may try to force gravity. Considering some abstract Hilbert space $\mathcal{H}_\Sigma$ associated to a 3-dimensional boundary $\Sigma = \partial M$, one expects amplitude maps

$$\begin{aligned} \rho_M &: \mathcal{H}_\Sigma \to \mathbb{C} \\ |\psi\rangle &\to \int_{(e,A)|_\Sigma=\psi} \mathcal{D}e\mathcal{D}A \; e^{\frac{i}{\hbar}S[e,A]} \,, \end{aligned} \tag{I.2.1}$$

where the integration over field configurations is such that the fields respect the boundary data characterizing $|\psi\rangle$. Such a generalized path-integral construction accounts for the fact that the underlying manifold $M$ may or may not have a simply-connected boundary, and it makes no assumption on the causal character[4] of the boundary surface. It makes minimal use of geometrical and metric-derived concepts, since one expects that in a

[4]Here and throughout this work the term *causal character* will always refer to the type of induced metric on the hypersurface in question. If the induced metric is Riemannian the surface is deemed space-like; if it is Lorentzian it is called time-like. If the surface is orthogonal to a null direction, in which case the induced metric is degenerate, the surface is termed light-like.

quantum theory the metric - a classical solution to the field equations - is dissolved into its quanta, from which it is unrecoverable[5]. Thus equation (I.2.1) generalizes the usual QFT situation where one has initial and final space-like hypersurfaces. Operational and conceptual concerns for such a framework have been extensively discussed by Oekl in what he termed the General Boundary Formalism [26–28].

As written, equation (I.2.1) is entirely formal. There are a number of very different quantum gravity proposals which all share this foundation, most eminently spin-foams, causal dynamical triangulations and perturbative gravity. Their point of departure is precisely in giving meaning to the abstract symbols above. Considering the conceptual discussion of the previous section, the spin-foam approach assumes that the absence of a-priori structures is a defining characteristic of GR. It follows: 1) that the theory ought not to be perturbative around a fixed background, departing from perturbative gravity; 2) that one ought to avoid Wick-rotating the action as not all metrics admit the procedure, thus insisting in a Lorentzian path-integral. The manner in which spin-foam models ascribe substance to the algebraic and path-integral structures of equation (I.2.1) is the subject of the following.

### I.2.2 The Holst action and $BF$ quantization

Rather than starting from the tetrad action of equation (I.1.1), one may consider the more general Holst action [29] (in full analogy with the $\theta$-term in QCD, see [30, Ch. 13])

$$S[\theta, A] = \int_M (\star + \gamma)\,(\theta_I \wedge \theta_J) \wedge F^{IJ}[A]\,, \qquad \text{(I.2.2)}$$

which differs from the former via the inclusion of the so-called Holst term, whose relative weight is controlled by the Immirzi parameter $\gamma$. From a classical point of view, and in the absence of matter, this additional term is inconsequential. The equations of motion determine the torsion to vanish $T^I := \mathrm{d}\theta^I + A^I{}_J \wedge \theta^J = 0$, and one may use Bianchi's identity relating torsion and the curvature tensor to show the Holst term to be identically zero. On the other hand, the Holst action is of fundamental importance to LQG, with which one may want to make contact through the spin-foam approach. Indeed, the canonical analysis of the tetrad action leads to very complicated constraints [31], which have been shown to simplify under a particular canonical transformation [32],

$$A \mapsto (\star \pm \gamma)\, A\,,$$

where the Hodge dual acts on the $\mathfrak{so}(1,3)$ vector space. The choice $\gamma = i$ amounts to a projection onto the self- (or anti-self-) dual part of a complexified $\mathfrak{so}(1,3)_{\mathbb{C}}$ connection[6], resulting in what is known as the Ashtekar variables [33], to be supplemented by reality conditions. One may alternatively take $\gamma$ to be real, in which case the resulting object is a real $\mathfrak{so}(3)$ connection, no longer Lorentz; still, under an appropriate choice of gauge and restriction to a space-like hypersurface, this variable can still be used to describe Lorentzian canonical quantum gravity [34]. It has thus become customary in the literature to allow $\gamma$ to be a free parameter of the theory, and it further figures in important predictions of LQG for the area spectrum [35, 36] and black hole entropy [37]. This freedom is captured correctly by the Holst action [38], which we then take as the action to quantize.

[5] In much the same way, one cannot expect to reconstruct a classical field configuration from $n$-particle states in a QFT.

The manner in which spin-foam models prescribe meaning to the path-integral (I.2.1) is by first constructing the partition function for a much simpler action known as $BF$-theory [39, 40]

$$S[B, A] = \int_M B_{IJ} \wedge F^{IJ}[A]\,, \quad B \in \Lambda^2 T^* M \otimes \mathfrak{g}\,, \tag{I.2.3}$$

[6] This is possible because the Hodge map in a 4-dimensional Lorentzian vector space $V$ acts on $\Lambda^2 V$ as a complex structure $\star^2 = -1$, a rather fortunate coincidence.

Table I.1: Cells of a 2-complex dual to a 4d triangulation

| 2-complex $\Delta^*$ | triangulation $\Delta$ |
|---|---|
| vertex $v$ | simplex $\sigma$ |
| edge $e$ | tetrahedron $\tau$ |
| face $f$ | triangle $t$ |

which reduces to the Holst action under the assumption of *simplicity* $B_{IJ} = (\star + \gamma)\ \theta_I \wedge \theta_J$. By quantizing such a purely topological action, and later enforcing the constraints at the quantum level, one sidesteps many of the difficulties with defining a path-integral for GR. As is customary, we will take the gauge group to be the double cover of the Lorentz group, i.e. $G = \mathrm{SL}(2,\mathbb{C})$. The $B$ field can be formally integrated, resulting in a Dirac delta which enforces the connection to be flat:

$$\int \mathcal{D}B\mathcal{D}A\, e^{i\int_M\ B_{IJ}\wedge F^{IJ}} = \int \mathcal{D}A\,\delta\left(F[A]\right)\,.$$

Leaving the matter of dealing with the constraints for later, we proceed by introducing a triangulation of the base manifold, i.e. a homeomorphism from $M$ to a geometrical realization of a simplicial complex $\Delta$. From $\Delta$ one can construct the Poincaré dual 2-complex $\Delta^*$ by mapping $(d-n)$-cells to $n$-cells, as in table I.1. A famous result in smooth geometry is that one can recover a connection from all parallel transports $\gamma \mapsto g_\gamma(A)$ on curves $\gamma$. The partition function of the discretized theory over $\Delta^*$ then takes the form

$$Z(\Delta^*) = \int \prod_e \mathrm{d}g_e \prod_f \delta\left(\prod_{e\in\partial f} g_e\right)\,, \tag{I.2.4}$$

where we have associated a group element $g_e$ to every edge $e \in \Delta^*$.

Although $\mathrm{SL}(2,\mathbb{C})$ is non-compact, it is locally compact, and functions in $L^2(G)$ admit a Plancherel decomposition into unitary irreducible representations [41, 42]. The reader is referred to appendix B for a review of the representation theory of $\mathrm{SL}(2,\mathbb{C})$, as well as for the associated notation which will be used throughout this monograph. The representations appearing in the Plancherel decomposition are known as the *principal series representations*, labeled by two "spins" $\chi = (\nu, \rho), \nu \in \mathbb{N}, \rho \in \mathbb{R}$. This allows for a formal decomposition of the delta function, which up to conventional pre-factors can be shown to read (cf. appendix B)

$$\delta(g) = \sum\!\!\!\!\!\!\int_{\nu,\rho} (\nu^2 + \rho^2)\,\mathrm{Tr}\left[D^\chi(g)\right] ,$$

for $D^\chi(g)$ a representation matrix. This equation can now be used to massage equation (I.2.4) in a series of steps which take into consideration the combinatorial structure of the simplicial complex, as follows:

$$\begin{aligned} Z(\Delta^*) &= \int \prod_e \mathrm{d}g_e \prod_f \left( \sum\!\!\!\!\!\!\int_{\nu,\rho} (\nu^2+\rho^2)\,\mathrm{Tr}\left[\prod_{e\in\partial f} D^\chi(g_e)\right]\right) \\ &= \sum\!\!\!\!\!\!\int_{\chi\to f} \int \prod_e \mathrm{d}g_e \prod_f \left( (\nu_f^2+\rho_f^2)\,\mathrm{Tr}\left[\prod_{e\in\partial f} D^{\chi_f}(g_e)\right]\right) \\ &= \sum\!\!\!\!\!\!\int_{\chi\to f} \prod_f (\nu_f^2+\rho_f^2)\,\mathrm{Tr}_f \left[\prod_e \left(\int \mathrm{d}g_e \prod_{f \text{ s.t. } e\in\partial f} D^{\chi_f}(g_e)\right)\right] . \end{aligned} \tag{I.2.5}$$

In passing to the second line we have exchanged the order of the product and the sum by summing over assignments of labels to faces $\chi \to f$. By $\mathrm{Tr}_f$ we mean that one contracts the indices of the product of matrices according to the combinatorics of the faces, i.e. matrices associated to the same face are contracted together (these manipulations become clearer in a diagrammatic language, which will follow.). By virtue of the left-right invariance of the Haar measure, we identify in the last line of (I.2.5) a projector[7] onto the

[7] Strictly speaking, due to non-compact nature of the Lorentz group, the map is only a pseudo-projector unless the Haar measure is regularized.

invariant subspace of a tensor product of representations associated to all faces $f$ sharing the same edge $e$,

$$\pi_e : \bigotimes_{f \text{ s.t. } e \in \partial f} \mathcal{H}^{\chi_f} \mapsto \mathrm{Inv}\left( \bigotimes_{f \text{ s.t. } e \in \partial f} \mathcal{H}^{\chi_f} \right)$$

$$\pi_e = \int_{\mathrm{SL}(2,\mathbb{C})} \mathrm{d}g_e \prod_{f \text{ s.t. } e \in \partial f} D^{\chi_f}(g_e) =:$$ 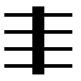 ,

where we have introduced a diagrammatic representation of the four matrices associated to the four faces sharing an edge[8], as well as the associated group integration. Thus, denoting by $d_f = \nu_f^2 + \rho_f^2$ the "dimension" of the representation,

$$Z(\Delta^*) = \sum\!\!\!\!\!\!\int_{\chi \to f} \prod_f d_f \, \mathrm{Tr}_f \Big[$$ 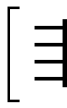 $$_{e,\{f\}} \Big] = \sum\!\!\!\!\!\!\int_{\chi \to f} \prod_f d_f \prod_v$$ 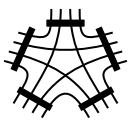 $_v$ . (I.2.6)

Equation (I.2.6) is composed of a sum over assignments of spin labels to every face (each of which is diagrammatically represented by a link), a product of face amplitudes $d_f$, and a product of vertex diagrams. The vertex diagrams are to be tiled together according to the combinatorics of $\Delta^*$ (and thus also $\Delta$, there being one such diagram per 4-simplex), in such a manner that one ends up with closed loops of links, i.e. matrix traces[9]. The vertex diagram admits a further refinement once one realizes that an orthonormal basis for $\mathrm{Inv}\left(\bigotimes_{f \text{ s.t. } e \in \partial f} \mathcal{H}^{\chi_f}\right)$ is given by intertwiners

$$\iota : \bigotimes_{f \text{ s.t. } e \in \partial f} \mathcal{H}^{\chi_f} \to \mathbb{C}\,, \quad \iota \circ \rho(g) v = \iota(v)\,,$$

[8] I remind the reader that the simplicial triangulation consists of 4-simplices and their subcells. Each 4-simplex has a boundary comprising of five tetrahedra, each of which containing four triangular faces.

[9] It is implicit in the construction that the manifold one starts with is boundary-less.

for which we can also assign a diagram as

$$\equiv = \sum_{\iota} |\iota\rangle\langle\iota| = \sum_{\iota} \frac{1}{\ominus_{\iota}} \ni^{\iota}\in \,,$$

where the melon-like drawing denotes a normalization factor such that $\pi_e \circ \pi_e = \pi_e$. It follows that the partition function (I.2.6) can equivalently be written as

$$Z(\Delta^*) = \sum\!\!\!\!\!\!\!\int_{\chi\to f,\, \iota\to e} \prod_f d_f \prod_e \ominus^{-1}_{\iota,\{\chi\}} \prod_v \bigstar_{v,\{\iota\},\{\chi\}} \,, \tag{I.2.7}$$

having introduced the $\mathrm{SL}(2,\mathbb{C})$ $15j$ symbol $\bigstar$. This is the canonical form of a spin-foam partition function, it being a state-sum of weights associated to the different cells of a triangulation.

Putting aside for now any concerns regarding the convergence of the above sum, equation (I.2.7) allows us to come full circle and interpret the objects of the prescription given in (I.2.1). The fact that the partition function is so neatly organized in terms of cells of the underlying simplicial complex allows one to perform an induction step, and to propose that it can define a transition amplitude provided some of the intertwiner and spin labels are regarded as boundary data (rather than them all being contracted against each other). This we do, accordingly:

1. The triangulation is now allowed to have a boundary $\partial\Delta$;
2. To every tetrahedron $\tau \in \partial\Delta$ one assigns the Hilbert space $\mathcal{H}_\tau = \bigoplus_{\chi_i} \mathrm{Int}\left(\bigotimes_{i=1}^{4} \mathcal{H}^{\chi_i}, \mathbb{C}\right)$;
3. The amplitude map $\rho_\Delta : \bigotimes_\tau \mathcal{H}_\tau \to \mathbb{C}$ is given by equation (I.2.7), where the sum over spins and intertwiners is restricted so as to conform with the boundary data characterizing a state $|\psi\rangle \in \bigotimes_\tau \mathcal{H}_\tau$.

I should remark that point 1 requires one to make an educated choice of basis for the traces figuring in the spin-foam amplitude, insofar as different such choices will lead to different kinds of induced boundary states, which should themselves have a physical and operational meaning. I will come back to this point later on.

As the simplest possible example of this framework, in the case where $\Delta$ is composed of a single 4-simplex $\sigma$, the amplitude map takes a set of five intertwiners, each labeled by an intertwiner label $\iota$ and four spins $\chi$, to the vertex amplitude

$\rho_\sigma :$ 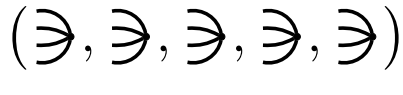 $\mapsto$ 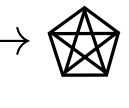 .

Note that these objects inherit the combinatorial properties of $\Delta$: each intertwiner diagram is dual to a tetrahedron with four faces, and each $15j$ diagram is dual to a 4-simplex with five boundary tetrahedra. Still, the five variables characterizing an intertwiner are not enough to uniquely determine a tetrahedron (otherwise characterized by its six edge lengths), from where the fuzzy character of quantum theory can be argued to manifest itself [39].

It is worthwhile to take a step back and appreciate what has been achieved. Starting from classical $BF$-theory, all geometric objects have been diluted into purely combinatorial and algebraic ones - a foam of spins, whence "spin-foam". If a method to deform this quantum theory into a gravitational one via the simplicity constraints is devised, one will have achieved a representation of quantum gravity without any reference to geometry. In this scenario a geometric space-time would not be fundamental, but rather emergent from more foundational structures.

### I.2.3 Simplicity constraints and current spin-foam models

Following the discretization philosophy above, one can naturally associate to every triangle $t \in \Delta$ a bivector as

$$b_t^{IJ} = \int_{t\in\Delta} B^{IJ} \, .$$

It has been shown [43] that the bivectors $b_t^{IJ}$ in a tetrahedron $\tau \in \Delta$ can be enforced to be either simple or the Hodge duals of simple bivectors through the following so-called *linear simplicity constraints*[10]:

$$\begin{cases} \exists\, n_I^\tau \in \mathbb{R}^4 \;\text{ s.t. }\; n_I^\tau b_t^{IJ} = 0\,, \;\; \forall t \in \tau \quad \Leftrightarrow \quad b_t^{IJ} = \theta_t^I \wedge \theta_t^J\,, \\ \exists\, n_I^\tau \in \mathbb{R}^4 \;\text{ s.t. }\; n_I^\tau (*b_t)^{IJ} = 0\,, \;\; \forall t \in \tau \quad \Leftrightarrow \quad b_t^{IJ} = \star(\theta_t^I \wedge \theta_t^J)\,. \end{cases}$$

Going back to $BF$ theory (I.2.3), one sees that the Holst action of equation (I.2.2) is recovered if one imposes

$$\frac{1}{\gamma^2+1}(\gamma - \star)B^{IJ} = \theta^I \wedge \theta^J\,, \tag{I.2.8}$$

and as such, still in a discretized classical context, one arrives at the constraints

$$\forall \tau\,, \;\; \exists\, n_I^\tau \;\text{ s.t. }\; n_I^\tau \left[\frac{1}{\gamma^2+1}(\gamma - \star) b_t^{IJ}\right] = 0\,. \tag{I.2.9}$$

Note that this form of the constraints requires picking an orthogonal vector $n_\tau$ at each tetrahedron, a choice which assigns to them a particular causal character: the tetrahedron can either be space-, time- or light-like. This is interpreted as additional necessary information for the theory.

The various spin-foam models now differ[11] in the way the constraints (I.2.9) are applied at the quantum level, and in which states are assigned to the boundary. With regards to the former, the fact that the vector space of Lorentzian bivectors $\Lambda^2\mathbb{R}^{1,3}$ is isometric to the dual lie algebra $\mathfrak{sl}(2,\mathbb{C})^*$ of the relevant symmetry group, which by itself admits a canonical Poisson structure induced by the Lie bracket, suggests that one should use that identification to define a phase space for bivectors associated to a tetrahedron. This both fixes a Poisson structure for bivectors, which will lead to commutators at the quantum level, and provides a quantum formulation for the simplicity constraints. Baez [44] has proceeded in this manner for a Riemannian signature in the context of geometric quantization, obtaining tetrahedron Hilbert spaces which are $\mathfrak{so}(4)$ analogs of the $\mathcal{H}_\tau$ defined above. An overview of some current models follows.

[10] This terminology differentiates the linear constraints from the *quadratic simplicity constraints* [43], which have historically preceded them in the literature.

[11] Spin-foam proposals employing methods other than those outlined in this work exist in the literature, e.g. the Reisenberger and the Gambini-Pullin models. A good reference is the review by Perez [40].

*(a) The Barrett-Crane (BC) model*

The BC model [45,46] was one of the first to be proposed. It is constructed for the case $\gamma = 0$, i.e. for when the action to be quantized does not contain a Holst term. The quantum constraints are formulated by restricting to the preferred time-like normal vector $n_\tau = \hat{e}_0$, thus making all the tetrahedra space-like. It suggests the correspondence

$$\begin{aligned} \beta_{\text{BC}} :\ \Lambda^2\mathbb{R}^{1,3} &\to \mathfrak{sl}(2,\mathbb{C}) \\ b_t^{IJ} &\to J^{IJ}\,, \end{aligned}$$

which, together with (I.2.9), leads to a condition on rotation generators

$$\mathcal{C}^i = L^i \sim 0\,.$$

These constraints are closed under the Lie bracket $[\mathcal{C}^i, \mathcal{C}^j] = i\epsilon^{ij}{}_k\mathcal{C}^k$, so they are first-class and can be imposed strongly on the SU(2) canonical $|(\nu,\rho);\ j,m\rangle$ basis of $\mathcal{H}^\chi$ (see appendix B), obtaining states $|(0,\rho);0,0\rangle$. Referring back to the $BF$ spin-foam amplitude, simplicity is imposed at the boundary of *every* 4-simplex, such that the vertex amplitude is simply given by restriction of the spins in equation (I.2.7). The boundary states are characterized by four $\rho > 0$ labels at every tetrahedron, forming the so-called Barrett-Crane intertwiners $|\iota_{BC}\rangle$

$$|\iota_{\text{BC}}\rangle\langle\iota_{\text{BC}}|_{\rho_1,\ldots,\rho_4} = \int_{\text{SL}(2,\mathbb{C})} \mathrm{d}g\,\prod_{i=1}^{4} D^{(0,\rho_i)}(g)\,.$$

An area operator can be constructed from the simple bivectors of equation (I.2.8),

$$A_{\mathrm{BC}}^2 = \frac{1}{2}\langle - \star b_t,\, - \star b_t \rangle = -\frac{C_1}{2} = \frac{\rho^2+4}{4} > 0\,.$$

*(b) The Engle-Pereira-Rovelli-Livine (EPRL) model*

The EPRL model [47] is defined for a general Immirzi parameter, but it still considers the same space-like tetrahedra with $n_\tau^I = \hat{e}_0$. It prescribes a different Poisson structure for the bivectors through the correspondence

$$\begin{aligned} \beta_{\mathrm{EPRL}} :\ \Lambda^2\mathbb{R}^{1,3} &\to \mathfrak{sl}(2,\mathbb{C}) \\ b_t^{IJ} &\to \gamma \star J^{IJ}\,, \end{aligned}$$

which is $\star$-dual to the one considered in the BC model, and weighted[12] by $\gamma$. Such a correspondence can be motivated from a covariant phase space formalism perspective [48, 49]. It leads to the simplicity condition

$$\mathcal{C}^i = K^i + \gamma L^i \sim 0\,. \tag{I.2.10}$$

This time the constraints do not close, and are thus second-class. The proposal of [47] is to implement them via the "master constraint" procedure [50], which defines $\mathcal{M} = \mathcal{C}_i\mathcal{C}^i$ and imposes it by finding those states which have a minimal eigenvalue. One arrives at the spin labels

$$\begin{cases} \nu = \sqrt{4j(j+1)-2} \sim 2j\,, \\ \rho = \gamma\nu\,. \end{cases}$$

Simplicity is again applied at the boundary of every 4-simplex by restriction of the allowed labels. The resulting boundary states are the EPRL

[12] In the original EPRL paper the authors started from a Holst action which differs from the one presented here by an overall factor of $\gamma^{-1}$. The above formulation of the EPRL correspondence is thus not usual in the literature, although something similar had already appeared in [37].

intertwiners $|\iota_{EPRL}\rangle$

$$\sum_{\iota} |\iota_{\text{EPRL}}\rangle\langle\iota_{\text{EPRL}}|_{\iota,j_1,\ldots,j_4} = \int_{\text{SL}(2,\mathbb{C})} \mathrm{d}g \prod_{i=1}^{4} D^{(2j_i,2\gamma j_i)}(g)\,.$$

each associated to a tetrahedron, and each depending on an SU(2) spin $j$ and on an intertwiner label $\iota$, a structure which is very reminiscent of LQG spin-networks [31]. The area operator is still defined with the simple bivectors (I.2.8), and one finds

$$\begin{aligned} A^2_{\text{EPRL}} &= \frac{1}{(\gamma^2+1)^2}\frac{\langle(\gamma-\star)b_t,\,(\gamma-\star)b_t\rangle}{2} \\ &= \frac{\gamma^2}{(\gamma^2+1)^2}\frac{(1-\gamma^2)C_1+2\gamma C_2}{2} \\ &= \gamma^2 j(j+1) > 0\,, \end{aligned}$$

after applying the simplicity condition (I.2.10) in passing to the second line, thus obtaining the usual LQG area spectrum.

*(c) The coherent state formulation*

An alternative suggestion was put forth in [51] for the boundary states of the EPRL model. It makes use of a set of coherent states [52] on the Hilbert spaces of unitary irreducible SU(2) representations $\mathcal{H}^j$, i.e. states of the form

$$|j,g\rangle = D^j(g)|j,m_r\rangle\,, \quad g\in \text{SU}(2)\,, \qquad \text{(I.2.11)}$$

where $|j,m_r\rangle$ denotes a fixed reference $L^3$ eigenstate, usually taken to be either one of $|j,\pm j\rangle$ with maximal/minimal weight $m_r=\pm j$. They satisfy a completeness relation on $\mathcal{H}^j$,

$$\mathbb{1}_j = (2j+1)\int_{\text{SU}(2)} \mathrm{d}g\,|j,g\rangle\langle j,g| =: \int_{\text{SU}(2)} \mathrm{d}g\ \text{—∘∘—}\,, \qquad \text{(I.2.12)}$$

and as such they form a basis for $\mathcal{H}^\chi$. Such coherent states are represented by a line with a node $|j,g\rangle =$ —∘, as above. They admit a geometrical

interpretation through the correspondence

$$\langle j, g|\sigma_i|j, g\rangle = \left([\pm\,\pi(g)\hat{e}_3]_i\right)^{2j} \ , \quad \pm\,\pi(g)\hat{e}_3 \in S^2 \ ,$$

where $\pi$ denotes the spin homomorphism $\pi : \mathrm{SL}(2,\mathbb{C}) \to \mathrm{SO}^+(3,1)$ (cf. section III.3.1). The prescription of [51] is to use these as the boundary states of the theory, in which case the EPRL vertex amplitude reads

$$\Big[\quad\Big]_{\mathrm{EPRL}} = \int_{\mathrm{SL}(2,\mathbb{C})} \prod_{\substack{a,b=1\\a<b}}^{5} \mathrm{d}g_a\, d_{j_a b} \cdot \langle j_{ab}, n_{ab}|D^{\chi_{ab}}(g_a)^\dagger D^{\chi_{ab}}(g_b)|j_{ab}, n_{ba}\rangle \ , \tag{I.2.13}$$

with $\chi_{ab} = (2j_{ab}, 2\gamma j_{ab})$ and $d_j = 2j+1$. It is to be implicitly understood that the boundary states in equation (I.2.13) refer to those states in $\mathcal{H}^\chi$ induced from the SU(2) basis. Note moreover that, as defined, the left-right invariance of the Haar measure dictates that the amplitude diverges due to an extraneous $\mathrm{SL}(2,\mathbb{C})$ integration, so it is common to regularize it by performing only four out of five integrations. One is justified in thinking of each coherent state as representing an orthogonal vector for each space-like triangle of each boundary tetrahedron, as this is indeed what one finds in a semiclassical analysis [51, 53].

*(d) The Conrady-Hnybida (CH) extension*

The CH extension [54] extrapolates the EPRL procedure for different choices of orthogonal vectors $n_\tau$, and hence for different choices of causal characters of tetrahedra and triangles. It makes use of the coherent state construction for the boundary of the theory, specifying the signature of the triangle faces by referring to the homogeneous spaces (cf. appendix C)

$$\mathrm{SU}(2)/\mathrm{U}(1) \simeq S^2 \ , \quad \mathrm{SU}(1,1)/\mathrm{U}(1) \simeq H^\pm \ ,$$
$$\mathrm{SU}(1,1)/\left(e^{iK_2} \times \mathbb{Z}_2\right) \simeq H^{\mathrm{sl}} \ ,$$

where $H^\pm$ and $H^{\mathrm{sl}}$ denote the two- and one-sheeted 2-hyperboloids, respectively. For the sake of brevity I will omit the derivation, and list only the final results of the extension. The reader can find in appendix A information about the representation theory of $\mathrm{SU}(1,1)$ as well as all necessary notation.

1. Space-like triangles $ab$ are labeled by $\rho_{ab} = \gamma \nu_{ab}$. They may be cells of space- or time-like tetrahedra;

   - If the tetrahedron is space-like, the boundary states are the EPRL ones discussed above, and $\nu_{ab} = 2j_{ab}$. The triangle is characterized by an element of $S^2$, with area $A_{ab}^2 = \gamma^2 j_{ab}(j_{ab}+1)$.

   - If the tetrahedron is time-like (orthogonal to a canonical normal $n_\tau = \hat{e}_3$), the boundary states $|\psi_{ab}\rangle$ are constructed from the usual basis for $\mathrm{SU}(1,1)$ representations in the discrete series, i.e.

   $$\begin{gathered} L^3|k,m\rangle = m|k,m\rangle\,, \\ k \in -\frac{\mathbb{N}}{2}\,, \quad m \in q(-k+\mathbb{N}^0)\,, \quad q = \pm\,, \\ |\psi_{ab}\rangle = D^j(g)|k_{ab}, -q_{ab}k_{ab}\rangle\,, \quad g \in \mathrm{SU}(1,1)\,, \end{gathered}$$

   and one takes $\nu_{ab} = -2k_{ab}$. The $q_{ab}$ parameter specifies whether the geometric vector is in the future- or past-pointing hyperboloid $H^\pm$. There is an associated area spectrum given by $A_{ab}^2 = \gamma^2 k_{ab}(k_{ab}+1) > 0$.

2. Time-like triangles $ab$ are necessarily cells of time-like tetrahedra, and they are labeled by $\nu_{ab} = -\gamma \rho_{ab}$;

   - Boundary states $|\psi_{ab}\rangle$ are constructed from eigenstates of a non-compact generator of $\mathrm{SU}(1,1)$ in the continuous series of its representations, i.e.

$$
\begin{aligned}
&K^1|j,\lambda,\sigma\rangle = \lambda|j,\lambda,\sigma\rangle\,,\\
&j = -\frac{1}{2} + is\,, \quad s\in\mathbb{R}\,, \quad \lambda\in\mathbb{C}\,, \quad \sigma = \pm\,,\\
&|\psi_{ab}\rangle = D^j(g)|j_{ab}, s_{ab}, +\rangle\,, \quad g\in \mathrm{SU}(1,1)\,,
\end{aligned}
$$

and one takes $\rho_{ab} = -2\sqrt{s_{ab}^2 + 1/4} \sim -2s_{ab}$. The triangle is characterized by an element of the one-sheeted hyperboloid $H^{\mathrm{sl}}$. The corresponding area reads $A_{ab}^2 = -\gamma^2\left(s_{ab}^2 + 1/4\right) < 0$.

Having prescribed states for the three types of boundary triangles, the vertex amplitude is constructed as before, paying attention to the different types of triangle interfaces that might be allowed. It is useful to fix terminology. These are 1) *achronal*, for space-like interfaces of space-like tetrahedra; 2) *parachronal*, for time-like interfaces of time-like tetrahedra; 3) *orthochronal*, for space-like interfaces of time-like tetrahedra; and 4) *heterochronal*, for space-like interfaces of tetrahedra with different causal characters. The generic amplitude reads simply

$$
\left[\;\;\right]_{\mathrm{EPRL-CH}} = \int_{\mathrm{SL}(2,\mathbb{C})} \prod_{\substack{a,b=1\\ a<b}}^{5} \mathrm{d}g_a\, d_{\chi_{ab}} \cdot \langle\psi_{ab}|D^{\chi_{ab}}(g_a)^\dagger D^{\chi_{ab}}(g_b)|\psi_{ba}\rangle\;,
$$

with $\chi_{ab}$ and $|\psi_{ab}\rangle$ standing for the appropriate labels and states, once a choice of causal character is made at the boundary. The multiplicative factor $d_\chi$ depends on the relevant spin, and its explicit form can be consulted in [54]. The same observations as for equation (I.2.13) regarding regularization and notation apply.

### I.2.4 Remarks on the state of affairs for the spin-foam programme

I have up to now merely reported on the formal aspects of spin-foams. A discussion of their physical consequences, shortcomings and difficulties is in order.

*(a) On the BC and EPRL Poisson structures*

The BC-type (after accounting for a causal character extension as done in [55]) and EPRL-type models are inherently different in that they propose different algebraic structures for the theory, as mentioned above. Unlike their classical counterparts, they are not simply related by a $\gamma \to 0$ limit. There is an extensive discussion in the literature as to which one is preferable [48, 56–59], though EPRL-type models seem to be favored due to their affinity with LQG. It is frequently claimed that the BC model fails to induce the correct graviton propagator [60], that its boundary states are too constrained [61] (as they allow for a single possible intertwiner once representations have been fixed), and that the shared faces of different tetrahedra are uncorrelated [43]. An apology of the BC model in response to this criticism can be found in [62]. I will exclusively focus on EPRL-type constructions in the remainder of this work.

*(b) On the subject of boundary states*

Insofar as a quantum theory should assign probabilities to states at the boundary, the choice of boundary states has operational importance. On one hand, they determine the Hilbert space structure at the boundary, and thus which observables can in principle be defined and eventually measured. On the other, they induce different dynamics. On the latter point, and for the particular case of coherent boundary states, the choice of reference state used in the construction clearly affects the vertex amplitude (though it is inconsequential if there is no boundary). The CH prescription for boundary states is but one of possible many. I will argue in chapter II that one can refine the choice of states associated to time-like faces.

*(c) On the semiclassical limit*

In the absence of experimental evidence for quantum gravity, the semiclassical correspondence with general relativity is an important requirement. It is a general result [53, 63–65] that EPRL-CH models recover a form of discretized GR known as Regge calculus when all spins are scaled to asymptotic infinity. This statement comes with a number of caveats, all mentioned in the relevant literature. First, the vertex amplitude asymptotes to a cosine of Regge actions, although one might expect a single exponential - this is known as the *cosine problem*. Second, the variables of the theory are by construction areas rather than lengths, which is a departure from discretized GR [66]. Third, the semiclassical limit includes not only the simplices of Regge calculus, but also other objects - including degenerate simplices and the so-called vector geometries - which are generally undesirable [67]. Fourth, for triangulations with more than one simplex (as the famous $\Delta_3$ triangulation [68]), one finds that the curvature at the bulk is exponentially suppressed in the semiclassical limit - this is known as the *flatness problem* [69]. Fifth, and finally, the asymptotics of time-like faces remains incomplete, in that there is not yet a closed-form expression for the vertex amplitude in that regime. I will expand on all these points in chapter III.

*(d) On computational feasibility*

Prediction requires computation, but spin-foam amplitudes are notoriously difficult to evaluate numerically for more than one vertex [70, 71]. There have recently been a number of proposals for efficient computational algorithms, such as the complex critical points algorithm [72], Lefschetz thimbles [73], the `sl2c-foam` package [74], the hybrid-algorithm proposal [75], and the effective spin-foam framework [76–78]. I will discuss a contribution to this goal in chapter IV, further providing additional insight on the structure of EPRL-type amplitudes.

*(e) On triangulation dependence and the continuum limit*

The spin-foam approach substantiates the path-integral for quantum gravity by introducing a discretization of the underlying manifold. Evidently, as the models stand, their predictions are directly dependent on the properties of the triangulation. It is clear that the discretization cannot be a manifestation of a minimal scale structure for space-time, as the expectation value of the area operator - which would in principle be the means by which one would determine the size of that minimal structure - depends itself on the number of simplices one assumes, it being therefore self-referential[13]. It is furthermore reasonable to doubt whether a discretization by simplices is fundamentally more appropriate than one using other types of polytopes. There is moreover the obstacle that 4-dimensional topologies do not admit a classification [79], rendering an eventual state-sum over topologies problematic. There are two main competing viewpoints on this subject: one suggesting that the triangulation serves as a regulator, which should be managed from a renormalization perspective (see [80] and references therein); and another suggesting that one ought to sum over all possible triangulations (this is the perspective of group field theory) [81]. There is, to the best of my knowledge, no definite answer on this matter. My personal speculation, stemming from the observation that discrete quantum toy models become trivial if the classical limit is taken before the continuum limit [82] (as seems to be the case for spin-foams and the flatness problem) together with results suggesting that the quantum simplicity constraints require an increasingly smaller $\gamma$ for finer triangulations [77] (consequently corresponding to smaller area eigenvalues), is that a renormalization procedure based on a flowing Immirzi parameter might be preferable; this would hypothetically allow one to identify a regime in which the predictions of the theory would be invariant under a further refinement of the triangulation, removing the discretization dependence but conserving the discrete properties of the quantum theory; the author of [83] conjectures on general grounds that this should happen at a fixed point with a second order phase transition, where the correlation length would diverge. A successful renormalization scheme for spin-foams is yet to be proposed, although initial efforts have been made in this direction: tensor-network renormalization [84–86] and the fusion basis [87] are two such examples.

[13] Who shaves the barber who shaves those who do not shave themselves?

### *(f) On the matter of matter*

Matter matters, as the saying goes [88]; this is however beyond the scope of this monograph, and it remains a difficult problem. There is a broad range of proposals for coupling fermionic matter to spin-foams. e.g. [89–92]. That I mention this - even in the absence of any meaningful insights on the topic - is supposed to impress upon the reader the importance I believe the subject to have.

# II. New Boundary States for Lorentzian Spin-foams

*In fact the opposition of instinct and reason is mainly illusory. Instinct, intuition, or insight is what first leads to the beliefs which subsequent reason confirms or confutes; but the confirmation, where it is possible, consists, in the last analysis, of agreement with other beliefs no less instinctive.*

— Bertrand Russel, Our Knowledge of the External World (1914)

As trivial an observation as it may be, the fact that the spin-foam construction purports to describe geometrical structures by employing strictly algebraic objects implies that those objects have already by themselves a certain geometrical affinity. One might then expect a further study of the underlying geometry to shed some light in the whole affair. As of the time of writing, there exist two manners in which the geometrical aspects of spin-foams have been understood: on one hand, there is a wealth of literature in the description of boundary states in terms of spinors and twistors, going by the name of *twisted geometries* [93–98]; on the other, and as mentioned in section I.2.3, the introduction of coherent state boundaries has allowed for a geometrical interpretation of the spin-foam semiclassical limit [53, 63–65].

Reiterating the considerations I have made at the end of chapter I, the states one chooses to assign to the spin-foam boundary are of the utmost importance in defining the model and its predictions. Such a choice is, however, not a settled matter. The twisted geometry picture allows for a rigorous derivation of the spin labels that ought to figure in the model, but it makes no further claims on which boundary states to pick. The CH prescription [54] for boundary coherent states has been successful insofar as it allows for the semiclassical limit of the models to be explored - giving weight to the notion that such states should be preferred - but these states are neither unequivocally necessary nor unique. Indeed, the specific case of time-like boundaries in the CH model seems particularly pathological, as it has been shown that the amplitude has a number of undesirable features in the asymptotic regime [64, 65] (cf. chapter III), with the unfortunate consequence that an explicit expression for the asymptotic amplitude is yet to be obtained. Given that the CH proposal for time-like states is manifestly atypical when compared to all other boundary cases (being constructed from *generalized* eigenstates of a non-compact $\mathrm{SU}(1,1)$ generator, rather than bona-fide elements of the relevant Hilbert space), one may wonder whether the model's peculiar behavior could be cured with a more refined choice of states at the boundary. It further complicates matters that the spinorial approach of [98] suggests different spin labels from the CH model, deriving a $\gamma$-independent area spectrum in disagreement with [54].

How can one then improve on the choice of boundary states for spin-foams? A possible approach, and the one I shall here advocate for, is to invert the process of determination: to postulate the desirable geometric structures one expects to find, and to see in which manner those structures might determine a prescription for the algebraic states one should use. The purpose of this chapter is to show that such a programme can be implemented, and that its successful application to the bothersome case of the space-like hyperboloid yields a new proposal for coherent states constructed from the continuous series of $\mathrm{SU}(1,1)$ representations, different from the ones of [54] and with more amenable properties for asymptotic analysis. The manner in which the states are obtained is through a chain of correspondences, succinctly represented in figure II.1. Using complexified quaternions as a geometric dictionary, I will start by proposing spinor states with the right geometrical properties adapted to the homogeneous spaces of $\mathrm{SU}(1,1)$; such states will then induce certain realizations of unitary and irreducible representations of the group, in turn suggesting particular coherent states associated to hypersurfaces in $\mathbb{R}^{1,2}$. Besides yielding new boundary states, this approach 1) strengthens the CH prescription for the different series of $\mathrm{SU}(1,1)$ representations, as it re-derives its results from a different perspective; 2) allows us to propose a new toy model for 3d quantum Lorentzian gravity with space-like boundaries. The entirety of this chapter is based on [99, 100].

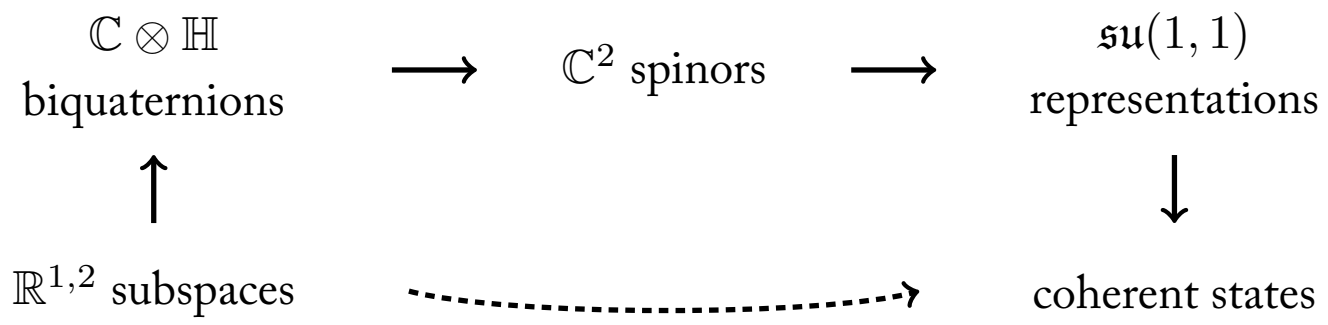

Figure II.1: A schematic representation of the relationships established throughout chapter II.

Since much of this work is inspired by the twisted geometries framework for LQG and spin-foams, a general review of the subject now follows.

## II.1 Prologue: twistors and spinors in spin-foams

It has long been known that the phase space of LQG on a fixed graph admits a twistorial $\mathbb{T} \simeq \mathbb{C}^2 \oplus \mathbb{C}^2$ description [93, 94, 101]. The twistorial formalism paints an elegant and compelling picture of the kinematics of LQG, in that the quantum states of gravity, as well as its geometry, can be encoded simply in terms of tuples of complex numbers with very little extra structure. As I will review below, the basic idea hinges on the identification between the classical phase space of a graph link $T^*\mathrm{SU}(2)$ with twistor space $\mathbb{T}$ as Poisson manifolds; this is achieved via a sympletic quotient $\mathbb{T}//\mathcal{C}$ with an "area-matching" constraint $\mathcal{C}$.

The twistor framework is of course not exclusive to LQG, and it admits a natural extension to spin-foam models - which are what concerns us presently - at the very least because the boundary states of the EPRL model and the LQG states are reminiscent of one another. Indeed, the formalism has been successfully refined in a covariant manner so as to start from two twistor copies, upon which one imposes the simplicity and area-matching constraints to obtain the desired $T^*\mathrm{SU}(2)$ quotient [95–97]; formally

$$\mathbb{T}^2 \overset{\mathcal{C}_{\text{area}}}{\to} T^*\mathrm{SL}(2,\mathbb{C}) \overset{\mathcal{C}_{\text{simple}}}{\to} T^*\mathrm{SU}(2)\,,$$

where $\mathcal{C}_{\text{area}}$ denotes an area-matching constraint (now adapted to $\mathbb{T}^2$), and $\mathcal{C}_{\text{simple}}$ is an appropriate formulation of the simplicity constraints of the EPRL model. These results have paved the way for later works concerned with describing states associated to hypersurfaces with different causal characters [98, 102].

### II.1.1 Spinors for space-like tetrahedra

Although the covariant twistor $\mathbb{T}^2 \to T^*\mathrm{SL}(2,\mathbb{C})$ characterization is beyond the purpose of this chapter, it will prove useful for what follows to go over the spinor $\mathbb{T} \to T^*\mathrm{SU}(2)$ formulation of the EPRL space-like boundary states in the spin-foam context. I remind the reader that, in the coherent state description, such states are objects of the type

$$|j,g\rangle = D^j(g)|j,m_r\rangle\,, \quad g \in \mathrm{SU}(2)\,,$$

i.e. actions of unitary irreducible representations of $\mathrm{SU}(2)$ on some reference state $|j,m_r\rangle$; the diffeomorphism $\mathrm{SU}(2)/U(1) \simeq S^2$ is then enough to associate to each of these states an element of the 2-sphere. A trivial observation is that a spinor $|z\rangle = (z_1,z_2) \in \mathbb{C}^2$ has enough degrees of freedom to model $S^2$. Defining the vector[14]

$$|z\rangle \to v_z = \langle z|\sigma|z\rangle\ , |v_z|^2 = \langle z|z\rangle = v_i v^i\,,$$

[14] Here and in the following I will always take Dirac's bracket notation to stand for the standard $\mathbb{C}^2$ inner product. All other necessary pairings will be denoted in the same manner, but with additional operators explicitly represented, e.g. $\langle u|\sigma_3|v\rangle$ for the $\mathrm{SU}(1,1)$-compatible pairing. It follows that $\langle u|$ will always stand for complex transposition.

where $\sigma = (\sigma_1, \sigma_2, \sigma_3)$ denotes a tuple of Pauli matrices, one easily sees that $v_z \in S^2$ if it is taken to have unit norm. There is a natural action $g \triangleright |z\rangle = |gz\rangle$ of $g \in \mathrm{GL}(2, \mathbb{C})$ on $\mathbb{C}^2$, and the subgroup preserving the vector norm and orientation is clearly SU(2)

$$|g \triangleright v_z|^2 = \langle z|g^\dagger g|z\rangle = \langle z|z\rangle \ , \qquad g \in \mathrm{SU}(2) \, .$$

Since SU(2) acts transitively on the sphere, any unit-norm spinor can be written in terms of a reference state. Introducing the notation

$$|+\rangle = \left(\begin{smallmatrix} 1 \\ 0 \end{smallmatrix}\right), \quad |-\rangle = \left(\begin{smallmatrix} 0 \\ 1 \end{smallmatrix}\right),$$

one can define e.g. $|z\rangle = g_z|\pm\rangle$, for some $g_z \in \mathrm{SU}(2)$, and recover the mapping to the sphere using the spin homomorphism $g \mapsto \pi(g)$ (cf. section III.3.1)

$$v_z^i = \langle \pm|g_z^\dagger \sigma^i g_z|\pm\rangle = \pi(g_z^\dagger)_j^{\ i} \langle \pm|\sigma^j|\pm\rangle = \pm\pi(g_z)^i{}_j \hat{e}_3^j \in S^2 \, ;$$

this is the usual coherent state geometrical association, but now strictly written in the defining representation of SU(2).

The spinor formalism proceeds as follows. Given a space-like tetrahedron with space-like faces, one associates to each triangle a spinor (for now allowed to have arbitrary non-vanishing norm). Each vertex in the tetrahedron's combinatorial graph 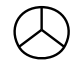 is thus decorated, and each link is labeled by a pair of spinors $(|z\rangle, |w\})\in \mathbb{C}^2 \oplus \mathbb{C}^2$,

$|z\rangle$ 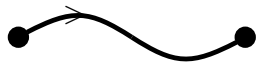 $|w\}$ ,

where the dual state $|w\} = Q|w\rangle$ was introduced, together with an orientation. The antilinear map $Q$ denotes the SU(2) quaternionic structure[15] $Q^2 = -1$ on $\mathbb{C}^2$,

$$\begin{aligned} Q :\ & \mathbb{C}^2 \to \mathbb{C}^2 \\ & |z\rangle \mapsto -i\sigma_2|\overline{z}\rangle \, , \end{aligned} \tag{II.1.1}$$

[15] $Q$ is usually denoted by $J$ in the spin-foam literature, e.g. [45]. I prefer $Q$ for "quaternionic", just as later I will use $R$ for the real structure.

which has the useful property of acting on $\mathrm{GL}(2,\mathbb{C})$ matrices $g$ as $QgQ^{-1} = g^{A\dagger}$, with $g^A$ standing for the adjugate operation. It furthermore serves as an orthogonality transformation, in that $\langle z|z\} = 0$. A "parallel transport" can be constructed from the two spinors,

$$g(z,w) = \frac{|w\}\langle z| - |w\rangle\{z|}{\sqrt{\langle z|z\rangle\,\{w|w\}}} \in \mathrm{SU}(2)\,, \tag{II.1.2}$$

which can be shown to be an element of $\mathrm{SU}(2)$ as it commutes with $Q$ and is constructed so as to have unit determinant. It may be interpreted as a parallel transport insofar as it satisfies

$$g(z,w)|z\rangle = \left(\frac{\langle z|z\rangle}{\{w|w\}}\right)^{\frac{1}{2}} |w\}\,, \tag{II.1.3}$$

and it transforms as $g(h_1 z, h_2 w) = h_2 g(z,w) h_1^{-1}$ for $h_1, h_2 \in \mathrm{SU}(2)$.

The two classes of objects so defined, i.e. the vectors $v_z, v_w$ and the group elements $g(z,w)$, are somewhat reminiscent of elements of the phase space $T^*\mathrm{SU}(2) \simeq \mathrm{SU}(2) \otimes \mathfrak{su}(2)$, since $\mathfrak{su}(2) \simeq \mathbb{R}^3 \supset S^2$ as vector spaces. One could hope then that supplementing $\mathbb{T}$ with a Poisson structure and an appropriate constraint might result in the desired identification; this is indeed the case. Equation (II.1.3) suggests considering $\mathcal{C} = \langle z|z\rangle - \{w|w\} \sim 0$, the so-called *area-matching* constraint. With regards to the Poisson structure, one takes the canonical symplectic structure $\omega = i\mathrm{d}z_i \wedge \mathrm{d}\overline{z}^i \in \Omega^2(\mathbb{C}^4)$ given by

$$\omega(\partial_i, \overline{\partial}_j) = i\delta_{ij}\,, \quad \omega(\partial_i, \partial_j) = 0\,, \quad \omega(\overline{\partial}_i, \overline{\partial}_j) = 0\,,$$

having introduced a basis $\{\partial_1, \partial_2\} \oplus \{\overline{\partial}_1, \overline{\partial}_2\}$ of $\mathbb{C}^2 \oplus \overline{\mathbb{C}}^2$ (and fixing the notational meaning of $\overline{\mathbb{C}}^2$). It induces Poisson brackets on coordinate functions,

$$\{\overline{z}_i, z_j\} = i\delta_{ij}\,,$$

all other brackets vanishing. Under the imposition of $\mathcal{C} \sim 0$, one can prove the Poisson relations

$$\begin{aligned}
&\{v_z^i, v_z^j\} = \epsilon^{ij}_{\ \ k} v_z^k \,, \\
&\{g_{ab}(z,w), g_{cd}(z,w)\} = 0 \,, \\
&\{g_{ab}(z,w), v_z^i\} = i g_{ab}(z,w)\, \sigma^i \,,
\end{aligned} \tag{II.1.4}$$

which match the Kirillov-Kostant-Souriau Poisson structure [103] on $T^*\mathrm{SU}(2)$. The theorem of [94] is that there exists a symplectomorphism

$$\Big( \mathbb{C}^2 \oplus \overline{\mathbb{C}}^2 \setminus \{\langle z|z\rangle = 0\} \Big) // \mathcal{C} \quad \simeq \quad T^*\mathrm{SU}(2) \setminus |v| = 0 \,, \tag{II.1.5}$$

where the double quotient signifies that elements of the left space both commute with and satisfy the area-matching constraint $\mathcal{C}$.

## II.2 From complexified quaternions to spinors

Having reviewed the standard lore, we continue with the first stated goal of deriving spinor states which can be put in correspondence with the surfaces of transitivity of $\mathrm{SU}(1,1)$. The relevant subspaces in $\mathbb{R}^{1,2}$ are the two-sheeted time-like hyperboloid $H^{\pm}$ and the one-sheeted space-like hyperboloid $H^{\mathrm{sl}}$ (defined in appendix C). As previously mentioned, this correspondence will be achieved by resorting to biquaternions as a geometric dictionary, which I now introduce.

### II.2.1 The algebra of biquaternions

The biquaternions, or complexified quaternions $\mathbb{H}_{\mathbb{C}}$, are a complex division algebra containing elements of the form

$$q = \alpha + i\beta_i k^i \,, \quad \alpha \,, \beta_i \in \mathbb{C} \,,$$

where $k^i$ are the three quaternionic roots of $-1$, satisfying $k_i k_j = -\delta_{ij} + \epsilon_{ij}^{\ \ k} k_k$. I still denote by $i$ the usual imaginary unit. Due to the quaternionic units the algebra is non-commutative, and one needs to distinguish between

left- and right-multiplication. There exist two conjugation operations, one complex $\bar{\cdot}$ and one quaternionic $\cdot^c$,

$$\overline{q} = \overline{\alpha} - i\overline{\beta}_i k^i\,, \qquad q^c = \alpha - i\beta_i k^i\,,$$

which are clearly commutative among themselves. Note that quaternionic conjugation is distributive under an order inversion. One has projections onto the scalar $S(q)$ and vector $V(q)$ (or quaternionic) parts

$$S(q) = \frac{q + q^c}{2} = \alpha\,, \qquad V(q) = \frac{q - q^c}{2} = i\beta_i k^i\,,$$

and there exists a complex-linear inner-product

$$\langle q, q' \rangle = S(q q'^c) = \alpha\alpha' - \beta^i \beta'_i\,,$$

inducing a pseudo-norm

$$|q|^2 = S(q q^c) = \alpha^2 - \beta^i \beta_i \; \in \mathbb{C}\,.$$

The aforementioned properties endow the biquaternions with a complex Minkowski structure, in that there is a vector space isometry

$$\left(\mathbb{H}_{\mathbb{C}}\,, \langle\cdot,\cdot\rangle\right) \to \left(\mathbb{C}^4, \langle\cdot,\cdot\rangle_\eta\right)\,, \quad \langle u, v\rangle_\eta = u^T \eta v\,,$$
$$q \mapsto \alpha \hat{e}_0 + \beta_i \hat{e}^i\,.$$

Of course, such complex quaternions offer more degrees of freedom than are necessary to describe real Minkowski space. To remedy this we can introduce a constraint, e.g.

$$q = \overline{q}^c \;\Leftrightarrow\; \alpha, \beta_i \in \mathbb{R}\,,$$

such that the isometry above yields genuinely real Minkowski vectors. Finally, it is straightforward to see that the maps

$$L_x\,, |x|^2 = 1 :\; q \mapsto x q \overline{x}^c$$

leave the norm of $q$ invariant, take real Minkowski vectors to real Minkowski vectors, and satisfy the composition law $L_{xy} = L_x L_y$, thus corresponding to Lorentz transformations.

Since left- and right-multiplication are linear, one might hope that the biquaternionic algebra admits a realization as a matrix algebra. This is indeed the case: consider the mapping

$$\begin{aligned} \mathbb{H}_{\mathbb{C}} &\to \mathbb{C}^{2\times 2} \\ (1, k_i) &\mapsto (\mathbb{1}, -i\sigma_i) \ , \end{aligned}$$

making use of the standard Pauli matrices $\sigma_i$. A general biquaternion then takes the form

$$q = \alpha + i\beta_i k^i \mapsto X_q = \begin{pmatrix} \alpha + \beta_3 & \beta_1 - i\beta_2 \\ \beta_1 + i\beta_2 & \alpha - \beta_3 \end{pmatrix} ,$$

and all the operations defined above extend to this representation: a dictionary between biquaternions and their matrix formulation can be found in table II.1, where the quaternionic structure $Q$ defined in equation (II.1.1) again makes an appearance. I will make extensive use of this correspondence.

### II.2.2 Spinors and the homogeneous spaces of the Lorentz group

The biquaternionic technology just discussed can now be used to probe the geometric content of spinor states. I will start by describing the strategy for the well-known space-like sphere case, to show that the objects of section II.1.1 can too be motivated from the biquaternionic perspective; the remaining $H^{\pm}$ and $H^{\mathrm{sl}}$ cases will follow. For ease of notation I will refer to both a quaternion proper and to its matrix representation by the letter $q$.

*(a) The space-like sphere $S^2$*

Consider the original EPRL choice where all tetrahedra are taken to be orthogonal to the time direction, and thus fully contained in $\mathbb{R}^3 \subset \mathbb{R}^{1,3}$. All unit normal vectors to triangles are space-like and elements of the sphere. Before seeking a correspondence between spinors and biquaternions we shall reduce the dimensionality of $q \in \mathbb{C}^{2\times 2}$ by requiring $\det q = 0$. Table II.1 shows this implies $|q|^2 = 0$, or in terms of biquaternionic components

$$\beta_1^2 + \beta_2^2 + \beta_3^2 = \alpha^2\,, \tag{II.2.1}$$

i.e. we restrict to null biquaternions. We are of course interested in real components $\alpha, \beta_i \in \mathbb{R}$, from where it must be that $q = q^\dagger$. The constraints $\det q_z = 0$ and $q_z = q_z^\dagger$ are clearly solved by $q_z = |z\rangle\langle z|$, and thus one can model space-like vectors of $\mathbb{R}^3$ through the sequence of bijections (note that there is a phase redundancy)

$$\begin{array}{ccccc} \mathbb{C}^2/\mathrm{U}(1) & \to & \mathbb{C}^{2\times 2}_{\substack{\det q = 0 \\ q = q^\dagger}} & \to & \left(\mathbb{R}^3, \eta_{(3)}\right) \\ |z\rangle & \mapsto & q_z = |z\rangle\langle z| & \mapsto & v_z^i = \dfrac{1}{2}\mathrm{Tr}\left[\sigma^i q_z\right]\,. \end{array}$$

Table II.1: Correspondence between biquaternions and complex matrices

| $\mathbb{H}_{\mathbb{C}}$ | $\mathbb{C}^{2\times 2}$ |
|---|---|
| $q \mapsto q^c$ | $X_q \mapsto X_q^A$ |
| $q \mapsto \overline{q}$ | $X_q \mapsto Q X_q Q^{-1}$ |
| $\lvert q\rvert^2 = S(qq^c)$ | $\lvert X_q\rvert^2 = \det X_q$ |

| $\mathbb{H}_{\mathbb{C}}$ | $\mathbb{C}^{2\times 2}$ |
|---|---|
| $S(q) = \frac{q+q^c}{2}$ | $S\left(X_q\right) = \frac{1}{2}\mathrm{Tr}X_q$ |
| $V(q) = \frac{q-q^c}{2}$ | $V\left(X_q\right) = \frac{\sigma^i}{2}\mathrm{Tr}\left[\sigma_i X_q\right]$ |
| $\langle q, q'\rangle = S(qq'^c)$ | $\langle X_q, X_{q'}\rangle = \frac{1}{2}\mathrm{Tr}\left[X_q X_{q'}^A\right]$ |

Here $\eta_{(3)} = \mathrm{diag}(-1,-1,-1)$ denotes the induced metric on $\mathbb{R}^3 \subset \mathbb{R}^{1,3}$ defined by

$$\eta_{(3)}(v_z, v_w) := \langle V(q_z), V(q_w)\rangle = -{v_z}^i v_{w\,i}\,,$$

and equation (II.2.1) guarantees that $|v_z|^2 = -\left(\frac{1}{2}\mathrm{Tr} q_z\right)^2$. The spinorial construction of section II.1.1 can be recovered through a suitable rescaling, and the mapping to the sphere reads

$$|z\rangle \text{ s.t. } \langle z|z\rangle^2 = 1 \;\mapsto\; v_z^i = \langle z|\sigma^i|z\rangle \in S^2 \,.$$

*(b) The time-like two-sheeted hyperboloid $H^{\pm}$*

Consider now a tetrahedron orthogonal to the $\hat{e}_3$ direction, implying it is fully contained in $\mathbb{R}^{1,2} \subset \mathbb{R}^{1,3}$. Suppose further that the triangular face is space-like, i.e. its normal vector is an element of the future- or past-pointing hyperboloid $H^{\pm}$. We would like to modify the above procedure in such a manner that the induced metric has signature $(+,-,-)$. This can be achieved by first permuting the components of $q$ so that the projection operator maps to the intended subspace; right-multiplication by the third unit $k^3$ gives

$$\begin{aligned} qk^3 &= -i\beta_3 + i\left(\beta_2 k_1 - \beta_1 k_2 - i\alpha k_3\right)\,, \\ |V(qk^3)|^2 &= \alpha^2 - \beta_1^2 - \beta_2^2\,, \end{aligned}$$

as desired. The construction then follows as before: we impose $\det q = 0$ and $q = q^\dagger$, but we right-rotate the quaternion $q \mapsto q(-i\sigma_3)$. The sequence of bijections

$$\begin{aligned} \mathbb{C}^2/\mathrm{U}(1) \;&\to\; \mathbb{C}^{2\times 2}_{\substack{\det q = 0 \\ \sigma_3 q \sigma_3 = -q^\dagger}} \;\to\; \left(\mathbb{R}^{1,2}_{|\cdot|^2 \geq 0,\, v^1 \geq 1},\, \eta_{(1,2)}\right) \\ |z\rangle \;&\mapsto\; q_z = -i|z\rangle\langle z|\sigma_3 \;\mapsto\; v_z^i = \frac{i}{2}\mathrm{Tr}\left[\varsigma^i q_z\right] \end{aligned} \tag{II.2.2}$$

yields the subspace of future-pointing[16] time-like vectors in $\mathbb{R}^{1,2} \subset \mathbb{R}^{1,3}$. The notation $\varsigma$ stands for a new tuple of Pauli matrices,

$$\varsigma := (\sigma_3, i\sigma_2, -i\sigma_1) = \sigma_3(\mathbb{1}, \sigma_1, \sigma_2)\,, \tag{II.2.3}$$

[16] The map only covers the future-pointing hyperboloid, as it can be checked that $v_z^1 \geq 1$. The lower hyperboloid can be obtained by taking the symmetric of the components in equation (II.2.2).

which is necessary in order to get the right vector components $v_z = (\alpha, \beta_1, \beta_2)$. This tuple can also be seen as a vector of $\mathrm{SU}(1,1)$ generators in the defining representation; we will later make use of this fact. The induced metric takes the form

$$\eta_{(1,2)}(v_z, v_w) := \langle V(q_z), V(q_w) \rangle = v_z^i v_w^j \eta_{(1,2)ij} \,,$$

with $\eta_{(1,2)} = \mathrm{diag}(1, -1, -1)$, and the norm is constrained by equation (II.2.1) to satisfy

$$|v_z|^2 = \beta_3^2 = \left(\frac{i}{2}\mathrm{Tr} q_z\right)^2 \geq 0 \,,$$

since $\beta_3$ was chosen to be real; all vectors so constructed are indeed time-like. The Minkowski metric is also naturally associated to $\varsigma$, since

$$\varsigma^i \varsigma^j = \eta_{(1,2)}^{ij} - i\epsilon^{ijk} \eta_{(1,2)kl} \varsigma^l \,. \tag{II.2.4}$$

Finally, the mapping from spinors to the two-sheeted hyperboloid reads

$$|z\rangle \text{ s.t. } \langle z|\sigma_3|z\rangle^2 = 1 \;\mapsto\; v_z^i = \pm \langle z|\sigma_3 \varsigma^i|z\rangle \in H^{\pm} \,. \tag{II.2.5}$$

Note that this construction has the right symmetry properties, since the vector norm is invariant under the natural action of $\mathrm{SU}(1,1)$ on $\mathbb{C}^2$,

$$|v_{gz}|^2 = \langle z|g^\dagger \sigma_3 g|z\rangle^2 = \langle z|\sigma_3|z\rangle^2 \,, \quad g \in \mathrm{SU}(1,1) \,.$$

*(c) The space-like one-sheeted hyperboloid* $H^{\mathrm{sl}}$

A triangle face in $\mathbb{R}^{1,2} \subset \mathbb{R}^{1,3}$ can alternatively be time-like, and its normal vector will be a space-like element of the one-sheeted hyperboloid $H^{\mathrm{sl}}$. As equation (II.2.1) shows, the induced vector norm $\alpha^2 - \beta_1^2 - \beta_2^2$ will be positive as long as $\beta_3$ remains real; it must thus be made purely imaginary, and this requires amending the hermiticity condition. One should consider instead

$$k^3 q k^3 = -\bar{q} \quad \Leftrightarrow \quad \alpha, \beta_1, \beta_2 \in \mathbb{R} \,, \; \beta_3 \in i\mathbb{R} \,,$$

which in the matrix formulation reads $\sigma_3 q \sigma_3 = QqQ^{-1}$. The real structure $R^2 = 1$ associated with $\mathrm{SU}(1,1)$

$$R: \mathbb{C}^2 \to \mathbb{C}^2$$
$$|z\rangle \mapsto \sigma_1|\overline{z}\rangle$$

naturally emerges, and the constraint has the equivalent form $q = RqR^{-1}$. Note that $R$ commutes with $g \in \mathrm{SU}(1,1)$, i.e. $Rg = gR$.

The solutions to $\det q = 0$, $q = RqR^{-1}$ are less immediate than in the previous cases. Because $q$ is singular, it can be written as the exterior product of two elements of $\mathbb{C}^2$, i.e $q = |x\rangle\langle y|$. The second constraint then implies $R|x\rangle = \lambda_y|x\rangle$ and $R|y\rangle = \lambda_x|y\rangle$ with $\lambda_x\overline{\lambda}_y = 1$ some complex numbers depending on $x$ and $y$, respectively. Up to multiplicative factors we are thus interested in eigenstates of the real structure, and these are given in terms of a general spinor as

$$|z^\pm\rangle = \frac{1}{\sqrt{2}}\left(|z\rangle \pm R|z\rangle\right) \, ,$$

with eigenvalues $R|z^\pm\rangle = \pm|z^\pm\rangle$. Hence we might propose two classes of solutions to the constraints,

$$\begin{cases} q = |z^\pm\rangle\langle z^\pm| \, , \\ q = \alpha|z^\mp\rangle\langle z^\pm| \, , \end{cases}$$

where $\alpha$ is some linear map anticommuting with the real structure. One can however quickly convince themselves that the first alternative is overconstrained, leading exclusively to null vectors[17] in $\mathbb{R}^{1,2}$. We are left with the second option, and the choice $\alpha = \sigma_3$ will prove to be particularly useful in simplifying calculations.

[17] Unsurprisingly so, since it is hermitian, thus satisfying the constraints associated to both space- and time-like vectors.

Before proceeding we address the cumbersomeness of the states $|z^{\pm}\rangle$ by performing a change of variables

$$\sigma_3|z^-\rangle = \begin{pmatrix} z_1 - \overline{z}_2 \\ \overline{z}_1 - z_2 \end{pmatrix} \mapsto |z^1\rangle := \begin{pmatrix} z_1 \\ \overline{z}_1 \end{pmatrix} ,$$

$$|z^+\rangle = \begin{pmatrix} z_1 + \overline{z}_2 \\ z_2 + \overline{z}_1 \end{pmatrix} \mapsto |z^2\rangle := \begin{pmatrix} z_2 \\ \overline{z}_2 \end{pmatrix} .$$

This justifies having picked $\alpha = \sigma_3$, which was done so that $|z^1\rangle$ and $|z^2\rangle$ have the same qualitative structure. In this manner the two complex degrees of freedom of a Weyl spinor have been split across two Majorana spinors - i.e. spinors which are invariant under the real structure, known as the charge conjugation operator in QFT - whose components are constrained to be related by complex conjugation. This signifies a substantial departure from the previous $S^2$ and $H^{\pm}$ cases, which could be described by a single Weyl spinor. Note moreover that, since the real structure commutes with $\mathrm{SU}(1,1)$, the natural action $g \triangleright |z^{1,2}\rangle = g|z^{1,2}\rangle$ takes Majorana spinors to Majorana spinors.

I can now state the sequence of bijections yielding space-like vectors. Considering again the right-product with the third quaternionic unit, they read

$$\begin{array}{ccccc} (\mathbb{C}_{\text{Majorana}} \oplus \mathbb{C}_{\text{Majorana}})/\mathrm{U}(1) & \to & \mathbb{C}^{2\times 2}_{\substack{\det q = 0 \\ RqR^{-1}=q}} & \to & \left(\mathbb{R}^{1,2}_{|\cdot|^2 \leq 0}, \eta_{(1,2)}\right) \\ \left(|z^1\rangle,\, |z^2\rangle\right) & \mapsto & q_z = -i|z^1\rangle\langle z^2|\sigma_3 & \mapsto & v_z^i = \dfrac{i}{2}\mathrm{Tr}\left[\varsigma^i q_z\right] . \end{array}$$

with the vector norm given by

$$|v_z|^2 = \beta_3^2 = \left(\frac{i}{2}\mathrm{Tr} q_z\right)^2 \leq 0 ,$$

which is again invariant under the $\mathrm{SU}(1,1)$ action, as intended. The one-sheeted hyperboloid can hence be written in terms of two Majorana spinors

as

$$|z^1\rangle,\ |z^2\rangle \ \text{s.t.}\ \langle z^2|\sigma_3|z^1\rangle^2 = -1 \ \mapsto\ v^i_z = \langle z^2|\sigma_3\varsigma^i|z^1\rangle \in H^{\text{sl}}\,. \tag{II.2.6}$$

## II.3 Spinorial realizations of SU$(1,1)$ representations

It was proven in [101, 104] that the parametrization of SU$(2)$ elements in terms of spinors (II.1.2) could be used to derive the Bargmann-Fock measure [105] on holomorphic functions from the Haar measure. It stands to reason that a similar result ought to be achievable for SU$(1,1)$. Indeed, it turns out that the spinorial objects considered above have enough structure not only to define a spinor measure, but also to induce particular realizations of the different series of unitary and irreducible representations of both SU$(1,1)$ and its Lie algebra. The representations obtained in this manner will be shown to match the proposals of Conrady and Hnybida [54] for space- and time-like boundaries in time-like tetrahedra, thus constituting an alternative derivation of their prescription.

### II.3.1 The discrete series from $H^{\pm}$ Weyl spinors

*(a) Poisson structure*

Recall from section II.2.2.(b) that the two-sheeted hyperboloid $H^{\pm}$ can be modeled on a Weyl spinor $|z\rangle \in \mathbb{C}^2$ as per equation (II.2.5). In the spirit of Bargmann-Fock quantization[18] we consider the larger space $\mathbb{C}^4 \simeq \mathbb{C}^2 \oplus \overline{\mathbb{C}}^2$, together with a basis $\{\partial_1, \partial_2\} \oplus \{\overline{\partial}_1, \overline{\partial}_2\}$. A general element in $\mathbb{C}^4$ can be written as $u = u^+ + u^-$, where $u^+ \in \mathbb{C}^2$ and $u^- \in \overline{\mathbb{C}}^2$. Complex conjugation takes $u^+ \in \mathbb{C}^2$ to $\overline{u}^+ \in \overline{\mathbb{C}}^2$, and vice-versa. The discussion of section II.2.2.(b) suggests that $\mathbb{C}^2$ should be equipped with the indefinite

[18]The well-known Bargmann-Fock quantization of an harmonic oscillator starts by picking a complex structure in order to complexify the phase space. Complex linear combinations of the canonical variables $(q, p)$ then lead to creation and annihilation operators. Since here we already start with a complex vector space, we consider simply a direct sum with its complex conjugate.

hermitian form

$$\langle \cdot, \cdot \rangle_{\mathbb{C}^2} : \mathbb{C}^2 \times \mathbb{C}^2 \to \mathbb{C}$$
$$(u^+, v^+) \mapsto \langle u^+, v^+ \rangle_{\mathbb{C}^2} = u^{+\dagger}\sigma_3 v^+ \,,$$

which is invariant under the natural action of $\mathrm{SU}(1,1)$ on $\mathbb{C}^2$. There is a standard construction for assigning a symplectic structure given such a pairing [105]: one defines

$$\begin{aligned} i\omega(u^-, v^+) &= \langle \overline{u}^-, v^+ \rangle_{\mathbb{C}^2} \,, \\ \omega(u^+, v^+) &= \omega(u^-, v^-) = 0 \,, \end{aligned}$$

and the requirement of antisymmetry induces another hermitian form in $\overline{\mathbb{C}}^2$ as $\langle u^-, v^- \rangle_{\overline{\mathbb{C}}^2} = -u^{+\dagger}\sigma_3 v^+$. The symplectic form thus reads

$$\omega = i\sigma_3^{ij}\, \mathrm{d}z_i \wedge \mathrm{d}\overline{z}_j \,, \tag{II.3.1}$$

differing from the standard symplectic structure on $\mathbb{C}^2 \oplus \overline{\mathbb{C}}^2$ by the inclusion of the third Pauli matrix. It induces Poisson brackets on coordinate functions

$$\{\overline{z_i}, z_j\} = i\sigma_{3ij} \,,$$

all other brackets vanishing.

*(b) Lie algebra realization*

Consider the geometric vector components of equation (II.2.2), now denoted by $v^i = \frac{1}{2}\langle z|\sigma_3 \varsigma^i|z\rangle$,

$$v^1 = \frac{1}{2}(|z_1|^2 + |z_2|^2) \,, \quad v^2 = \frac{1}{2}(z_1\overline{z}_2 + \overline{z}_1 z_2) \,, \quad v^3 = \frac{i}{2}(z_1\overline{z}_2 - \overline{z}_1 z_2) \,,$$

with vector norm

$$\begin{aligned} |v|^2 &= (v^1)^2 - (v_2)^2 - (v_3)^2 \\ &= \frac{1}{4}(|z_1|^2 - |z_2|^2)^2 \geq 0 \,. \end{aligned}$$

A straightforward calculation employing the identity (II.2.4) shows that

$$\{v^i, v^j\} = -\epsilon^{ijk}\eta_{kl}v^l\,, \tag{II.3.2}$$

where for the rest of this chapter $\eta := \eta_{(1,2)}$. The functions $v^i$ induce Hamiltonian vector fields $X^i$ through the symplectic form $\omega(X^i, \cdot) = \mathrm{d}v^i(\cdot)$, which explicitly read

$$X^1 = -\frac{i}{2}\left(z_1\partial_1 - z_2\partial_2 + \overline{z}_2\overline{\partial}_2 - \overline{z}_1\overline{\partial}_1\right)\,,$$

$$X^2 = -\frac{i}{2}\left(z_2\partial_1 - z_1\partial_2 + \overline{z}_1\overline{\partial}_2 - \overline{z}_2\overline{\partial}_1\right)\,,$$

$$X^3 = -\frac{1}{2}\left(z_2\partial_1 + z_1\partial_2 + \overline{z}_2\overline{\partial}_1 + \overline{z}_1\overline{\partial}_2\right)\,.$$

and which satisfy the Lie bracket relations $[X^i, X^j] = \epsilon^{ijk}\eta_{kl}X^l$. The reader may recognize in this equation the Lie brackets of SU$(1,1)$, suggesting the vector fields can be thought of as its generators; the algebra has however too many degrees of freedom. Recalling the orthogonal decomposition $\mathbb{C}^2 \oplus \overline{\mathbb{C}}^2$ - which was itself used in the construction of the symplectic structure -, we therefore choose to project the vectors onto the first factor $\mathbb{C}^2$. Under the identification $(X^1, X^2, X^3) \mapsto (iL^3, iK^1, iK^2)$ (which includes imaginary units to match the physics literature [106]), one gets the vector fields

$$L^3 = \frac{1}{2}\left(z_2\partial_2 - z_1\partial_1\right)\,, \quad K^1 = \frac{1}{2}\left(z_1\partial_2 - z_2\partial_1\right)\,, \qquad K^2 = \frac{i}{2}\left(z_1\partial_2 + z_2\partial_1\right)\,, \tag{II.3.3}$$

together with the algebra

$$[L^3, K^1] = iK^2\,, \quad [L^3, K^2] = -iK^1\,, \quad [K^1, K^2] = -iL^3\,,$$

which the reader may recognize as the Lie brackets of $\mathfrak{su}(1,1)$. The vector fields so constructed thus constitute a spinorial representation of the Lie algebra. In this realization the Casimir element $Q$ takes the form

$$\begin{aligned} Q &= (L^3)^2 - (K^1)^2 - (K^2)^2 \\ &= \frac{1}{4}\left[2(z_1\partial_2 + z_2\partial_2) + (z_1\partial_2 + z_2\partial_2)^2\right] , \end{aligned} \tag{II.3.4}$$

and one can define the objects

$$L^{\pm} = K^1 \pm iK^2 \,, \quad L^{+} = -z_2\partial_1 \,, \quad L^{-} = z_1\partial_2 \,,$$

which will soon be shown to act as creation and annihilation operators, respectively.

*(c) Group representations*

The $\mathfrak{su}(1,1)$ generators obtained above are all holomorphic derivations depending on coordinate functions $z_1$ and $z_2$. This suggests seeking representations of SU(1, 1) on the ring of formal power series $\mathbb{C}[[z_1, z_2]]$. Indeed, the algebra representation (II.3.3) corresponds to the usual group representation in terms of operators on a function space

$$\begin{gathered} D : \mathrm{SU}(1,1) \to \mathcal{O}\left(\mathbb{C}[[z_1, z_2]]\right) \\ D(g)f(z_1, z_2) = f(g^{-1} \triangleright (z_1, z_2)) \,, \quad g \triangleright (z_1, z_2) = g\left(\begin{smallmatrix} z_1 \\ z_2 \end{smallmatrix}\right) , \end{gathered}$$

as can be retroactively checked by setting $(L^3, K^1, K^2) = \frac{1}{2}(\varsigma_1, \varsigma_2, \varsigma_3)$ in the defining representation, i.e.

$$L^3 = D'(\sigma_3/2) \,, \quad K^1 = D'(i\sigma_2/2) \,, \quad K^2 = D'(-i\sigma_1/2) \,,$$

where $D'(X)$ stands for the induced Lie algebra representation. For example, considering $K^1$ one finds

$$\begin{aligned} D'\left(\frac{i\sigma_2}{2}\right) f(z_1, z_2) &:= -i\frac{\mathrm{d}}{\mathrm{d}t}|_{t=0} D\left(e^{i\frac{i\sigma_2}{2}t}\right) f(z_1, z_2) \\ &= \frac{1}{2}\left(z_1\frac{\partial f}{\partial z_2} - z_2\frac{\partial f}{\partial z_1}\right) , \end{aligned}$$

which is exactly the action of $K^1$ on $f(z_1, z_2)$ according to (II.3.3).

According to Schur's lemma [105], in any irreducible representation the Casimir operator is proportional to the identity. Hence in searching for irreducible representations it is useful to consider the eigenfunctions of the Casimir element $Q$. Clearly, any solution to

$$(z_1\partial_2 + z_2\partial_2)f(z_1, z_2) = 2kf(z_1, z_2)\,, \qquad \text{(II.3.5)}$$

will also be an eigenfunction of $Q$ by virtue of equation (II.3.4). But Euler's theorem on homogeneous functions dictates that every maximal smooth solution to such an equation is an homogeneous function of degree $2k \in \mathbb{Z}$, i.e. $f(\alpha z_1, \alpha z_2) = \alpha^{2k} f(z_1, z_2)$. The eigenfunctions of $Q$ are therefore the homogeneous series

$$Qf_k(z_1, z_2) = k(k+1)f_k(z_1, z_2)\,.$$

*(d) Hermitian inner product and unitarity*

We are interested in unitary representations, and this requires finding an invariant hermitian form on $\mathbb{C}[[z_1, z_2]]$. Our strategy will be to derive such a form from the $L^2(\mathrm{SU}(1,1))$ inner product, since 1) the Haar measure is left- and right-invariant and 2) $\mathrm{SU}(1,1)$ can be parametrized in terms of the spinor $|z\rangle$. With the parametrization

$$g(z_1, z_2) = \begin{pmatrix} z_1 & z_2 \\ \overline{z}_2 & \overline{z}_1 \end{pmatrix}\,,$$

the Haar measure takes the form $\mathrm{d}g(z) = \pi^{-2}\delta(\langle z|\sigma_3|z\rangle - 1)\, Dz_1 Dz_2$ [42], where $Dz_i := \frac{i}{2}\mathrm{d}z_i \wedge \mathrm{d}\overline{z}_i$. Let $f_k, g_{k'}$ be two homogeneous functions of degree $2k, 2k'$, respectively.

Then

$$\begin{aligned}
&\frac{1}{\pi^2}\int Dz_1 Dz_2\ \delta(\langle z|\sigma_3|z\rangle - 1)\overline{f_k(z_1,z_2)}g_{k'}(z_1,z_2)\\
&= \frac{1}{2\pi^2}\int \mathrm{d}\theta_1\mathrm{d}\theta_2\mathrm{d}r\ \frac{r}{(r^2-1)^2}\Theta(r-1)\\
&\qquad \cdot \overline{f}_k\left(\frac{re^{i\theta_1}}{\sqrt{r^2-1}},\frac{e^{i\theta_2}}{\sqrt{r^2-1}}\right)g_{k'}\left(\frac{re^{i\theta_1}}{\sqrt{r^2-1}},\frac{e^{i\theta_2}}{\sqrt{r^2-1}}\right)\\
&= \frac{i}{2\pi}\delta_{kk'}\int_{D^1} Dz\,(1-|z|^2)^{-2k-2}\overline{f}_k(1,z)g_{k'}(1,z)\,,
\end{aligned}$$

where $z_i =: \rho_i e^{i\theta_i}, r := \rho_1/\sqrt{\rho_1^2-1}, z^{-1} := re^{i\theta_1}$ and $D^1$ denotes the unit disk in $\mathbb{C}$. One can then define the inner product

$$\langle f,g\rangle := \frac{1}{\pi}\int_{D^1} Dz\,(1-|z|^2)^{-2k-2}\overline{f}(1,z)g(1,z)\,, \qquad \text{(II.3.6)}$$

which is by construction invariant, i.e. $\langle D(g)f, D(g)h\rangle = \langle f,g\rangle$ and well-defined for $k \leq -1$. Following Bargmann [107] we extend this inner product to the case $k=-1/2$ by setting[19]

$$\langle f,g\rangle_{-\frac{1}{2}} := \lim_{k\to-\frac{1}{2}} \frac{-2k-1}{\pi}\int_{D^1} Dz\,(1-|z|^2)^{-2k-2}\overline{f}(1,z)g(1,z)\,.$$

The ring $\mathbb{C}[[z_1,z_2]]$ can now be restricted the Hilbert space $\mathcal{B}_k(D^1)$ of holomorphic functions on the disk.

The invariant pairing of equation (II.3.6) is precisely the inner product on the discrete series of unitary representations of $\mathrm{SU}(1,1)$ [107], showing the manner in which the one-sheeted hyperboloid is associated to the discrete series. The representation $D(g)f(z_1,z_2) = f(g^{-1}(z_1,z_2))$ reduces to the more common multiplier representation derived by Bargmann [107] on the disk $\psi(z) := f(1,z)$ when homogeneity is taken into account:

$$D(g)\psi(z) = (\overline{\alpha}-\beta z)^{2k}\psi\left(\frac{\alpha z-\overline{\beta}}{\overline{\alpha}-\beta z}\right)\,,\quad g = \left(\begin{smallmatrix}\alpha & \beta\\ \overline{\beta} & \overline{\alpha}\end{smallmatrix}\right). \qquad \text{(II.3.7)}$$

[19] That our method glosses over the lowest spin representation is a consequence of the fact that $k=-1/2$ is absent from the Fourier series of square-integrable functions on the group, cf. (A.4.1).

One furthermore has from the above law that

$$D(\mathbb{1})\psi(z) = D(e^{4\pi i L^3})\psi(z) = e^{-4i\pi k}\psi(z)\,,$$

and requiring single-valuedness restricts $k$ to be half-integer, and thus to lie in the range $k = -\frac{1}{2} - \frac{\mathbb{N}^0}{2}$. The Casimir element is therefore non-negative

$$Q \sim k(k+1) \geq 0\,,$$

and the sign of $D(-\mathbb{1}) = \pm\mathbb{1}$ is controlled by whether $k$ takes an integral or definite half-integral value.

*(e) The $L^3$ eigenbasis*

Consider the functions

$$f_{k,m}(z_1,z_2) = \frac{1}{\sqrt{\gamma_{k,m}}} z_1^{k-m} z_2^{k+m} \quad \left( \underset{D^1}{\rightarrow} \; \psi_m(z) = \frac{1}{\sqrt{\gamma_{k,m}}} z^{k+m} \right)\,, \tag{II.3.8}$$

where $\gamma_{k,m} = \frac{\Gamma(-2k-1)\Gamma(1+k+m)}{\Gamma(m-k)}$, which are clearly homogeneous of degree $2k$. They are also eigenfunctions of $L^3$, since

$$L^3 f_{k,m} = m f_{k,m}\,.$$

The requirement that $\psi_m$ reduces to a polynomial on $D^1$ implies that the magnetic index takes the range $m = -k + \mathbb{N}^0$. Together with the transformation rule of equation (II.3.7), this characterizes the *positive series* $\mathcal{D}_k^+$ as defined by [107]. The functions $\psi_m$ are moreover orthonormal under (II.3.6), as can be shown using the integral representation of the Beta function, and hence they constitute an orthonormal basis for $\mathcal{D}_k^+$. The ladder operators act as

$$L^{\pm} f_{k,m} = \sqrt{(m \pm k \pm 1)(m \mp k)} f_{k,m\pm 1}$$

and $L^-$ annihilates the lowest weight state with $m = -k$.

The *negative series* $\mathcal{D}_k^-$ can be obtained in a similar manner. There is an obvious outer automorphism of the Lie algebra (II.3.3) leaving the brackets invariant, namely the one induced by taking $(z_1, z_2) \mapsto (z_2, z_1)$. Denoting the automorphism by $P$ (for parity, to match the terminology of [108]), it reads

$$PL^3P^{-1} = -L^3\,, \quad PK^1P^{-1} = -K^1\,, \quad PK^2P^{-1} = K^2\,, \quad \text{(II.3.9)}$$

The parity-transformed algebra can be integrated to a group representation

$$D(g)f(z_1, z_2) \mapsto f(\overline{g}^{-1} \triangleright (z_1, z_2))\,,$$

such that

$$D(g)\psi(z) = (\alpha - \overline{\beta}z)^{2k}\psi\left(\frac{\overline{\alpha}z - \beta}{\alpha - \overline{\beta}z}\right)\,,$$

agreeing with the original findings of [107] for $\mathcal{D}_k^-$. One can then see from equation (II.3.8) that under $PL^3P^{-1}$ the magnetic index takes the values $m = k - \mathbb{N}^0$, and this is indeed the negative series. We collect both positive and negative series under the notation $\mathcal{D}_k^q$, with $q = \pm$.

The representations just constructed are known to be irreducible. A standard proof of this fact using the ladder operators can be found in [109].

### II.3.2 The continuous series from $H^{\mathrm{sl}}$ Majorana spinors

*(a) Poisson structure*

Turning now to the one-sheeted hyperboloid, recall that the discussion of section II.2.2.(c) resulted in a set of spinors $|z_{1,2}\rangle$ and a mapping to $H^{\mathrm{sl}}$ (II.2.6). In searching for SU(1, 1) representations we again consider a 4-dimensional complex space, but we pick an orthogonal decomposition adapted to the Majorana spinors, i.e. $\mathbb{C}^4 \simeq (\mathbb{C}^2)^1 \oplus (\mathbb{C}^2)^2$ with a choice of basis

$$\{\partial_1^1, \partial_2^1\} \oplus \{\partial_1^2, \partial_2^2\} := \{\partial_1, \overline{\partial}_1\} \oplus \{\partial_2, \overline{\partial}_2\}\,,$$

so that the upper index denotes the factor. A general element $u \in (\mathbb{C}^2)^1 \oplus (\mathbb{C}^2)^2$ takes the form $u = u^1 + u^2$ for $u^i \in (\mathbb{C}^2)^i$, and there is a linear involution commuting the factors

$$T : \mathbb{C}^4 \mapsto \mathbb{C}^4$$
$$T\,\partial_i^1 = \partial_i^2 \,.$$

The first component is equipped with an indefinite hermitian pairing

$$\langle \cdot, \cdot \rangle_{(\mathbb{C}^2)^1} : (\mathbb{C}^2)^1 \times (\mathbb{C}^2)^1 \to \mathbb{C}$$
$$(u^1, v^1) \mapsto \langle u^1, v^1 \rangle_{(\mathbb{C}^2)^1} = u^{1\dagger}\sigma_3 v^1 \,,$$

which is invariant under the action of $\mathrm{SU}(1,1)$. Taking inspiration from before, we consider the symplectic form[20]

$$i\omega(u_i^2, v_j^1) = \langle T\overline{u_i^2}, v_j^1 \rangle_{(\mathbb{C}^2)^1} \,,$$
$$\omega(u^1, v^1) = \omega(u^2, v^2) = 0 \,,$$

where antisymmetry requires that $\langle u^2, v^2 \rangle_{(\mathbb{C}^2)^2} = u^{2\dagger}\sigma_3 v^2$. Note that $\omega$ has the same structure as the symplectic form defined in equation (II.3.1), but now with an additional involution map $T$ (which was complex conjugation itself in the former case). Explicitly it takes the form

$$\omega = i(\mathrm{d}z_1 \wedge \mathrm{d}\overline{z}_2 - \mathrm{d}\overline{z}_1 \wedge \mathrm{d}z_2) \,, \tag{II.3.10}$$

leading to the only non-vanishing Poisson brackets[21]

$$\{\overline{z}_i^1, z_j^2\} = i\sigma_{3ij} \,. \tag{II.3.11}$$

[20] This was once revealed to me in a dream.

[21] For the reader's convenience I reiterate that the double-index notation is to be understood as

$$z_1^1 := z_1 \,, \quad z_2^1 := \overline{z}_1 \,, \quad z_1^2 = z_2 \,, \quad z_2^2 = \overline{z}_2 \,.$$

*(b) Lie algebra realization*

The vector components $v^i = \frac{1}{2}\langle z^2|\sigma_3\varsigma^i|z^1\rangle$ determined by equation (II.2.6) read

$$v^1 = \frac{1}{2}(z_1\bar{z}_2 + \bar{z}_1 z_2)\,, \quad v^2 = \frac{1}{2}(z_1 z_2 + \bar{z}_1\bar{z}_2)\,, \quad v^3 = \frac{i}{2}(z_1 z_2 - \bar{z}_1\bar{z}_2)\,,$$

with norm

$$\begin{aligned} |v|^2 &= (v^1)^2 - (v_2)^2 - (v_3)^2 \\ &= \frac{1}{4}(z_1\bar{z}_2 - \bar{z}_1 z_2)^2 \leq 0\,, \end{aligned}$$

which is non-positive as intended. The component functions satisfy the same Poisson algebra as equation (II.3.2),

$$\{v^i, v^j\} = -\epsilon^{ijk}\eta_{kl}v^l\,, \tag{II.3.12}$$

which can now be recognized as being analogous to the $\mathfrak{su}(1,1)$ Lie algebra. Using the symplectic form (II.3.10), the functions $v^i$ induce the Hamiltonian vector fields

$$\begin{aligned} X^1 &= -\frac{i}{2}\left(z_1\partial_1 + z_2\partial_2 - \bar{z}_1\bar{\partial}_1 - \bar{z}_2\bar{\partial}_2\right)\,, \\ X^2 &= -\frac{i}{2}\left(\bar{z}_1\partial_1 + \bar{z}_2\partial_2 - z_1\bar{\partial}_1 - z_2\bar{\partial}_2\right)\,, \\ X^3 &= -\frac{1}{2}\left(\bar{z}_1\partial_1 + \bar{z}_2\partial_2 + z_1\bar{\partial}_1 + z_2\bar{\partial}_2\right)\,, \end{aligned}$$

with Lie brackets given by $[X^i, X^j] = \epsilon^{ijk}\eta_{kl}X^l$. As we did before, we proceed by projecting onto the first factor $\mathbb{C}^2_1$ and subsequently identifying $(X^1, X^2, X^3) \mapsto (iL^3, iK^1, iK^2)$. One obtains in this manner the generators

$$\begin{gathered} L^3 = \frac{1}{2}\left(\bar{z}\bar{\partial} - z\partial\right)\,, \quad K^1 = \frac{1}{2}\left(z\bar{\partial} - \bar{z}\partial\right)\,, \\ K^2 = \frac{i}{2}\left(z\bar{\partial} + \bar{z}\partial\right)\,, \end{gathered} \tag{II.3.13}$$

having done away with the subscript $z_1 \mapsto z$, since it is for now irrelevant. There is again an immediate outer automorphism induced by the mapping $(z,\overline{z}) \mapsto (\overline{z}, z)$,

$$PL^3P^{-1} = -L^3\,, \quad PK^1P^{-1} = -K^1\,, \quad PK^2P^{-1} = K_2\,,$$

which is precisely the parity map of equation (II.3.9). The Casimir operator in this realization reads

$$\begin{aligned} Q &= (L^3)^2 - (K^1)^2 - (K^2)^2 \\ &= \frac{1}{4}\left[2(z\partial + \overline{z}\overline{\partial}) + (z\partial + \overline{z}\overline{\partial})^2\right]\,, \end{aligned} \tag{II.3.14}$$

and the ladder operators are given by

$$L^+ = -\overline{z}\partial\,, \quad L^- = z\overline{\partial}\,.$$

*(c) Group representations*

It is once more the case that the Lie algebra representation (II.3.13) induces a group representation. We take as carrier space the ring of formal power series $\mathbb{C}[[z,\overline{z}]]$, and one can retroactively check that the representation

$$\begin{gathered} D : \mathrm{SU}(1,1) \to \mathcal{O}\left(\mathbb{C}[[z,\overline{z}]]\right) \\ D(g)f(z,\overline{z}) \mapsto f(g^{-1} \triangleright (z,\overline{z}))\,, \quad g \triangleright (z,\overline{z}) = g\left(\begin{smallmatrix} z \\ \overline{z} \end{smallmatrix}\right)\,, \end{gathered}$$

does lead to the realization of equation (II.3.13). Regarding the eigenfunctions of the Casimir element (II.3.14), note that any solution to

$$(z\partial + \overline{z}\overline{\partial})f(z,\overline{z}) = (-2j-2)f(z,\overline{z})$$

is also an eigenfunction[22] of $Q$. The differential equation (II.3.5) of the previous subsection is analogous to the above up to the fact that $z$ and $\overline{z}$ are complex conjugated. One is thus led to consider homogeneous functions as possible solutions, and indeed the homogeneous series $f_j(rz, r\overline{z}) = r^{-2j-2} f_j(z, \overline{z})$, $r \in \mathbb{R}$ solve the differential problem. Hence the $Q$ eigenfunctions satisfy

$$Qf_j = j(j+1)f_j \, .$$

[22] Although one could have chosen to pick $2j$ as the degree of homogeneity as in [107], our choice is to match the convention of [108]. The eigenvalue of $Q$ agrees in both cases, and once one fixes $j = -\frac{1}{2} + is$ both conventions are related by complex conjugation.

*(d) Hermitian inner product and unitarity*

We again consider the Haar measure in terms of two complex variables $z_1, z_2$. Let $f_j(z_1)$ and $g_{j'}(z_1)$ be two homogeneous functions as above, having reinstated the lower index of $z_1$. Then

$$\begin{aligned} &\frac{1}{\pi^2} \int Dz_1 Dz_2 \; \delta(\langle z|\sigma_3|z\rangle - 1) \overline{f}_j(z_1) g_{j'}(z_1) \\ &= \frac{1}{\pi} \int \mathrm{d}\theta r \mathrm{d}r \; \Theta(r-1) \overline{f}_j \left(re^{i\theta}\right) g_{j'} \left(re^{i\theta}\right) \\ &= \frac{1}{\pi} \int \mathrm{d}\theta \int_0^\infty \mathrm{d}k \, e^{2k[(j'+\overline{j})+1]} \overline{f}_j \left(e^{i\theta}\right) g_{j'} \left(e^{i\theta}\right) \, , \end{aligned} \tag{II.3.15}$$

where $z_1 =: re^{i\theta}$ and $k := \ln r$. Equation (II.3.15) contains an integral representation of the Dirac delta when the exponent of the last line is purely imaginary. Indeed, setting

$$j + \frac{1}{2} = is \, , \quad s \in \mathbb{R}^+ \, ,$$

the whole expression reduces to

$$\begin{aligned} &\frac{1}{\pi^2} \int Dz_1 Dz_2 \; \delta(\langle z|\sigma_3|z\rangle - 1) \overline{f}_j(z_1) g_{j'}(z_1) \\ &\qquad\qquad \sim \delta(s-s') \int_{S^1} \mathrm{d}\theta \, \overline{f}_j \left(e^{i\theta}\right) g_{j'} \left(e^{i\theta}\right) \, , \end{aligned}$$

suggesting one ought to define the inner product

$$\langle f, g\rangle = \frac{1}{2\pi} \int_{S^1} \mathrm{d}z \, \overline{f}(z) g(z) \tag{II.3.16}$$

over the circle. The formal series $\mathbb{C}[[z,\overline{z}]]$ can be restricted to functions on $S^1$ with period $p$, and upon completion through the norm induced by (II.3.16) one gets the Hilbert space of square-integrable functions $L^2_p(S^1)$.

The invariant pairing defined in (II.3.16) agrees exactly with the inner product on the continuous series of unitary representations of SU$(1,1)$ [107], and hence the representations associated to the $H^{\rm sl}$ Majorana spinors do correspond to the continuous series. The representation $D(g)f(z,\overline{z}) = f(g^{-1}(z,\overline{z}))$ matches the multiplier representation defined in [41, Ch. VII] through the homogeneity property on $\psi(\theta) = f(e^{i\theta})$,

$$D(g)\psi(\theta) = |\overline{\alpha}e^{i\theta} - \beta e^{-i\theta}|^{-2j-2}\psi\left(\arg\frac{\overline{\alpha}e^{i\theta} - \beta e^{-i\theta}}{|\overline{\alpha}e^{i\theta} - \beta e^{-i\theta}|}\right) ,$$
$$g := \left(\begin{smallmatrix}\alpha & \beta \\ \overline{\beta} & \overline{\alpha}\end{smallmatrix}\right) .$$

With respect to the labels $j = -\frac{1}{2} + is$ the Casimir operator acts as

$$Q \sim -\left(\frac{1}{4} + s^2\right) < 0 ,$$

and as expected its spectrum is always negative.

*(e) The $L^3$ eigenbasis*

The functions

$$f_{j,m}(z) = e^{i\varphi_{j,m}} z^{-j-1-m}\overline{z}^{-j-1+m} \qquad \left(\underset{S^1}{\rightarrow} \ \psi_m(\theta) = e^{i\varphi_{j,m}} e^{-2i\theta m}\right)$$

with $e^{i\varphi_{j,m}} = \left(\frac{\Gamma(m-\overline{j})}{\Gamma(m-j)}\right)^{\frac{1}{2}}$ are homogeneous of degree $-2j-2$, and satisfy

$$L^3 f_{j,m} = m f_{j,m} .$$

The range of the magnetic index can be determined by requiring the representation property $D^j(\mathbb{1}) = \mathbb{1}_j$, i.e.

$$D^j(e^{4i\pi L^3}) f_{j,m} = e^{4i\pi m} f_{j,m} \stackrel{!}{=} f_{j,m} ,$$

by virtue of which $m$ must take integer and half-integer values. This splits the continuous series into two classes of representations: the *integer series* $\mathcal{C}_j^0$, for which

$$m = \mathbb{Z}\,, \quad D(-\mathbb{1})\psi_m(\theta) = \psi_m(\theta)\,, \quad \psi(\theta+\pi) = \psi(\theta)\,,$$

and the *half-integer* series $\mathcal{C}_j^{\frac{1}{2}}$, characterized by

$$m = \frac{1}{2} + \mathbb{Z}\,, \quad D(-\mathbb{1})\psi_m(\theta) = -\psi_m(\theta)\,, \quad \psi(\theta+\pi) = -\psi(\theta)\,.$$

The states $\psi_m$ in $\mathcal{C}_j^0$ and $\mathcal{C}_j^{\frac{1}{2}}$ thus respectively constitute an orthonormal basis for $L^2_\pi(S^1)$ and $L^2_{2\pi}(S^1)$, as is known from Fourier analysis. The phase $e^{i\varphi_{j,m}}$ was chosen such that the ladder operators have the same expression[23] as in the discrete series, i.e.

$$L^\pm f_{j,m} = \sqrt{(m \pm j \pm 1)(m \mp j)}\, f_{j,m\pm 1}\,.$$

The parity operator acts by conjugation $z \mapsto \overline{z}$, so

$$P f_{j,m} = e^{i\varphi_{j,m}} e^{-i\varphi_{j,-m}} f_{j,-m} = e^{i\pi m} f_{j,-m}\,,$$

confirming that $P^2 = \mathbb{1}$ is an involution. Finally, irreducibility for each $\mathcal{C}^\delta$, $\delta = 0, \frac{1}{2}$ can be proven following the same strategy as for the discrete case.

### II.3.3 Twistor gauge reduction for Majorana spinors

The symplectomorphism of equation (II.1.5) was originally derived with spinor states adapted to the space-like sphere [94], and a generalization of the statement to SU(1,1) was afterwards obtained by [98]. However, since the spinor states I propose for $H^{\text{sl}}$ are substantially different from the ones of this latter work (a circumstance which will be discussed shortly), it is worthwhile to show how the $T^*\text{SU}(1,1)$ phase structure can be made to arise from the Majorana spinors introduced above. All necessary ingredients to do so are already in place.

[23] Under $k \mapsto j$. Note that $[(m \pm j \pm 1)(m \mp j)]^{\frac{1}{2}} = |m \mp \overline{j}|$ for the continuous series. The chosen $e^{i\varphi_{j,m}}$ phase is a solution to the recursion relation $(m - \overline{j})e^{i\varphi_{j,m}} = |m - \overline{j}|e^{i\varphi_{j,m+1}}$.

We again consider a graph link, together with two sets of Majorana spinors,

$(|z^1\rangle, |z^2\rangle)$ 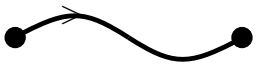 $(|w^2\rangle, |w^1\rangle)\,.$

A parallel transport can be constructed as

$$g(z,w) = \frac{|w^1\rangle\!\langle z^1|\sigma_3 + |w^2\rangle\!\langle z^2|\sigma_3}{\sqrt{\langle w^2|\sigma_3|w^1\rangle\,\langle z^2|\sigma_3|z^1\rangle}}\,,$$

taking source spinors to target spinors

$$g(z,w)|z^1\rangle = \sqrt{\frac{\langle z^2|\sigma_3|z^1\rangle}{\langle w^2|\sigma_3|w^1\rangle}}\,|w^2\rangle\,, \quad g(z,w)|z^2\rangle = \sqrt{\frac{\langle z^2|\sigma_3|z^1\rangle}{\langle w^2|\sigma_3|w^1\rangle}}\,|w^1\rangle\,,$$

and transforming as $g(h_1 z, h_2 w) = h_2 g(z,w) h_1^{-1}$ for $h \in \mathrm{SU}(1,1)$. This suggests considering an area-matching constraint

$$\mathcal{C} = \langle w^2|\sigma_3|w^1\rangle - \langle z^2|\sigma_3|z^1\rangle \sim 0\,,$$

which has the same structure as the area constraint used in section II.1.1 insofar as each of the bilinears corresponds to the norm of the associated geometrical vector, as per equation (II.2.6). Upon enforcing the constraint it is straightforward to check that $g(z,w)$ commutes with the real structure and that $\det g(z,w) = 1$, from where $g(z,w) \in \mathrm{SU}(1,1)$ in the constraint hypersurface. Regarding the Poisson brackets we employ (II.3.11), and it was already established in (II.3.12) that then $\{v_z^i, v_z^k\} = -\epsilon^{ijk}\eta_{kl}v_z^l$, where $v_z^i := \langle z^2|\sigma_3\varsigma^i|z^1\rangle$ are functions dual through $\omega$ to a complete set of left-invariant vector fields. Straightforward computations show[24] that

[24] That the Poisson commutator of group elements vanishes can be shown by developing the brackets and making use of the identity

$$\sigma_2 = -i\frac{|z^1\rangle\!\langle \overline{z}^2| - |z^2\rangle\!\langle \overline{z}^1|}{\langle z^2|\sigma_3|z^1\rangle} = -i\frac{|w^1\rangle\!\langle \overline{w}^2| - |w^2\rangle\!\langle \overline{w}^1|}{\langle w^2|\sigma_3|w^1\rangle}\,.$$

on the constraint hypersurface

$$\{g_{ab}(z,w), g_{cd}(z,w)\} = 0\,, \quad \{g_{ab}(z,w), v_z^i\} = -i g_{ab}(z,w)\,\varsigma^i\,,$$

analogously to (II.1.4) but with $\mathrm{SU}(1,1)$ generators $\varsigma^i$; $\mathcal{C}$ moreover commutes with $g(z,w)$ and $v_z$. This all establishes the symplectomorphism

$$\left[\,(\mathbb{C}^2)^1 \oplus (\mathbb{C}^2)^2 \setminus \{\langle z^2|\sigma_3|z^1\rangle = 0\}\,\right] // \mathcal{C} \quad \simeq \quad T^*\mathrm{SU}(1,1) \setminus |v| = 0\,,$$

which serves as an $\mathrm{SU}(1,1)$ version of the original findings of [94].

As mentioned above, the Majorana spinors for $H^{\text{sl}}$ proposed here are substantially different from the Weyl spinors of Rennert [98], which are claimed to also characterize the space-like hyperboloid. A careful reading of [98] reveals however that these latter objects are inherently adapted to $H^{\pm}$; this is particularly clear from eqs. (3.135-3.136) of that work, which - once a missing square exponent is fixed - show the induced vectors have positive square norm.

## II.4 Boundary coherent states

Although a relationship was established between the $H^{\pm}$, $H^{\text{sl}}$ spinors and the discrete and continuous series of $\mathrm{SU}(1,1)$, respectively, we have not yet determined which particular states in those series should be associated to the boundaries of spin-foam models. Doing so requires some criteria, and the following seem reasonable: 1) that in the semiclassical limit - where geometry is to be recovered - the model should effectively depend on the spinor states defined above; and 2) that that dependence ought to be weighted by the parameter which controls the degree of classicality, as in any path-integral formalism. The coherent states originally defined for the EPRL model do satisfy these requirements, and it is useful to see just how so before continuing with the $\mathrm{SU}(1,1)$ formalism. We will consider a 3-dimensional model for simplicity.

The spin-foam vertex amplitude for Riemannian 3d quantum gravity with SU(2) coherent states (I.2.11) $|j, n\rangle := D^j(n)|j, j\rangle$ reads [110]

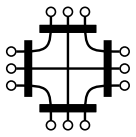

$$= \int_{\mathrm{SU}(2)} \prod_{a=1}^{4} \mathrm{d}g_a \prod_{a \neq b} \langle j_{ab}, n_{ab}| D^{j_{ab}}(g_a)^\dagger D^{j_{ab}}(g_b) |j_{ab}, n_{ba}\rangle \ ,$$

it being a product of 6 coherent state pairings. The unitary representations of SU(2) satisfy the Clebsch-Gordan isomorphism

$$I : \bigotimes_{i=1}^{2j} \mathcal{H}^{\frac{1}{2}} \simeq V \oplus \mathcal{H}^j \ , \tag{II.4.1}$$

where $\mathcal{H}^j$ is the supporting Hilbert space of the fundamental representation, and $V$ is a reducible representation. Since $D^{\otimes 2j}(L^3)|+\rangle^{\otimes 2j} = j$, it follows that $I \circ |+\rangle^{\otimes 2j} = |j, j\rangle$. Thus each of the SU(2) inner products appearing in the amplitude can be written in the defining representation as

$$\begin{aligned} \langle j_{ab}, n_{ab}| D^{j_{ab}}(g_a)^\dagger D^{j_{ab}}(g_b) |j_{ab}, n_{ba}\rangle &= \langle 1/2, n_{ab}| g_a^\dagger g_b |1/2, n_{ba}\rangle^{2j_{ab}} \\ &= \langle + | n_{ab}^\dagger g_a^\dagger g_b n_{ba} | + \rangle^{2j_{ab}} \ , \end{aligned}$$

or, setting $|z_{ab}\rangle := n_{ab}|+\rangle$,

$$\langle j_{ab}, n_{ab}| D^{j_{ab}}(g_a)^\dagger D^{j_{ab}}(g_b) |j_{ab}, n_{ba}\rangle = \langle z_{ab} | g_a^\dagger g_b | z_{ba}\rangle^{2j_{ab}} \ .$$

The Riemannian 3d model can therefore be thought of as a convolution of SU(2)-invariant pairings of Weyl spinors, weighted by the spin label. Since the classical regime is attained when all spins are scaled uniformly to infinity (cf. chapter III), our proposed criteria are satisfied. The crux of the argument is then the formal correspondence

$$D^j(g)|j, j\rangle \sim g|+\rangle^{2j} \sim |z\rangle^{2j} \ , \tag{II.4.2}$$

which is guaranteed by equation (II.4.1).

In searching for appropriate SU(1, 1) coherent states, it is clear that the criteria outlined in the paragraphs above can be satisfied if those coherent states behave in a similar manner to (II.4.2), adapted to the right spinor states. The obvious obstacle is that unlike SU(2) the SU$(1, 1)$ group is non-compact, and a standard argument [111] shows that all unitary representations must be infinite-dimensional (indeed this was confirmed in section II.3); hence a relationship between the unitary representations and the defining one as in (II.4.1) can never hold. It happens to be the case, however, that there is a sense in which such a correspondence is possible, and this is the subject of what follows.

### II.4.1 Coherent states for $H^{\pm}$ Weyl spinors

In the following, in order to streamline notation as much as possible, I denote by

$$[u|v] := \overline{u}^T \sigma_3 v\,, \quad u, v \in \mathbb{C}^2\,,$$

the SU$(1, 1)$-invariant inner product on $\mathbb{C}^2$; I retain however $\cdot^\dagger$ for complex transposition, and set the convention that $|v] = |v\rangle$, while $[v| = \langle v|\sigma_3$.

Focusing first on the defining representation, we would like to write any unit-norm Weyl spinor as an SU$(1, 1)$ action on a reference state in $\mathbb{C}^2$, i.e.

$$g|\chi] \stackrel{!}{=} \left(\begin{smallmatrix} z_1 \\ z_2 \end{smallmatrix}\right), \quad g \in \mathrm{SU}(1,1)\,, \quad [z|z]^2 = 1\,,$$

which is clearly possible if $|\chi] = |\pm]$. The $|\pm]$ are eigenstates of $L^3$ in the defining representation, and indeed the eigenstates of higher representations (II.3.8) can be expressed as powers of $|\pm]$ spinor monomials

$$f_{k,m}(w_1, w_2) = \frac{(-1)^{k+m}}{\sqrt{\gamma_{k,m}}}[+|w]^{k-m}[-|w]^{k+m}\,. \tag{II.4.3}$$

This suggests that we should consider coherent states constructed from $L^3$ eigenstates, i.e. coherent states of the form $D^k(g)|k, m_r\rangle$ for some reference $m_r$, in which case

$$D^k(g) f_{k,m_r} = \frac{(-1)^{k+m_r}}{\sqrt{\gamma_{k,m_r}}}[g\cdot +|w]^{k-m_r}[g\cdot -|w]^{k+m_r}\,,$$

and the geometric spinors $|g \cdot \pm]$ - the ones which were put in correspondence with the two-sheeted hyperboloid - directly figure in the representation function.

Looking at the functions (II.4.3), there exists a distinguished and particularly simple lowest weight state $\psi_{-k}(w) = \gamma_{k,-k}^{-1/2} \cdot 1$. Consider then the states $D^k(g)|k,-k\rangle$ which, according to the transformation law (II.3.7), can be expressed using the binomial series as[25]

$$\begin{aligned} D^k(g)\psi_{-k}(w) &= \gamma_{k,-k}^{-1/2}(\overline{\alpha} - \beta w)^{2k} \\ &= \gamma_{k,-k}^{-1/2} \sum_{l \geq 0} \binom{2k}{l} \overline{\alpha}^{2k-l}(-\beta)^l w^l \end{aligned}$$

whenever $|\beta w/\overline{\alpha}| < 1$. Using the explicit form of the inner product in $\mathcal{D}_k^+$ (II.3.6), and setting $\mathrm{d}\omega_w := Dw\,(1-|w|^2)^{-2k-2}$, one has for the pairing of two coherent states

$$\begin{aligned} &\langle k,-k|D^{k\dagger}(g')D^k(g)|k,-k\rangle \\ &= \frac{\alpha'^{2k}\overline{\alpha}^{2k}}{\gamma_{k,-k}} \sum_{l,t \geq 0} \binom{2k}{t}\binom{2k}{l}\left(\frac{-\overline{\beta}'}{\alpha'}\right)^t \left(\frac{-\beta}{\overline{\alpha}}\right)^l \frac{1}{\pi}\int_{D^1} \mathrm{d}\omega_w\, \overline{w}^t w^l \\ &= \sum_{l \geq 0} \binom{2k}{l}^2 \binom{-2k+l-1}{l}^{-1} (\alpha'\overline{\alpha})^{2k-l}(\overline{\beta}'\beta)^l \\ &= (\alpha'\overline{\alpha} - \overline{\beta}'\beta)^{2k}\,. \end{aligned}$$

Hence the following identity involving the defining representation holds[26],

$$\langle k,-k|D^{k\dagger}(g')D^k(g)|k,-k\rangle = [g' \cdot -|g \cdot -]^{2k}\,, \tag{II.4.4}$$

[25] It is actually sufficient to consider the action of the maximal compact subgroup generated by $L^3$, in which case the convergence of the binomial series is assured inside the disk $|z| < 1$.

[26] It is quite interesting to note that equation (II.4.4) is formally similar to a CFT two-point function (recall $k$ is always negative). The matrix coefficients of (II.4.4) appear in the definition of the 3d coherent spin-foam vertex for a time-like boundary, which *very* heuristically might suggest a relationship to AdS-3/CFT-2 duality.

and our choice of coherent states satisfies the requirements outlined earlier. The formal identification $|k,-k\rangle \sim |-]^{2k}$ is still valid for the discrete series of $\mathrm{SU}(1,1)$, but now only in the sense of equation (II.4.4). Note that the coherent states $D^k(g)|k,-k\rangle$ are precisely the ones proposed in [54] for space-like boundaries.

### II.4.2 Coherent states for $H^{\mathrm{sl}}$ Majorana spinors

Turning now to the space-like hyperboloid, and thus to Majorana-type spinors, we require

$$g|\chi_1] \overset{!}{=} \left(\begin{smallmatrix} z_1 \\ \bar{z}_1 \end{smallmatrix}\right), \quad g|\chi_2] \overset{!}{=} \left(\begin{smallmatrix} z_2 \\ \bar{z}_2 \end{smallmatrix}\right), \quad g \in \mathrm{SU}(1,1)\,, \quad [z^2|z^1]^2 = -1\,,$$

which is solved by $|\chi_1] = |l^+]$ and $|\chi_2] = i|l^-]$, having defined

$$|l^\pm] := \frac{1}{\sqrt{2}}(|+] \pm |-])\,.$$

The elements $|l^\pm]$ are not $L^3$ eigenstates, but rather eigenstates of $K^2$ in the defining representation. In full analogy with the discrete series case just discussed, it is reasonable to expect that an identity similar to that of equation (II.4.4) may hold, but seemingly requiring coherent states constructed from $K^2$. I shall show that such an identity can be found. It is however known [108] that $K^2$, being a non-compact operator with continuous spectrum, does not have eigenstates in $\mathcal{C}_j^\delta$; moving forward with the construction requires first pointing out a few facts about generalized eigenstates.

*(a) Gelfand triple for $\mathcal{C}_j^\delta$*

I will follow the treatment of the subject given by Lindblad in [108], remarking that our conventions for the $\mathfrak{su}(1,1)$ generators and unitary representations match exactly those provided in that work.

Given the Hilbert space $\mathcal{C}_j^\delta$, it is possible to define a dense subspace $\mathcal{D}$ of "rapidly decreasing sequences"

$$\mathcal{D} = \left\{ \sum_m c_m |j,m\rangle \;\middle|\; \forall n \in \mathbb{N}\,, \lim_{|m|\to\infty} m^n c_m = 0 \right\}\,,$$

with a certain topology cf. [108]. Then there is a Gelfand triple $\mathcal{D} \subset \mathcal{C}_j^\delta \subset \mathcal{D}'$, where $\mathcal{D}'$ is the space of continuous functionals on $\mathcal{D}$ and $\mathcal{C}_j^\delta$ (identified with its dual) is dense in $\mathcal{D}'$. The functional-analytical nuances of the construction are such that a spectral theorem can be applied to self-adjoint operators with continuous spectra on such a triple. For the purposes of this work it is sufficient to state that the nuclear spectral theorem guarantees that $K^2$ (by virtue of being self-adjoint and continuous in $\mathcal{D}$ and leaving it invariant) has a complete set of generalized eigenvectors in $\mathcal{D}'$. This is meant in the sense that

$$F_{\lambda,\sigma}(K^{2\dagger}\psi_m) = \lambda \langle j, m | j, \lambda \rangle \ , \quad F_{\lambda,\sigma} \in \mathcal{D}' \ , \quad \psi_m \in \mathcal{D},$$
$$\langle \psi, \varphi \rangle_{\mathcal{C}_j^\delta} = \sum_\sigma \int \mathrm{d}\lambda \ \langle \psi | j, \lambda, \sigma \rangle \langle j, \overline{\lambda}, \sigma | \varphi \rangle \ , \tag{II.4.5}$$

with $\sigma$ standing for the degeneracy of the distribution at $\lambda$. It is crucial for the following to note that $\lambda \in \mathbb{C}$ is not required to be real since - as pointed out by Lindblad himself - while $K^2$ is self-adjoint in $\mathcal{D}$ its extension to $\mathcal{D}'$ is not. That a complex-conjugated eigenvalue $\overline{\lambda}$ is necessary in the completeness relation (II.4.5) follows however from the self-adjoint property of $K^2$ in $\mathcal{C}_j^\delta$,

$$\begin{aligned}
\langle K^{2\dagger}\psi, \varphi \rangle_{\mathcal{C}_j^\delta} &= \sum_\sigma \int \mathrm{d}\lambda \, F_{\lambda,\sigma}(K^{2\dagger}\psi) \overline{F_{\overline{\lambda},\sigma}(\varphi)} \\
&= \sum_\sigma \int \mathrm{d}\lambda \, F_{\lambda,\sigma}(\psi) \, \lambda \, \overline{F_{\overline{\lambda},\sigma}(\varphi)} \\
&= \langle \psi, K^{2\dagger}\varphi \rangle_{\mathcal{C}_j^\delta} \ .
\end{aligned}$$

Since we will be interested in determining matrix coefficients of the type $\langle j, \overline{\lambda'}, \sigma' | D^j(g) | j, \lambda, \sigma \rangle$, it is important to note that such objects may not be by themselves well-defined; they are generally to be understood as distributions, i.e.

$$F_{\lambda,\sigma}(D^{j\dagger}(g)\psi) = \sum_{\sigma'} \int \mathrm{d}\lambda' \ \langle \psi | j, \lambda', \sigma' \rangle \langle j, \overline{\lambda'}, \sigma' | D^j(g) | j, \lambda, \sigma \rangle \ .$$

*(b) The generalized $K^2$ eigenbasis*

With the above preliminaries established, we proceed to finding a complete set of generalized eigenstates of $K^2$. Inspired by the discussion of section II.4.1, we consider powers of $[l^{\pm}|w]$ monomials, with $|w]$ a general Majorana spinor, and define[27]

$$f^{\sigma}_{j,\lambda}(w) = \alpha_j \left|[l^-|w]\right|^{\bar{j}-i\lambda} \left|[l^+|w]\right|^{\bar{j}+i\lambda} \\ \cdot\ \mathrm{sgn}^{\sigma}\Im\left([l^+|w][l^-|w]\right)\mathrm{sgn}^{2\delta}\Re\left([l^-|w]\right)\ , \\ |w] := \left(\begin{smallmatrix} w \\ \bar{w}\end{smallmatrix}\right) \in \mathbb{C}^2\ , \quad \alpha_j := 2^j\ .$$

These functions are homogeneous of degree $-2j-2$, and the reduction to $|w| = 1$ reads

$$\psi_{\lambda,\sigma}(\theta) = \frac{1}{2}|\cos\theta|^{\bar{j}-i\lambda}\,|\sin\theta|^{\bar{j}+i\lambda}\,\mathrm{sgn}^{\sigma}\left[\cos\theta\sin\theta\right]\,\mathrm{sgn}^{2\delta}\left[\cos\theta\right]\ .$$

It is straightforward to verify that

$$Q\psi_{\lambda,\sigma} = j(j+1)\psi_{\lambda,\sigma}\ , \\ K^2\psi_{\lambda,\sigma} = \lambda\psi_{\lambda,\sigma}\ ,$$

as intended. The sgn functions were introduced in order to control the parity behavior of $\psi_{\lambda,\sigma}$ as the argument moves around the circle of integration $\theta \in [0, 2\pi)$. Indeed, one has

$$P\psi_{\lambda,\sigma} = (-1)^{\sigma}\psi_{\lambda,\sigma}\ , \quad \sigma \in \{0,1\}\ , \\ \psi_{\lambda,\sigma}(\theta+\pi) = (-1)^{2\delta}\psi_{\lambda,\sigma}(\theta)\ , \quad \delta \in \{0, \frac{1}{2}\}\ ,$$

so that the states diagonalize $P^2 = \mathbb{1}$ (recall $[P, K^2] = 0$) and satisfy the periodicity property of $\mathcal{C}^{\delta}_j$. One can explicitely derive the integral identity

$$\langle j, m|j, \lambda, \sigma\rangle = \frac{1}{2\pi}\left(\frac{\Gamma(m-j)}{\Gamma(m-\bar{j})}\right)^{\frac{1}{2}}\int_0^{\pi/2}\left(e^{2i\theta m} + (-1)^{\sigma}e^{-2i\theta m}\right) \\ \cdot\,(\cos\theta)^{\bar{j}-i\lambda}(\sin\theta)^{\bar{j}+i\lambda}\,\mathrm{d}\theta\ ,$$

[27] $z^w := e^{w\ln z}$ for $z, w \in \mathbb{C}$ is defined in terms of the principal branch $\arg z \in [-\pi, \pi)$ of the logarithm.

recovering a result of [108]. Making use of this last equation it is possible to verify completeness and orthogonality,

$$\sum_\sigma \int_{\mathbb{R}+ix} \mathrm{d}\lambda \, \langle j,m|j,\lambda,\sigma\rangle \, \langle j,\overline{\lambda},\sigma|j,n\rangle = \delta_{m,n}\,, \quad x\in\mathbb{R}\,,$$
$$\sum_m \langle j,\overline{\lambda'},\sigma'|j,m\rangle \, \langle j,m|j,\lambda,\sigma\rangle = \delta(\lambda-\lambda')\,, \quad \Im\lambda = \Im\lambda'\,.$$

In short, there is a family of orthonormal[28] bases

$$\{\, |j,\lambda+ix,\sigma\rangle \,|\, \lambda\in\mathbb{R}\,,\sigma\in\{0,1\}\}_{x\in\mathbb{R}}$$

although the $\sigma$ labels are not orthogonal among themselves.

*(c) A proposal for $K^2$ coherent states*

The reader might remember that there was a suggestive choice of reference state in the construction of coherent states for the discrete series - they were the lowest- (or highest-) weight states, for which the corresponding functions considerably simplified. The fact that the continuous series does not terminate in either direction invalidates applying the same criterion. An alternative requirement, which I have used in [65], is to consider those states which minimize the variance of the generators $F^i := (L^3, K^1, K^2)$,

$$\langle\Delta|F^i|\rangle := \langle F^i F_i\rangle - \langle F^i\rangle\,\overline{\langle F_i\rangle} = -s^2 - \frac{1}{4} + |\lambda|^2\,,$$

i.e. those states lying in the circle $|\lambda|^2 = s^2 + \frac{1}{4}$.

In determining matrix coefficients it is enough to consider the maximal subgroup generated by $L^3$. The general expression (A.3.2) has already been computed in [108] for arbitrary complex eigenvalues. The result is a well-defined meromorphic function of $\lambda$ and $\lambda'$, except when $\Delta\lambda := \lambda - \lambda' = 0$ - fortune dictates this is precisely our case of interest. The expression can however easily be regularized by taking[29] $\Gamma(\pm i\Delta\lambda) \mapsto \Gamma(\pm[i\Delta\lambda + \epsilon])$,

[28] Note that there is no orthogonality in $\sigma$.

[29] This regularization was already proposed in the original paper [108].

and we define

$$\langle j,\overline{\lambda},\sigma|D^j(e^{-i\alpha L^3})|j,\lambda',\sigma'\rangle_{\text{reg}} =$$

$$\lim_{\epsilon\to 0}\frac{1}{2\pi}\left\{\frac{\Gamma(\frac{-\bar{j}+\sigma-i\lambda}{2})\Gamma(\frac{-j+\sigma'+i\lambda'}{2})}{\Gamma(\frac{-j+\sigma+i\lambda}{2})\Gamma(\frac{-\bar{j}+\sigma'-i\lambda'}{2})}\Gamma(i\Delta\lambda+\epsilon)\psi_-(\alpha)\right.$$

$$\left.+(-1)^{2\delta}\frac{\Gamma(\frac{-\bar{j}+2\delta+(-1)^{2\delta}\sigma+i\lambda}{2})\Gamma(\frac{-j+2\delta+(-1)^{2\delta}\sigma'-i\lambda'}{2})}{\Gamma(\frac{-j+2\delta+(-1)^{2\delta}\sigma-i\lambda}{2})\Gamma(\frac{-\bar{j}+2\delta+(-1)^{2\delta}\sigma'+i\lambda'}{2})}\Gamma(-i\Delta\lambda-\epsilon)\psi_+(\alpha)\right\}$$

$$\cdot\ \cos\frac{\pi}{2}(i\Delta\lambda+\sigma-\sigma')\,,$$

where $\alpha\in[-\pi,\pi]$ and $f_\pm(\alpha)$ takes the form

$$\psi_\pm(\alpha)=\cos\left(\frac{\alpha}{2}\right)^{-2j-2}\left|2\tan\frac{\alpha}{2}\right|^{\pm i\Delta\lambda}$$

$$\cdot\ {}_2F_1\left(j+1\pm i\lambda, j+1\mp i\lambda', 1\pm i\Delta\lambda; -\tan^2\frac{\alpha}{2}\right)\text{sgn}^{\sigma-\sigma'}\alpha\,.$$

Resorting to a well-known identity for the hypergeometric function ${}_2F_1$ with repeated coefficients, and setting $\lambda=ij$, it is straightforward to see that

$$\begin{aligned}\langle j,\overline{\pm ij},\sigma|D^j(e^{-i\alpha L^3})|j,\pm ij,\sigma\rangle^{\delta=0}_{\text{reg}}\\ &=\lim_{\epsilon\to 0}\frac{\Gamma(\epsilon)+\Gamma(-\epsilon)}{2\pi}\cos\left(\frac{\alpha}{2}\right)^{-2j-2}\left(1+\tan^2\frac{\alpha}{2}\right)^{2j+1}\\ &=-\frac{\gamma}{\pi}\cos^{2j}\frac{\alpha}{2}\,,\end{aligned}\tag{II.4.6}$$

with $\gamma$ the Euler-Mascheroni constant (not to be confused with the Immirzi parameter).

One has to take care in generalizing the discussion to an arbitrary $g\in\text{SU}(1,1)$, since intermediate computation steps may diverge. Our strategy will therefore be to resort to equation (II.4.6) as much as possible. To that end note first that

$$\left[\sum_m e^{-i\alpha m}\langle j,\overline{ij},\sigma|j,m\rangle\langle j,m|j,ij,\sigma\rangle\right]_{\text{reg}}=-\frac{\gamma}{\pi}\cos^{2j}\frac{\alpha}{2}$$

by virtue of completeness of $L^3$ eigenstates $|j,m\rangle$ in $\mathcal{C}^0_j$. Moreover, under the parametrization $g = e^{-i\alpha L^3}e^{-itK^1}e^{-iuK^2}$ (see section A.2 of the appendix) one has

$$\langle j,\overline{ij},\sigma|D^j(g)|j,ij,\sigma\rangle_{\text{reg}}$$
$$= e^{ju}\left[\sum_m e^{-i\alpha m}\langle j,\overline{ij},\sigma|j,m\rangle\langle j,m|D^j(e^{-itK^1})|j,ij,\sigma\rangle\right]_{\text{reg}} .$$

The matrix elements of the subgroup generated by $K^1$ were computed by Lindblad in a subsequent paper [112], and massaging equation (A.3.3) the identity

$$\langle j,m|D^j(e^{-itK^1})|j,ij,\sigma\rangle = \langle j,m|j,ij,\sigma\rangle\,(1+i\sinh t)^{\frac{j+m}{2}}\,(1-i\sinh t)^{\frac{j-m}{2}}$$

can be obtained. Thus, defining $\phi = \arg\left(\cosh\frac{t}{2} + i\sinh\frac{t}{2}\right)$,

$$\langle j,\overline{ij},\sigma|D^j(g)|j,ij,\sigma\rangle_{\text{reg}}$$
$$= \left|\cosh\frac{t}{2} + i\sinh\frac{t}{2}\right|^{2j} e^{ju}\left[\sum_m e^{-im(\alpha-2\phi)}\langle j,\overline{ij},\sigma|j,m\rangle\langle j,m|j,ij,\sigma\rangle\right]_{\text{reg}}$$
$$= -\frac{\gamma}{\pi}\left|\cosh\frac{t}{2} + i\sinh\frac{t}{2}\right|^{2j}\cos^{2j}\left(\frac{\alpha}{2}-\phi\right)e^{ju} .$$

One can see at once that $D^{j\dagger}(g) = D^j(g^{-1})$, and a rewriting of the last line of the previous equation guarantees the principal result of this section[30], namely

$$\langle j,\overline{ij},\sigma|D^{j\dagger}(g')D^j(g)|j,ij,\sigma\rangle_{\text{reg}} = -\frac{\gamma}{\pi}[g'\cdot l^+|g\cdot l^-]^{2j} , \qquad \text{(II.4.7)}$$

with respect to which we may formally write $|j,ij,\sigma\rangle \sim |l^-]^{2j}$.

[30] The reader may wonder whether the asymmetry of equation (II.4.7) justifies considering the alternative pairing, where the $ij$ eigenstate appears on the left. This matter is discussed in [100], but omitted here for simplicity.

A comment on the regularization proposal leading to equation (II.4.7) is due. It is undeniable that regularizing a diverging object is less of a science than it is an art: the procedure is not unique, and strictly speaking one can only say with certainty that without it the matrix coefficients are undefined. Still, the circumstance that this specific regularization of generalized eigenstates is particularly simple, and that it leads - as per (II.4.7) - to the same qualitative behavior as that of the discrete series coefficients (II.4.4) (and indeed the SU$(2)$ ones) lends credence to the choice made. There is a parallel to be made with the well-known Feynman $i\epsilon$ regularization of the eigenfunctions of the momentum operator in QFT, ultimately legitimized by the empirical success of the theory. For what concerns spin-foams, in the absence of even the prospect of a similar experimental verification in the near future, an appeal to plausibility will have to suffice.

## II.5 A new model for Lorentzian 3d quantum gravity

As stated at the beginning of this chapter, much of the motivation behind the construction of continuous series coherent states had to do with clarifying the difficulties plaguing the EPRL model with time-like triangles. Finding ourselves in the possession of such states with analogous properties to those ones used for other types of boundaries, we can now make use of them in a 3-dimensional toy model as a first approximation to the 4-dimensional theory. The obstacles found in the asymptotic analysis of the vertex amplitude with time-like triangles [64, 65] will reappear in a simpler form, consequently clarifying their causes and possible solutions.

### II.5.1 The vertex amplitude

Three basic ingredients are needed in defining the model. First observe that there exists a Plancherel formula for smooth compactly-supported functions on SU$(1,1)$ (A.4.1), leading to a Fourier decomposition of the

Dirac delta distribution as

$$\delta(g) = \sum_{\delta=0,\frac{1}{2}} \int_{-\infty}^{\infty} \mathrm{d}s\, s \tanh^{1-4\delta}(\pi s)\, \mathrm{Tr}\left[D^{j(\delta)}(g)\right] + \sum_{q=\pm} \sum_{2k=-1}^{-\infty} (-2k-1)\mathrm{Tr}\left[D^{k(q)}(g)\right] ,$$

where the notation and conventions used throughout the chapter have been kept. Secondly recall that - as a purely topological theory - the tetradic action for $2+1$ gravity agrees with the 3d version of unconstrained $BF$ theory,

$$S[\theta, A] = \int_M (\star\theta)_{IJ} F^{IJ} \;\leftrightarrow\; S[B, A] = \int_M B_{IJ} F^{IJ} ,$$

where $I = 0, 1, 2$, $\dim M = 3$, and the gauge group is taken to be the spin group of isometries of 3d Minkowski space, i.e. $\mathrm{SU}(1,1)$. The same reasoning as the one leading to equation (I.2.5) then results in the *formal* partition function

$$\begin{aligned} Z(\Delta^*) = \sum_{\delta\to f} \int_{j\to f} \mathrm{d}s \prod_f \left[s_f \tanh^{1-4\delta_f}(\pi s_f)\right] \\ \cdot \mathrm{Tr}_f \left[\prod_e \Big( \int \mathrm{d}g_e \prod_{f \text{ s.t. } e\in\partial f} D^{j_f(\delta_f)}(g_e)\Big)\right] \\ + \sum_{q\to f} \sum_{k\to f} \prod_f (-2k_f - 1)\, \mathrm{Tr}_f \left[\prod_e \left( \int \mathrm{d}g_e \prod_{f \text{ s.t. } e\in\partial f} D^{k_f(q_f)}(g_e)\right)\right] ,\end{aligned}$$

with a discretization based on a dual 2-complex as in table II.2; any concerns regarding convergence or pragmatics are for now to be boldly ignored (it is worthile to mention at this stage that this partition function has previosuly made an appearence in the literature [113, 114]).

Table II.2: Cells of a 2-complex dual to a 3d triangulation

| 2-complex $\Delta^*$ | triangulation $\Delta$ |
|---|---|
| vertex $v$ | tetrahedron $\tau$ |
| edge $e$ | triangle $t$ |
| face $f$ | triangle-edge $\epsilon$ |

Thirdly, it is necessary to pick a basis with which to take traces, and it is at this stage that a novelty is introduced. Since the various matrix coefficients are orthonormal among themselves (A.4.2), the coherent states of the previous sections constitute orthonormal bases for their respective Hilbert spaces. Indeed for the discrete series $\mathcal{D}_k^q$ one has the completeness relation

$$\int \mathrm{d}g\; D^{k(q)}(g)|k,-qk\rangle\langle k,-qk|D^{k(q)\dagger}(g) = \frac{\mathbb{1}_{k(q)}}{-2k-1}\,,$$

while for the continuous series $\mathcal{C}_j^\delta$ it holds that[31]

$$\int \mathrm{d}g\; D^{j(\delta)}(g)|j,\lambda,\sigma\rangle\langle j',\overline{\lambda}',\sigma|D^{j(\delta)\dagger}(g) = \frac{\mathbb{1}_{j(\delta)}\delta(j-j')\delta(\lambda-\lambda')}{s\tanh^{1-4\delta}\pi s}\,. \tag{II.5.1}$$

Accordingly, set $|qk,g\rangle := D^{k(q)}(g)|k,-qk\rangle$, $|j,g\rangle := D^{j(0)}(g)|j,ij,0\rangle$ and $|j,g] := D^{j(0)}(g)|j,\overline{ij},0\rangle$ for the remainder of the chapter. Introduce the diagram

$$ {}^{n}\;\cdots\;{}^{n'} := d_k\,\langle qk,n|D^{k(q)}(g)|qk,n'\rangle\;,\quad d_k = -2k-1\,, \tag{II.5.2}$$

for a pairing of discrete series coherent states. For a continuous series pairing, let

$$ \cdots\;{}^{n'} := d_s\,\mathcal{C}_{n,gn'}\;[j,n|D^{j(0)}(g)|j,n'\rangle\,,\quad d_s = s\tanh\pi s\,. \tag{II.5.3}$$

[31] The two Dirac deltas in equation (II.5.1) reflect the facts that 1) the matrix coefficients of the continuous series are not square-integrable [107], and 2) the states $|j,\lambda,\sigma\rangle$ are only *generalized* eigenstates.

The term $\mathcal{C}_{n,n'}$ is a function of the boundary data

$$\mathcal{C}_{n,n'} := e^{s[l_n^+|l_{n'}^+]^2}\,, \quad |l_n^\pm] := n|l^\pm]\,,$$

which I have appended to the continuous series pairings ex post facto. Its inclusion will prove fundamental in guaranteeing a well-behaved semi-classical limit, as it corresponds to an otherwise absent Gaussian implementation of a gluing constraint between the edges $n$ and $n'$. I will come back to this point in the context of asymptotic analysis.

The amplitude I wish to propose now follows. Pick 12 group elements $n_{ab} \in \mathrm{SU}(1,1)$ and 6 spin labels $j_{ab} = j_{ba}$ or $k_{ab} = k_{ba}$ (and in that case also $q_{ab} = q_{ba}$) as boundary data, one for each edge. The vertex amplitude is constructed from a convolution of the diagrams (II.5.2) and (II.5.3), following the combinatorics of a tetrahedron. For example, the amplitude for a tetrahedron with all edges time-like is given by

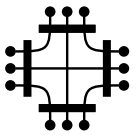

$$= \int \prod_{a=1}^{3} \mathrm{d}g_a \prod_{a<b} d_{k_{ab}} \cdot \langle qk_{ab}, n_{ab}| D^{k_{ab}(q)_{ab}}(g_a)^\dagger D^{k_{ab}(q)_{ab}}(g_b) |qk_{ab}, n_{ba}\rangle$$
$$= \int \prod_{a=1}^{3} \mathrm{d}g_a \prod_{a<b} d_{k_{ab}} [g_a \cdot (-q)_{ab} | g_b \cdot (-q)_{ba}]^{2k_{ab}}\,.$$

where $|\pm_{ab}] := n_{ab}|\pm]$. The amplitude for a tetrahedron with all edges space-like, reads

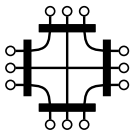

$$= \int \prod_{a=1}^{3} \mathrm{d}g_a \prod_{a<b} d_{s_{ab}} \mathcal{C}_{g_a n_{ab}, g_b n_{ba}} \cdot [j_{ab}, n_{ab}| D^{j_{ab}(0)}(g_a)^\dagger D^{j_{ab}(0)}(g_b) |j_{ab}, n_{ba}\rangle$$
$$= \int \prod_{a=1}^{3} \mathrm{d}g_a \prod_{a<b} \left(-\frac{\gamma}{\pi} d_{s_{ab}} e^{s_{ab}[g_a \cdot l_{ab}^+ | g_b \cdot l_{ba}^+]^2}\right) [g_a \cdot l_{ab}^+ | g_b \cdot l_{ba}^-]^{-1+2is_{ab}}\,. \tag{II.5.4}$$

A general vertex with space- and time-like edges involves combinations of all types of pairings. On all amplitudes we set $g_4 = \mathbb{1}$ in order to regularize the Haar integral, as usual. A number of remarks on the vertex amplitude as defined above are necessary:

1. The correlation between causal character and representation series is inverted relative to the 4-dimensional case: in the latter the vertex amplitude of a space-like boundary makes use of the discrete series, and the time-like one uses the continuous series [54]; this is because for us the boundary state $|j_{ab}, n_{ab}\rangle$ turns out to be asymptotically *parallel* to a triangle edge (as shall soon be confirmed), while in 4 dimensions the same data would label the *normal* vector to a triangle.

2. The space-like model is defined only for $\delta = 0$, which is necessary for (II.4.7) to hold. While in the time-like model the sign $q$ can be shown to index the upper and lower hyperboloid, the label $\delta$ is immaterial for vectors in the one-sheeted hyperboloid, so I take it to be geometrically inconsequential.

3. The Dirac distributions appearing in the completeness relation (II.5.1) have been discarded in the definition of (II.5.4), but strictly speaking they would have to be considered when using the completeness relation in a complex with more than one vertex. How this should be done is unclear, and it is by itself an interesting question for further study.

4. Although the time-like amplitude is straightforward and likely well-known, the space-like amplitude introduces two novelties: the usage of $|j, ij, 0\rangle$ coherent states satisfying (II.4.7) and the ad-hoc inclusion of the function $\mathcal{C}$. The chosen notation is supposed to be suggestive, and indeed $\mathcal{C}$ is intended as a Gaussian constraint on the amplitude - some form of additional constraint turns out to be necessary to recover geometry in the asymptotic regime, as will be argued in what follows, and $\mathcal{C}$ plays that role in a particularly simple manner.

### II.5.2 Asymptotic analysis of the space-like model

The proposal of (II.5.4) as a vertex amplitude for space-like boundaries is predicated on its formal similarity to its time-like counterpart, as well as its desirable asymptotic behavior. To see how we shall follow protocol [53,63–65] and resort to a stationary phase approximation of the amplitude for large spins $s$; the reader is referred to Hörmander's theorem in chapter III for details on the procedure.

The vertex amplitude with uniformly scaled spins $\Lambda s_{ab}$ can be rewritten as an exponential integral,

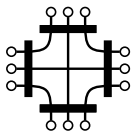

$$= \left(\frac{\gamma\Lambda}{\pi}\right)^6 \int \prod_{a=1}^{3} \mathrm{d}g_a \prod_{a<b} \frac{s_{ab}\tanh(\pi\Lambda s_{ab})}{\langle l^+_{ab}|g^\dagger_a\sigma_3 g_b|l^-_{ba}\rangle} e^{\Lambda S_{ab}} ,$$

with an action given by

$$S_{ab} = 2is_{ab}\ln\langle l^+_{ab}|g^\dagger_a\sigma_3 g_b|l^-_{ba}\rangle + s_{ab}\,\langle l^+_{ab}|g^\dagger_a\sigma_3 g_b|l^+_{ba}\rangle^2 .$$

When $\Lambda \to \infty$ the integral is dominated by stationary contributions $\delta_g S_{ab} = 0$ with maximal real part. Observe to that end that $[g_a \cdot l^+_{ab}|g_b \cdot l^-_{ba}] \in \mathbb{R}$, so that the maximum of $\Re S_{ab}$ is attained at

$$\Re S_{ab} = 0 \quad \Leftrightarrow \quad \langle l^+_{ab}|g^\dagger_a\sigma_3 g_b|l^+_{ba}\rangle = 0 \;\wedge\; \langle l^+_{ab}|g^\dagger_a\sigma_3 g_b|l^-_{ba}\rangle > 0 ,$$

which implies $g_b|l^+_{ba}\rangle = \vartheta_{ab} g_a|l^+_{ab}\rangle$; acting with the real structure $R$ then shows that $\vartheta_{ab} \in \mathbb{R}$, and the second equation above enforces $\vartheta_{ab} > 0$. To find the stationary phase we pick adapted coordinates in the group manifold (see section III.2.1), i.e. coordinates $x_I$ for which

$$\partial_I g = \frac{i}{2}\varsigma_I g , \quad \mathrm{d}g = (4\pi)^{-2}\,\mathrm{d}x^1 \wedge \mathrm{d}x^2 \wedge \mathrm{d}x^3 ;$$

it then follows that

$$\partial_I^{(b)} \sum_{a<b} S_{ab} = 0 \quad \Leftrightarrow \quad \sum_{a|a\neq b} s_{ab}\epsilon_{ab} \left[ \frac{\langle l_{ab}^+ | g_a^\dagger \sigma_3 \varsigma_I g_b | l_{ba}^- \rangle}{\langle l_{ab}^+ | g_a^\dagger \sigma_3 g_b | l_{ba}^- \rangle} \right.$$
$$\left. + \, i \, \langle l_{ab}^+ | g_a^\dagger \sigma_3 g_b | l_{ba}^+ \rangle \, \langle l_{ab}^+ | g_a^\dagger \sigma_3 \varsigma_I g_b | l_{ba}^+ \rangle \right] = 0 \, .$$

The symbol $\epsilon_{ab}$ above stands for a sign depending on the assumed orientation of the pairings in the amplitude. With the conventions of equation (II.5.4), $\epsilon_{ab}$ is negative whenever $a > b$. Note that by symmetry the derivatives with respect to the parameters of $g_a$ are redundant. Taken together, the stationary and reality conditions may be written in the simple form

$$\Re S_{ab} = 0 \quad \Leftrightarrow \quad g_b | l_{ba}^+ \rangle = \vartheta_{ab} g_a | l_{ab}^+ \rangle \, , \tag{II.5.5}$$

$$\forall b \, , \; \partial_I^{(b)} \sum_{a|a\neq b} S_{ab} = 0 \quad \Leftrightarrow \quad \forall b \, , \; \sum_{a|a\neq b} s_{ab} \, \langle l_{ba}^+ | g_b^\dagger \sigma_3 \varsigma^I g_b | l_{ba}^- \rangle = 0 \, , \tag{II.5.6}$$

which the informed reader may already recognize as the gluing and closure conditions, respectively. The need for the constraint $\mathcal{C}$ should now be clear: since the object $[g_a \cdot l_{ab}^+ | g_b \cdot l_{ba}^-]$ is always real by virtue of the structure of $\mathrm{SU}(1,1)$, in the absence of $\mathcal{C}$ there would be no real part of the action to maximize, and therefore no gluing condition (II.5.5).

Regarding closure (II.5.6), the spin homomorphism $\pi : \mathrm{SU}(1,1) \to \mathrm{SO}_0(1,2)$ allows us to write (with indices contracted according to $\eta$ and $P_{1,2} = \mathrm{diag}(1,-1,-1)$; cf. chapter III)

$$\sum_{a|a\neq b} s_{ab} \, \langle l_{ba}^+ | g_b^\dagger \sigma_3 \varsigma^I g_b | l_{ba}^- \rangle = \sum_{a|a\neq b} \sum_{\mu=0}^{2} s_{ab} \left[ P_{1,2} \pi(n_{ba}^\dagger g_b^\dagger) \right]_\mu^{\;I} \langle l_{ba}^+ | \sigma^\mu | l_{ba}^- \rangle$$
$$= -i \sum_{a|a\neq b} s_{ab} \left[ \pi(g_b n_{ba}) \hat{e}_2 \right]^I \, ,$$

such that equation (II.5.6) becomes

$$\forall b\,, \sum_{a|a\neq b} s_{ab}\left[\pi(n_{ba})\hat{e}_2\right] = 0\,. \tag{II.5.7}$$

Since $\mathrm{SO}(1,2)$ acts transitively on the one-sheeted space-like hyperboloid $H^{\mathrm{sl}}$ and $\hat{e}_2 \in H^{\mathrm{sl}}$, the vectors $v_{ba} := \pi(n_{ba})\hat{e}_2$ are all elements of $H^{\mathrm{sl}}$, and the whole hyperboloid is covered by all possible such vectors. Turning to the gluing condition, note first that (II.5.5) also implies $g_b|l^-_{ba}\rangle = \vartheta^{-1}_{ab}\, g_a|l^-_{ab}\rangle$, from where

$$g_b|l^+_{ba}\rangle\!\langle l^-_{ba}|g_b^\dagger = g_a|l^+_{ab}\rangle\!\langle l^-_{ab}|g_a^\dagger \quad \Leftrightarrow \quad \pi(g_b)v_{ba} = \pi(g_a)v_{ab}\,, \tag{II.5.8}$$

through the same reasoning as above.

Equations (II.5.8) and (II.5.7) are well-known in the asymptotic analysis of spin-foam models, and they afford a geometrical interpretation for the dominant configurations in the vertex amplitude. Minkowski's theorem on convex polyhedra (see appendix D) guarantees that, for all $v_{a(b)}$ not colinear (fixing $b$), equation (II.5.7) holds if and only if there exists (up to rigid body motions) a triangle with edge vectors $v_{a(b)}$ and edge lengths $s_{a(b)}$. The vertex amplitude is thus suppressed if the boundary data is not in correspondence with four geometrical triangles at the boundary. In its turn the gluing equation (II.5.8) dictates that the amplitude is dominated by configurations in which the triangles are $\mathrm{SO}(1,2)$-rotated into coinciding edges. I show in appendix E that the gluing equations admit two kinds of solutions when the boundary data so allows: one either recovers a degenerate tetrahedron (i.e. a triangle) or a proper tetrahedron (and its reflected counterpart). It is in this sense that one can make the claim that the space-like vertex amplitude induces geometricity in the semiclassical regime.

Tying it all up, we can apply Theorem III.1 of chapter III and obtain an explicit expression for the asymptotic amplitude when the boundary data is that of a tetrahedron with space-like edges, reading

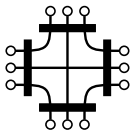

$$
= e^{i\frac{\pi}{4}} \frac{\Lambda^{\frac{3}{2}}\gamma^6}{(2\pi)^{\frac{15}{2}}} \left[\prod_{a<b} s_{ab} \tanh(\pi\Lambda s_{ab})\right] \cdot \left(\frac{1}{H_{\mathbb{1}}^{1/2}} + \frac{e^{\sum_{a\neq b}(-1+2i\Lambda s_{ab})\theta_{ab}}}{H_{\theta}^{1/2}}\right) + \mathcal{O}\left(\Lambda^{\frac{1}{2}}\right) ,
$$

which is valid if the Hessian $H$ of the action is non-singular at the critical points $\mathbb{1}, \theta$ (see appendix E for the explicit critical points). The parameter $\theta_{ab} := \ln \vartheta_{ab}$ is the external dihedral angle between faces $a, b$, with the latter defined such that its sign agrees with equation (E.4),

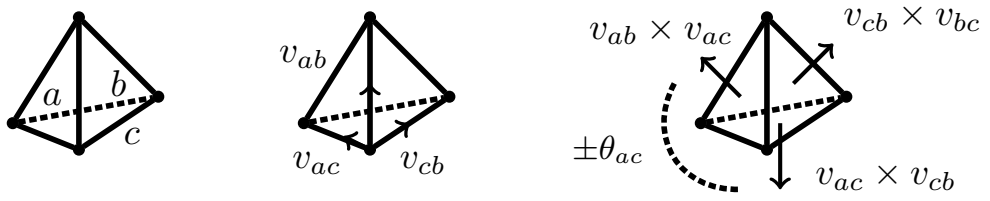

.

I would like to add a few remarks on closing this chapter.

1. Reiterating an earlier observation, the inclusion of the constraint $\mathcal{C}_{n,n'} = e^{s[l_n^+|l_{n'}^+]^2}$ is unjustified in the context of the usual spin-foam prescription: its presence does not follow from a direct manipulation of the path integral for 3d quantum gravity. Rather surprisingly, however, had we not defined the vertex amplitude with this additional constraint the stationary point equations would not be sufficient to determine isolated critical points - which would *strictly* invalidate the application of Hörmander's theorem, and *heuristically* lead to dominant configurations not required to coherently glue. This state of affairs is ultimately a consequence of the mapping from $\mathrm{SU}(1,1)$ to $H^{\mathrm{sl}}$ involving matrix coefficients of the type $[l^+|g|l^-]$, which are always real (while for the $\mathrm{SU}(2)$ model or the time-like $\mathrm{SU}(1,1)$ one the analogous objects are complex). Moreover it seems unlikely for the lack of sufficient constraints to be a consequence of our regularization prescription (II.4.7), given that any sensible model should always recover a $[l^+|g|l^-]$ structure (as does the 4d EPRL-CH model, even though it uses different boundary states), at the very least due to the discussion of section II.2.2.(c). On the other hand there does not seem to be any physical reason to expect that space-like boundary data should be less constrained. Just why does the space-like amplitude exhibit this odd behavior remains unclear.

2. The expected Regge action [115] for a space-like tetrahedron $iS = i\sum_{a\neq b} s_{ab}\theta_{ab}$ figures in the asymptotic formula, but so does an additional imaginary term $i\mathfrak{I} := -\frac{1}{2}\sum_{a\neq b}\theta_{ab}$. The presence of $\mathfrak{I}$ is another peculiarity of the model, and it can be traced back to the real part of the complex spin $j = -1/2 + is$. The most immediate interpretation is to assume the model describes a tetrahedron with complex lengths given by $-ij = s + i/2$; however odd, this would be in agreement with the space-like area spectrum (now understood as a length), which reads $A^2 = -(s^2 + 1/4)$ according to [54] and to this very chapter. The imaginary part of such lengths would be fixed, and small compared to the spins in the regime where gravity is expected to be recovered. Whether the presence of a small imaginary component on space-like lengths leads to interesting consequences is a compelling question for a future time. The second possibility is to interpret $\mathfrak{I}$ as part of the amplitude's measure. Since the measure already depends on spins and angles via the Hessian determinant, doing so should simply be a matter of convention. The exact same phenomenon is already present in the time-like CH amplitude [64, 65].

3. It will be shown in chapter III that the Conrady-Hnybida model for the 4d vertex amplitude with time-like triangles is similarly unconstrained as is the 3d space-like model proposed in this chapter.

This suggests that the CH amplitude might itself benefit from additionally imposed constraints. The manner in which this is to be done deserves further investigation.

4. The explicit formulas for both the amplitude and its asymptotics allow for numerically studying configurations involving space- and time-like boundaries, a research direction which has remained unexplored due to the obstacles of the CH time-like amplitude. This opens the door to e.g. 1) comparative studies between the spin-foam framework and that of Causal Dynamical Triangulations [22], potentially bridging the two approaches; and 2) explorations of cosmological scenarios requiring both space- and time-like regions, as is the case for the FRW universe.

5. Earlier $(2+1)$ state-sums of the Ponzano-Regge type (in that the boundary data consists solely of spin labels) have been constructed; the associated amplitudes correspond to tetrahedra with entirely time-like [116, 117] or entirely space-like edges [118]. The model here described extends these proposals to mixed edges, and the coherent-state formulation allows for a straightforward generalization to higher-valent polyhedra.

6. The Majorana spinors discussed in the first part of the present chapter provide an alternative description of the $\mathrm{SU}(1,1)$ phase space. It remains to elevate the construction to the level of twistors and $T^*\mathrm{SL}(2,\mathbb{C})$ in the style of [98], which would then hopefully reduce to $T^*\mathrm{SU}(1,1)$ under the weight of appropriately-defined simplicity constraints; to do so constitutes an interesting research avenue.

# III. Asymptotics of 4d Spin-foams with Time-like Polygons

*'Well, I mean, yes idealism, yes the dignity of pure research, yes the pursuit of truth in all its forms, but there comes a point I'm afraid where you begin to suspect that the entire multidimensional infinity of the Universe is almost certainly being run by a bunch of maniacs.'*

— Frankie Mouse, in The Hitchhiker's Guide to the Galaxy (Douglas Adams, 1979)

It is now time to leave the relatively comfortable domain of 3-dimensional gravity and turn to the much less-trivial problem of proper 4-dimensional Lorentzian gravity, which we believe to describe nature. Such an increment in subject matter inevitably demands a proportional decrease in control - and to some extent rigor - of the relevant objects of the spin-foam description. The results of the following two chapters are therefore less strict than the ones preceding them, but perhaps no less relevant if they are taken as a tentative initial exploration of the problems they tackle.

As to what concerns the present chapter, the problem at hand is to study the semiclassical limit of the Conrady-Hnybida model, succinctly reviewed in section I.2.3.(d), with a particular emphasis on the case in which some of the boundary faces are time-like. That this sector of the model is particularly problematic had already been observed by Han and Liu in [64], who were the first to analyze its asymptotics. In the wake of their results Steinhaus and I believed their conclusions could be pushed further [65]: by generalizing to other boundary faces beyond triangles, by making a more judicious choice of boundary states leading to a better-behaved amplitude, by relaxing some assumptions, and finally by pointing out the existence of previously disregarded additional critical configurations. I shall now report on our conclusions, with the caveat that there are differences of convention relative to the original paper.

## III.1 Definition of the model

*(a) Vertex amplitude*

We consider an EPRL-CH-type vertex amplitude generalized to arbitrary polytopes. Given such a 4d polytope, its dual graph is composed of a single vertex and of as many edges as there are boundary polyhedra. The standard example is given by the graph dual to a 4-simplex 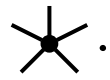. Label as before each ordered pair of graph edges by a pair of ordered letters $ab$. To each such polyhedral face $ab$ a boundary state $\psi_{ab}$ is assigned as follows:

1. Space-like faces $ab$ are labeled by $\rho_{ab} = \gamma\nu_{ab}$. They may be cells of space- or time-like polyhedra;

   - If the polyhedron is space-like, the face is characterized by an element of $S^2$. Set $\nu_{ab} = 2j_{ab}$ and $\psi_{ab} = \mathcal{I}^\chi \circ D^j(n_{ab})|j_{ab}, j_{ab}\rangle$, for $n_{ab} \in \mathrm{SU}(2)$. $\mathcal{I}^\chi$ denotes the mapping of subgroup representations to principal series representations $\mathcal{D}_\chi$ as described in section B.2 of the appendix. $j$ labels the unitary irreducible representations of $\mathrm{SU}(2)$.

   - If the polyhedron is time-like (orthogonal to a canonical normal $\hat{e}_3$) the face is characterized by an element of the future- or past-pointing hyperboloid $H^\pm$. Set $\nu_{ab} = -2k_{ab}$ and $\psi_{ab} = \mathcal{I}^\chi \circ D^{k_{ab}(\tau_{ab})}(n_{ab})|k_{ab}, -\tau_{ab}k_{ab}\rangle$, for $n_{ab} \in \mathrm{SU}(1,1)$ and $\tau_{ab} = \pm$ labeling the hyperboloid. $k$ labels the discrete series $\mathcal{D}^q_k$ of $\mathrm{SU}(1,1)$ representations.

2. Time-like faces $ab$ are necessarily cells of time-like polyhedra, and they are labeled by $\nu_{ab} = -\gamma\rho_{ab}$;

- The polygonal face is characterized by an element of $H^{\text{sl}}$. Set $\rho_{ab} = 2s_{ab}$ and $\psi_{ab} = \mathcal{I}^{\chi} \circ D^{j_{ab}(0)}(n_{ab})|j_{ab}, ij_{ab}, 0\rangle$, for $n_{ab} \in \text{SU}(1,1)$. There is a costate reading $\tilde{\psi}_{ba} = \mathcal{I}^{\chi} \circ D^{j_{ba}(0)}(n_{ba})|j_{ba}, \overline{ij}_{ba}, 0\rangle$. $j$ labels the continuous series $\mathcal{C}_j^{\delta}$ of $\text{SU}(1,1)$ representations.

Among the above prescriptions, the restriction on representations and the definition of boundary states associated to space-like polyhedra is exactly as originally discussed in [54]. The boundary states for time-like polyhedra are a new proposal, as is taking the continuous label to be $\rho = 2s$ (rather than $\rho = -2s$); the latter simplifies both the boundary states and their analysis, and it should be inconsequential for end-results given that $(\chi, -\chi)$ constitute equivalent representations [42]. The vertex amplitude for the model then reads

$$\mathcal{A} = \int_{\text{SL}(2,\mathbb{C})} \prod_{a=1}^{n} \mathrm{d}g_a \delta(g_n) \prod_{a<b} \langle D^{\chi_{ab}}(g_a)\psi_{ab}, D^{\chi_{ab}}(g_b)\psi_{ba}\rangle \ , \tag{III.1.1}$$

where $n$ denotes the number of boundary polyhedra to the polytope, and $\delta(g_n)$ is included for regularization as usual. The bracket $\langle\cdot,\cdot\rangle$ denotes the inner product in $\mathcal{D}_{\chi}$ as in equation (B.1.1). $\chi_{ab}$ and $\psi_{ab}$ stand for the labels and states discussed above, for a given choice of causal character and boundary data. Observe that the structure of the amplitude is that of a product of pairings of boundary states, each of which is either of the achronal, heterochronal, parachronal or orthochronal type[32].

[32] I remind the reader that the terminology is as follows: achronal for space-like interfaces of space-like polyhedra; parachronal for time-like interfaces of time-like polyhedra; orthochronal for space-like interfaces of time-like polyhedra; an heterochronal, for space-like interfaces of space- and time-like polyhedra.

*(b) Boundary states*

We shall need explicit expressions for the $\mathrm{SL}(2,\mathbb{C})$ coherent states $\psi_{ab}$ as homogeneous functions. To that end consider first the $\mathbb{C}^2$ inner products

$$\langle u|v\rangle := u^\dagger v\,, \quad [u|v] := u^\dagger \sigma_3 v\,,$$

which are invariant under the natural action of $\mathrm{SU}(2)$ and $\mathrm{SU}(1,1)$ respectively. We keep the convention that $|v\rangle = |v]$ and $[v| = \langle v|\sigma_3$, and define as before

$$|+\rangle := \left(\begin{smallmatrix}1\\0\end{smallmatrix}\right), \quad |-\rangle := \left(\begin{smallmatrix}0\\1\end{smallmatrix}\right), \quad |l^\pm\rangle := \frac{1}{\sqrt{2}}(|+\rangle \pm |-\rangle)\,, \quad |z\rangle := \left(\begin{smallmatrix}z_1\\z_2\end{smallmatrix}\right).$$

The manner in which states of unitary irreducible representations of $\mathrm{SU}(2)$ and $\mathrm{SU}(1,1)$ map to states in $\mathcal{D}_\chi$ is discussed at length in section B.2 of the appendix. Starting with the $\mathrm{SU}(2)$ maximal-weight state $\mathcal{I}^\chi|j,j\rangle$, equation (B.2.4) together with the explicit expression for the group's matrix elements found in [53][33] shows

$$F^\chi_{j,j}(z) = (2j+1)^{\frac12}\left(\frac{\Gamma(j+\frac{\nu}{2}+1)\Gamma(j-\frac{\nu}{2}+1)}{\Gamma(2j+1)}\right)^{\frac12} \cdot \langle z|z\rangle^{i\frac{\rho}{2}-1-j}\,\langle z|+\rangle^{j+\nu/2}\,(-\langle -|z\rangle)^{j-\nu/2}\,.$$

Turning to the $\mathrm{SU}(1,1)$ subgroup and the discrete series of representations $\mathcal{D}^q_k$, equations (B.2.8) and (A.3.1) give for $\mathcal{I}^\chi|k,-qk\rangle$ with $q=\tau$ that

$$F^{\chi,\tau}_{k,-\tau k}(z) = (-2k-1)^{\frac12}\,\frac{\tau^{k+\frac{\nu}{2}}}{(\frac{\nu}{2}+k)!}\left(\frac{\Gamma(\frac{\nu}{2}+k+1)\Gamma(\frac{\nu}{2}-k)}{\Gamma(-2k)}\right)^{\frac12} \cdot \Theta\left(\tau[z|z]\right)\left(\tau[z|z]\right)^{i\frac{\rho}{2}-1-k}\,[\tau|z]^{k-\frac{\nu}{2}}\,[z|-\tau]^{k+\frac{\nu}{2}}\,, \quad \frac{\nu}{2} \geq -k\,,$$

[33] Using a convention where the representation acts on the left by $g^T$, the formula reads

$$D^j_{mn}(g) = \left(\frac{\Gamma(j+m+1)\Gamma(j-m+1)}{\Gamma(j+n+1)\Gamma(j+m+1)}\right)^{\frac12} \cdot \sum_l \binom{j+n}{l}\binom{j-n}{j+m-l} g_{11}^l g_{12}^{j+m-l} g_{21}^{j+n-l} g_{22}^{l-m-n}\,.$$

and it will be enough to consider $\frac{\nu}{2} \geq -k$.

The homogeneous functions associated with the continuous series $\mathcal{C}_j^\delta$ of $\mathrm{SU}(1,1)$ representations require a lengthier discussion. One obtains first for the matrix coefficients (A.3.3) with $|j, ij, 0\rangle$ that

$$D^{j(0)}_{m,ij,0}(g) = \frac{\Gamma(0)}{-2\sin\pi j}\left(\frac{1}{\Gamma(-m-j)\Gamma(m-\overline{j})} + [m \to -m]\right) \cdot \left(\frac{\Gamma(m-j)}{\Gamma(m-\overline{j})}\right)^{\frac{1}{2}} (g_{11}-g_{12})^{j+m}(\overline{g}_{11}-\overline{g}_{12})^{j-m}\,.$$

It should not be surprising that a $\Gamma(0)$ divergence appears in the equation above, since a similar behavior was observed (and dealt with) in section II.4.2.(c); as discussed then, that such a divergence may occur is a consequence of employing generalized eigenstates. Note also that the regularization procedure proposed in chapter II cannot be directly applied, since the mapping to $\mathcal{D}_\chi$ requires mixed coefficients $D^j_{m\lambda}$ as per equation (B.2.9). In the absence (at this point in time) of a better approach, we shall simply factor the diverging term in all following calculations. With regard to the costate $|j, \overline{ij}, 0\rangle$, we work around the presence of an hypergeometric function in its expression as follows: observe first that $\overline{ij} = ij + i$, and furthermore that there exists [108] a ladder operator acting as

$$K^+ = L^3 + K^1\,, \quad K^+|j,\lambda,\sigma\rangle := i(\overline{j}+i\lambda)|j,\lambda+i,(\sigma+1)_{\mathrm{mod}\,2}\rangle\,;$$

one can therefore argue

$$\begin{aligned}
&\langle j,m|D^{j(0)}(g)|j,\overline{ij},0\rangle \\
&\quad = \frac{i}{2j+1}\langle j,m|D^{j(0)}(g)(D'^{j(0)}(L^3)+D'^{j(0)}(K^1))|j,ij,1\rangle \\
&\quad = \frac{1}{2j+1}\frac{\mathrm{d}}{\mathrm{d}t}\bigg|_{t=0}\left[D^{j(0)}_{m,ij,1}(ge^{itL^3}) + D^{j(0)}_{m,ij,1}(ge^{itK^1})\right]\,,
\end{aligned}$$

and conclude that

$$D^{j(0)}_{m,\overline{ij},0}(g) = \frac{\Gamma(0)}{2i\sin\frac{\pi}{2}(1-2j)}\left(\frac{1}{\Gamma(-m-j)\Gamma(m-\bar{j})} - [m\to -m]\right)$$
$$\cdot\left(\frac{\Gamma(m-j)}{\Gamma(m-\bar{j})}\right)^{\frac{1}{2}}\frac{i}{2j+1}\left[(j+m)\frac{g_{11}+g_{12}}{g_{11}-g_{12}} - (j-m)\frac{\overline{g}_{11}+\overline{g}_{12}}{\overline{g}_{11}-\overline{g}_{12}}\right]$$
$$\cdot\,(g_{11}-g_{12})^{j+m}(\overline{g}_{11}-\overline{g}_{12})^{j-m}\,.$$

Equation (B.2.9) can now be used to obtain the functions in $\mathcal{D}_\chi$, which read

$$F^{\chi,\tau}_{j,0,ij,0}(z) = A^j_{\frac{\tau\nu}{2}}\,\Theta\left(\tau[z|z]\right)\left(\tau[z|z]\right)^{i\rho/2-1-j}\left(\tau[z|l^-]\right)^{j+\frac{\nu}{2}}\left(\tau[l^-|z]\right)^{j-\frac{\nu}{2}}\,,$$

$$F^{\chi,\tau}_{j,0,\overline{ij},0}(z) = \tilde{A}^j_{\frac{\tau\nu}{2}}\,\Theta\left(\tau[z|z]\right)\left(\tau[z|z]\right)^{i\rho/2-1-j}\left(\tau[z|l^-]\right)^{j+\frac{\nu}{2}}\left(\tau[l^-|z]\right)^{j-\frac{\nu}{2}}$$
$$\cdot\left[\left(j+\frac{\nu}{2}\right)\frac{[z|l^+]}{[z|l^-]} - \left(j-\frac{\nu}{2}\right)\frac{[l^+|z]}{[l^-|z]}\right]\,,$$

with coefficients

$$A^j_m := (s\tanh\pi s)^{1/2}\left(\frac{\Gamma(m-j)}{\Gamma(m-\bar{j})}\right)^{\frac{1}{2}}\frac{2^j\Gamma(0)}{-2\sin\pi j}$$
$$\cdot\left(\frac{1}{\Gamma(-m-j)\Gamma(m-\bar{j})} + [m\to -m]\right)\,,$$

$$\tilde{A}^j_m := \frac{i\,(s\tanh\pi s)^{1/2}}{2j+1}\left(\frac{\Gamma(m-j)}{\Gamma(m-\bar{j})}\right)^{\frac{1}{2}}\frac{2^j\Gamma(0)}{2i\sin\frac{\pi}{2}(1-2j)}$$
$$\cdot\left(\frac{1}{\Gamma(-m-j)\Gamma(m-\bar{j})} - [m\to -m]\right)\,.$$

Having derived the image under $\mathcal{I}^\chi$ of all reference states, obtaining the boundary coherent states is a simple matter. Since the map $\mathcal{I}^\chi$ commutes with SU(2) and SU(1,1) representations in the sense of equations (B.2.5) and (B.2.2), one can simply act with an element $n_{ab}$ of the relevant subgroup on the homogeneous functions defined above. Proceeding in this manner, and denoting generally $|\cdot_{ab}\rangle := |n_{ab}\,\cdot\rangle$, one finally finds

1. for a space-like polygon orthogonal to $S^2$, with $n_{ab} \in \mathrm{SU}(2)$:

$$\psi^{j}_{ab}(z) = (2j+1)^{\frac{1}{2}} \langle z|z\rangle^{j(i\gamma-1)-1} \langle z|+_{ab}\rangle^{2j} \; ; \qquad \text{(III.1.2)}$$

2. for a space-like polygon orthogonal to $H^\tau$, with $n_{ab} \in \mathrm{SU}(1,1)$:

$$\psi^{k,\tau}_{ab}(z) = (-2k-1)^{\frac{1}{2}}\, \Theta\left(\tau[z|z]\right) \left(\tau[z|z]\right)^{-k(i\gamma+1)-1} [\tau_{ab}|z]^{2k} \; ;$$

3. for two time-like polygons orthogonal to $H^{\mathrm{sl}}$, with $n_{ab}$, $n_{ba} \in \mathrm{SU}(1,1)$:

$$\psi^{s,\tau}_{ba}(z) = A^{j}_{-\tau\gamma s} \Theta\left(\tau[z|z]\right) \left(\tau[z|z]\right)^{-\frac{1}{2}} \cdot \left(\tau[z|l^-_{ba}]\right)^{(i-\gamma)s-\frac{1}{2}} \left(\tau[l^-_{ba}|z]\right)^{(i+\gamma)s-\frac{1}{2}} \; ,$$

$$\tilde{\psi}^{s,\tau}_{ab}(z) = \tilde{A}^{j}_{-\tau\gamma s} \Theta\left(\tau[z|z]\right) \left(\tau[z|z]\right)^{-\frac{1}{2}} \cdot \left(\tau[z|l^-_{ab}]\right)^{(i-\gamma)s-\frac{1}{2}} \left(\tau[l^-_{ab}|z]\right)^{(i+\gamma)s-\frac{1}{2}} \cdot \left[\left((i-\gamma)s - \frac{1}{2}\right) \frac{[z|l^+_{ab}]}{[z|l^-_{ab}]} - \left((i+\gamma)s - \frac{1}{2}\right) \frac{[l^+_{ab}|z]}{[l^-_{ab}|z]}\right] .$$

This concludes the characterization of the spin-foam vertex amplitude.

## III.2 The asymptotic problem

I shall follow the general procedure applied in [53, 63]. There exists the following asymptotic theorem due to Hörmander [119, Theorem 7.7.5]:

**Theorem III.1** (Hörmander I). *Let $S(x)$ be smooth and complex-valued in a neighborhood $K$ of $x_0 \in \mathbb{R}^n$, such that $\Im S \geq 0$, $\Im S(x_0) = 0$, $S'(x_0) = 0$ and* $\det S''(x_0) \neq 0$. *Consider furthermore $u(x) \in \mathcal{C}_0^\infty(K)$. Then*

$$\int dx\; u(x) e^{i\lambda S(x)} = \left(\frac{2\pi i}{\lambda}\right)^{n/2} \frac{u(x_0)e^{i\lambda S(x_0)}}{\sqrt{\det S''(x_0)}} + \mathcal{O}\left(\lambda^{-n/2-1}\right) .$$

Defining $\Omega_{ab} = \overline{g_a \triangleright \psi_{ab}} \cdot g_b \triangleright \psi_{ba}$ for special linear matrices $g_a, g_b$, we explicitly write the inner product of equation (III.1.1) as an integral,

$$A_v = \int_{\mathrm{SL}(2,\mathbb{C})} \prod_{a=1}^{n} \mathrm{d}g_a \delta(g_n) \prod_{a<b} \int_{\mathbb{C}P} \omega(z_{ab}) \Omega_{ab}(z_{ab}, g_a, g_b)\,, \quad \text{(III.2.1)}$$

where $\omega(z_{ab})$ is the integration measure defined in equation (B.1.1), and subsequently bring $\Omega_{ab}$ into the generic form of the exponential of an action, $\Omega_{ab} = f_{ab}\, e^{\Lambda S_{ab}}$. We are then interested in the critical points of $S = \sum_{a<b} S^{\nu}_{ab}$, and these are characterized firstly by a "reality condition",

$$\Re\, S_{ab}(z_{ab}, g_a, g_b) = 0,\; \forall\, a, b\,,$$

and secondly by the critical point conditions

$$\begin{cases} \delta_{z_{ab}} S_{ab}(z_{ab}, g_a, g_b) = 0,\; \forall\, a, b\,, \\ \sum_{b\neq a} \delta_{g_a} S_{ab}(z_{ab}, g_a, g_b) = 0,\; \forall\, a\,. \end{cases}$$

Note that one need only vary the action with respect to the holomorphic spinor and group variables $z_{ab}$ and $g_a$, since the action is constrained to be purely imaginary [63].

Generally one ought to consider every possible type of interface between polyhedra, as per the prescription of the previous section. The calculation for achronal interfaces has already been carried out in great detail by Barrett et al. in [53], so I will refrain from repeating it here. The case of orthochronal interfaces has also been thoroughly discussed in [63]. While the remaining two possibilities - parachronal and heterochronal - were analyzed in [63] and [64], respectively, we found that revisiting these cases proved useful in further understanding the structure of the model. The present analysis will therefore restrict itself to the last two situations: the case of a time-like interface between two time-like polyhedra and the one of a space-like interface between time- and space-like polyhedra. I will simply state the remaining results when needed.

### III.2.1 Adapted coordinates on the group manifold

*A note on notation: henceforth Minkowski- and Euclidean-type inner products will frequently appear back-to-back. In order to avoid confusion I will use Greek indices $\mu = 0, 1, 2, 3$ to refer to Minkowski contractions, while capitalized latin indices $I = 1, 2, 3$ will be reserved for Euclidean contractions on $\mathbb{C}^3$. The vertical placement of indices will only be of consequence for Greek letters. A sum will explicitly be written for Minkowski contractions whenever the summation involves only an index subset. I moreover define $\sigma_\mu = (\mathbb{1}, \sigma_1, \sigma_2, \sigma_3)_\mu$, such that the Pauli matrices with lower indices stand for the conventional ones.*

Performing the Haar integrals in equation (III.2.1), as well as computing the derivatives determining the critical points, requires one to choose coordinates $\phi_i : U_i \in G \to \mathbb{C}^3$ on the group manifold $G = \mathrm{SL}(2, \mathbb{C})$. However, since we are interested in making use of Theorem III.1 rather than analytically evaluating the integral, we can substantially simplify the discussion by implicitly picking useful coordinates and explicitly specifying only the values of the derivatives in those coordinates.

With a slight abuse of notation (I omit the dependence on $\phi_i$), let $\mathrm{d}g$ be the exterior derivative of the preimage of a chart,

$$\mathrm{d}g : \mathbb{C}^3 \to T_g G \ .$$

The differential and right-multiplication $R_g$ for matrix groups satisfies $\mathrm{d}R_g^{-1} X = X g^{-1}$ for $X \in TG$, $g \in G$, and hence the map $\mathrm{d}g g^{-1}$ must take values in the Lie algebra $\mathfrak{g} \simeq T_e G$. One may think of this object as a 1-form[34] in $\mathbb{C}^3$ with values in $\mathfrak{g}$, such that it admits an expansion in terms

[34] This amounts to a coordinate representation of the usual Maurer-Cartan 1-form.

of $\sigma_I$ generators

$$\mathrm{d}gg^{-1} = \frac{i}{2}\sigma_I\Omega^I\,, \quad \Omega^I \in T^*_g\mathbb{C}^3 \simeq \mathbb{C}^3\,. \tag{III.2.2}$$

Generally one then sees that coordinate derivatives of $g \in G$ may always be written as $\partial_I g = \frac{i}{2}\Omega^J{}_I\sigma_J g$, for $\Omega^J{}_I$ a matrix of coefficients dependent on the choice of charts. A particular simple choice of complex coordinates $x^I$ is that in which the matrix of coefficients reduces to the identity,

$$\partial_I g = \frac{i}{2}g\sigma_I\,,$$

and this is the choice we will make for coordinates on the special linear group whenever $g_a$ is associated to a space-like polyhedron $a$. If the element under consideration is instead associated to a time-like polyhedron $b$, we pick adapted coordinates $y^I$ for $g_b$ such that

$$\partial_I g = \frac{i}{2}g\varsigma_I\,,$$

where $\varsigma_I$ was defined in (II.2.3). While not strictly necessary, this second set of coordinates will make the asymptotic analysis clearer.

Besides derivative terms, the only object in Theorem III.1 which depends on the choice of charts is the Haar measure $\mathrm{d}\mu(g)$. But this too can be identified without explicitly defining the maps $g = g(x^I)$ or $g = g(y^I)$. Indeed, note that the 1-form of equation (III.2.2) is right-invariant. We may thus construct a measure on $G$ by taking the trace of its sixth exterior power, which is bi-invariant by virtue of the cyclicity property of the trace operator. It follows from that equation that

$$\begin{aligned}\mathrm{d}\mu(g) &= N\mathrm{Tr}\left[(\mathrm{d}gg^{-1})^{\wedge 3} \wedge (\mathrm{d}gg^{-1})^{\dagger\wedge 3}\right] \\ &= N\frac{2\cdot 3!^2\cdot 6^2}{2^6}|\det\Omega_w|^2\mathrm{d}w^1\wedge\mathrm{d}\overline{w}^1\wedge \ldots \wedge \mathrm{d}w^3\wedge\mathrm{d}\overline{w}^3\end{aligned} \tag{III.2.3}$$

where $N$ is a normalization factor to be fixed, $w^I$ are some coordinates and $\Omega_w$ denotes the matrix coefficients of $\Omega$ in those same coordinates.

The Haar measure on locally-compact groups is known to be unique up to a multiplicative factor, so the above measure *is* the Haar measure; it remains to determine $N$. An often-used convention in the spin-foam literature [74,120] is the one of Rühl, for which the Haar measure reads [42]

$$\mathrm{d}\mu(g)_R = \frac{(2\pi)^{-4}}{|a_{22}|^2}\left(\frac{i}{2}\right)^3 \mathrm{d}a_{12}\wedge \mathrm{d}\overline{a}_{12}\wedge \mathrm{d}a_{21}\wedge \mathrm{d}\overline{a}_{21}\wedge \mathrm{d}a_{22}\wedge \mathrm{d}\overline{a}_{22}\,,$$
$$g = \left(\begin{smallmatrix} a_{11} & a_{12} \\ a_{21} & a_{22}\end{smallmatrix}\right).$$

The $N$ factor of equation (III.2.3) can be made to agree with the convention of Rühl by direct comparison. Letting $w^I$ now stand for Rühl's coordinates one finds $|\det \Omega_R|^2 = 2^4|a_{22}|^{-2}$. On the other hand, it is clear that $|\det \Omega_x|^2 = |\det \Omega_y|^2 = 1$ and hence, requiring $\mathrm{d}\mu(g) = \mathrm{d}\mu(g)_R$,

$$N_R = \frac{-i\cdot(2\pi)^{-4}}{3!^2\cdot 6^2\cdot 2^2}.$$

This fixes the normalization we will use throughout. In terms of the adapted coordinates $x^I$ and $y^I$, the Haar measure reads simply

$$\begin{aligned}\mathrm{d}\mu(g) &= \frac{1}{(4\pi)^4}\mathrm{d}\Re x^1\wedge\ldots\wedge\mathrm{d}\Re x^3\wedge\mathrm{d}\Im x^1\wedge\ldots\wedge\mathrm{d}\Im x^3\\ &= \frac{1}{(4\pi)^4}\mathrm{d}\Re y^1\wedge\ldots\wedge\mathrm{d}\Re y^3\wedge\mathrm{d}\Im y^1\wedge\ldots\wedge\mathrm{d}\Im y^3\,.\end{aligned}$$

### III.2.2 Critical point equations

Having established useful coordinates on the special linear group, I proceed to derive the critical point equations for the two cases of interest.

*(a) Heterochronal interfaces*

Consider the product $\Omega_{ab} = \overline{g_a \triangleright \psi^j_{ab}}\cdot g_b \triangleright \psi^{k,\tau}_{ba}$ describing the interface between space-like and time-like polyhedra. For the amplitude to be well-defined the $\chi$ labels of both states must agree, from where $j_{ab} = -k_{ab}$. Introducing a uniform scaling parameter $\Lambda$, and defining

$$\Omega_{ab} =: f_{ab}(z_{ab}, g_a, g_b)e^{\Lambda S_{ab}(z_{ab},g_a,g_b)}\,,$$

the functions $f_{ab}$ and $S_{ab}$ read

$$f_{ab}(z_{ab}, g_a, g_b) = \frac{\sqrt{4\Lambda^2 j_{ab}^2 - 1}\;\Theta\left(\tau_{ba}[g_b^{-1} z_{ab}|g_b^{-1} z_{ab}]\right)}{\tau_{ba}\,\langle g_a^{-1} z_{ab}|g_a^{-1} z_{ab}\rangle\,[g_b^{-1} z_{ab}|g_b^{-1} z_{ab}]}\,,$$
$$S_{ab} = (i\gamma + 1) j_{ab} \ln \tau_{ba} \frac{[g_b^{-1} z_{ab}|g_b^{-1} z_{ab}]}{\langle g_a^{-1} z_{ab}|g_a^{-1} z_{ab}\rangle} + 2 j_{ab} \ln \frac{\langle +_{ab}|g_a^{-1} z_{ab}\rangle}{\tau_{ba}[\tau_{ba}|g_b^{-1} z_{ab}]}\,. \tag{III.2.4}$$

The reality condition $\Re S_{ab} = 0$ implies

$$\ln\left|\frac{[g_b^{-1} z_{ab}|g_b^{-1} z_{ab}]}{\langle g_a^{-1} z_{ab}|g_a^{-1} z_{ab}\rangle}\left(\frac{\langle +_{ab}|g_a^{-1} z_{ab}\rangle}{[\tau_{ba}|g_b^{-1} z_{ab}]}\right)^2\right| - \gamma \arg\left(\tau_{ba}\frac{[g_b^{-1} z_{ab}|g_b^{-1} z_{ab}]}{\langle g_a^{-1} z_{ab}|g_a^{-1} z_{ab}\rangle}\right) = 0$$
$$\Leftrightarrow \begin{cases} |g_a^{-1} z_{ab}\rangle = \lambda_{ab}|+_{ab}\rangle \\ |g_b^{-1} z_{ab}\rangle = \lambda_{ba}|\tau_{ba}\rangle \end{cases}, \quad \lambda_{ab}, \lambda_{ba} \in \mathbb{C}\,. \tag{III.2.5}$$

A general variation of the action with respect to the spinor variable yields

$$\delta_{z_{ab}} S_{ab} = (i\gamma - 1)\, j_{ab}\left(\frac{\tau_{ba}[\tau_{ba}|g_b^{-1}}{\lambda_{ba}} - \frac{\langle +_{ab}|g_a^{-1}}{\lambda_{ab}}\right)\delta|z_{ab}\rangle\,, \tag{III.2.6}$$

having substituted-in the solution to the reality condition. Regarding the special linear group, we consider the adapted coordinates discussed above, such that

$$\delta g_a = \frac{i}{2} g_a \epsilon_I \sigma^I\,, \quad \delta g_b = \frac{i}{2} g_b \epsilon_I \varsigma^I\,,$$

for some small complex displacement $\epsilon_I$. After applying the reality condition one gets

$$\begin{aligned} \delta_{g_a} S_{ab} &= -\frac{\epsilon_I}{2}(i+\gamma) j_{ab}\,\langle +_{ab}|\sigma^I +_{ab}\rangle\,, \\ \delta_{g_b} S_{ab} &= \frac{\epsilon_I}{2}(i+\gamma) j_{ab} \tau_{ba}[\tau_{ba}|\varsigma^I \tau_{ba}]\,. \end{aligned} \tag{III.2.7}$$

*(b) Parachronal interfaces*

Take now the boundary states $\tilde{\psi}_{ab}^{s,\tau_{ab}}$ and $\psi_{ba}^{s,\tau_{ba}}$ to be those associated to time-like faces, and set $\Omega_{ab} = \overline{g_a \triangleright \tilde{\psi}_{ab}^{s,\tau_{ab}}} \cdot g_b \triangleright \psi_{ba}^{s,\tau_{ba}}$. As before the spins

$s_{ab} = s_{ba}$ must agree, and expanding

$$\Omega_{ab} =: f_{ab}(z_{ab}, g_a, g_b) e^{\Lambda S_{ab}(z_{ab}, g_a, g_b)}\,, \tag{III.2.8}$$

the action takes the form

$$S_{ab} = (i+\gamma) s_{ab} \ln \frac{\tau_{ba}[l^-_{ba}|g_b^{-1} z_{ab}]}{\tau_{ab}[l^-_{ab}|g_a^{-1} z_{ab}]} + (i-\gamma) s_{ab} \ln \frac{\tau_{ba}[g_b^{-1} z_{ab}|l^-_{ba}]}{\tau_{ab}[g_a^{-1} z_{ab}|l^-_{ab}]}\,. \tag{III.2.9}$$

It is a crucial remark that, unlike all other interface types, the action for parachronal interfaces is purely imaginary; no reality condition, which would otherwise refine the critical configurations, need be enforced. The exact same behavior had already been observed in the analog 3d model proposed in section II.5, where additional ad-hoc gluing constraints were necessary in order to obtain a reasonable semiclassical limit. We shall indeed see that the present amplitude is similarly insufficiently constrained.

Proceeding with the variations of the action, they read

$$\delta_{z_{ab}} S_{ab} = (i+\gamma) s_{ab} \left( \frac{[l^-_{ba}|g_b^{-1}}{[l^-_{ba}|g_b^{-1} z_{ab}]} - \frac{[l^-_{ab}|g_a^{-1}}{[l^-_{ab}|g_a^{-1} z_{ab}]} \right) \delta|z_{ab}\rangle$$

for the spinor variable, as well as

$$\delta_{g_a} S_{ab} = -\frac{\epsilon_I}{2}(1 - i\gamma) s_{ab} \frac{[l^-_{ab}|\varsigma^I g_a^{-1} z_{ab}]}{[l^-_{ab}|g_a^{-1} z_{ab}]}$$

for the group elements. Following [64], in the absence of a reality condition we parametrize the spinor variable in terms of a $\{|l^\pm_{ab}\rangle\}$ basis of $\mathbb{C}^2$, i.e.

$$|g_a^{-1} z_{ab}\rangle = \alpha_{ab} \left( |l^+_{ab}\rangle + \beta_{ab} |l^-_{ab}\rangle \right)\,, \quad \alpha_{ab}, \beta_{ab} \in \mathbb{C}\,, \tag{III.2.10}$$

and analogously for $|g_b^{-1} z_{ab}\rangle$. Under these expansions the previous equations take the form

$$\delta_{z_{ab}} S_{ab} = \left( \frac{[l^-_{ba}|g_b^{-1}}{\alpha_{ba}} - \frac{[l^-_{ab}|g_a^{-1}}{\alpha_{ab}} \right) \delta|z_{ab}\rangle\,, \tag{III.2.11}$$

$$\begin{aligned}\delta_{g_a} S_{ab} &= -\frac{\epsilon_I}{2}(i+\gamma)s_{ab}\left(-i[l^-_{ab}|\varsigma^I l^+_{ab}] - i\beta_{ab}[l^-_{ab}|\varsigma^I l^-_{ab}]\right)\,, \\ \delta_{g_b} S_{ab} &= \frac{\epsilon_I}{2}(i+\gamma)s_{ab}\left(-i[l^-_{ba}|\varsigma^I l^+_{ba}] - i\beta_{ba}[l^-_{ba}|\varsigma^I l^-_{ba}]\right)\,.\end{aligned} \tag{III.2.12}$$

For completeness we also expand the pre-factor function of equation (III.2.8) in terms of (III.2.10), finding

$$\begin{aligned}f_{ab}(z_{ab}, g_a, g_b) = \frac{1}{2}\overline{\tilde{A}}^{-\frac{1}{2}+i\Lambda s_{ab}}_{-\tau_{ab}\gamma\Lambda s_{ab}} A^{-\frac{1}{2}+i\Lambda s_{ab}}_{-\tau_{ba}\gamma\Lambda s_{ab}} \\ \cdot\, \Theta(\tau_{ab}\,\Re\beta_{ab})\Theta(\tau_{ba}\,\Re\beta_{ba})\,|\alpha_{ab}\alpha_{ba}|^{-2}\,(\tau_{ab}\Re\beta_{ab})^{-\frac{1}{2}}\,(\tau_{ba}\Re\beta_{ba})^{-\frac{1}{2}} \\ \cdot\, \left[(2\Lambda s_{ab} - i)\,\Im\beta_{ab} - 2\gamma\Lambda s_{ab}\,\Re\beta_{ab}\right].\end{aligned} \tag{III.2.13}$$

All critical point equations follow from the variations above.

## III.3 Geometrical formulation of critical point equations

We are now in possession of the *algebraic* critical point equations with which to study the dominant configurations of the spin-foam amplitude. These equations can, however, be brought into a more amenable form, which will later on be useful in formulating a *geometrical* understanding of the semiclassical limit; this is the main subject of the present section.

### III.3.1 The spin homomorphism

The attentive reader will already have identified in equations (III.2.7) and (III.2.12) the very same type of structures that in chapter II were shown to determine geometrical vectors. Indeed, the objects $\langle +_{ab}|\sigma^I +_{ab}\rangle$, $[\tau_{ab}|\varsigma^I \tau_{ab}]$ and $[l^-_{ab}|\varsigma^I l^+_{ab}]$ are nothing but associations between the chosen coherent states at the boundary and the homogeneous spaces of the Lorentz group[35]. An explicit characterization can be achieved through the spin

[35] That such objects should correspond to homogeneous spaces can be argued as follows. Note first that there is a natural isometry between the algebras of interest and the relevant vector spaces $\mathfrak{su}(2) \simeq \mathbb{R}^3$ and $\mathfrak{su}(1,1) \simeq \mathbb{R}^{1,2}$, using the inner product on the algebras given by the Killing form $\langle X, Y\rangle = \frac{1}{2}\mathrm{Tr}[XY]$. The adjoint action of each group on its own algebra then constitutes a norm-preserving linear mapping of 3-vectors, which can be checked

homomorphism

$$\begin{aligned}\pi:\ \mathrm{SL}(2,\mathbb{C}) &\to \mathrm{SO}_0^+(1,3)\\ g\sigma_\mu g^\dagger &= \pi(g)^\nu{}_\mu \sigma_\nu\,,\end{aligned}$$

where the usual conventions with Minkowski indices apply. This map is 2-to-1, and clearly restricts to the SU(2) and SU(1,1) subgroups via reduction to $(\sigma_1,\sigma_2,\sigma_3)$ and $(\mathbb{1},\sigma_1,\sigma_2)$, respectively.

It will be useful for the remainder of the chapter to establish the image under $\pi$ of a certain transformation of a general group element. Let then $g, i^{1-\eta_{\mu 0}}\sigma_\mu \in \mathrm{SL}(2,\mathbb{C})$ (no sum over $\mu$). We can argue

$$\begin{aligned}\pi\left[\left(i^{1-\eta_{\mu 0}}\sigma_\mu\right) g^\dagger \left(i^{1-\eta_{\nu 0}}\sigma_\nu\right)\right]^\alpha{}_\beta\,\sigma_\alpha &= \sigma_\mu g^\dagger \sigma_\nu\sigma_\alpha\sigma_\nu g\sigma_\mu\\ &= \sigma_\mu g^\dagger (\Sigma_\nu)_\beta{}^\alpha \sigma_\alpha g\sigma_\mu\,,\end{aligned}$$

where $\Sigma_\nu$ is a $4\times 4$ diagonal matrix which depends on the index $\nu$ as

$$\Sigma_0 = \mathbb{1}\,,\quad (\Sigma_i)_0{}^0 = (\Sigma_i)_i{}^i = 1\,,\quad (\Sigma_i)_j{}^j = -1\,,\ i\neq j\,,$$

such that $(\Sigma_\nu)_\alpha{}^\beta\sigma_\beta = \sigma_\nu\sigma_\alpha\sigma_\nu$. Making use of the quaternionic structure $Q$ defined in equation (II.1.1), note that $Q\sigma_\mu Q^{-1} = P_\mu{}^\alpha\sigma_\alpha$, with $P$ the parity map $P = \mathrm{diag}(1,-\vec{1})$. Carrying on,

$$\begin{aligned}&\pi\left[\left(i^{1-\eta_{\mu 0}}\sigma_\mu\right) g^\dagger \left(i^{1-\eta_{\nu 0}}\sigma_\nu\right)\right]^\alpha{}_\beta\sigma_\alpha\\ &\qquad= (\Sigma_\nu)_\beta{}^\alpha P_\alpha{}^\gamma \sigma_\mu Q g^{-1}\sigma_\gamma {g^{-1}}^\dagger Q^{-1}\sigma_\mu\\ &\qquad= (\Sigma_\nu P)_\beta{}^\alpha \pi(g^{-1})^\gamma{}_\alpha (P\Sigma_\mu)_\gamma{}^\delta\sigma_\delta\\ &\qquad= (\Sigma_\nu P\pi(g)P\Sigma_\mu)_\beta{}^\alpha\sigma_\alpha\,,\end{aligned}$$

and denoting by $(R_v)^\mu{}_\nu = \delta^\mu{}_\nu - \frac{2v^\mu v_\nu}{\langle v,v\rangle}$ a reflection in $\mathbb{R}^{1,3}$ with respect to the hyperplane orthogonal to $v$, one finally finds

$$\pi\left[\left(i^{1-\eta_{\mu 0}}\sigma_\mu\right) g^\dagger \left(i^{1-\eta_{\nu 0}}\sigma_\nu\right)\right]^\alpha{}_\beta\,\sigma_\alpha = (R_\mu\pi(g)R_\nu)_\beta{}^\alpha\sigma_\alpha\,. \qquad \text{(III.3.1)}$$

---

to further preserve orientation. Since $\langle +|\sigma^I +\rangle$, $[\tau|\varsigma^I\tau]$ and $[l^-|\varsigma^I l^+]$ are by themselves elements of $S^2$, $H^\tau$ and $H^{\mathrm{sl}}$, respectively, and knowing that the action of SU(2) and SU(1,1) on these spaces is transitive and injective up to $\mathbb{Z}^2$, it must be that $\langle +_{ab}|\sigma^I +_{ab}\rangle$, $[\tau_{ab}|\varsigma^I\tau_{ab}]$ and $[l^-_{ab}|\varsigma^I l^+_{ab}]$ cover the surfaces of transitivity of the Lorentz group.

Equation (III.3.1) describes reflections on the image of the spin homomorphism in terms of a transformation of special linear matrices. As a first application, it can be used to derive a correspondence between the boundary states of the theory and geometrical spaces. Define $P_3$ and $P_{1,2}$ to be the restrictions of $P$ to the subspaces $\mathbb{R}^3, \mathbb{R}^{1,2} \subset \mathbb{R}^{1,3}$. That correspondence, for every type of polygon, is as follows:

*(a) Space-like polygons in space-like polyhedra*

The relevant object in equation (III.2.7) is $\langle +_{ab}|\sigma^I +_{ab}\rangle = \langle +|n^\dagger_{ab}\sigma^I n_{ab}|+\rangle$. In terms of Minkowski components $\sigma^\mu = (\mathbb{1}, -\sigma_1, -\sigma_2, -\sigma_3)$,

$$\begin{aligned}\langle +|n^\dagger_{ab}\sigma^\mu n_{ab}|+\rangle &= \pi(n^\dagger_{ab})_\nu{}^\mu \, \langle +|\sigma^\nu|+\rangle \\ &= (P\pi(n_{ab})P)^\mu{}_\nu(\delta^\nu_0 - \delta^\nu_3) \ .\end{aligned}$$

It is well-known that $\pi$ has the block form $\pi(n_{ab}) = \text{diag}\left(1, \pi_{\text{SU}(2)}(n_{ab})\right)$ when restricted to the $\text{SU}(2)$ subgroup. The previous equation therefore implies

$$\sum_{\mu=1}^{3} \langle +_{ab}|\sigma^\mu +_{ab}\rangle \, \hat{e}_\mu = P_3\pi_{\text{SU}(2)}(n_{ab})\, \hat{e}_3 \ ,$$

suggesting the definition

$$\vec{n}_{ab} := \pi_{\text{SU}(2)}(n_{ab}) \ (-\hat{e}_3) \in S^2 \subset \mathbb{R}^3 \ . \tag{III.3.2}$$

The vector components thus satisfy[36]

$$\langle +_{ab}|\sigma^I +_{ab}\rangle = -n^I_{ab} \ . \tag{III.3.3}$$

*(b) Space-like polygons in time-like polyhedra*

Take now $[\tau_{ab}|\varsigma^I \tau_{ab}] = \langle \tau|n^\dagger_{ab}(\mathbb{1}, \sigma_1, \sigma_2)^I n_{ab}|\tau\rangle$ from (III.2.7), for which

$$\begin{aligned}\langle \tau|n^\dagger_{ab}\sigma^\mu n_{ab}|\tau\rangle &= \pi(n^\dagger_{ab})_\nu{}^\mu \, \langle \tau|\sigma^\nu|\tau\rangle \\ &= (P\pi(n_{ab})P)^\mu{}_\nu(\delta^\nu_0 - \tau\delta^\nu_3) \ .\end{aligned}$$

Since the homomorphism satisfies $\pi(n_{ab}) = \text{diag}\left(\pi_{\text{SU}(1,1)}(n_{ab}), 1\right)$ when restricted to $\text{SU}(1,1)$, it must be that

$$\sum_{\mu=0}^{2} \tau \, \langle \tau_{ab}|\sigma^\mu \tau_{ab}\rangle \, \hat{e}_\mu = P_{1,2}\pi_{\text{SU}(1,1)}(n_{ab})\, \tau\hat{e}_0 \ ,$$

We can then define

$$\vec{n}_{ab} := \pi_{\text{SU}(1,1)}(n_{ab})\, \tau\hat{e}_0 \in H^\tau \subset \mathbb{R}^{1,2} \ , \tag{III.3.4}$$

for which the vector components read

$$\tau[\tau_{ab}|\varsigma^I \tau_{ab}] = n^I_{ab} \ . \tag{III.3.5}$$

[36] I have chosen to define $\vec{n}_{ab}$ in terms of $-\hat{e}_3$ rather than $\hat{e}_3$ because the bivector equation (III.3.20) suggests this is the right association.

*(c) Time-like polygons in time-like polyhedra and* $n_{ab} \in \mathrm{SU}(1,1)$
Consider $[l^-_{ab}|\varsigma^I l^+_{ab}] = \langle l^-|n^\dagger_{ab}(\mathbb{1},\sigma_1,\sigma_2)^I n_{ab}|l^+\rangle$ from equation (III.2.12) and

$$\begin{aligned}\langle l^-|n^\dagger_{ab}\sigma^\mu n_{ab}|l^+\rangle &= \pi(n^\dagger_{ab})_\nu{}^\mu\,\langle l^-|\sigma^\nu|l^+\rangle \\ &= (P\pi(n_{ab})P)^\mu{}_\nu(i\delta^\nu_2 - \delta^\nu_3)\,.\end{aligned}$$

Like before one has

$$\sum_{\mu=0}^{2} i[l^-_{ab}|\varsigma^\mu\, l^+_{ab}]\hat{e}_\mu = P_{1,2}\pi_{\mathrm{SU}(1,1)}(n_{ab})\,\hat{e}_2\,,$$

from where we set

$$\vec{n}_{ab} := \pi_{\mathrm{SU}(1,1)}(n_{ab})\,(-\hat{e}_2) \in H^{\mathrm{sl}} \subset \mathbb{R}^{1,2} \tag{III.3.6}$$

with vector components

$$-i[l^-_{ab}|\varsigma^I l^+_{ab}] = n^I_{ab}\,. \tag{III.3.7}$$

With regards to $[l^-_{ab}|\varsigma^I l^-_{ab}]$ the same arguments lead to

$$\sum_{\mu=0}^{2} [l^-_{ab}|\varsigma^\mu\, l^-_{ab}]\hat{e}_\mu = P_{1,2}\pi_{\mathrm{SU}(1,1)}(n_{ab})\,(\hat{e}_0 - \hat{e}_1)\,,$$

and thus

$$\begin{aligned}\vec{m}_{ab} &:= \pi_{\mathrm{SU}(1,1)}(n_{ab})(\hat{e}_0 - \hat{e}_1) \in \mathcal{C}^+ \subset \mathbb{R}^{1,2}\,,\\ [l^-_{ab}|\varsigma^I l^-_{ab}] &= m^I_{ab}\,,\end{aligned} \tag{III.3.8}$$

where $\mathcal{C}^+$ denotes the future-pointing light-cone. The symbol $\vec{m}_{ab}$ labels the null vector and distinguishes it from $\vec{n}_{ab}$.

The three items above fix the meaning of the notation $\vec{n}_{ab}, \vec{m}_{ab}$ (contingent on the causal character of the polygon $ab$) from here on.

### III.3.2 A complex structure on $\Lambda^2\mathbb{R}^{1,3}$ and $\mathbb{C}^3$ duality

There exists another group homomorphism which will figure naturally in the analysis of critical point configurations, yielding a second interpretation for the boundary data in geometrical terms. It reads [121, Ch. 8]

$$\begin{aligned}\rho:\ &\mathrm{SL}(2,\mathbb{C}) \to \mathrm{SO}(3,\mathbb{C})\\ &g\sigma_I g^{-1} = \rho(g)^J{}_I\sigma_J,\end{aligned}$$

mapping from the special linear group to the complex *orthogonal* group, the latter acting naturally on $\mathbb{C}^3$. Since what concerns us presently is the subject of Minkowski geometry, an identification between the complex vector space and some real Minkowski space is required. The immediate candidate is of course the second exterior power of Minkowski space $\Lambda^2\mathbb{R}^{1,3}$, which matches $\mathbb{C}^3$ in real dimensions.

Introduce to this end the notion of a complex structure $J$ on a real inner product vector space $\mathbb{R}^{2n}$,

$$J: \ \mathbb{R}^{2n} \to \mathbb{R}^{2n}, \quad J^2 = -1,$$

and declare it to be compatible with that inner product $\langle J\cdot, J\cdot\rangle = \langle\cdot,\cdot\rangle$. Note that if $\{e_a, e_{n+a}\}$ is a basis for $\mathbb{R}^{2n}$, then clearly so is $\{e_a, Je_a\}$, as must be $\langle v, Jv\rangle = 0$. There is then a canonical correspondence

$$\begin{aligned} \delta_J: \ (\mathbb{R}^{2n}, J) &\to \mathbb{C}^n \\ e_a + Je_a &\mapsto (1+i)e_a, \end{aligned}$$

where complex scalar multiplication in $\mathbb{C}^n$ is defined in terms of $J$. This I name $\delta_J$ for *duality*. Finding ourselves in the position of having to choose a complex structure with which to identify $\Lambda^2\mathbb{R}^{1,3}$ and $\mathbb{C}^3$, we may note that there exists a natural endomorphism on $\Lambda^2\mathbb{R}^{1,3}$ satisfying our requirements; it is the Hodge map $\star: \ \Lambda^2\mathbb{R}^{1,3} \to \Lambda^2\mathbb{R}^{1,3}$. It is worthwhile remarking that the Hodge star can only serve as a complex structure because the metric signature is Lorentzian and because we consider specifically the second exterior power. Our duality finally takes the form

$$\begin{aligned} \delta_\star: \ \Lambda^2\mathbb{R}^{1,3} &\to \mathbb{C}^3 \\ \theta_I + \star\theta_I &\mapsto (1+i)\vartheta_I. \end{aligned}$$

where $\{\theta_I, \star\theta_I\}$ and $\{\vartheta_I\}$ are bases for their respective spaces.

Having established the desired vector-space identification, the aforementioned correspondence between boundary states and bivectors can be achieved. Start by picking a basis[37] for $\Lambda^2\mathbb{R}^{1,3}$,

$$\theta_I := \hat{e}_0 \wedge \hat{e}_I, \quad \star\theta_1 = \hat{e}_2 \wedge \hat{e}_3, \quad \star\theta_2 = \hat{e}_3 \wedge \hat{e}_1, \quad \star\theta_3 = \hat{e}_1 \wedge \hat{e}_2,$$

[37] I use the convention that $\epsilon_{0123} = 1$ and $\epsilon^{0123} = -1$.

yielding a dual basis of $\mathbb{C}^3$ given by $\{\vartheta_I := \delta_\star \theta_I\}$. I will again describe each type of polygon separately. The strategy consists in analyzing similar objects to the ones of section III.3.1, but taken as vector components of $\mathbb{C}^3$ elements:

*(a) Space-like polygons in space-like polyhedra and* $n_{ab} \in \mathrm{SU}(2)$
Take once again $\langle +_{ab}|\sigma^I +_{ab}\rangle = -n^I_{ab}$, and form the combination $-n^I_{ab}\vartheta_I$. Its pre-image under the duality is

$$\delta_\star^{-1}\left(-n^I_{ab}\vartheta_I\right) = -n^I_{ab}\theta_I = (0, \vec{n}_{ab}) \wedge (1, \vec{0})\,,$$

where in the last equality $\vec{n}_{ab} \in S^2$ is defined as in (III.3.2). The resulting bivector is orthogonal to both the time-like direction and the plane orthogonal to $\vec{n}_{ab}$, thus admitting an interpretation in terms of a space-like polygon in a space-like polyhedron.

*(b) Space-like polygons in time-like polyhedra and* $n_{ab} \in \mathrm{SU}(1,1)$
Consider now the object $\tau[\tau_{ab}|\sigma^I \tau_{ab}] = (\xi \vec{n}_{ab})^I$, where $\xi$ is a complex matrix such that $\xi^I{}_J \varsigma^J = \sigma^I$, i.e.

$$\xi = \begin{pmatrix} & & i \\ & -i & \\ 1 & & \end{pmatrix}\,.$$

Then one finds

$$\delta_\star^{-1}\left(\xi^I{}_J n^J_{ab}\vartheta_I\right) = \delta_\star^{-1}(n^3_{ab} i\vartheta_1 - n^2_{ab} i\vartheta_2 + n^1_{ab}\vartheta_3) = (\vec{n}_{ab}, 0) \wedge (\vec{0}, 1)\,, \quad \text{(III.3.9)}$$

with $\vec{n}_{ab} \in H^\tau$ as in (III.3.4). Through the same argument as above the resulting bivector describes a space-like polygon in a time-like polyhedron.

*(c) Time-like polygons in time-like polyhedra and* $n_{ab} \in \mathrm{SU}(1,1)$
Regarding $-i[l^-_{ab}|\sigma^I l^+_{ab}] = (\xi \vec{n}_{ab})^I$ and $[l^-_{ab}|\sigma^I l^-_{ab}] = (\xi \vec{m}_{ab})^I$ one finds at once

$$\begin{aligned} \delta_\star^{-1}\left(\xi^I{}_J n^J_{ab}\vartheta_I\right) &= (\vec{n}_{ab}, 0) \wedge (\vec{0}, 1)\,, \\ \delta_\star^{-1}\left(\xi^I{}_J m^J_{ab}\vartheta_I\right) &= (\vec{m}_{ab}, 0) \wedge (\vec{0}, 1)\,, \end{aligned} \quad \text{(III.3.10)}$$

this time with $\vec{n}_{ab} \in H^{\mathrm{sl}}$ and $\vec{m}_{ab} \in C^+$ as per (III.3.6) and (III.3.8). Each bivector can be understood as describing a space-like or null polygon in a time-like polyhedron, respectively.

A pair of final observations is necessary to conclude our discussion. The first has to do with the duality between the $\pi$ and $\rho$ homomorphisms under

$\delta_\star$. To that end, note first that

$$\begin{aligned}\pi(g)^{\wedge 2}\theta_I &= \pi(g)^\alpha{}_0 \pi(g)^\beta{}_I \ \hat{e}_\alpha \wedge \hat{e}_\beta \\ &= \pi(g)^\alpha{}_0 \pi(g)^\beta{}_I \left[ (\delta^0_\alpha \delta^J_\beta - \delta^0_\beta \delta^J_\alpha)\theta_J + \epsilon_{0\alpha\beta}{}^J (\star\theta)_J \right] \\ &\overset{\delta_\star}{\to} \pi(g)^\alpha{}_0 \pi(g)^\beta{}_I \left( \delta^0_\alpha \delta^J_\beta - \delta^0_\beta \delta^J_\alpha + i\epsilon_{0\alpha\beta}{}^J \right) \vartheta_J \,.\end{aligned}$$

Direct computation moreover shows
$\delta^0_\alpha \delta^J_\beta - \delta^0_\beta \delta^J_\alpha + i\epsilon_{0\alpha\beta}{}^J = \frac{1}{2}\mathrm{Tr}\left[\sigma_\beta (P\sigma)_\alpha \sigma^J\right]$, whence

$$\begin{aligned}\delta_\star \left[ \left( \pi(g)^{\wedge 2} \right)^J{}_I \right] &= \pi(g)^\alpha{}_0 \pi(g)^\beta{}_I \frac{1}{2}\mathrm{Tr}\left[\sigma_\beta (P\sigma)_\alpha \sigma^J\right] \\ &= \pi(g)^\alpha{}_0 \pi(g)^\beta{}_I \frac{1}{2}\mathrm{Tr}\left[\sigma_\beta Q \sigma_\alpha Q^{-1} \sigma^J\right] \\ &= \frac{1}{2}\mathrm{Tr}\left[g\sigma_I g^\dagger Q g g^\dagger Q^{-1} \sigma^J\right] \\ &= \frac{1}{2}\mathrm{Tr}\left[g\sigma_I g^{-1} \sigma^J\right] \,.\end{aligned}$$

Appealing to the definition of $\rho$ and to the orthonormality of Pauli matrices under $\frac{1}{2}\mathrm{Tr}$, one arrives at the identity

$$\delta_\star \circ \pi(g)^{\wedge 2} = \rho(g) \circ \delta_\star \,, \tag{III.3.11}$$

which will allow us to equivalently formulate critical configurations either in terms of complex vectors or in terms of real bivectors and their respective rotations.

The second remark concerns the symmetric bilinear in $\mathbb{C}^3$ restricted to the complex vectors of points 1., 2. and 3. above. Although the $\delta_\star$ map cannot constitute an isomorphism (the bilinear in $\mathbb{C}^3$ is complex, while the one in $\mathbb{R}^{1,3}$ is real), it still yields the correct Minkowski inner product for vectors in each of the homogeneous spaces. Indeed one has for $\vec{n}_{ab}, \vec{n}_{ba} \in S^2$ that

$$\langle n^I_{ab}\vartheta_I, n^I_{ba}\vartheta_I \rangle_{\mathbb{C}^3} = \langle \vec{n}_{ab}, \vec{n}_{ba} \rangle_{\mathbb{R}^3}$$

simply by virtue of the components $n^I_{ab}, n^I_{ba}$ being real. Perhaps more interesting is to note that for $\vec{n}_{ab}, \vec{n}_{ba}$ either in $H^\tau$ or $H^{\mathrm{sl}}$ one has

$$\langle (\xi \vec{n}_{ab})^I \vartheta_I, (\xi \vec{n}_{ba})^I \vartheta_I \rangle_{\mathbb{C}^3} = \vec{n}^T_{ab} \, (\xi^T \xi) \, \vec{n}_{ba} = \langle \vec{n}_{ab}, \vec{n}_{ba} \rangle_{\mathbb{R}^{1,2}} \ ,$$

since $\xi^T \xi = \eta_{(1,2)}$ yields the Minkowski metric. Thus the $\xi$ matrix, which appears naturally from the mapping of bivectors to complex vectors in equations (III.3.9), (III.3.10), takes the analog role of square root of the Minkowski metric.

With all preliminary tools established, I now proceed to geometrically develop the critical point equations.

### III.3.3 The closure constraints

Recall that part of the stationarity condition is obtained from the variation with respect to group elements, and that for each polyhedron there is the constraint

$$\sum_{b \neq a} \delta_{g_a} S_{ab} = 0 \, , \, \forall \, a \, . \tag{III.3.12}$$

Consider once more the general form of the spin-foam amplitude of equation (III.2.1). For each choice of polyhedron $a$ one has a product of $\Omega_{ab}$ functions, one for each polygon labeled by $ab$. Since there is one group integration for each polyhedron, the concrete form of the critical point equations obtained from (III.3.12) will depend on the causal character of each interface $ab$.

#### *(a) Space-like polyhedra*

If we take $a$ to be a space-like polyhedron, then every face $ab$ must be space-like. The relevant interfaces can only be heterochronal or achronal. The critical point equations (III.2.7) for the former type were obtained in the previous section, and the result of [53] for the latter type has the exact same form (this is not surprising, since the variations with respect to $g_a$ should only depend on the states $\psi_{ab}$ and not $\psi_{ba}$). Equation (III.3.12)

then implies

$$\forall \text{ s.l. } a\,, \quad \sum_b j_{ab} \left\langle +_{ab} | \sigma^I +_{ab} \right\rangle = 0\,,$$

which, by virtue of equation (III.3.3), can be rewritten as

$$\forall \text{ s.l. } a\,, \quad \sum_b j_{ab} \vec{n}_{ab} = 0\,, \tag{III.3.13}$$

with $\vec{n}_{ab} = \pi_{\mathrm{SU}(2)}(n_{ab})\,(-\hat{e}_3)$ an element of the sphere $S^2$ associated to the face $ab$. Stationarity with respect to group variables thus induces a closure condition for space-like polyhedra.

*(b) Time-like polyhedra*

Alternatively we may take $a$ to label a time-like polyhedron. Every interface $ab$ will then either be space-like, and thus orthochronal or heterochronal, or time-like, and thus parachronal. The equations for the variation of the action with respect to the group for orthochronal interfaces obtained in [63] have the same form as (III.2.7). Collecting equations (III.2.7) and (III.2.12) we find, for every term in the sum of (III.3.12),

$$\forall \text{ t.l. } a\,, \quad \sum_{b\colon \text{s.l.}ab} j_{ab} \tau_{ab} [\tau_{ab} | \varsigma^I \tau_{ab}] + \sum_{b\colon \text{t.l.}ab} -i s_{ab} \left( [l^-_{ab} | \varsigma^I l^+_{ab}] + \beta_{ab} [l^-_{ab} | \varsigma^I l^-_{ab}] \right) = 0\,.$$

Making use of equations (III.3.5), (III.3.7) and (III.3.8), and separating the real and imaginary parts, the above equation takes the form

$$\forall \text{ t.l. } a\,, \quad \sum_{b\colon \text{s.l.}ab} j_{ab} \vec{n}_{ab} + \sum_{b\colon \text{t.l.}ab} s_{ab} \left( \vec{n}_{ab} + \Im \beta_{ab} \vec{m}_{ab} \right) = 0\,,$$

$$\forall \text{ t.l. } a\,, \quad \sum_{b\colon \text{t.l.}ab} s_{ab} \Re \beta_{ab} \vec{m}_{ab} = 0\,, \tag{III.3.14}$$

where the various $\vec{n}_{ab}$ are defined in section III.3.1 and are either elements of $H^\tau$ or $H^{\mathrm{sl}}$ (depending on the causal character of the polygon $ab$), while $\vec{m}_{ab} \in C^+$.

The comparatively strange form - due to the presence of a null vector $\vec{m}_{ab}$ and the complex parameter $\beta_{ab}$ - of the resulting closure conditions demands a couple of remarks.

1. On $\Re\beta_{ab}$: it is claimed in [64] that it would be desirable to establish a condition that would make the real part of $\beta_{ab}$ vanish, so as to constrain the dominant configurations. The proposal of that work is that each tetrahedron should consist of at least one space-like triangle, in which case equation (III.3.14) sets $\Re\beta_{ab} = 0$. This however explicitly excludes triangulations used in CDT [22], which require tetrahedra consisting entirely of time-like triangles (as in the "3-2" 4-simplex). Believing this to be undesirable, I alternatively propose to define the model with the same $\tau_{ab} = \pm$ for all time-like polygons $ab$; this has no effect on the geometric interpretation of the boundary states, but the Heaviside thetas in equation (III.2.13) then determine that the signs of all $\Re\beta_{ab}$ should agree. Equation (III.3.14) thus dictates that a sum of similarly-oriented null vectors must vanish, which is only possible if $\Re\beta_{ab} = 0$.

2. On $\Im\beta_{ab}$: it was Bommer [122] who first pointed out that the combination $\vec{n}^{\beta}_{ab} := \vec{n}_{ab} + \Im\beta_{ab}\vec{m}_{ab}$ constitutes an element of $H^{\text{sl}}$ for any choice of $\beta_{ab}$. That it is so can be argued from equations (III.3.7) and (III.3.8), since

$$\vec{n}^{\beta}_{ab} = \pi_{\text{SU}(1,1)}(n_{ab})\left[-\hat{e}_2 + \Im\beta_{ab}(\hat{e}_0 - \hat{e}_1)\right] \in H^{\text{sl}} . \quad \text{(III.3.15)}$$

   From this point of view the critical point equations still determine a family of closure conditions for time-like polyhedra indexed by $\Im\beta_{ab}$,

$$\forall \text{ t.l. } a\,, \quad \sum_{b:\,\text{s.l.}ab} j_{ab}\vec{n}_{ab} + \sum_{b:\,\text{t.l.}ab} s_{ab}\vec{n}^{\beta}_{ab} = 0\,. \quad \text{(III.3.16)}$$

### III.3.4 The gluing constraints

We now turn to the algebraic equations derived from the $z_{ab}$ variations,

$$\delta_{z_{ab}} S_{ab}(z_{ab}, g_a, g_b) = 0, \; \forall\, a, b\,,$$

which I will show can be rephrased as constraints on Minkowski bivectors.

*(a) Heterochronal interfaces*

The solution to the reality condition of equation (III.2.5) determines

$$\lambda_{ab} g_a |+_{ab}\rangle = \lambda_{ba} g_b |\tau_{ba}\rangle\,, \tag{III.3.17}$$

while it follows from equation (III.2.6) that

$$\frac{1}{\bar{\lambda}_{ab}} g_a^{-1\dagger} |+_{ab}\rangle = \frac{\tau_{ba}}{\bar{\lambda}_{ba}} g_b^{-1\dagger} \sigma_3 |\tau_{ba}\rangle\,. \tag{III.3.18}$$

The adjoint of the latter equation applied to the first yields

$$\begin{aligned} \langle +_{ab}| g_a^{-1} \sigma_I g_a |+_{ab}\rangle &= \tau_{ba} [\tau_{ba}| g_b^{-1} \sigma_I g_b |\tau_{ba}] \\ \Rightarrow\; \rho(g_a)_I{}^J \langle +_{ab}|\sigma_J +_{ab}\rangle &= \rho(g_b)_I{}^J \tau_{ba} [\tau_{ba}|\sigma_J \tau_{ba}] \\ \Leftrightarrow\; -\rho(g_a)\vec{n}_{ab} &= \rho(g_b)\xi\vec{n}_{ba}\,, \end{aligned} \tag{III.3.19}$$

having used in the last line the results of section III.3.1. To this last equation we may apply the $\delta_\star$ duality, which according to (III.3.11) and the discussion of section III.3.2 results in

$$\pi(g_a)^{\wedge 2}(0, \vec{n}_{ab}) \wedge (1, \vec{0}) = \pi(g_b)^{\wedge 2}(\vec{n}_{ba}, 0) \wedge (\vec{0}, 1)\,, \tag{III.3.20}$$

describing the Lorentz rotation of two Minkowski bivectors into each other. An equivalent characterization is given by (III.3.19) itself, which refers to $\mathrm{SO}(3,\mathbb{C})$ rotations of $\mathbb{C}^3$ vectors. Both equations may be interpreted as identifying a space-like polygon $ab$ in a space-like polyhedron $a$ with a space-like polygon $ba$ in a time-like polyhedron $b$.

*(b) Parachronal interfaces*

The spinor equation (III.2.11) states

$$\frac{1}{\overline{\alpha}_{ab}} g_a^{-1\dagger}\sigma_3|l_{ab}^-\rangle = \frac{1}{\overline{\alpha}_{ba}} g_b^{-1\dagger}\sigma_3|l_{ba}^-\rangle \tag{III.3.21}$$

while the parametrization of equation (III.2.10) implies

$$\alpha_{ab} g_a \left(|l_{ab}^+\rangle + \beta_{ab}|l_{ab}^-\rangle\right) = \alpha_{ba} g_b \left(|l_{ba}^+\rangle + \beta_{ba}|l_{ba}^-\rangle\right) . \tag{III.3.22}$$

The same argument as above leads to (recall we have established $\Re\beta_{ab} = 0$)

$$\begin{aligned}
[l_{ab}^-|g_a^{-1}\sigma_I g_a \left(|l_{ab}^+] + i\Im\beta_{ab}|l_{ab}^-]\right) &= [l_{ba}^-|g_b^{-1}\sigma_I g_b \left(|l_{ba}^+] + i\Im\beta_{ba}|l_{ba}^-]\right) \\
\Rightarrow\ \rho(g_a)_I{}^J [l_{ab}^-|\sigma_J(-i|l_{ab}^+] + \Im\beta_{ab}|l_{ab}^-]) & \\
&= \rho(g_b)_I{}^J [l_{ba}^-|\sigma_J \left(-i|l_{ba}^+]\Im\beta_{ba}|l_{ba}^-]\right) \\
\Leftrightarrow\ \rho(g_a)\xi\vec{n}_{ab}^{\beta} &= \rho(g_b)\xi\vec{n}_{ba}^{\beta} ,
\end{aligned}$$

having again resorted to section III.3.1 and the definition of equation (III.3.15). Considering the $\delta_\star$ duality, one has equivalently

$$\pi(g_a)^{\wedge 2}(\vec{n}_{ab}^{\beta}, 0) \wedge (\vec{0}, 1) = \pi(g_b)^{\wedge 2}(\vec{n}_{ba}^{\beta}, 0) \wedge (\vec{0}, 1) , \tag{III.3.23}$$

for both $\vec{n}_{ab}^{\beta}, \vec{n}_{ba}^{\beta} \in H^{\mathrm{sl}}$ parametrized by $\Im\beta_{ab}, \Im\beta_{ba}$, respectively. Once more the resulting constraint describes the identification of two time-like faces among each other.

### III.3.5 Dihedral angles

In exploring the critical point equations it remains to further ascribe a geometrical interpretation to the complex parameters $\lambda_{ab}$ and $\alpha_{ab}$.

*(a) Heterochronal interfaces*

Consider the action of the quaternionic structure $Q$ on (III.3.17) and (III.3.18), which gives

$$\begin{aligned}
\overline{\lambda}_{ab} g_a^{-1\dagger}|-_{ab}\rangle &= -\overline{\lambda}_{ba} g_b^{-1\dagger}\sigma_3|-\tau_{ba}\rangle , \\
\frac{1}{\lambda_{ab}} g_a|-_{ab}\rangle &= \frac{\tau_{ba}}{\lambda_{ba}} g_b|-\tau_{ba}\rangle .
\end{aligned}$$

Factoring out the state $|{-\tau_{ba}}\rangle$, both equations imply

$$g_a^{-1} g_b \sigma_3 g_b^\dagger g_a^{-1\dagger} |{-_{ab}}\rangle = -\tau_{ba} \left| \frac{\lambda_{ab}}{\lambda_{ba}} \right|^{-2} |{-_{ab}}\rangle \,. \tag{III.3.24}$$

The same argument applied to the original equations (III.3.17), (III.3.18) yields a second identity

$$g_a^{-1} g_b \sigma_3 g_b^\dagger g_a^{-1\dagger} |{+_{ab}}\rangle = \tau_{ba} \left| \frac{\lambda_{ab}}{\lambda_{ba}} \right|^{2} |{+_{ab}}\rangle \,. \tag{III.3.25}$$

Since $\{|\pm_{ab}\rangle\}$ is a basis of $\mathbb{C}^2$, the eigensystem (III.3.24),(III.3.25) uniquely determines the matrix $g_a^{-1} g_b \sigma_3 g_b^\dagger g_a^{-1\dagger}$. Note that

$$\begin{aligned} n^I_{ab} L_I |\pm_{ab}\rangle &= -\frac{1}{2} \sum_{\mu,\nu=1}^{3} \hat{e}_3^\mu \, \pi_{\mathrm{SU}(2)}(n_{ab})^\nu{}_\mu \, \sigma_\nu |\pm_{ab}\rangle \\ &= -\frac{1}{2} n_{ab} \sigma_3 n_{ab}^\dagger n_{ab} |\pm\rangle \\ &= \mp \frac{1}{2} |\pm_{ab}\rangle \,, \end{aligned}$$

where $\vec{n}_{ab} \in S^2$ is defined as per equation (III.3.2) and $L_I = \sigma_I/2$ denotes the usual $\mathrm{SU}(2)$ generators. This is enough to find

$$g_a^{-1} g_b \sigma_3 g_b^\dagger g_a^{-1\dagger} = -\tau_{ba} \exp\left( 2 \ln \left| \frac{\lambda_{ba}}{\lambda_{ab}} \right|^2 n^I_{ab} L_I \right) 2 n^I_{ab} L_I \,, \tag{III.3.26}$$

as the matrix on the right-hand side satisfies the same eigensystem. The left-hand side, on the other hand, has the form of a spin homomorphism; massaging both we finally arrive at

$$\begin{aligned} &[\pi\left(g_a^{-1}\right) \pi\left(g_b\right) \hat{e}_3]^\mu \sigma_\mu \\ &\qquad = -\tau_{ba} \left( \sigma_0 \cosh \ln \left| \frac{\lambda_{ba}}{\lambda_{ab}} \right|^2 + n^I_{ab} \sigma_I \sinh \ln \left| \frac{\lambda_{ba}}{\lambda_{ab}} \right|^2 \right) n^I_{ab} \sigma_I \,, \end{aligned}$$

from where we infer

$$\sinh \ln \left| \frac{\lambda_{ab}}{\lambda_{ba}} \right|^2 = \tau_{ba} \left\langle \pi(g_a) \hat{e}_0, \, \pi(g_b) \hat{e}_3 \right\rangle \,.$$

This last equation can equivalently be written

$$\cosh\left(\ln\left|\frac{\lambda_{ab}}{\lambda_{ba}}\right|^2 - i\tau_{ba}\frac{\pi}{2} + 2ik\pi\right) = \frac{\langle N_a^{\hat{e}_0},\, N_b^{\hat{e}_3}\rangle}{||N_a^{\hat{e}_0}||\,||N_b^{\hat{e}_3}||}\,, \quad k \in \mathbb{Z}\,, \tag{III.3.27}$$

with the convention that $||\hat{e}_3|| = i$, $N_a^{\hat{e}_0} := \pi(g_a)\hat{e}_0$ and $N_b^{\hat{e}_3} := \pi(g_b)\hat{e}_3$.

The reader is now directed to the appendix, where a general discussion on Minkowski geometry can be found; in section D.2 the concept of Lorentzian $\mathbb{R}^{1,1}$ angle is introduced, containing an imaginary part which keeps track of the number of light-cone crossings between the vectors which determine it. The definition is such that equation (III.3.27) implies

$$\pm\phi_{N_a^{\hat{e}_0},N_b^{\hat{e}_3}} = \ln\left|\frac{\lambda_{ab}}{\lambda_{ba}}\right|^2 - i\tau_{ba}\frac{\pi}{2}\,,$$

the sign of the angle being contingent on convention. I will follow the spirit of [53] and define $\phi_{ab} := \Re\phi_{N_a^{\hat{e}_0},N_b^{\hat{e}_3}}$ to be the parameter of the Lorentz transformation which 1) keeps the Minkowski plane $\mathcal{P} = \text{span}\left\{N_a^{\hat{e}_0}, N_b^{\hat{e}_3}\right\}$ fixed and 2) transforms the reflection of $N_a^{\hat{e}_0}$ - through the single null line which separates the hyperboloids of $N_a^{\hat{e}_0}$ and $N_b^{\hat{e}_3}$ in $\mathcal{P}$ - into $N_b^{\hat{e}_3}$. In order to do so observe first that it follows from equation (III.3.20) that[38]

$$\begin{aligned} N_b^{\hat{e}_3} &= \langle \pi(g_a)(0,\vec{n}_{ab}) \wedge N_a^{\hat{e}_0}, \pi(g_b)(\vec{n}_{ba},0)\rangle \\ \Rightarrow \left(N_a^{\hat{e}_0} \wedge N_b^{\hat{e}_3}\right) &= \langle N_a^{\hat{e}_0}, \pi(g_b)(\vec{n}_{ba},0)\rangle \left(\pi(g_a)(0,\vec{n}_{ab}) \wedge N_a^{\hat{e}_0}\right)\,, \end{aligned} \tag{III.3.28}$$

from where one sees that $\mathcal{P}$ is spanned by $N_a^{\hat{e}_0}$ and $\pi(g_a)(0,\vec{n}_{ab})$; in particular, the null rays $c^{\pm}$ on $\mathcal{P}$ admit the representation

$$c^{\pm} = N_a^{\hat{e}_0} \pm \pi(g_a)(0,\vec{n}_{ab})\,.$$

and one of these rays separates $N_a^{\hat{e}_0}$ and $N_b^{\hat{e}_3}$ on $\mathcal{P}$. Two unit-norm vectors on the plane which sum to a null vector must be reflections of each other

[38] Here $\langle\cdot,\cdot\rangle$ is understood as the inner product on the exterior algebra of Minkowski space.

along that same null direction; we may conclude that the desired reflection map is one of

$$R_\pm N_a^{\hat{e}_0} = \pm\pi(g_a)(0, \vec{n}_{ab}) \, ,$$

the sign of which will be fixed in the following. I demand that there exists $D_{ab} = D_{ab}(\phi_{ab}) \in \mathrm{SL}(2,\mathbb{C})$ such that

$$\begin{aligned} \pi(D_{ab}) R_\pm \pi(g_a)\hat{e}_0 &= \pi(g_b)\hat{e}_3 \\ \Leftrightarrow \;\; \pm\pi(D_{ab})\pi(g_a)(0, \vec{n}_{ab}) &= \pi(g_b)\hat{e}_3 \, , \end{aligned} \tag{III.3.29}$$

for one of either $R_\pm$, and moreover

$$\pi(D_{ab})^{\wedge 2} \star \left( N_a^{\hat{e}_0} \wedge N_b^{\hat{e}_3} \right) = \star \left( N_a^{\hat{e}_0} \wedge N_b^{\hat{e}_3} \right) \, . \tag{III.3.30}$$

By virtue of the identity (III.3.28), the requirement of (III.3.30) can be rewritten as

$$\begin{aligned} \pi(D_{ab})^{\wedge 2} \star \left( \pi(g_a)(0, \vec{n}_{ab}) \wedge N_a^{\hat{e}_0} \right) &= \star \left( \pi(g_a)(0, \vec{n}_{ab}) \wedge N_a^{\hat{e}_0} \right) \\ \overset{\delta_\star}{\Rightarrow} D_{ab} g_a \left( n_{ab}^I \sigma_I \right) g_a^{-1} D_{ab}^{-1} &= g_a \left( n_{ab}^I \sigma_I \right) g_a^{-1} \, , \end{aligned}$$

such that, for some $\phi_{ab} \in \mathbb{R}$ and up to a sign to which $\pi$ is ignorant,

$$D_{ab} = \pm g_a \exp\left( \phi_{ab} n_{ab}^I L_I \right) g_a^{-1} \, .$$

Substituting this last expression into (III.3.29) one finds

$$g_a^{-1} g_b \sigma_3 g_b^\dagger g_a^{-1\dagger} = \pm \exp\left( 2\phi_{ab} n_{ab}^I L_I \right) \, 2 n_{ab}^I L_I \, ,$$

and a comparison with (III.3.26) yields $\phi_{ab} = \ln \left| \frac{\lambda_{ba}}{\lambda_{ab}} \right|^2$, fixing moreover $R_{-\tau_{ba}}$ as the correct reflection map. This all characterizes

$$\phi_{N_a^{\hat{e}_0}, N_b^{\hat{e}_3}} = \ln \left| \frac{\lambda_{ba}}{\lambda_{ab}} \right|^2 + i\tau_{ba} \frac{\pi}{2} \tag{III.3.31}$$

as the oriented dihedral angle between the hypersurfaces normal to $N_a^{\hat{e}_0}, N_b^{\hat{e}_3}$.

*(b) Parachronal interfaces*

The parachronal case is analogous. The action of the quaternionic structure on equations (III.3.21) and (III.3.22) shows

$$\frac{1}{\alpha_{ab}} g_a |l_{ab}^-\rangle = \frac{1}{\alpha_{ba}} g_b |l_{ba}^-\rangle \,,$$
$$\overline{\alpha}_{ab} g_a^{-1\dagger} \sigma_3 \left( |l_{ab}^+\rangle + i\Im\beta_{ab} |l_{ab}^-\rangle \right) = \overline{\alpha}_{ba} g_b^{-1\dagger} \sigma_3 \left( |l_{ba}^+\rangle + i\Im\beta_{ba} |l_{ba}^-\rangle \right) \,.$$

Together with (III.3.21) and (III.3.22) this implies

$$g_a^{-1} g_b \sigma_3 g_b^\dagger g_a^{-1\dagger} \sigma_3 |l_{ab}^-\rangle = \frac{\alpha_{ba}\overline{\alpha}_{ab}}{\alpha_{ab}\overline{\alpha}_{ba}} |l_{ab}^-\rangle \,,$$
$$g_a^{-1} g_b \sigma_3 g_b^\dagger g_a^{-1\dagger} \sigma_3 \left( |l_{ab}^+\rangle + i\Im\beta_{ab} |l_{ab}^-\rangle \right) = \frac{\alpha_{ab}\overline{\alpha}_{ba}}{\alpha_{ba}\overline{\alpha}_{ab}} \left( |l_{ba}^+\rangle + i\Im\beta_{ba} |l_{ba}^-\rangle \right) \,.$$

According to the definition (III.3.15) of $\vec{n}_{ab}^\beta$,

$$\begin{aligned}
&(n_{ab}^\beta)^I F_I^\dagger |l_{ab}^\pm\rangle \\
&\quad = \frac{1}{2} \sum_{\mu,\nu=0}^{2} \left( -\hat{e}_2 + \Im\beta_{ab}(\hat{e}_0 - \hat{e}_1) \right)^\mu \pi_{\mathrm{SU}(1,1)}(n_{ab})^\nu{}_\mu \, \sigma_\nu \sigma_3 |l_{ab}^\pm\rangle \\
&\quad = \frac{1}{2} n_{ab} \left[ -\sigma_2 + \Im\beta_{ab}(\sigma_0 - \sigma_1) \right] n_{ab}^\dagger \sigma_3 n_{ab} |l^\pm\rangle \\
&\quad = \mp \frac{i}{2} \left( |l_{ab}^\pm\rangle \pm i\Im\beta_{ab}(1 \pm 1) |l_{ab}^\mp\rangle \right) \,,
\end{aligned}$$

where $F_I = \varsigma_I/2$ denotes the usual $\mathrm{SU}(1,1)$ generators and $\sigma_3 \varsigma_I \sigma_3 = \varsigma_I^\dagger$. Succinctly,

$$(n_{ab}^\beta)^I F_I^\dagger |l_{ab}^-\rangle = \frac{i}{2} |l_{ab}^-\rangle \,,$$
$$(n_{ab}^\beta)^I F_I^\dagger \left( |l_{ab}^+\rangle + i\Im\beta_{ab} |l_{ab}^-\rangle \right) = -\frac{i}{2} \left( |l_{ab}^+\rangle + i\Im\beta_{ab} |l_{ab}^-\rangle \right) \,.$$

Since $\left( |l_{ab}^+\rangle + i\Im\beta_{ab} |l_{ab}^-\rangle \right)$ and $|l_{ab}^-\rangle$ are linearly independent for any $\beta_{ab}$, it follows that

$$g_a^{-1} g_b \sigma_3 g_b^\dagger g_a^{-1\dagger} \sigma_3 = \exp\left[ 4\arg\frac{\alpha_{ba}}{\alpha_{ab}} (n_{ab}^\beta)^I F_I^\dagger \right] \,. \qquad \text{(III.3.32)}$$

Identifying the spin homomorphism on the left, and expanding the right-hand side, one arrives at

$$
\begin{aligned}
&[\pi\left(g_a^{-1}\right)\pi\left(g_b\right)\hat{e}_3]^\mu\sigma_\mu \\
&\qquad = \left[\sigma_3\cos\left(2\text{arg}\frac{\alpha_{ba}}{\alpha_{ab}}\right) + \sigma_3(n^\beta_{ab})^I\varsigma_I\sin\left(2\text{arg}\frac{\alpha_{ba}}{\alpha_{ab}}\right)\right] ,
\end{aligned}
$$

and hence

$$
\cos\left(2\text{arg}\frac{\alpha_{ba}}{\alpha_{ab}} + 2k\pi\right) = \frac{\langle\pi(g_a)\hat{e}_3,\, \pi(g_b)\hat{e}_3\rangle}{||\hat{e}_3||\,||\hat{e}_3||} , \quad k\in\mathbb{Z} .
$$

Just as in the heterochonal case, the above equation identifies the parameter $2\text{arg}\frac{\alpha_{ba}}{\alpha_{ab}}$ as an Euclidean angle between the vectors $N_a^{\hat{e}_3}$ and $N_b^{\hat{e}_3}$. Keeping the convention for oriented angles, there must exist $D_{ab} = D(\phi_{N_a^{\hat{e}_3},N_b^{\hat{e}_3}}) \in \text{SL}(2,\mathbb{C})$ such that

$$
\pi(D_{ab})\pi(g_a)\hat{e}_3 = \pi(g_b)\hat{e}_3 , \tag{III.3.33}
$$

and moreover

$$
\pi(D_{ab})^{\wedge 2}\star\left(N_a^{\hat{e}_3}\wedge N_b^{\hat{e}_3}\right) = \star\left(N_a^{\hat{e}_3}\wedge N_b^{\hat{e}_3}\right) . \tag{III.3.34}
$$

Resorting to identity (III.3.23) one can see that

$$
\begin{aligned}
N_b^{\hat{e}_3} &= \langle N_a^{\hat{e}_3}\wedge\pi(g_a)(\vec{n}^\beta_{ab},0),\pi(g_b)(\vec{n}^\beta_{ba},0)\rangle \\
\Rightarrow\star\left(N_a^{\hat{e}_3}\wedge N_b^{\hat{e}_3}\right) &= \langle N_a^{\hat{e}_3},\pi(g_b)(\vec{n}^\beta_{ba},0)\rangle\star\left(\pi(g_a)^{\wedge 2}(\vec{0},1)\wedge(\vec{n}^\beta_{ab},0)\right) ,
\end{aligned}
$$

and thus equation (III.3.34) implies via the $\delta_*$ duality

$$
D_{ab}g_a\left[(\xi\vec{n}^\beta_{ab})^I\sigma_I\right]g_a^{-1}D_{ab}^{-1} = g_a\left[(\xi\vec{n}^\beta_{ab})^I\sigma_I\right]g_a^{-1} .
$$

This is enough to argue that any transformation stabilizing the $N_a^{\hat{e}_3}, N_b^{\hat{e}_3}$ plane must take the form

$$
\begin{aligned}
D_{ab} &= \pm g_a\exp\left[\phi_{N_a^{\hat{e}_3},N_b^{\hat{e}_3}}(\xi\vec{n}^\beta_{ab})^I\frac{\sigma_I}{2}\right]g_a^{-1} \\
&= \pm g_a\exp\left[\phi_{N_a^{\hat{e}_3},N_b^{\hat{e}_3}}(n^\beta_{ab})^I F_I^\dagger\right]g_a^{-1} ,
\end{aligned} \tag{III.3.35}
$$

for some $\phi_{N_a^{\hat{e}3},N_b^{\hat{e}3}} \in \mathbb{R}$. Substituting (III.3.35) in equation (III.3.33) one arrives at

$$g_a^{-1} g_b \sigma_3 g_b^\dagger g_a^{-1\dagger} \sigma_3 = \exp\left[2\phi_{N_a^{\hat{e}3},N_b^{\hat{e}3}} (n_{ab}^\beta)^I F_I^\dagger\right] ,$$

and direct comparison with equation (III.3.32) finally establishes

$$\phi_{N_a^{\hat{e}3},N_b^{\hat{e}3}} = 2\mathrm{arg}\frac{\alpha_{ba}}{\alpha_{ab}} \tag{III.3.36}$$

to be the oriented dihedral angle between the hypersurfaces normal to $N_a^{\hat{e}_3}$ and $N_b^{\hat{e}_3}$.

## III.4 Induced geometry of critical configurations

Although the discussion of the previous two sections has been extensive, its purpose can be succinctly summarized in a couple of simple equations. For the reader's convenience I reproduce here those very same equations which characterize the critical points of the vertex amplitude. They are

$$\forall \text{ s.l. } a\,, \sum_b j_{ab}\vec{n}_{ab} = 0\,, \quad \forall \text{ t.l. } a\,, \sum_{b\text{: s.l.}ab} j_{ab}\vec{n}_{ab} + \sum_{b\text{: t.l.}ab} s_{ab}\vec{n}_{ab}^\beta = 0\,, \tag{III.4.1}$$

coming from (III.3.13), (III.3.16),

$$\begin{aligned} &\pi(g_a)^{\wedge 2} \star \left[(0,\vec{n}_{ab}) \wedge (1,\vec{0})\right] \\ &\qquad\qquad = \pi(g_b)^{\wedge 2} \star \left[(\vec{n}_{ba},0) \wedge (\vec{0},1)\right] , \quad a \text{ s.l.}, b \text{ t.l.}, ab \text{ s.l.}\,, \\ &\pi(g_a)^{\wedge 2} \star \left[(\vec{n}_{ab}^\beta,0) \wedge (\vec{0},1)\right] \\ &\qquad\qquad = \pi(g)^{\wedge 2} \star \left[(\vec{n}_{ba}^\beta,0) \wedge (\vec{0},1)\right] , \quad a \text{ t.l.}, b \text{ t.l.}, ab \text{ t.l.}\,, \end{aligned} \tag{III.4.2}$$

according to (III.3.20), (III.3.23), and finally

$$\begin{aligned} \phi_{N_a^{\hat{e}0},N_b^{\hat{e}3}} &= \ln\left|\frac{\lambda_{ba}}{\lambda_{ab}}\right|^2 + i\tau_{ba}\frac{\pi}{2}\,, \quad a \text{ s.l.}, b \text{ t.l.}, ab \text{ s.l.}\,, \\ \phi_{N_a^{\hat{e}3},N_b^{\hat{e}3}} &= 2\mathrm{arg}\frac{\alpha_{ba}}{\alpha_{ab}}\,, \quad a \text{ t.l.}, b \text{ t.l.}, ab \text{ t.l.}\,, \end{aligned} \tag{III.4.3}$$

as per equations (III.3.31), (III.3.36). All vectors - which are uniquely (with the exception of $\vec{n}^{\beta}_{ab}$, which is parametrized by $\Im\beta_{ab}$) determined by the boundary data - have been defined in the analysis of section III.3.1.

### III.4.1 Geometrical reconstruction

Recall that the aforementioned equations single out those points which are assumed to dominate the integral defining the amplitude of the theory. It is well-known that EPRL-type models behave in such a way that those configurations amount to geometrical constructions, and we are now in a position to see just how.

*I will always assume that the boundary data at every polyhedron is non-degenerate*, i.e. that the vectors $\{\vec{n}_{ab}\}_{b\neq a}$ (or $\{\vec{n}^{\beta}_{ab}\}_{b\neq a}$) at every $a$ are not all collinear; if non-degeneracy is not assumed the geometrical meaning of the critical points is diluted. Given such data at $a$, Minkowski's theorem for convex polyhedra, which I have generalized to Minkowski space in appendix D, Theorem D.2, states that the closure equation (III.4.1) uniquely determines a convex polyhedron in $\mathbb{R}^3$ or $\mathbb{R}^{1,2}$ up to isometries, in such a way that $\{\vec{n}_{ab}\}_{b\neq a}$ (or $\{\vec{n}^{\beta}_{ab}\}_{b\neq a}$) are the outward normals to the faces and $\{j_{ab}\}_{b\neq a}$ (or $\{s_{ab}\}_{b\neq a}$) correspond to the areas of those faces[39]. Our first observation is therefore that the amplitude is dominated by configurations which yield 3-dimensional polyhedra of the adequate causal character.

While the closure constraints refer exclusively to 3-dimensional objects, the gluing equations (III.4.2) refer to bivectors of 4-dimensional Minkowski space. Interpreting those bivectors in terms of the reconstructed polyhedra requires embedding them in $\mathbb{R}^{1,3}$, which we do in the following manner. Let $P^s_a$ stand for a reconstructed space-like polyhedron $a$, and $P^t_a$ for a time-like one. Denote by $P^{s,t}_{ab}$ the face orthogonal to $\vec{n}_{ab}$ (or $\vec{n}^{\beta}_{ab}$).

[39] The reader is referred to appendix D for the definition of Minkowski polyhedra, volumes and areas.

Consider the affine embeddings

$$\begin{gathered} P_a^s \subset \mathbb{R}^3 \hookrightarrow \hat{P}_a^s \subset \mathbb{R}^{1,3} \\ \text{s.t. } \hat{P}_a^s \perp N_a^{\hat{e}_0}\,,\ \hat{P}_{ab}^s \perp \pi(g_a)(0,\vec{n}_{ab})\,, \end{gathered}$$

and

$$\begin{gathered} P_a^t \subset \mathbb{R}^{1,2} \hookrightarrow \hat{P}_a^t \subset \mathbb{R}^{1,3} \\ \text{s.t. } \hat{P}_a^t \perp N_a^{\hat{e}_3}\,,\ \hat{P}_{ab}^t \perp \begin{cases} \pi(g_a)(\vec{n}_{ab},0) \text{ if } P_{ab}^t \perp \vec{n}_{ab}\,,\ ab \text{ s.l.} \\ \pi(g_a)(\vec{n}_{ab}^\beta,0) \text{ if } P_{ab}^t \perp \vec{n}_{ab}^\beta\,,\ ab \text{ t.l.} \end{cases}\,, \end{gathered}$$

subject to the requirement that $N_a^{\hat{e}_0}, N_a^{\hat{e}_3}$ are outward-pointing for every $a$, $\pi(g_a)(0,\vec{n}_{ab})$, $\pi(g_a)(\vec{n}_{ab},0)$ and $\pi(g_a)(\vec{n}_{ab}^\beta,0)$ are outward-pointing whenever $b > a$, and moreover $-\pi(g_a)(0,\vec{n}_{ab})$, $-\pi(g_a)(\vec{n}_{ab},0)$ and $-\pi(g_a)(\vec{n}_{ab}^\beta,0)$ are outward-pointing whenever $a > b$[40]. In this manner equation (III.4.2) determines that

$$\begin{gathered} \hat{P}_{ab}^s \parallel \hat{P}_{ba}^t\,,\ A_{ab} = A_{ba}\,,\quad \text{for } ab \text{ s.l.} \\ \hat{P}_{ab}^t \parallel \hat{P}_{ba}^t\,,\ A_{ab} = A_{ba}\,,\quad \text{for } ab \text{ t.l.}\,, \end{gathered}$$

having denoted by $A_{ab}$ the area of the polygonal face $P_{ab}$.

Solutions to the critical point equations therefore describe different ways of gluing 3-dimensional polyhedra along their faces according to some fixed combinatorial structure, prescribed by the boundary data and by the definition of the vertex amplitude. The areas of glued faces are required to agree, but since the identification is done at the level of normal vectors not so their polygonal shape. The resulting entity goes by the name of *twisted geometry* in the literature [67, 93]. The angles of (III.4.3) finally correspond to the dihedral angles between embedded polyhedra.

[40] This requirement is necessary in order for the bivector constraints to describe gluing along antiparallel vectors; this involves embedding reflections of the original polyhedra. The same goal is attained in [53] by considering boundary states which are already opposite-pointing.

A great deal more can be said regarding the classification of possible solutions to the critical equations; a particularly exhaustive study for the space-like case was done in [67]. Still I would like to make three straightforward remarks:

1. It is quite crucial that the geometric reconstruction of time-like faces is not uniquely determined by the boundary data, since as repeatedly mentioned the vectors $\vec{n}^{\beta}_{ab}$ depend on $\Im\beta_{ab}$.

2. Consider an amplitude modeled on the combinatorics of a 4-simplex, and the subset of solutions for which the five normals $N_a$ span $\mathbb{R}^{1,3}$. Then this subset can only contain isometries of the 4-simplex, since any 4-simplex is uniquely determined by its edge lengths and these are fixed when applying Minkowski's theorem in reconstructing the boundary tetrahedra. If the amplitude is based on any other convex polytope the argument does not hold; simplices are in some sense ideally rigid.

3. There may generally be solutions for which all polyhedra are identified in a lower dimensional subset of $\mathbb{R}^{1,3}$. Such configurations are known as *vector geometries* [53, 63, 64]. It is interesting to note that vector geometries cannot exist whenever a polytope contains two polyhedra of different causal characters, for in that case the two polyhedra embed into different 3-dimensional subsets of Minkowski space.

### III.4.2 Symmetries of the solutions

A benefit of the geometrical interpretation is that it allows for a straightforward analysis of the symmetries of the solutions to the critical point equations. To see how, observe first that the spin-foam actions (III.2.4), (III.2.9) are invariant under the following transformations, using the terminology of [53]:

1. Lorentz: a global action of $g \in \mathrm{SL}(2,\mathbb{C})$ at every $g_a \mapsto g\, g_a$, afforded

by the properties of the Haar measure. This symmetry is gauge-fixed by the Dirac delta in (III.1.1).

2. Spin lift: a local transformation of $g_a \mapsto -g_a$.

3. Spinor rescaling: a local transformation at each face $ab$ taking $z_{ab} \mapsto \kappa z_{ab}$, with $\kappa \in \mathbb{C}^*$.

These symmetries carry over to the equations themselves[41]. Regarding the bona-fide symmetries of the dominant configurations, note that among the isometries of Minkowski space-time only reflections are not contemplated above. Let then $\pi(g_a)\hat{e}_x$, $x = 0, 3$ stand for a polyhedron normal $N_a^{\hat{e}_0}$ or $N_a^{\hat{e}_3}$. A general reflection $R_v$ along a hypersurface orthogonal to $v$ transforms the normal as[42]

$$\begin{aligned}\pi(g_a)\hat{e}_x &= \pi(g_a) \star (e_1 \wedge e_2 \wedge e_3) \\ &= - \star\, \pi(g_a)^{\wedge 3}(e_1 \wedge e_2 \wedge e_3) \\ &\overset{R_v}{\mapsto} - \star\, (R_v\, \pi(g_a))^{\wedge 3}\,(e_1 \wedge e_2 \wedge e_3) \\ &= R_v\, \pi(g_a) R_{\hat{e}_x} \hat{e}_x\,,\end{aligned}$$

where $e_1, e_2, e_3$ are linearly independent edge vectors of the polyhedron with common base. As was argued in equation (III.3.1), such a reflection of the image of the spin-homomorphism induces a transformation of its object; one thus sees that a reflection along $\hat{e}_\mu$ of a given solution induces on the special linear matrices $g_a$ the transformation

$$R_\mu :\; g_a \mapsto (i^{1-\eta_{\mu 0}}\sigma_\mu) g_a^{-1\dagger}(i^{1-\eta_{x0}}\sigma_x)\,,$$

[41] Note that the special linear group elements $\pm g$ map to the same $\pi(g)$ and $\rho(g)$. There is an apparent additional local symmetry of the bivector equations (III.4.2) in taking $\pi(g_a) \mapsto -\pi(g_a)$, but since $-\pi(g_a) \notin \mathrm{SO}_0^+(1,3)$ this Lorentz transformation does not lift to an $\mathrm{SL}(2,\mathbb{C})$ element.

[42] Here I make use of the identity $g^{\wedge j} = -\det g \star g^{\wedge(n-j)} \star$, which holds for $\eta$-orthogonal linear maps on the exterior algebra of a real vector space of dimension $n$ with inner product $\langle u, v\rangle = u^T \eta v$. The proof uses the properties of $\star$ to show first that $a \wedge g^{\wedge(n-j)} b = b \wedge \star (g^{-1})^{\wedge(n-j)} \star a$. Using $g^{\wedge n} b \wedge a = \det g\, b \wedge a$ the result then follows.

dependent on whether the polyhedron $a$ is space-like ($x = 0$) or time-like ($x = 3$). These transformed matrices do indeed correspond to a second solution of the critical point equations (which may be equivalent to the first, as so happens for vector geometries). It is worthwhile analyzing this behavior explicitly in a case-by-case basis:

*(a) Every polyhedron is space-like*

The interfaces must be all achronal. This case was discussed in [53], from where the algebraic gluing equations adapted to our notation read

$$g_a|+_{ab}\rangle = \frac{\varsigma_{ba}}{\varsigma_{ab}} g_b|+_{ba}\rangle\,,$$
$$g_a^{-1\dagger}|+_{ab}\rangle = \frac{\overline{\varsigma}_{ab}}{\overline{\varsigma}_{ba}} g_b^{-1\dagger}|+_{ba}\rangle\,,$$

where $\varsigma_{ab}$ is the proportionality parameter in $|g_a^{-1} z_{ab}\rangle = \varsigma_{ab}|+_{ab}\rangle$. Because every polygon and polyhedron is space-like, every normal is of the type $\hat{e}_0$. By making use of the global Lorentz gauge we may consider the reflection $R_0$, and the resulting transformation can be understood as a parity operation (this is precisely the transformation found in [53]). One may then check that if $\{g_a, \varsigma_{ab}\}$ is a solution for given states $\{|+_{ab}\rangle\}$ then so is $\{(g_a^\dagger)^{-1}, \overline{\varsigma}_{ba}\}$ for the same states.

At the level of the bivector equations, geometrically more explicit, the existence of the second solution follows from the particular structure of the vectors associated with the boundary data, which remains invariant under $R_0$:

$$\pi(g_a)^{\wedge 2}(0, \vec{n}_{ab}) \wedge (1, \vec{0}) = \pi(g_b)^{\wedge 2}(0, \vec{n}_{ba}) \wedge (1, \vec{0})$$
$$\Leftrightarrow \quad [R_0\pi(g_a)R_0]^{\wedge 2}\,(0, \vec{n}_{ab}) \wedge (1, \vec{0}) = [R_0\pi(g_b)R_0]^{\wedge 2}\,(0, \vec{n}_{ba}) \wedge (1, \vec{0})$$
$$\Leftrightarrow \quad \pi(g_a^{-1\dagger})^{\wedge 2}(0, \vec{n}_{ab}) \wedge (1, \vec{0}) = \pi(g_b^{-1\dagger})^{\wedge 2}(0, \vec{n}_{ba}) \wedge (1, \vec{0})\,,$$

having used (III.3.1) in the last line.

*(b) Every polygon is time-like*

In this case, the parachronal bivector algebraic equations are

$$g_a|l^-_{ab}\rangle = \frac{\alpha_{ab}}{\alpha_{ba}} g_b|l^-_{ba}\rangle\,,$$
$$g_a^{-1\dagger}\sigma_3|l^-_{ab}\rangle = \frac{\overline{\alpha}_{ab}}{\overline{\alpha}_{ba}} g_b^{-1\dagger}\sigma_3|l^-_{ba}\rangle\,,$$

and all polyhedral normals are space-like of the form $\hat{e}_3$. One can consider an $R_0$ reflection, and check that any solution $\{g_a, \alpha_{ab}\}$ induces a second one $\{g_a^{-1\dagger} i\sigma_3, \overline{\alpha}_{ab}\}$, in agreement with what was found in [63]. In terms of bivectors,

$$\pi(g_a)^{\wedge 2}(\vec{n}^{\beta}_{ab}, 0) \wedge (\vec{0}, 1) = \pi(g_b)^{\wedge 2}(\vec{n}^{\beta}_{ba}, 0) \wedge (\vec{0}, 1)$$
$$\Leftrightarrow \quad [R_0\pi(g_a)R_3]^{\wedge 2}\,(\vec{n}^{\beta}_{ab}, 0) \wedge (\vec{0}, 1) = [R_0\pi(g_b)R_3]^{\wedge 2}\,(\vec{n}^{\beta}_{ba}, 0) \wedge (\vec{0}, 1)$$
$$\Leftrightarrow \quad \pi(g_a^{-1\dagger} i\sigma_3)^{\wedge 2}(\vec{n}^{\beta}_{ab}, 0) \wedge (\vec{0}, 1) = \pi(g_b^{-1\dagger} i\sigma_3)^{\wedge 2}(\vec{n}^{\beta}_{ba}, 0) \wedge (\vec{0}, 1)\,.$$

*(c) There exist both space- and time-like polyhedra*

Then at least some of the faces $ab$ will be described by the heterochronal equations

$$g_a|+_{ab}\rangle = \frac{\lambda_{ba}}{\lambda_{ab}} g_b|\tau_{ba}\rangle\,,$$
$$g_a^{-1\dagger}|+_{ab}\rangle = \frac{\overline{\lambda}_{ab}}{\overline{\lambda}_{ba}} g_b^{-1\dagger}\tau\sigma_3|\tau_{ba}\rangle\,.$$

A reflection of the polytope induces different transformations on $g_a$ and $g_b$, since these elements are associated to normals of the type $\hat{e}_0$ and $\hat{e}_3$, respectively. The individual transformations are as before, and given one solution $\{g_a, g_b, \lambda_{ab}, \lambda_{ba}\}$ to the equations, a second one can be constructed as $\{(g_a^{-1})^\dagger, (g_b^{-1})^\dagger i\sigma_3, -i\tau\overline{\lambda}_{ba}, \overline{\lambda}_{ab}\}$. Regarding the bivector equation

$$\begin{aligned} & \pi(g_a)^{\wedge 2}(0,\vec{n}_{ab}) \wedge (1,\vec{0}) = \pi(g_b)^{\wedge 2}(\vec{n}_{ba},0) \wedge (\vec{0},1) \\ \Leftrightarrow\quad & \left[R_0\pi(g_a)R_0\right]^{\wedge 2}(0,\vec{n}_{ab}) \wedge (1,\vec{0}) = \left[R_0\pi(g_b)R_3\right]^{\wedge 2}(\vec{n}_{ba},0) \wedge (\vec{0},1) \\ \Leftrightarrow\quad & \pi(g_a^{-1\dagger})^{\wedge 2}(0,\vec{n}_{ab}) \wedge (1,\vec{0}) = \pi(g_b^{-1\dagger} i\sigma_3)^{\wedge 2}(\vec{n}_{ba},0) \wedge (\vec{0},1)\,. \end{aligned}$$

This discussion reiterates for the EPRL-CH model the well-known fact that any spin-foam critical point induces a second one, geometrically related to the first via a reflection of its underlying geometry (possibly coinciding with the first). The existence of a second solution is at the root of the so-called *cosine problem*, first described in the simplicial achronal analysis of [53]: as mentioned before, in case the spin-foam model is derived from a simplicial triangulation, the critical point equations can determine a unique simplex up to isometries; a given set of achronal boundary data then induces two solutions related by reflection, and thus the asymptotic expansion of the spin-foam amplitude evaluates formally to

$$A_v \sim e^{iS_c} + e^{-iS_c} \sim \cos S_c\,,$$

where $S_c$ denotes the action at critical points. I will have more to say on this subject in closing the chapter.

## III.5 The action at critical points and concluding remarks

Sizable as it was, the main preoccupation of the present chapter has been determining and examining the critical point equations of the vertex am-plitude III.1.1; the end goal should however be to employ those results in an explicit formula for the asymptotic vertex amplitude through Theorem III.1, as I did in section II.5.2 for the 3d model, and as obtained in [53, 63] for the achronal, ortho- and heterochronal cases. Regrettably, we are confronted with two difficult problems for the parachronal action: 1) its critical points are not isolated, since there is a line of dominant configurations parametrized by $\Im\beta_{ab}$; and 2) the function (III.2.13) which precedes the action exponential, here reproduced[43],

$$f_{ab}(z_{ab}, g_a, g_b) = \frac{1}{2}\overline{\widetilde{A}}_{-\gamma\Lambda s_{ab}}^{-\frac{1}{2}+i\Lambda s_{ab}} A_{-\gamma\Lambda s_{ab}}^{-\frac{1}{2}+i\Lambda s_{ab}} \Theta(\Re\beta_{ab})\Theta(\Re\beta_{ba})$$
$$\cdot \left|\alpha_{ab}\alpha_{ba}\right|^{-2} (\Re\beta_{ab})^{-\frac{1}{2}} (\Re\beta_{ba})^{-\frac{1}{2}} \left[(2\Lambda s_{ab} - i)\,\Im\beta_{ab} - 2\gamma\Lambda s_{ab}\,\Re\beta_{ab}\right],$$

contains a branch point at $\Re\beta_{ab} = \Re\beta_{ba} = 0$. Both 1) and 2) are in contradiction with the conditions of Hörmander's theorem, which is consequently non-applicable. Results for such cases have been obtained in one dimension [123, 124], where the integrand contains branch points which are allowed to coalesce with critical points of the exponential function; to my knowledge no general theorem exists for higher dimensional integrals.

One can still argue on heuristic grounds that the stationary points obtained throughout this chapter should dominate the integral, and indeed the action at critical points behaves as one would expect. The total vertex action for parachronal and heterochronal interfaces at a single stationary point reads

$$S = i\gamma\left[\sum_{ab}^{\text{hetero.}} j_{ab}\,\ln\left|\frac{\lambda_{ba}}{\lambda_{ab}}\right|^2 + \sum_{ab}^{\text{para.}} s_{ab}\,2\arg\frac{\alpha_{ba}}{\alpha_{ab}}\right]$$
$$+\,i\left[\sum_{ab}^{\text{hetero.}} j_{ab}\,2\arg\frac{\lambda_{ab}}{\lambda_{ba}} + \sum_{ab}^{\text{para.}} s_{ab}\,\ln\left|\frac{\alpha_{ba}}{\alpha_{ab}}\right|^2\right],$$

[43] Recall we have set $\tau = \pm$ for all $\tau_{ab}$. For simplicity I take now $\tau = +$.

and equivalently, referring to the discussion of section III.3.5,

$$S = i\gamma \left[ \sum_{ab}^{\text{hetero.}} j_{ab}\, \Re\phi_{N_a^{\hat{e}_0}, N_b^{\hat{e}_3}} + \sum_{ab}^{\text{para.}} s_{ab}\, \phi_{N_a^{\hat{e}_3}, N_b^{\hat{e}_3}} \right] + i \left[ \sum_{ab}^{\text{hetero.}} j_{ab} \ln \mathbb{W}_{ab} + \sum_{ab}^{\text{para.}} s_{ab} \ln \mathbb{W}_{ab} \right] . \tag{III.5.1}$$

The first term in the above equation is precisely the Regge action weighted by the Immirzi parameter $\gamma$, which is a function of dihedral angles between polyhedra and face areas given by the spins. The parachronal amplitude is in this sense in agreement with the remaining cases. The second term involves the objects $\mathbb{W}_{ab}$, whose definition depends on the type of interface as follows[44]

$$\mathbb{W}_{ab} = \tau_{ab} \left\langle \pi(g_a n_{ab})[\hat{e}_1 - i\hat{e}_2],\, \pi(g_b n_{ba})[\hat{e}_1 + i\tau_{ab}\hat{e}_2] \right\rangle \ , \ a \text{ s.l.}, b \text{ t.l.}, \ ,$$

$$\mathbb{W}_{ab} = \left\langle \pi(g_a n_{ab})[\hat{e}_0 + \hat{e}_1],\, \pi(g_b n_{ba})[\hat{e}_0 + \hat{e}_1] \right\rangle \ , \quad ab \text{ t.l.} \, .$$

Note that $\mathbb{W}_{ab}$ amounts to a scalar product between pairs of vectors which lie on identified polygonal faces in the reconstructed geometry - the notation is meant to be suggestive; surprisingly, the semiclassical action is sensitive to the relative twisting of polygons between identified tetrahedra:

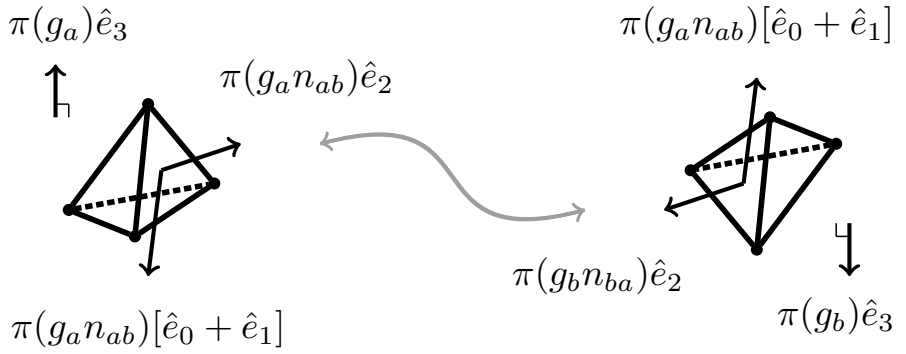

.

[44]Here the Minkowski bilinear $\langle \cdot, \cdot \rangle$ is complex-linear extended. The equations for $\mathbb{W}_{ab}$ follow straightforwardly from the algebraic gluing conditions (III.3.17) and (III.3.21) by observing that

$$\frac{\alpha_{ba}}{\alpha_{ab}} = [l^-_{ba}|g_b^{-1} g_a|l^+_{ab}] \, ,$$

$$\frac{\lambda_{ab}\overline{\lambda}_{ba}}{\lambda_{ba}\overline{\lambda}_{ab}} = \tau \left\langle +_{ab}|g_a^{-1} g_b|\tau_{ba}\right\rangle \left\langle -\tau_{ba}|g_b^{\dagger} g_a^{-1\dagger}|-_{ab}\right\rangle \, .$$

In the case of space-like interfaces the various $\mathbb{W}_{ab}$ will generally involve an imaginary unit, but its separate real and imaginary parts still admit the above interpretation.

In the absence of a final expression for the semiclassical parachronal amplitude, the principal result of the present chapter is in my judgment the development of analytical methods and insights for its study, hopefully useful once a better-behaved model is proposed. In closing I briefly restate the most important results, together with some final remarks.

*(a) New boundary states for time-like faces*

The parachronal amplitude defined in section III.1 and studied throughout this chapter is not the original one proposed by Conrady and Hnybida in [54], and for which [64] studied its asymptotics. The difference lies in the new boundary states $|j, \lambda = ij\rangle$, which I have initially proposed in the context of the 3d model of chapter III. The main benefit of using the new states is that the associated action (III.2.9) considerably simplifies; the original action discussed in [64] is given in terms of a hypergeometric function, which the authors had to first approximate before proceeding with the analysis. The usefulness of these states in both the 3- and 4-dimensional models reinforces their validity.

*(b) Minkowski's theorem in* $\mathbb{R}^{1,2}$

As far as I am aware, no theorem on the reconstruction of Minkowski convex polyhedra from its normal vectors and areas existed in the literature prior to [65]. The Minkowski theorem of appendix D allows for a geometrical understanding of Lorentzian spin-foams without appealing to the simplicial *bivector theorem* of Barrett [53], thus expanding the argument to general polytopes.

*(c) Fixing* $\Re\beta_{ab} = 0$ *for general polyhedra*

In attempting to constrain the parachronal action as much as possible, it was argued in [64] that $\Re\beta_{ab} = 0$ could be achieved if all tetrahedra of the simplicial model are required to contain at least one space-like triangle. This restriction is in direct conflict with the framework of Causal Dynamical Triangulations. I have shown it is possible to achieve the same goal without any demands on the causal character by instead demanding all parameters $\tau_{ab}$ to agree, which generalizes nicely to polytopes of higher valency; the value of $\tau_{ab}$ is geometrically inconsequential for time-like faces.

*(d) Remark 1: on the pathological behavior of time-like faces*

As I have frequently stated, the parachronal amplitude is particularly ill-behaved when compared to the remaining interfaces. On this point I must make two observations: 1) that the existence of a branch point in the integrand is common between the present model and the original one discussed in [64]; and 2) that the circumstance that the stationary points lie on a critical *line* (parametrized by $\Im\beta_{ab}$; also observed in [64] but argued to be due to a gauge freedom) is a phenomenon which also occurs in the 3d model (without the imposition of additional constraints). This makes it unlikely that the anomalous behavior is caused by the choice of boundary states; it seems instead more probable that it follows from the structure of the $\mathrm{SU}(1,1)$ continuous series, as suggested by the discussion of section II.5.2. It is not entirely out of the question that a more refined regularization procedure for continuous series states may address the branch point problem. It is also conceivable that additional ad-hoc constraints as in section II.5.2 could isolate the critical points. I believe these possibilities to be most promising in further developing 4d Lorentzian spin-foam models.

*(e) Remark 2: on complex dihedral angles and causality violations*

The notion of angle in $\mathbb{R}^{1,1}$ inevitably requires extending it to the complex plane (cf. appendix D). It is commonly accepted in the literature [115] that the Lorentzian Regge action should be privy to the imaginary part of dihedral angles, which keeps track of light-cone crossings; the action could then account for causality violations, i.e. configurations where more than one light-cone meet at a point. It may then seem strange that the Regge action (III.5.1) obtained from spin-foam asymptotics is entirely ignorant of such imaginary components. But one can heuristically see that it must be so: the spin-foam amplitude derives from a path-integral, and non-vanishing imaginary terms in the action must be comparatively suppressed. The inherent geometry of spin-foam models only manifests itself in the semiclassical regime, from where that geometry must be oblivious to the imaginary part of Lorentzian angles. Any eventual sensitivity of spin-foam models to causality violations must therefore manifest itself in the deep quantum regime, where some pre-geometric characterization is necessary.

### *(f) Remark 3: on simplicial shape matching and* $\mathbb{W}_{ab}$

It is well-known that among the possible solutions to the critical point equations there exist vector geometries. It is claimed in [125] that such vector geometries are characterized by a failure of shape-matching between two tetrahedra. On the other hand, I have argued that the semiclassical action is sensitive to relative axial rotations of neighboring tetrahedra. The additional $j_{ab} \ln \mathbb{W}_{ab}$ term was already present in Barrett's space-like analysis [53], but in a case by case basis: for a Lorentzian simplex solution it was shown to be a constant phase, and for an Euclidean simplex solution it was shown to be the Regge action for the simplex. The characterization of equation (III.5.1) contextualizes these results in a different light. Whenever the solution is a Lorentzian simplex the triangle shapes must match, so that the $\mathbb{W}_{ab}$ terms simply induce a phase factor; the overall semiclassical action is therefore the Regge action multiplied by the Immirzi parameter. When the solution is that of a vector geometry there are generally relative twists between tetrahedra, and in the case of an Euclidean simplex the $\ln \mathbb{W}_{ab}$ relate to its effective dihedral angles; since the proper dihedral angles $\phi_{N_a,N_b}$ vanish for a vector geometry, the overall action is in this case the Regge action without the Immirzi parameter.

*(g) Remark 4: on the cosine problem*

It is a frequent claim in the context of the EPRL simplicial model that the existence of two 4-simplex solutions together with their conjugated actions is problematic [53, 63], since the vertex amplitude then yields *the cosine* of the Regge action - not a simple exponential, as what would perhaps be desirable. On this topic I would like to point out that - since the vertex amplitude ultimately derives from a partition function, cf. the review of chapter I - it is somewhat expected that it contains *all possible* semiclassical configurations compatible with the boundary data. Indeed, the same boundary data is shared between a simplex and its reflection (see appendix E for a pictorial justification). If one insists on obtaining a partition function whose semiclassical limit involves a single simplex, it is therefore likely that the data at the boundary must be supplemented with additional information (e.g. an orientation), and the vertex amplitude must be made sensitive to that additional structure (see e.g. [126]).

# IV. Effective Gluing Constraints in Spin-Foam Models

*The centripetal force on our planet is still fearfully strong, Alyosha. I have a longing for life, and I go on living in spite of logic. Though I may not believe in the order of the universe, yet I love the sticky little leaves as they open in spring.*

— Ivan Karamazov, in The Brothers Karamazov (Fyodor Dostoevsky, 1880)

—— ▲ ——

That the spin-foam framework manages to conciliate a necessarily complex (and as of yet incomplete) description of quantum gravity with a rather concise and straightforward picture of quantum space-time as fuzzy geometry is certainly one of its most appealing features. The heuristic notion - afforded by the semiclassical analysis, as in chapters II and III - of a pre-geometric space-time of quantum shapes glued among themselves is a pervasive one; so pervasive, in fact, that the proposal of *effective spin-foams* [76–78] - one of the most prolific developments of the field in the recent past - is itself a direct implementation of that idea.

Effective spin-foams, usually taken as simplicial, start by having well-defined, proper 4-simplices at every vertex of the dual complex. Borrowing from the original formulation, the configuration variables are triangle areas, and neighboring tetrahedra are not required to match in shape; this is understood as a consequence of metric discontinuities or torsion [66, 127]. Shape-matching is only weakly enforced, and this is done via "gluing constraints", proposed to be Gaussian functions peaked on identified tetrahedra. Among the notions put forth by its authors is the idea that the effective implementation of such gluing constraints requires an inverse relationship between the Immirzi parameter $\gamma$ and the triangle spins $j$ [76], opening the door to a possibly running $\gamma = \gamma(j)$ in a renormalization context.

Insofar as effective spin-foams are still formulated as a quantum theory, they are not truly classical; in that they postulate bona-fide geometrical tetrahedra, they are not truly quantum. The proposal resides instead on a transitioning no-mans land, *navigating between Scylla and Charybdis*, as the original authors [76] so eloquently put it. The evident question then arises of whether it is possible to ground effective spin-foams - and in particular the notion of gluing constraints - on the complete quantum spin-foam model. Since the constraints are assumed to be peaked at shape-matching, doing so would require a more refined asymptotic analysis which would take into consideration further configurations beyond the critical ones. Succeeding in doing so would further confirm the heuristics discussed in the opening of this chapter.

I shall show that effective gluing constraints can be seen to arise from both SU(2) BF-theory and the EPRL model [75]; summarily that spin-foam tetrahedra are indeed *sticky*.

## IV.1 Gluing constraints

The original idea for the following definition of gluing constraints in full spin-foam models is due to Steinhaus in [75]. We restrict ourselves to spin-foams obtained from simplicial triangulations, such that the vertex amplitudes are associated to 4-simplices. The theories under analysis will all be space-like, in the sense that the boundary states are taken to be SU(2) coherent states.

### IV.1.1 General description

In the coherent state representation, the partition function associated to a 2-complex can be expressed as a product of coherent vertex amplitudes after doubling the gauge group integration and inserting resolutions of the identity on the representation spaces of that group, expressed in terms of coherent states (cf. chapter I). The partition function would then take the

schematic form

$$Z = \sum_{\{\chi_t\}} \int \prod_{(\sigma,\sigma')} \prod_t \dim \chi_t \int_{\mathrm{SU}(2)} \mathrm{d}h_t$$

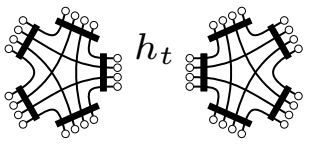

,

where $\chi_t$ refers to triangle $t$ representation labels, $(\sigma, \sigma')$ is a pair of simplices $\sigma, \sigma'$ and $h_t \in \mathrm{SU}(2)$ is the group element characterizing the coherent state associated to $t$. As one may see, the resulting vertex amplitudes are not all independent, as the same coherent data associated to a common edge is shared between a pair of vertices. There is however a straightforward way to gain better control over the mutual dependence of the vertices (denoted above by the script-size $h_t$). We can isolate the vertex amplitudes by inserting at the common tetrahedron a completeness relation in terms of coherent states (I.2.12),

$$Z = \sum_{\{\chi_t\}} \int \prod_{(\sigma,\sigma')} \prod_t \dim^2 \chi_t \int_{\mathrm{SU}(2)} \mathrm{d}h_t \mathrm{d}k_t$$

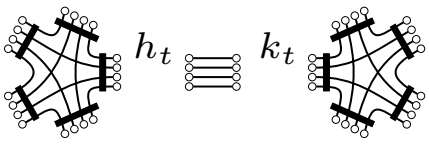

,

such that now every vertex amplitude carries its own coherent data at the boundary. In this manner we get for every (bulk) tetrahedron shared by a pair of 4-simplices a term which is represented by four lines in-between the amplitudes. It is a tensor product of pairings between coherent states associated to the triangles belonging to the pair of simplices, i.e.

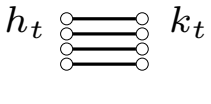

$$= \bigotimes_{t=1}^{4} \mathbb{1}_{\chi_t} \langle \chi_t, h_t | \chi_t, k_t \rangle \ , \qquad \text{(IV.1.1)}$$

where the brakets stand for the inner product on the relevant representation space $\mathcal{H}^{\chi_t}$.

As they stand, the objects of (IV.1.1) do not share any of the symmetry properties of the vertex amplitude. Since these are ultimately derived from the symmetries of the Haar measure, we may remedy this by proposing a refinement which involves an integration over the gauge group $G$ of the theory. We choose to do so, and consider instead

$$\mathcal{G}_\tau := \int_G \mathrm{d}g \bigotimes_{t=1}^{4} \mathbb{1}_{\chi_t} \langle Q \triangleright \chi_t, h_t | D^{\chi_t}(g_\tau) | \chi_t, k_t \rangle = {}^{h_t} \quad {}^{k_t} \,. \tag{IV.1.2}$$

It is to this object that we shall call the *gluing constraint* associated to the tetrahedron $\tau$. The map $Q$ is related to the quaternionic structure of equation (II.1.1); its action on coherent states will be explicitly discussed shortly. The inclusion of $Q$ follows the prescription of Barrett [53], and its purpose is to induce a geometric boundary vector with flipped sign, as shall be seen. Two remarks regarding the definition of (IV.1.2) are in order:

1. The particular symmetry properties of $\mathcal{G}_\tau$ depend on what one considers to be its domain of definition. Indeed, if one takes the domain of the constraints to be general group elements $h_t, k_t \in \mathrm{SU}(2)$, then the constraints are *invariant* under the action of $g, g' \in \mathrm{SU}(2)$ by virtue of the bi-invariance of the Haar measure, i.e. $\mathcal{G}_\tau(gh_t, g'k_t) = \mathcal{G}_\tau(h_t, k_t)$. On the other hand, if one considers the domain to be the coherent states themselves (defined up to a $\mathrm{U}(1)$ redundancy), then the gluing constraints are only invariant up to a phase factor (a property which extends to the coherent vertex amplitude), and this has been termed *covariance* in the literature [125]. The reason is that the group action on a coherent state results in a second coherent state up to a phase, as $\mathrm{SU}(2)/\mathrm{U}(1)$ is not closed under multiplication; it holds only that $|\mathcal{G}_\tau(g \triangleright \vec{h}_t, g' \triangleright \vec{k}_t)| = |\mathcal{G}_\tau(\vec{h}_t, \vec{k}_t)|$. In order to avoid having to deal with multiplicative phase factors we will always take the gluing constraints to be functions of $\mathrm{SU}(2)$.

2. Such objects have already made an appearance in the literature to some extent. The gluing constraint of equation (IV.1.2) involves two coherent intertwiners (not necessarily the same) associated to every bulk edge. In the special case when the two coherent intertwiners are the same, the gluing constraint becomes the norm of that coherent intertwiner. The asymptotics of the norm of coherent intertwiners has been studied in [51] for $G = \mathrm{SU}(2)$.

Having at hand a formal definition of the gluing constraints, we can now proceed to a more explicit characterization for two classes of models: $\mathrm{SU}(2)$ BF-theory and the EPRL model.

### IV.1.2 Gluing constraints for $\mathrm{SU}(2)$ $BF$ theory

As usual, we label the unitary and irreducible representations of $\mathrm{SU}(2)$ by spins $j \in \frac{\mathbb{N}}{2}$. The canonical basis states in $\mathcal{H}^j$ are the $L^3$ eigenstates $|j, m\rangle$ for (half-)integers $m = -j, \cdots, j$. We define a coherent state as the action of the group on the highest-weight state, i.e.

$$|j, h\rangle = D^j(h)|j, j\rangle\,, \quad h \in \mathrm{SU}(2)\,.$$

As pointed out in section I.2.3.(c), such coherent states admit an interpretation in terms of the sphere via the spin homomorphism $\pi$,

$$\langle j, h|\sigma_i|j, h\rangle = \left([\pi(h)\hat{e}_3]_i\right)^{2j}\,.$$

Using the standard Clebsh-Gordan isomorphism $|j, j\rangle \simeq |\frac{1}{2}, \frac{1}{2}\rangle^{2j}$, the quaternionic structure can be naturally made to act on $\mathcal{H}^j$ with $Q^2 = (-1)^{2j}$. The resulting coherent state then maps to

$$\langle Q \triangleright j, h|\vec{\sigma}|Q \triangleright j, h\rangle = \left([-\pi(h)\hat{e}_3]_i\right)^{2j}\,,$$

i.e. the symmetric of what would otherwise be the corresponding geometric vector. With all this in mind, the $\mathrm{SU}(2)$ gluing constraint for a tetrahedron $\tau$ is defined as

$$\mathcal{G}_\tau^{\mathrm{SU}(2)} := \int_{\mathrm{SU}(2)} \mathrm{d}g \bigotimes_{i=1}^{4} \langle Q \triangleright j_i, h_i\,| D^{j_i}(g)\,| j_i, k_i\rangle = {}^{h_i} \; \cdots \; {}^{k_i}\,, \tag{IV.1.3}$$

with $\mathrm{d}g$ standing for the normalized Haar measure.

The gluing constraints admit equivalent formulations in terms of related objects. Referring again to the Clebsh-Gordan map, one has just as well

$$\mathcal{G}_\tau^{\mathrm{SU}(2)} := \int_{\mathrm{SU}(2)} \mathrm{d}g \prod_{i=1}^{4} \langle Q h_i \,| g \,| k_i \rangle^{2 j_i} \;, \qquad \text{(IV.1.4)}$$

where $g$ is now in the defining representation and $|h\rangle := h|+\rangle$. In terms of the intertwiner basis

$$|j_1, .., j_4; \iota\rangle = \sum_{m_i} \begin{pmatrix} j_1 & j_2 & j_3 & j_4 \\ m_1 & m_2 & m_3 & m_4 \end{pmatrix}^{(\iota)} |j_1, m_1\rangle ... |j_4, m_4\rangle \,,$$

$$\int \mathrm{d}g \bigotimes_{i=1}^{4} D^{j_i}(g) = \sum_{\iota} \mathrm{dim}\iota \, |\{j_i\}; \iota\rangle\langle\{j_i\}; \iota| \,,$$

obtained from the canonical basis via the Wigner $4jm$ symbols, $\mathcal{G}_\tau^{\mathrm{SU}(2)}$ can also be written as

$$\mathcal{G}_\tau^{\mathrm{SU}(2)} = \sum_{\iota} \mathrm{dim}\iota \;\; \langle \{j_i\}, \{h_i\} | \{j_i\}; \iota\rangle \, \langle\{j_i\}; \iota | \{j_i\}, \{k_i\}\rangle$$

$$= \sum_{\iota} \mathrm{dim}\iota \; {}^{h_i}$$ 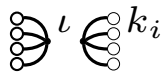 $${}^{k_i} \;;$$

the diagrams on the right-hand side are frequently termed *coherent intertwiners*. This latter characterization is particularly useful for numerical applications, where efficient methods for computing Wigner symbols are available [75].

### IV.1.3 Gluing constraints for the Lorentzian EPRL model

The underlying gauge group for the EPRL spin-foam is $\mathrm{SL}(2, \mathbb{C})$. The relevant representation theory is succinctly discussed in appendix B, to which the reader is referred. The explicit expression for an SU(2) coherent state as an homogeneous function in $\mathcal{D}^\chi$, $\chi = (2\gamma j, 2j)$ was already obtained in equation (III.1.2), reading

$$|j, h\rangle \sim \psi_h^j(z) = (2j+1)^{\frac{1}{2}} \, \langle z|z\rangle^{j(i\gamma - 1) - 1} \, \langle z|h\rangle^{2j} \;, \quad h \in \mathrm{SU}(2) \,,$$

where we persist with the notation $|h\rangle := h|+\rangle$. The action of the quaternionic structure is defined following [53][45]: setting $Q \triangleright \psi := \overline{\mathcal{A}\psi}$ for the intertwiner $\mathcal{A} : \mathcal{D}^\chi \to \mathcal{D}^{-\chi}$ of equation (B.1.2), one can show that

$$Q \triangleright \psi_h^j(z) = \mathcal{N}(2j+1)^{\frac{1}{2}} \langle z|z\rangle^{j(i\gamma-1)-1} \langle z|Qh\rangle^{2j} \ , \quad h \in \mathrm{SU}(2)\,,$$

with $\mathcal{N}$ a numerical factor depending on the choice of normalization of $\mathcal{A}$. It is common to take $\mathcal{N} = (-1)^{2j} e^{-i\arctan\gamma}$, $\gamma > 0$, a convention adopted in the aforementioned work. The gluing constraint for a tetrahedron $\tau$ thus reads[46]

$$\begin{aligned} \mathcal{G}_\tau^{\mathrm{SL}(2,\mathbb{C})}(h_i, k_i) = \int \mathrm{d}g \prod_{i=1}^4 \int_{\mathbb{C}P} \omega'(z_i) \\ \cdot \frac{\langle g z_i, g z_i\rangle^{j_i(i\gamma-1)-1}}{\langle z_i, z_i\rangle^{j_i(i\gamma+1)+1}} \langle g z_i, k_i\rangle^{2j_i} \langle Q h_i, z_i\rangle^{2j_i} \ , \end{aligned} \tag{IV.1.5}$$

where $\mathrm{d}g$ is the Haar measure in $\mathrm{SL}(2,\mathbb{C})$ as normalized in (B.3.1), and $\omega'(z)$ stands for a multiple of the $\mathcal{D}^\chi$ measure,

$$\omega'(z) = \overline{\mathcal{N}} \frac{2j+1}{\pi} \frac{i}{2} (z_1 \mathrm{d}z_2 - z_2 \mathrm{d}z_1) \wedge (\bar{z}_1 \mathrm{d}\bar{z}_2 - \bar{z}_2 \mathrm{d}\bar{z}_1)\,,$$

as per equation (B.1.1).

This concludes the definition of the gluing constraints for our cases of interest.

## IV.2 Asymptotics away from critical points

Equations (IV.1.3) and (IV.1.5) explicitly define the gluing constraints proposed formally in the beginning of this chapter. They implement the heuristics suggested above insofar as they ascribe to every vertex amplitude its own boundary data, and they account to the degree in which the simplices match by their dependence on that data; indeed, one can already see from (IV.1.1) that the gluing constraints ought to be peaked on configurations where the left and right coherent states agree.

[45] The usual notation uses a calligraphic $\mathcal{J}\psi = \overline{\mathcal{A}\psi}$.

[46] The integrand depends on $|g z_i\rangle$ rather than $|g^{-1} z_i\rangle$. The latter would be expected from the representation $D^\chi(g)\psi(z) = \psi(g^{-1}z)$. The Haar measure is however inversion-invariant for unimodular groups.

As mentioned before, the problem of extracting useful semiclassics from the gluing constraints can not be surmounted by Hörmander's first theorem III.1, as it makes no claims on the asymptotic value of integrals away from stationary configurations; but the critical points of the different $\mathcal{G}_\tau$ are a triviality. Fortunately, in the very same work where theorem III.1 was first formulated, Hörmander proves a second asymptotic theorem for integrals subject to free parameters, and this is precisely what is required for the problem at hand. Its application to the gluing constraints, besides clarifying their role in the context of the pseudo-quantum regime of effective spin-foams, serves as a first exploration of asymptotic analysis of spin-foams away from the critical points; more on this will be said later on.

### IV.2.1 Hörmander's second theorem

I shall make thorough use of Hörmander's second theorem [119, Theorem 7.7.12] on the asymptotic evaluation of integrals subject to free parameters, which I reproduce here for the reader's convenience.

**Theorem IV.1** (Hörmander II). *Let $S(x,y)$ be smooth and complex-valued in a neighborhood $K$ of $(0,0) \in \mathbb{R}^{n+m}$, such that $\Im S \geq 0$, $\Im S(0,0) = 0$, $S'_x(0,0) = 0$ and $\det S''_{xx}(0,0) \neq 0$. Consider furthermore $u \in C_0^\infty(K)$. Then*

$$\int dx\; u(x,y) e^{i\lambda S(x,y)} = \left(\frac{2\pi i}{\lambda}\right)^{n/2} \frac{u^0(y) e^{i\lambda S^0(y)}}{\sqrt{\left(\det S''_{xx}\right)^0(y)}} + \mathcal{O}\left(\lambda^{-n/2-1}\right), \tag{IV.2.1}$$

*where the superscript $f^0(y)$ denotes an $x$-independent residue in the residue class of $f(x,y) \bmod \mathcal{I}$, for $\mathcal{I}$ the ideal generated by the partial derivatives $\partial_{x^i} S$.*

To be clear, one says that $r$ is in the residue class $[a]_b$ if $a = r \bmod b$. Thus the possible $f^0(y)$ are defined by expansions of the type

$$f(x,y) = f^0(y) + \frac{\partial S}{\partial x^i}(x,y)q^i(x,y)\,, \tag{IV.2.2}$$

for $n$ smooth functions $q^i(x,y)$. The term $f^0$ is not unique, but another choice of representative will induce a correction of the same order as the error term in equation (IV.2.1). That $f(x,y)$ can be brought into this form near the origin is a consequence of the Malgrange preparation theorem [119], which can be thought of as a division theorem with remainder $f^0(y)$. As it stands, however, the explicit form of the coefficients in equation (IV.2.2) are difficult to obtain. We remedy this by resorting to yet another lemma of [119].

**Lemma IV.1.** *Let $b_j(x,y),\ j = 1,...,n$, be smooth and complex-valued in a neighborhood $K$ of $(0,0) \in \mathbb{R}^{n+m}$, such that $b_j(0,0) = 0$ and $\det \partial_{x^i} b_j \neq 0$. Then*

$$\mathcal{I}(b_1,...,b_n) = \mathcal{I}(x_1 - X_1(y),...,x_n - X_n(y))$$

*for some $X_j \in \mathcal{C}^\infty$ vanishing at the origin.*

Making use of the Malgrange preparation theorem (IV.2.2) $N$ times, and then applying lemma IV.1, equation (IV.2.2) may be written as

$$f(x,y) = \sum_{|\alpha|<N} f^\alpha(y)\,(x - X(y))^\alpha \operatorname{mod}\mathcal{I}^N\,, \tag{IV.2.3}$$

for some smooth $f^\alpha(y)$ near the origin and
$\mathcal{I}^N = \{\sum_{|\alpha|=N} s^\alpha\,(x - X(y))^\alpha \mid s^\alpha \in \mathcal{C}^\infty(K)\}$. Note that $\alpha = (\alpha_1,...,\alpha_n)$ stands for a multi-index set, and $|\alpha| = \sum_j \alpha_j$. Note furthermore that the $|\alpha| = 1$ term is an element of $\mathcal{I}^N$, and thus it can be omitted from equation (IV.2.3). This $x$-polynomial representation of $f(x,y)$ will be useful in what follows, as it allows a direct comparison with the function's Taylor series.

### IV.2.2 Asymptotics of SU(2) gluing constraints

The SU(2) case is the simplest among the ones we will study, so I start by deriving its asymptotic behavior first. As a function of the boundary data for a certain choice of spins, the gluing constraint of equation (IV.1.3) may be brought into the form

$$\mathcal{G}_{j_i}(h_i, k_i) = \int_{\mathrm{SU}(2)} \mathrm{d}g \, \exp\left\{ \sum_{i=1}^{4} \ln \langle Q \triangleright j_i, h_i \,|\, D^{j_i}(g) \,|\, j_i, k_i \rangle \right\} . \tag{IV.2.4}$$

*(a) Malgrange expansion*

In order to identify the residue in equation (IV.2.4), we first approximate the exponent (which I will refer to as the *action*) by a second order Taylor series using coordinates $g^I$, $I = 1, 2, 3$,

$$\begin{aligned} S(g, y) =& S(g_c, y) + \partial_I S(g_c, y)(g - g_c)^I \\ &+ \frac{1}{2} \partial^2_{IJ} S(g_c, y)(g - g_c)^I (g - g_c)^J + \mathcal{O}(g^3) \,, \end{aligned} \tag{IV.2.5}$$

where we let $y$ stand collectively for the boundary data and $g_c$ is a critical point of $S(g, y)$ at some particular configuration $y_c$ of the boundary, i.e. $\partial_I S(g_c, y_c) = 0$. The $N = 3$ Malgrange expansion (IV.2.3) of the same function reads

$$S(g, y) = S^0(y) + S^2_{IJ}(y)(g - X(y))^I (g - X(y))^J \mod \mathcal{I}^3 \,,$$

and, after matching monomials in $g$ and disregarding elements in $\mathcal{I}^3$, one finds

$$\begin{cases} S^2_{IJ}(y) = \frac{1}{2} H_{IJ} \,, \\ S^2_{IJ}(y) X^I(y) = -\frac{1}{2} \left( \partial_J S(g_c, y) - H_{IJ} g_c^I \right) \,, \\ S^0(y) = S(g_c, y) - \frac{1}{2} \partial^I S(g_c, y) \, \partial^J S(g_c, y) \, H^{-1}_{IJ} \,, \end{cases} \tag{IV.2.6}$$

where $H_{IJ} = \partial^2_{IJ} S(g_c, y)$ stands for the Hessian matrix. These equations uniquely specify a representative $S^0$ in the residue class $[S]_{\mathcal{I}}$.

*(b) Haar measure and coordinates in $SU(2)$*

Performing the Haar integral in equation (IV.2.4), as well as computing the derivatives appearing in (IV.2.6), requires one to choose coordinates $\phi_i : U_i \in G \to \mathbb{R}^3$ on the group manifold $G = \mathrm{SU}(2)$. This we do by resorting to the discussion on adapted coordinates of section III.2.1. Picking a chart $g^I$ such that

$$\partial_I g = \frac{i}{2}\sigma_I g \,,$$

it is straightforward to show that the normalized Haar measure reads

$$\mathrm{d}g = (4\pi)^{-2}\mathrm{d}g^1 \wedge \mathrm{d}g^2 \wedge \mathrm{d}g^3 \,.$$

*(c) Asymptotic formula*

Recall that through the Clebsch-Gordan isomorphism one has the identification $|j,j\rangle = |\frac{1}{2},\frac{1}{2}\rangle^{2j}$. The action of equation (IV.2.4) may thus be rewritten as

$$S(g,y) = \sum_{i=1}^{4} 2j_i \ln \langle Qh_i|\, g \,|\, k_i\rangle \ ,$$

where $|k_i\rangle = k_i|+\rangle$ and $|Qh_i\rangle = h_i|-\rangle$. A critical point is characterized by a vanishing $\partial_I S$ derivative, which reads

$$\begin{aligned}\partial_I S(g,y) &= i\sum_{i=1}^{4} j_i \frac{\langle Qh_i\,|\,\sigma_I g\,|\,k_i\rangle}{\langle Qh_i\,|\,g\,|\,k_i\rangle} \\ &= i\sum_{i=1}^{4} j_i \frac{\left[\pi(g)\vec{k}_i - \vec{h}_i - i\pi(g)\vec{k}_i \times \vec{h}_i\right]^{(I)}}{1-\pi(g)\vec{k}_i\cdot\vec{h}_i} \ ,\end{aligned} \qquad \text{(IV.2.7)}$$

having made repeated use of the spin homomorphism $\pi$ of section III.3.1. All vectors in the equation above are defined as

$$\vec{h}_i = \pi(h_i)\hat{e}_3 \,, \quad \vec{k}_i = \pi(k_i)\hat{e}_3 \,.$$

On the other hand, the real part of the action satisfies

$$\mathrm{Max}\,\Re S \leq \sum_{i=1}^{4} 2j_i \ln\,\mathrm{Max}\,|\,\langle Qh_i|\,g\,|\,k_i\rangle\,| = \sum_{i=1}^{4} 2j_i \ln 1\,,$$

so that $\Re S \leq 0$ (as required by theorem IV.1) and $\Re S = 0$ is attained at

$$g|k_i\rangle = e^{i\phi_i}|Qh_i\rangle\,, \quad \phi_i \in [0, 2\pi)\,. \tag{IV.2.8}$$

Under this condition, equation (IV.2.7) reduces to the expected closure relation for both sets $\{h_i\}$, $\{k_i\}$ of boundary data. By making use of the quaternionic structure $Q$ we may characterize $g$ through the eigensystem

$$\begin{cases} g|k_i\rangle = e^{i\phi_i}|Qh_i\rangle \\ g|Qk_i\rangle = -e^{-i\phi_i}|h_i\rangle \end{cases} \quad \Rightarrow \quad g = e^{-i\phi_i\vec{\sigma}\cdot\vec{h}_i}h_i(-i\sigma_2)k_i^\dagger\,,\ \forall i\,, \tag{IV.2.9}$$

and, using again the spin homomorphism, one can show

$$\pi(g)\vec{k}_i = -\vec{h}_i\,,$$

i.e. there are critical points if the boundary vectors can be rotated into each other.

Suppose now that equation (IV.2.9) admits at least one solution, labeled by $\{\hat{g}, \hat{\phi}_i\}$. Then it must be the case that, for any other solution $\{g, \phi_i\}$,

$$e^{-i(\phi_i-\hat{\phi}_i)\vec{\sigma}\cdot\vec{h}_i} = e^{-i(\phi_j-\hat{\phi}_j)\vec{\sigma}\cdot\vec{h}_j}\,, \quad i \neq j\,.$$

Making the simplifying assumption that *the boundary vectors are not all colinear*, the previous set of equations implies $\phi_i = \hat{\phi}_i \vee \phi_i = \hat{\phi}_i + \pi$ mod $2\pi$. For non-colinear boundary data, then, there exist two critical point configurations given by

$$g_c = \pm\, e^{-i\hat{\phi}_i\vec{\sigma}\cdot\vec{h}_i}h_i(-i\sigma_2)k_i^\dagger\,, \tag{IV.2.10}$$

and this holds for any $i = 1, ..., 4$.

Although the critical points $g_c$ were determined for boundary data $h_i, k_i$, the gluing constraints are only defined up to a global $\mathrm{SU}(2)$ gauge afforded by the bi-invariance of the Haar measure. The element $g_c$ in equation (IV.2.10) is thus actually gauge-dependent, and one is free to make the simplifying choice $g_c = \pm\mathbb{1}$ by appropriately rotating the boundary. A straightforward computation then fixes the coefficients of the Malgrange expansion for non-colinear arbitrary data as follows:

$$\begin{cases} H_{IJ}(y) = -\frac{1}{2}\sum_i j_i \left(\delta_{IJ} - V_I^i(y)V_J^i(y)\right) , \\ \partial_I S(y) = i j_i V_I^i(y) , \\ S^0(y) = S(\mathbb{1}, y) + \frac{1}{2} j_k j_l V_I^k(y)(H^{-1})^{IJ}(y)V_J^l(y) , \end{cases}$$

having further defined

$$\vec{V}_i(y) = \frac{\vec{k}_i - \vec{h}_i - i\vec{k}_i \times \vec{h}_i}{1 - \vec{k}_i \cdot \vec{h}_i} .$$

Finally, observe that, since we considered a series expansion of $S(g, y)$ to second order in equation (IV.2.5), the argument of the square root in theorem IV.1 approximates to $(\det H)^0(y) = (\det H)(\mathbb{1}, y)$. The resulting asymptotic expansion therefore reads

$$\begin{aligned} \mathcal{G}_{j_i}(h_i, k_i) \simeq & \frac{(1 + (-1)^{2\sum_i j_i}) \prod_i \langle Q h_i \,|\, k_i \rangle^{2 j_i}}{\sqrt{32\pi}\sqrt{-\det H}} \\ & \cdot \exp\left\{ \frac{1}{2} j_k j_l V_I^k (H^{-1})^{IJ} V_J^l \right\} , \end{aligned} \tag{IV.2.11}$$

where the prefactor $(-32\pi)^{-1/2} = (4\pi)^{-2} \cdot (-2\pi)^{3/2}$ is obtained from the normalization of the Haar measure and from the numerical factor of theorem IV.2.1, respectively.

We have thus obtained an explicit expression for the $\mathrm{SU}(2)$ gluing constraints for arbitrary boundary data. As expected, the result is proportional to an exponential function which is maximal when its argument vanishes, i.e. when $\vec{h}_i = -\vec{k}_i$. Before discussing this result further, however, we turn first to the Lorentzian theory.

### IV.2.3 Asymtpotics of EPRL gluing constraints

*Below the notation $\vec{u} \cdot \vec{v}$ will always refer to the Euclidean pairing.*

Recall the Lorentzian gluing constraints from equation (IV.1.5),

$$\mathcal{G}_{j_i}(h_i, k_i) = \int \mathrm{d}g \prod_{i=1}^{4} \int_{\mathbb{C}P} \omega'(z_i)$$
$$\cdot \frac{\langle gz_i, gz_i \rangle^{j_i(i\gamma-1)-1}}{\langle z_i, z_i \rangle^{j_i(i\gamma+1)+1}} \langle gz_i, k_i \rangle^{2j_i} \langle Qh_i, z_i \rangle^{2j_i} ,$$

to which the arguments laid down in section IV.2.2 apply with little modification. We start by rewriting the constraints in a form adapted to asymptotic analysis,

$$S(g, z, y) = \sum_{i=1}^{4} 2j_i \left( \ln \langle gz_i, k_i \rangle \langle Qh_i, z_i \rangle + \ln \frac{\langle gz_i, gz_i \rangle^{\frac{i\gamma-1}{2}}}{\langle z_i, z_i \rangle^{\frac{i\gamma+1}{2}}} \right) , \tag{IV.2.12}$$

$$u(g, z) = \mathrm{d}g \prod_{i=1}^{4} \frac{\omega'(z_i)}{\langle z_i, z_i \rangle \langle gz_i, gz_i \rangle} ,$$

$$\mathcal{G}_{j_i}(h_i, k_i) = \int_{\mathrm{SL}(2,\mathbb{C})} \int_{\mathbb{C}P} u(g, z) e^{S(g,z,y)} ,$$

with a slight abuse of notation in the definition of $u(g, z)$; I still refer by $y$ to the collection of external parameters $h_i, k_i$. Once more it is seen from (IV.2.12) that $\Re S \leq 0$, and the maximum is attained at

$$|gz_i\rangle = \lambda_i |k_i\rangle , \quad |z_i\rangle = \chi_i |Qh_i\rangle , \quad \lambda_i, \chi_i \in \mathbb{C} , \tag{IV.2.13}$$

moreover implying

$$|Qh_i\rangle = \frac{\lambda_i}{\chi_i} g^{-1} |k_i\rangle . \tag{IV.2.14}$$

The critical points of $S(g, z, y)$ may be straightforwardly characterized through the first derivatives in the spinor and group variables; as in section

III.2.1, we pick adapted real coordinates $g^I$, $I = 1, ..., 6$ for the special linear group such that

$$\partial_I g = \frac{i}{2} \Sigma_I g \,, \quad \Sigma_I = (\sigma, i\sigma)_I \,.$$

One then finds

$$\partial_I S = \sum_{i=1}^{4} i j_i \left[ \frac{i\gamma - 1}{2} \frac{\langle g z_i, (\Sigma_I - \Sigma_I^\dagger) g z_i \rangle}{\langle g z_i, g z_i \rangle} - \frac{\langle g z_i, \Sigma_I^\dagger k_i \rangle}{\langle g z_i, k_i \rangle} \right] ,$$

$$\partial_{z_i^a} S = 2 j_i \left[ \frac{\langle Q h_i, a \rangle}{\langle Q h_i, z_i \rangle} + \frac{i\gamma - 1}{2} \frac{\langle g z_i, g a \rangle}{\langle g z_i, g z_i \rangle} - \frac{i\gamma + 1}{2} \frac{\langle z_i, a \rangle}{\langle z_i, z_i \rangle} \right] ,$$

for $z_i^a$ the $a$th component $\langle a, z_i \rangle$. We remark that, due to the conjugation property of Wirtinger derivatives, if $\Re S = 0$ then $\partial_{z_i^a} S = 0 \Leftrightarrow \partial_{\bar{z}_i^a} S = 0$. Under equation (IV.2.13), a vanishing gradient $\partial S = 0$ reduces to the two equations

$$\sum_{i=1}^{4} j_i \vec{k}_i = 0 \,, \quad |Q h_i\rangle = \frac{\overline{\chi}_i}{\overline{\lambda}_i} g^\dagger |k_i\rangle \,, \qquad \text{(IV.2.15)}$$

and, as expected, one identifies a closure condition in the first equation above (holding equivalently for $\vec{h}_i$ by virtue of the second equation).

*(a) Rotation of the boundary data*

Equations (IV.2.14) and (IV.2.15) admit a similar treatment as that of the previous section. A general element $g \in \mathrm{SL}(2, \mathbb{C})$ can be polar-decomposed in terms of a pure boost $b = e^{\vec{\beta} \cdot \vec{\sigma}}$ and a unitary $a \in \mathrm{SU}(2)$ as $g = ba$. One may then combine those equations into the eigenvalue condition

$$b |k_i\rangle = \left| \frac{\lambda_i}{\chi_i} \right| |k_i\rangle \,,$$

from where it follows

$$\begin{cases} b|k_i\rangle = \left|\frac{\lambda_i}{\chi_i}\right| |k_i\rangle \\ b|Q k_i\rangle = \left|\frac{\chi_i}{\lambda_i}\right| |Q k_i\rangle \end{cases} \Rightarrow \quad b = e^{\ln\left|\frac{\lambda_i}{\chi_i}\right| \vec{k}_i \cdot \vec{\sigma}} \,, \ \forall i \,. \qquad \text{(IV.2.16)}$$

If one assumes that the boundary data corresponds to non-colinear vectors, then, through (IV.2.16), equations (IV.2.14) and (IV.2.15) imply

$$g|Qh_i\rangle = \frac{\lambda_i}{\chi_i}|k_i\rangle \quad g \in \mathrm{SU}(2)\,, \qquad \left|\frac{\lambda_i}{\chi_i}\right| = 1\,,$$

a result entirely analogous to the critical point condition (IV.2.8) of the $\mathrm{SU}(2)$ model, such that the discussion there can be immediately carried over. Given that $\mathrm{SU}(2)$ is a subgroup of $\mathrm{SL}(2,\mathbb{C})$, and the Haar measure induces an $\mathrm{SL}(2,\mathbb{C})$ symmetry at the level of the boundary data, we may again pick a convenient gauge with $g_c = \pm\mathbb{1}$.

*(b) Haar measure normalization*

The Haar measure for locally-compact groups is only unique up to a multiplicative constant. I follow the normalization of Rühl [42], for which the measure in adapted coordinates was already computed in section III.2.1. It reads

$$\mathrm{d}g = \frac{1}{(4\pi)^4}\mathrm{d}g^1 \wedge \ldots \wedge \mathrm{d}g^6$$

for the six real coordinates $g^I$.

*(c) Hessian matrix*

The $z$-integral in equation (IV.1.5) requires a choice of section in $\mathbb{C}^{2*}$, which has up to now remained unspecified. To simplify calculations I shall fix that section before taking derivatives. Since the set $\{|k_i\rangle, |Qk_i\rangle\}$ spans $\mathbb{C}^2$, we are free to pick

$$|z_i\rangle = |k_i\rangle + \beta|Qk_i\rangle\,,$$

where $\beta \in \mathbb{C}$; this choice considerably simplifies the discussion. Note that this is indeed a global section of the bundle $\mathbb{C}^{2*} \to \mathbb{C}\mathrm{P}$, since - under a judicious choice of the range of $\beta$ - it crosses every line through the origin only once. Such a choice further restricts the complex parameters to $\lambda_i = \pm 1$ (given $g_c = \pm\mathbb{1}$), and equation (IV.2.13) determines $\beta_c = 0$.

Below all relevant derivatives to the asymptotic analysis are listed (including only the symmetric part of second derivatives). Denoting by $S_c$ an evaluation at the critical point $S(g_c, \beta_c, y)$, one has

$$\partial_{\overline{\beta}_i} S_c = 0\,, \quad \partial_{\beta_i} S_c = -2\Theta_i\,, \quad \partial_I S_c^R = -i\Gamma_I\,, \quad \partial_I S_c^B = -i\gamma\Gamma_I\,,$$

$$\partial^2_{\overline{\beta}_i} S_c = 0\,, \quad \partial^2_{\beta_i} S_c = -2\frac{\Theta_i^2}{j_i}\,, \quad \partial^2_{\beta_i\overline{\beta}_i} S_c = -2j_i\,,$$

$$\partial^2_{I\beta_i} S_c^R = 0\,, \quad \partial^2_{I\beta_i} S_c^B = j_i(1-i\gamma)\kappa_{iI}\,, \quad \partial^2_{I\overline{\beta}_i} S_c^R = -ij_i\overline{\kappa}_{iI}\,,$$

$$\partial^2_{I\overline{\beta}_i} S_c^B = -i\gamma j_i\overline{\kappa}_{iI}\,, \quad \partial^2_{IJ} S_c^{RR} = -\frac{1}{2}\sum_i j_i\left(\delta_{IJ} - k_{iI}k_{iJ}\right)\,,$$

$$\partial^2_{IJ} S_c^{BR} = \frac{i}{2}\sum_i j_i\left(\delta_{IJ} - k_{iI}k_{iJ}\right)\,, \partial^2_{IJ} S_c^{RB} = \frac{i}{2}\sum_i j_i\left(\delta_{IJ} - k_{iI}k_{iJ}\right)\,,$$

$$\partial^2_{IJ} S_c^{BB} = \frac{2i\gamma - 1}{2}\sum_i j_i\left(\delta_{IJ} - k_{iI}k_{iJ}\right)\,.$$

Three objects were additionally defined in the equations above. They are 1) an auxiliary complex vector

$$\kappa_{iI} := \langle k_i|\sigma_I|Qk_i\rangle = \left[\pi(k_i)(\hat{e}_1 - i\hat{e}_2)\right]_I\ ;$$

2) a function which vanishes on closing tetrahedra, coming from the first group derivative,

$$\Gamma^I := \sum_i j_i k_i^I\ ;$$

and 3) a function which vanishes on gluing $\vec{h}_i = -\vec{k}_i$ triangles, coming from the first spinor derivative,

$$\Theta_i := j_i\frac{1 + \vec{h}_i\cdot\vec{k}_i}{\vec{h}_i\cdot\vec{\overline{\kappa}}_i}\,.$$

The capital indices $I$ were moreover split into rotation ($I = 1, 2, 3$) and boost ($I = 4, 5, 6$) subsets, denoted by $R$ and $B$ superscripts, such that now $I = 1, 2, 3$ in the equations above. The Hessian, which is a $14 \times 14$ symmetric matrix, has then the schematic structure

$$H(g_c, z_c, y) = \underbrace{\begin{pmatrix} H^{RR}_{gg} & H^{RB}_{gg} & H^{R}_{g\beta} & H^{R}_{g\overline{\beta}} \\ & H^{BB}_{gg} & H^{B}_{g\beta} & H^{B}_{g\overline{\beta}} \\ & & H_{\beta\beta} & H_{\beta\overline{\beta}} \\ & & & H_{\overline{\beta\beta}} \end{pmatrix}}_{3 \quad 3 \quad 4 \quad 4} \begin{matrix} \}3 \\ \}3 \\ \}4 \\ \}4 \end{matrix} .$$

*(d) Asymptotic formula*

Equations (IV.2.6) immediately generalize to the present case by including derivatives with respect to both sets of integration variables. The prefactor term $u(g, z)$ can be shown to be constant for our choice of section at the critical points. Hence the final expression for the gluing constraints in the asymptotic regime reads

$$\mathcal{G}_{j_i}(h_i, k_i) \simeq \overline{\mathcal{N}}^4_{j_i} \frac{(1 + (-1)^{2\sum_i j_i}) \prod_i (2j_i + 1) \langle Qh_i \mid k_i \rangle^{2j_i}}{32\pi\sqrt{-\det H}} \cdot \exp\left\{ V_\alpha (H^{-1})^{\alpha\beta} V_\beta \right\} , \quad \text{(IV.2.17)}$$

where $V_\alpha$ is a 14-component vector formally defined as

$$V = \Big( \underbrace{i\vec{\Gamma},}_{3} \quad \underbrace{i\gamma\vec{\Gamma},}_{3} \quad \underbrace{2\vec{\Theta},}_{4} \quad \underbrace{\vec{0}_4}_{4} \Big) ,$$

and $\overline{\mathcal{N}}^4_{j_i}$ is to be understood as a product of the phase $\mathcal{N}$ for every spin $j_i$. The coefficient $(32\pi)^{-1} = (4\pi)^{-4} \cdot (2\pi)^{-4} \cdot (2\pi)^7$ is obtained from the normalization of the Haar measure, from $u(g, z)$ and from the numerical factor of theorem IV.1, respectively. Observe that $V$ is sensitive to both closure and gluing through $\Gamma$ and $\Theta$, and that the Immirzi parameter figures in the exponential function through both $V$ and the Hessian matrix.

## IV.3 Numerical analysis

Equations (IV.2.11) and (IV.2.17) were the result of a straightforward application of Hörmander's second theorem to the gluing constraints. We have found, unsurprisingly, that they are maximal when evaluated on closing tetrahedra with identified face normal vectors (or which can be rotated into one-another). Before turning to a general discussion of the procedure and its conceivable applications, we find it important to numerically study these results. Doing so requires fixing conventions and settling on concrete boundary data.

*(a) Parametrization*

Our choice of parametrization on the SU(2) manifold is

$$g(\phi,\theta,\psi) = e^{-i\phi\frac{\sigma_3}{2}}\, e^{-i\theta\frac{\sigma_2}{2}}\, e^{-i\psi\frac{\sigma_3}{2}}\,,$$
$$-\pi \leq \phi \leq \pi,\ -\frac{\pi}{2} \leq \theta \leq \frac{\pi}{2},\ -2\pi \leq \psi \leq 2\pi\,,$$

with corresponding Haar measure

$$\mathrm{d}g = (4\pi)^{-2} \sin\theta \, \mathrm{d}\phi \mathrm{d}\theta \mathrm{d}\psi\,.$$

An SU(2) coherent state $|j, h\rangle$ can be parametrized by a spin $j$ and two Euler angles, where $h \equiv h(\phi,\theta) := h(\phi,\theta,-\phi) \in \mathrm{SU}(2)$. The element $h$ maps to a unit vector $\vec{h} \in S^2$ directly through the Euler angles, and generally via the $\mathbb{C}^2$ inner product $h_i = \langle h|\sigma_i|h\rangle$, as discussed above and in appendix C.

*(b) Boundary data*

As a prototypical case we consider constraints $\mathcal{G}(h_i, k_i)$ evaluated 1) on a bona-fide closed tetrahedron $h_i(\phi_i, \theta_i)$ and 2) on a perturbation $k_i = h_i(\phi_i, \theta_i + \delta\theta_i)$ of that same tetrahedron[47]. The tetrahedron $h_i$ is taken to

[47] Other types of boundary data were studied in [75].

be equilateral with spins $j_i = \Lambda$, with the explicit realization

$$\vec{h}_1 = (0,0,1)\ , \quad \vec{h}_2 = \frac{1}{3}\left(0, 2\sqrt{2}, -1\right)\ ,$$
$$\vec{h}_3 = \frac{1}{3}\left(\sqrt{6}, -\sqrt{2}, -1\right)\ , \quad \vec{h}_4 = \frac{1}{3}\left(-\sqrt{6}, -\sqrt{2}, -1\right)\ ,$$

for the face normal vectors. For plotting purposes we set $\delta\theta_i = 0, i \neq 1$ and allow only perturbations on $k_1$:

$$\vec{k}_1 = (\sin(\delta\theta_1), 0, \cos(\delta\theta_1))\ , \quad \vec{k}_i = \vec{h}_i\ ,\ i = 2,3,4\ .$$

*(c) Method for* $\mathcal{G}^{\mathrm{SU}(2)}$
Comparing the asymptotic formula to the full $\mathrm{SU}(2)$ gluing constraints requires computing the latter. This was done by Asante in [75] using the representation of equation (IV.1.4) and the Julia package `Cuba`[48] for multidimensional numerical integration.

*(d) Method for* $\mathcal{G}^{\mathrm{SL}(2,\mathbb{C})}$
The full constraints (IV.2.17) for the Lorentzian EPRL model involve fourteen real improper integrations, coming from a six-dimensional integration over $\mathrm{SL}(2,\mathbb{C})$ and two-dimensional real integrations over $\mathbb{C}P$ for each of the four nodes. Since these integrals are known to be difficult to evaluate numerically, we have instead resorted to the `sl2cfoam-next` package[49] developed in [74]. In order to apply the algorithm we first formulated the constraints in terms of $\mathrm{SU}(2)$ *boosted intertwiners*, following an idea of [129]. Diagrammatically, this is done through the set of equalities

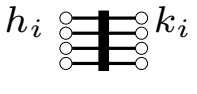 $=$ 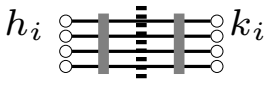 $= \sum_{\iota\,,\,\iota'} \dim\iota\, \dim\iota'$ 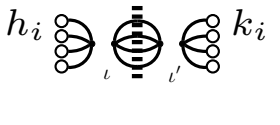 ,   (IV.3.1)

where the grey boxes represent $\mathrm{SU}(2)$ integrations and the dashed box stands for a non-compact integration; explicitly the $\mathrm{SL}(2,\mathbb{C})$ Haar measure

[48] github.com/giordano/Cuba.jl, based on [128].

[49] The package yields only real amplitudes; we thus compute only the absolute value of the gluing constraints.

of the left-hand side is Cartan-decomposed as

$$\mathrm{d}g = \mathrm{d}u_1 \mathrm{d}\mu(r) \mathrm{d}u_2 \,, \quad \mathrm{d}\mu(r) = \frac{1}{4}\pi \sinh^2 r \, \mathrm{d}r \,, \quad 0 \leq r < \infty \,,$$

with $\mathrm{d}u_{1,2}$ the normalized measure on $\mathrm{SU}(2)$. The melon-like diagram on the right-hand side of (IV.3.1) can be natively computed in `sl2cfoam-next`, and the coherent intertwiner diagrams are bona-fide $\mathrm{SU}(2)$ objects[50]. The Immirzi parameter is fixed to $\gamma = 0.123$.

*(e) Numerical results*

Having listed all necessary preliminary objects, the following figures IV.1 and IV.2 represent our numerical findings. We have performed computations for the absolute value of the full gluing constraints at different spin values $j_i = \Lambda$, and we have moreover compared the constraints to our asymptotic formulas at spins $\Lambda = 50$.

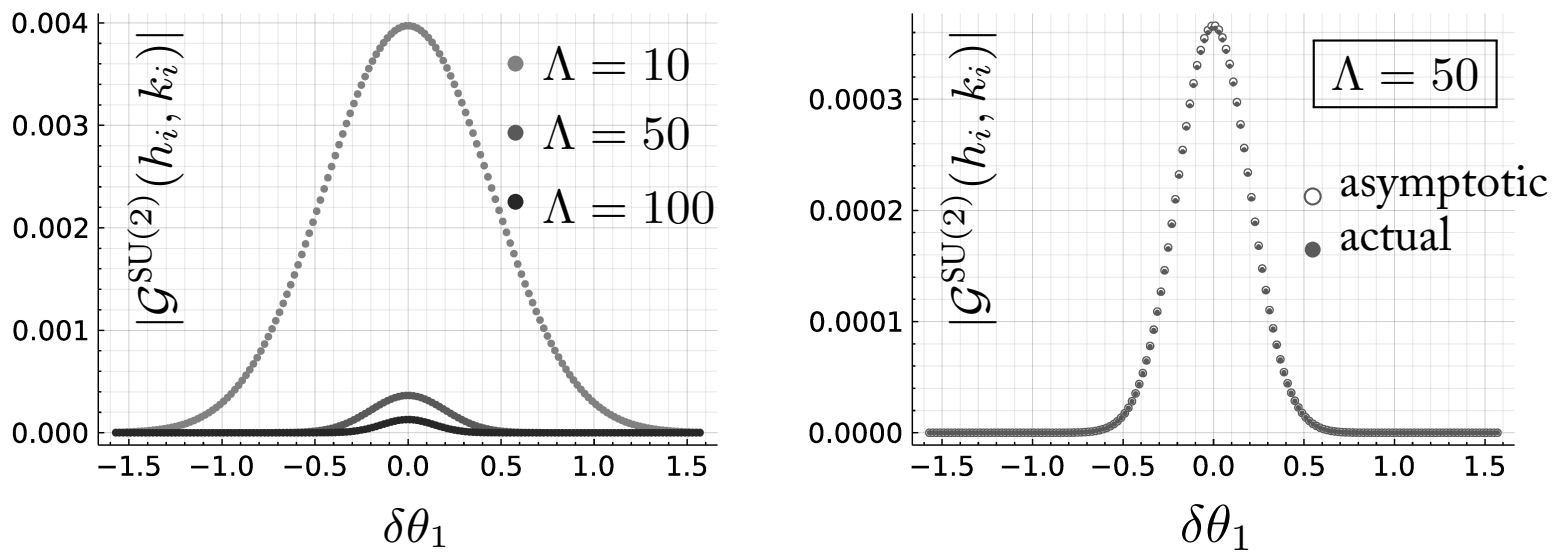

Figure IV.1: $\mathrm{SU}(2)$ gluing constraints evaluated on a tetrahedron and its deformation in terms of an Euler angle $\delta\theta_1$. The right panel shows the Gaussian behavior for both the actual amplitude and its asymptotic approximation.

A number of features can be observed from figures IV.1 and IV.2. One sees first of all confirmed the heuristic expectation that the gluing constraints should take the form of Gaussian functions peaked at configurations associated to identified geometric tetrahedra. If one insists in approaching the effective spin-foams framework, the perturbations $\delta\theta$ under consideration must be required to satisfy closure; this was studied in [75], yielding qualitatively similar results. Secondly, the agreement between the asymptotic formulae and the full constraints for both $\mathrm{SU}(2)$ and $\mathrm{SL}(2,\mathbb{C})$ seems to be quite satisfactory, even more so for the former. It is conceivable that the comparatively worse matching in the Lorentzian case is due to approximations which must be made in `sl2cfoam`, chiefly perhaps the truncations on infinite spin sums [74]. We have furthermore observed that the discrepancy between the peaks of the actual and the asymptotic $\mathcal{G}^{\mathrm{SL}(2,\mathbb{C})}$ functions increasingly worsens for larger values of the Immirzi parameter $\gamma$. As mentioned in [75], we believe this to be due to numerical instabilities in the asymptotic formula: the inverse Hessian matrix appearing in the exponential of equation (IV.2.17) seems to be extremely sensitive to small changes of the matrix coefficients.

[50]Note that equation (IV.3.1) implicitly involves the isomorphism of equation (B.2.3) and the identity (B.2.5).

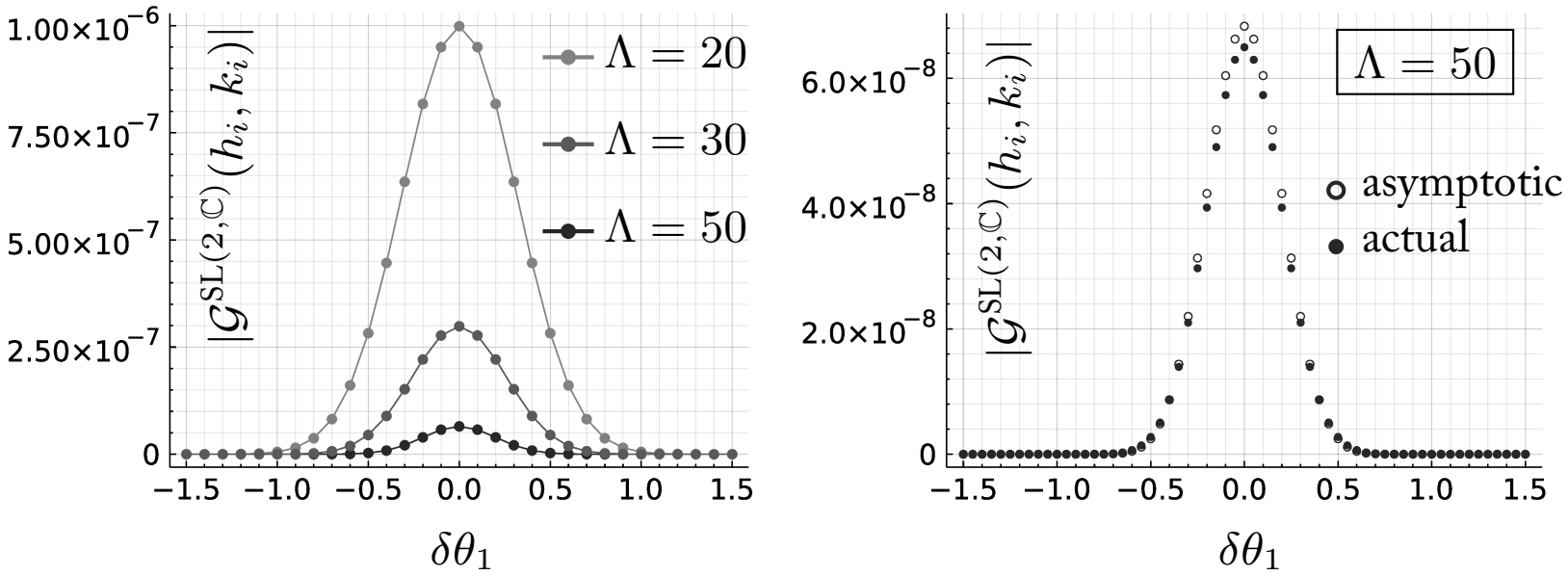

Figure IV.2: EPRL gluing constraints with $\gamma = 0.123$ evaluated on a tetrahedron and its deformation in terms of an Euler angle $\delta\theta_1$. The right panel shows the Gaussian behavior for both the actual amplitude and its asymptotic approximation.

## IV.4 Final remarks

The formulae (IV.2.11) and (IV.2.17) are to my knowledge the first *analytically explicit* asymptotic expressions for spin-foam objects generally away from their critical points. I close the chapter with a number of observations on the matter:

*(a) Generalization to the vertex amplitude*
Insofar as they correctly capture the qualitative behavior of the gluing constraints, our semiclassical formulae serve as a first proof of concept that there is more to be explored in the semiclassical analysis of spin-foams beyond the dominant configurations found in [53, 63–65]. Although certainly harder to technically develop, it seems quite reasonable to conjecture that the methods of this chapter can be applied to the full vertex amplitude: beyond conceivably yielding manageable functions with which to perform computations in a quantum-to-classical transitioning regime, such a more general semiclassical approximation has the potential to clarify the the dynamics of the model at the interfaces between different critical configurations, e.g. between Regge geometries and vector geometries. This possibility should be explored in future research.

*(b) Relation to "complex critical points"*
There exists a connection between the asymptotics discussed in this chapter and the recent proposal of *complex critical points* [72, 130]. The latter seeks to describe precisely the same structures as the former - the behavior of spin-foam objects away from critical points. The concrete techniques implemented by the complex critical point program are however slightly different, involving an analytical continuation of the action domain into the complex plane, and with a special affinity towards numerical computations: indeed the algorithm of [72, 130] yields only a discrete set of configurations away from the real critical points, and the problem of computing the Hessian matrix for a general complex remains unaddressed - the scaling behavior of the full amplitude is therefore unknown.

It is claimed in [72] and latter works that complex critical points (e.g. for the spin-foam amplitude of the $\Delta_3$ triangulation [68]) constitute "non-suppressed dominant contributions" with non-vanishing curvature, which is then argued to solve the famous *flatness problem* at finite spins. It should be noted however that the reason why such "complex configurations" were initially disregarded in [53,63–65] is that they are indeed suppressed relative to the "real critical points" - that much is the content of Hörmander's first theorem III.1. In the simple case of the gluing constraints studied in this chapter, the complex configurations would correspond to the points laying slightly away from the Gaussian peaks of figures IV.1 and IV.2, where $\Re S < 0$. This is to say that the claim that the bulk curvature of the $\Delta_3$ complex vanishes for arbitrarily large spins should still hold, as the configurations with $\Re S = 0$ will always comparatively dominate (even is slightly, when relative to points with $\Re S$ close to zero); it is my opinion that the flatness problem is as such not yet entirely out of the picture[51].

*(c) On the proposal of a hybrid algorithm and many-vertex asymptotics*

The gluing constraints discussed throughout this chapter were originally introduced in [75] in the context of a proposal for a hybrid algorithm, which would compute the amplitude of a general cell-complex by switching to the asymptotic approximation whenever profitable. This evidently requires having a working methodology for the asymptotics of complexes with more than one vertex. The hope was that, having disentangled the boundary data of the different vertices through the gluing constraints, one would then be able to apply the well-known single-vertex semiclassical approximation locally at each vertex, evaluating the gluing constraints on those configurations.

Subsequent numerical calculations have unfortunately shown that this procedure fails to recover the correct scaling behavior of the full amplitude. I can attempt a justification in hindsight: for one to apply the asymptotic approximation locally it is required that the full expression for the total amplitude itself admits such an approximation. Both theorems III.1 and IV.1 demand that the Hessian matrix is non-singular, which can only happen if there are no redundant integrations in the integral defining the relevant object; but by their very nature the gluing constraints must introduce such redundancies. Hence, while it can be argued that the parametrization of spin-foam models emphasizes to some degree some of its local properties, they cannot be helpful when it comes to approximating the amplitude for large spins.

[51] See [82], where it is argued that the flatness problem might be caused by insisting on taking the classical limit before the continuum limit.

The problem of redundant gauge orbits is, in fact, present in any spin-foam model. Since the quantization procedure described in chapter I does not involve any gauge-fixing, the resulting amplitude for a general complex will contain a number of redundant Lie group $G$ integrations (which moreover diverge if $G$ is non-compact). Performing a stationary phase analysis *explicitly*, in the spirit of this chapter, thus requires manually discarding all redundant integrals, a procedure which is strongly dependent on the combinatorics of the cell-complex at hand. Each gauge-fixed vertex has then the potential to involve more or less integrations, and the critical point equations must depend on this structure[52]. Deriving a semiclassical approximation for a general cell-complex is therefore a complicated problem, even at the level of the otherwise straightforward SU(2) $BF$ theory. I believe it to be likely that more refined methods of asymptotic approximations of integrals - allowing for singular Hessian matrices - are necessary; see [131, Theorem 3] and follow-up works[53] for an early analysis of such problems.

[52] This is entirely analogous to the perturbative path-integral quantization of quantum gauge theories, which *do* require gauge-fixing because one essentially performs a stationary phase approximation around the free theory.

[53] I was made aware of these papers by Tim Hoffmann.

# Overview and Outlook

The common thread throughout this monograph is the problem of defining a complete amplitude for 4-dimensional vacuum Lorentzian gravity with space- and time-like boundaries, as well as that of understanding its putative agreement with the classical theory. In the spirit of a conclusion, I would like to contextualize our results with respect to the current state of the spin-foam approach to quantum gravity; much of it will be from a personal and subjective perspective, unavoidably so.

I shall start with a succinct review of what was achieved. The relationship between spinors and the 2-dimensional homogeneous spaces of the Lorentz group was explored in chapter II. This has allowed for an extension of existing results on the symplectomorphism between constrained Weyl spinors and the phase spaces of $\mathrm{SU}(2)$ and $\mathrm{SU}(1,1)$ to a new symplectomorphism based on Majorana spinors, the space-like hyperboloid and $T^*\mathrm{SU}(1,1)$. These observations have in turn informed the proposal of a new coherent vertex amplitude for Lorentzian (2+1)-dimensional quantum gravity with both space- and time-like boundary edges, which was moreover shown to asymptotically match the Regge action.

Chapter III was in turn concerned with the asymptotic analysis of the Conrady-Hnybida extension to the EPRL model, with a particular emphasis on parachronal (i.e. time-like) interfaces. A final formula for the asymptotic amplitude for such interfaces was unfortunately not attained; still, I believe the benefit of my analysis lies in further clarifying the role and structure of many of the objects which turn out to be necessary for the semiclassical approximation. Among these: new boundary coherent states were proposed, leading to a better behaved time-like action; a Minkowski-type theorem for Lorentzian convex polyhedra was proven; it was shown that time-like data does not induce a unique reconstructed 4-simplex, instead determining a whole family of such simplices; and a geometrical interpretation was assigned to every term of the semiclassical action.

The problem of performing an asymptotic analysis away from the critical points was considered in chapter IV. The objects under study were gluing constrains, defined in analogy to their namesakes from effective spin-foams. The initial hope was that such gluing constraints could be useful for obtaining an asymptotic approximation of the amplitude for a whole simplicial complex, as they would separate the boundary data of every simplex; further scrutiny later showed that this proposal would require more refined asymptotic methods, allowing in particular for redundant integration directions. The gluing constraints have however proven to be by themselves ideal toy models on which to apply the asymptotic methods of Hörmander - this much was done, and the explicit analytical results seem to match the full gluing constraints quite satisfactorily, even for boundary data which does not allow for critical points.

***

Turning to the current status of the field, one ought first recognize that which *is* understood - and that which works. The first fundamental problem for spin-foams is that of obtaining a quantum amplitude for vacuum gravity: a map from whatever degrees of freedom are considered relevant to the field of complex numbers [39]. The two main contemporary answers to this problem are the simplicial Barrett-Crane [45] and EPRL [47] models (and their respective extensions [54, 55] to various causal structures). They both derive from the quantization of a topological theory, and their principal difference is in the manner in which the simplicity constraints are enforced. Both models are well-defined and reasonably well-understood whenever the tetrahedra they consider are assumed to be space-like. Their space-like vertex amplitude is finite and can be numerically evaluated for arbitrary boundary data; efficient numerical methods have in particular been developed for the EPRL model [74]. Simple gauge- (but not diffeomorphism-) invariant observables can be defined and evalu-ated [89]. The single-vertex large-spin asymptotic limit of the space-like BC and EPRL models is well-known, and it yields in certain cases the Regge action for discretized gravity [53]; initial results for larger complexes exist [130]. First steps in exploring the discretization-(in)dependence of spin-foams have been taken [86]. Powerful heuristics can be provided by analog models, namely in the context of "effective spin-foams" [77]. Beyond the amplitude for pure gravity, a cosmological term can be included e.g. by considering quantum-group deformations of SU(2) [132]. Proposals have been made for coupling bosonic and fermionic matter [90] to spin-foams by postulating an appropriate weight for the amplitude.

Much is likewise *not* known. Starting with the aforementioned foundational goal of defining a reasonable amplitude for pure gravity, the cohesive picture of the previous paragraph is tainted by numerous obstacles. The asymptotic behavior of more than a single vertex remains mysterious, insofar as it is not known how to generally obtain the global Hessian matrix for the complex. The one-vertex amplitude is dominated not only by 4-simplex configurations (desirable, according to our classical expectations) but also by vector geometries - perhaps due to areas rather than lengths being the fundamental degrees of freedom in spin-foam models. An analytical formula for the asymptotic amplitude with time-like faces is not known, and it may well turn out that the EPRL amplitude for such boundaries is itself not finite. Since these basic difficulties have not yet been surmounted, most consequential aspects of Lorentzian spin-foams are likewise yet to be fully understood. The comparatively complicated nature of the extensions to time-like boundaries has hampered the type of proliferation of numerical results which has been observed in the space-like case. The discretization-dependence of Lorentzian 4d models is still to be addressed, and it is unlikely that anything meaningful about the real world can be said while it remains so; finding the continuum limit of spin-foam models remains a substantial and unsolved puzzle. The possibility of defining measurable empirical quantities is probably predicated on having the coupling between spin-foams and matter under control, but such questions are only now starting to be considered in 4 dimensions.

***

It is against this backdrop that our research was carried out. While the foundational problem which inspired it - that of defining a complete amplitude and its semiclassical limit - has not yet been surmounted, I believe our conclusions to be useful in eventually approaching a solution.

I conclude with a personal perspective on what need still be understood in the future (and laboring not to repeat the remarks which close each chapter), exclusively focusing on the matters which title this thesis: Lorentzian spin-foams and their semiclassical limit. I believe it essential to define a well-behaved EPRL-type vertex amplitude for time-like polygons, i.e. to refine the Conrady-Hnybida proposal. I expect this to require - much like the 3-dimensional model of chapter III did - both a better understanding of the divergences related to the generalized states of the continuous series of $\mathrm{SU}(1,1)$ as well as a prescription for further constraining the amplitude. Once and if such a goal is achieved, the parachronal vertex amplitude will conceivably stand on the same footing as the space-like with regards to its semiclassical behavior. This would set a solid foundation for proceeding with more complex concerns; still the 3-dimensional amplitude may itself be useful for exploring such matters, e.g. observables and their interaction with the causal structure of the model.

On the subject of asymptotics, having a semiclassical formula for extended simplicial complexes is definitely desirable. Although major strides have recently been made regarding numerical characterizations of critical points of many-vertex spin-foams, it is still not known how to efficiently compute the Hessian determinant of the whole complex - and this is strictly necessary in order to obtain the correct scaling and oscillatory behavior of the amplitude. Until this is addressed, there is still room to further strengthen the claim that many-vertex spin-foam models recover the Regge action. Pursuing the original intent of the hybrid algorithm discussed in chapter IV - following the heuristic notion of "localizing" the asymptotic behavior at each vertex - seems to me to be a rather promising avenue; as mentioned previously, there exist asymptotic theorems for redundant integrations, and these ought to be explored. It seems moreover reasonable that the generalized asymptotic analysis of chapter IV may be applicable to the vertex amplitude, for which a better understanding of its behavior for arbitrary boundary data would certainly be useful.

I cannot resist the opportunity of laying down two final conjectures on what the future may hold for the spin-foam approach. Text-length constraints dictate they will have to be expanded on elsewhere. I would argue 1) that - by virtue of containing a curvature 3-form, to be integrated over 3-surfaces dual to simplicial edges - higher-gauge theory models will eventually correct the presence of vector geometries, and restitute the role of length variables in spin-foams; 2) that - since the amplitude ought to be oblivious to both the coarseness of the discretization as well as the valency of polyhedra in an appropriate regime - renormalization methods should eventually contemplate both scaling directions.

# *Appendix A* : Facts and Conventions on SU(1,1)

## A.1 Unitary irreducible representations

The unitary irreducible representations of $\mathrm{SU}(1,1)$ were classified for the first time by Bargmann in [107], and the analysis of generalized eigenstates was carried out by Lindblad in [108]. The following collection of facts is lifted directly from the two authors; we follow the conventions of the latter.

The $\mathfrak{su}(1,1)$ algebra is spanned by the generators $F^i = (L^3, K^1, K^2)$, defined in the fundamental representation with the standard Pauli matrices as $\varsigma^i/2 = (\sigma_3/2, i\sigma_2/2, -i\sigma_1/2)$. The respective subgroups read

$$e^{i\alpha L^3} = \begin{pmatrix} e^{i\frac{\alpha}{2}} & 0 \\ 0 & e^{-i\frac{\alpha}{2}} \end{pmatrix}, \quad e^{itK^1} = \begin{pmatrix} \cosh\frac{t}{2} & i\sinh\frac{t}{2} \\ -i\sinh\frac{t}{2} & \cosh\frac{t}{2} \end{pmatrix},$$

$$e^{iuK^2} = \begin{pmatrix} \cosh\frac{u}{2} & \sinh\frac{u}{2} \\ \sinh\frac{u}{2} & \cosh\frac{u}{2} \end{pmatrix},$$

and the Casimir element is given by $Q^2 = (L^3)^2 - (K^1)^2 - (K^2)^2$. There are two families of unitary irreducible representations characterized by the eigenvalues of $Q$ and $L^3$, called the *discrete* and *continuous series*. Regarding the first, the Hilbert space $\mathcal{D}_k^q$ is spanned by the orthonormal states

$$Q|k,m\rangle = k(k+1)|k,m\rangle\,, \quad k \in -\frac{\mathbb{N}}{2}\,,$$
$$L^3|k,m\rangle = m|k,m\rangle\,, \quad m \in q(-k+\mathbb{N}^0)\,, \quad q = \pm\,.$$

For the continuous series, an orthonormal basis for the Hilbert space $\mathcal{C}_j^\delta$ is given by

$$Q|j,m\rangle = j(j+1)|j,m\rangle\,, \quad j = -\frac{1}{2} + is\,, \quad s \in \mathbb{R}^+\,,$$
$$L^3|j,m\rangle = k|j,m\rangle\,, \quad m \in \delta + \mathbb{Z}\,, \quad \delta \in \{0, \frac{1}{2}\}\,.$$

An alternative orthonormal basis of $\mathcal{C}_j^\delta$ can be obtained from generalized eigenstates of the non-compact operator $K^2$. The eigenstates satisfy

$$Q|j,\lambda,\sigma\rangle = j(j+1)|j,\lambda,\sigma\rangle\,, \quad j=-\frac{1}{2}+is\,, \quad s\in\mathbb{R}^+\,,$$
$$K^2|j,\lambda,\sigma\rangle = \lambda|j,m\rangle\,, \quad \lambda\in\mathbb{C}\,,$$
$$P|j,\lambda,\sigma\rangle = (-1)^\sigma|j,\lambda,\sigma\rangle\,, \quad \sigma\in\{0,1\}\,,$$

where $P$ is an outer automorphism of the Lie algebra taking $(L^3,K^1,K^2)\mapsto(-L^3,-K^1,K^2)$. They are complete and orthonormal in the sense that

$$\sum_\sigma\int_{\mathbb{R}+i\alpha}\mathrm{d}\lambda\;\langle j,m|j,\lambda,\sigma\rangle\,\langle j,\overline{\lambda},\sigma|j,n\rangle = \delta_{m,n}\,, \quad \alpha\in\mathbb{R}\,,$$
$$\sum_m\langle j,\overline{\lambda'},\sigma'|j,m\rangle\,\langle j,m|j,\lambda,\sigma\rangle = \delta(\lambda-\lambda')\,, \quad \Im\lambda=\Im\lambda'\,,$$

and indeed there is a family of bases $\{\,|j,\lambda+i\alpha,\sigma\rangle\mid\lambda\in\mathbb{R}\,,\sigma\in\{0,1\}\}_\alpha$ indexed by $\alpha\in\mathbb{R}$.

## A.2 Parametrization and Haar measure

Among the possible parametrizations of the group the following (coherently normalized) are useful for this work [42,112]

1) $g=e^{i\alpha L^3}e^{itK^1}e^{iuK^2}\,, \quad \mathrm{d}g=(4\pi)^{-2}\cosh t\,\mathrm{d}\alpha\,\mathrm{d}t\,\mathrm{d}u\,,$

   with $0\le\alpha<4\pi\,,\ -\infty<t,u<\infty$;

2) $g=e^{i\alpha L^3}e^{iuK^2}e^{i\beta L^3}\,, \quad \mathrm{d}g=(4\pi)^{-2}\sinh u\,\mathrm{d}\alpha\,\mathrm{d}t\,\mathrm{d}\beta\,,$

   with $0\le\alpha<4\pi\,,\ 0\le\beta<2\pi\,,\ u\ge0$;

3) $g=\begin{pmatrix}\alpha & \beta\\ \overline{\beta} & \overline{\alpha}\end{pmatrix}\,, \quad \mathrm{d}g=\pi^{-2}\delta(|\alpha|^2-|\beta|^2-1)D\alpha D\beta\,,$

   $D\alpha=\frac{i}{2}\mathrm{d}\alpha\wedge\mathrm{d}\overline{\alpha}\,, \quad m.m.\ D\beta\,;$

4) $g=\begin{pmatrix}\alpha & \beta\\ \overline{\beta} & \overline{\alpha}\end{pmatrix}\,, \quad \mathrm{d}g=(2\pi)^{-2}|c|^{-1}\mathrm{d}a\,\mathrm{d}b\,\mathrm{d}c\,,$

with $a = \dfrac{(\alpha - \overline{\alpha}) - (\beta - \overline{\beta})}{2i}$ , $b = \dfrac{(\overline{\alpha} - \alpha) + (\overline{\beta} - \beta)}{2i}$ ,
$c = \dfrac{(\alpha + \overline{\alpha}) - (\beta + \overline{\beta})}{2}$ .

## A.3 Formulas for matrix coefficients

We accept the convention that the representation acts on the relevant set of functions by left translation $g \triangleright f(z) = f(g^{-1}z)$.

*(a) Discrete series*

Define $\Theta_{mn} = \frac{1}{(m-n)!}\left(\frac{\Gamma(m+1+k)\Gamma(m-k)}{\Gamma(n+1+k)\Gamma(n-k)}\right)^{\frac{1}{2}}$ and set $D^{k(+)}_{mn}(\alpha,\beta) := \langle j, m|D^{k(+)}(g)|j,n\rangle$ where $g = g(\alpha,\beta)$ is parametrized as in A.2.3). Then by [107] one has for $\mathcal{D}^+_k$ that

$$\begin{aligned} D^{k(+)}_{mn}(\alpha,\beta) &= \Theta_{mn}\overline{\alpha}^{-m-n}\beta^{m-n} \\ &\cdot {}_2F_1(-k-n, 1-n+k, 1+m-n; -\beta\overline{\beta})\,, \quad m \geq n\,, \end{aligned} \tag{A.3.1}$$

$$\begin{aligned} D^{k(+)}_{mn}(\alpha,\beta) &= (-1)^{m-n}\Theta_{nm}\overline{\alpha}^{-m-n}\overline{\beta}^{n-m} \\ &\cdot {}_2F_1(-k-m, 1-m+k, 1+n-m; -\beta\overline{\beta})\,, \quad n \geq m\,. \end{aligned}$$

The matrix coefficients $D^{k(-)}_{mn}(\alpha,\beta)$ of the negative series $\mathcal{D}^-_k$ can be obtained from the above equations by making use of the relation $D^{k(+)}_{mn}(\alpha,\beta) = (-1)^{m-n}\overline{D^{k(-)}_{-m,-n}}(\alpha,\beta)$ [107].

*(b) Continuous series*

The following is due to [108, 112]. Setting $\Delta\lambda = \lambda - \lambda'$ and $\alpha \in [-\pi, \pi]$, the matrix coefficients in the $K^2$ eigenbasis read

$$
\begin{aligned}
&\langle j, \overline{\lambda}, \sigma | D^j(e^{-i\alpha L^3}) | j, \lambda', \sigma' \rangle_{\text{reg}} = \\
&\lim_{\epsilon \to 0} \frac{1}{2\pi} \Bigg\{ \frac{\Gamma(\frac{-\overline{j}+\sigma-i\lambda}{2})\Gamma(\frac{-j+\sigma'+i\lambda'}{2})}{\Gamma(\frac{-j+\sigma+i\lambda}{2})\Gamma(\frac{-\overline{j}+\sigma'-i\lambda'}{2})} \Gamma(i\Delta\lambda)\psi_-(\alpha) \\
&+ (-1)^{2\delta} \frac{\Gamma(\frac{-\overline{j}+2\delta+(-1)^{2\delta}\sigma+i\lambda}{2})\Gamma(\frac{-j+2\delta+(-1)^{2\delta}\sigma'-i\lambda'}{2})}{\Gamma(\frac{-j+2\delta+(-1)^{2\delta}\sigma-i\lambda}{2})\Gamma(\frac{-\overline{j}+2\delta+(-1)^{2\delta}\sigma'+i\lambda'}{2})} \Gamma(-i\Delta\lambda)\psi_+(\alpha) \Bigg\} \\
&\qquad \cdot \cos\frac{\pi}{2}(i\Delta\lambda + \sigma - \sigma')\,,
\end{aligned}
\tag{A.3.2}
$$

with

$$
\begin{aligned}
f_\pm(\alpha) = {} & \cos\left(\frac{\alpha}{2}\right)^{-2j-2} \left|2\tan\frac{\alpha}{2}\right|^{\pm i\Delta\lambda} \\
& \cdot {}_2F_1\left(-\overline{j} \pm i\lambda, -\overline{j} \mp i\lambda', 1 \pm i\Delta\lambda; -\tan^2\frac{\alpha}{2}\right) \operatorname{sgn}^{\sigma-\sigma'}\alpha\,.
\end{aligned}
$$

For the coefficients in a mixed basis, and defining $x = (1 + i\sinh t)/2$,

$$
\begin{aligned}
\langle j, m | D^j(e^{-itK^1}) | j, \lambda, \sigma \rangle = {} & \left(\frac{\Gamma(m-j)}{\Gamma(m-\overline{j})}\right)^{\frac{1}{2}} \\
& \cdot \frac{2^{j-1}\Gamma(-j-i\lambda)}{i^\sigma \sin\frac{\pi}{2}(-j+\sigma+i\lambda)} \left(f_m(x) + (-1)^\sigma f_{-m}(\overline{x})\right)\,,
\end{aligned}
\tag{A.3.3}
$$

where

$$
f_m(x) = \frac{\overline{x}^{\frac{m+i\lambda}{2}} x^{\frac{m-i\lambda}{2}}}{\Gamma(-m-j)\Gamma(m+1-i\lambda)} {}_2F_1(m-j, m-\overline{j}, m+1-i\lambda; x)\,.
$$

## A.4 Harmonic analysis

There exists a Plancherel formula for $L^2(\mathrm{SU}(1,1))$. According to [42,133, 134], given any function $f \in \mathcal{C}_0^\infty$,

$$f(\mathbb{1}) = \sum_\delta \int_{-\infty}^{\infty} \mathrm{d}s\, s \tanh(\pi s)^{1-4\delta} \mathrm{Tr}_{\mathcal{C}_j^\delta} \left[ \int_{\mathrm{SU}(1,1)} \mathrm{d}g\, f(g) D^j(g) \right]$$
$$+ \sum_q \sum_{2k=-1}^{-\infty} (-2k-1) \mathrm{Tr}_{\mathcal{D}_k^q} \left[ \int_{\mathrm{SU}(1,1)} \mathrm{d}g\, f(g) D^k(g) \right] , \tag{A.4.1}$$

where the Haar measure is as above. Observe that the lowest $k = -\frac{1}{2}$ discrete representation is absent from the decomposition of $f$. Setting $f(g) := \int \mathrm{d}h \overline{f}_1(h) f_2(gh)$ it follows from (A.4.1) that

$$\int \mathrm{d}g\, \overline{f}_1(g) f_2(g) = \sum_\delta \int_{-\infty}^{\infty} \mathrm{d}s\, s \tanh(\pi s)^{1-4\delta} (\overline{c}_1)^{j(\delta)}_{\lambda\sigma,\lambda'\sigma'} (c_2)^{j(\delta)}_{\lambda\sigma,\lambda'\sigma'}$$
$$+ \sum_q \sum_{2k=-1}^{-\infty} (-2k-1) (\overline{c}_1)^{k(q)}_{mm'} (c_2)^{k(q)}_{mm'} ,$$

where all lower repeated indices are appropriately contracted, and the Fourier coefficients read

$$(c_i)^{k(q)}_{mm'} = \int \mathrm{d}g\, f_i(g) D^{k(q)}_{mm'}(g) ,$$
$$(c_i)^{j(\delta)}_{\lambda\sigma,\lambda'\sigma'} = \int \mathrm{d}g\, f_i(g) D^{j(\delta)}_{\lambda\sigma,\lambda'\sigma'}(g) .$$

The coefficients for the continuous series could of course also be written with respect to the $L^3$ eigenbasis. Yet another consequence of equation

(A.4.1) are the orthogonality relations

$$\begin{aligned}
\int \mathrm{d}g\, \overline{D}^{k(q)}_{mn}(g) D^{k'(q')}_{m'n'}(g) &= \frac{\delta_{q,q'}\delta_{k,k'}}{-2k-1}\, \delta_{m,m'}\delta_{n,n'}\,, \\
\int \mathrm{d}g\, \overline{D}^{j(\delta)}_{m,n}(g) D^{j'(\delta')}_{m'n'}(g) &= \frac{\delta_{\delta,\delta'}\delta(j-j')}{s\tanh(\pi s)^{1-4\delta}}\, \delta_{m,m'}\delta_{n,n'}\,, \\
\int \mathrm{d}g\, \overline{D}^{j(\delta)}_{\lambda\sigma,\mu\epsilon}(g) D^{j'(\delta')}_{\lambda'\sigma',\mu'\epsilon'}(g) &= \frac{\delta_{\delta,\delta'}\delta(j-j')}{s\tanh(\pi s)^{1-4\delta}}\, \delta(\lambda-\lambda')\delta(\mu-\mu')\,,
\end{aligned} \tag{A.4.2}$$

which hold for the matrix coefficients of the discrete and continuous series in the $L^3$ eigenbasis, and the coefficients of the continuous series in the $K^2$ eigenbasis, respectively.

# *Appendix B* : Facts and Conventions on SL(2,C)

This appendix contains a number of useful identities regarding the special linear group and its representations.

## B.1 Unitary irreducible representations

The algebra $\mathfrak{sl}(2,\mathbb{C})$ is spanned by the generators $J^i = \sigma^i/2$ and $K^i = i\sigma^i/2$ in the fundamental representation, with commutation relations

$$[J^i, J^j] = i\epsilon^{ijk}J^k\,, \quad [K^i, K^j] = -\epsilon^{ijk}K^k\,, \quad [J^i, K^j] = i\epsilon^{ijk}K^k\,;$$

it can equivalently be obtained by the complexification of either $\mathfrak{su}(2)$ or $\mathfrak{su}(1,1)$. The representations of $\mathrm{SL}(2,\mathbb{C})$ are constructed on the space $\mathcal{D}_{(n_1,n_2)}$ of homogeneous functions of two complex variables [41, 42, 135], i.e. functions $F : \mathbb{C}^2 \to \mathbb{C}$ satisfying

$$F(\lambda z_1, \lambda z_2) = \lambda^{n_1-1}\overline{\lambda}^{n_2-1}F(z_1, z_2), \quad \lambda \in \mathbb{C}\,,$$

such that the representation $D^{(n_1,n_2)}$ acts on functions through the usual multiplication by inverse in $\mathbb{C}^2$,

$$\begin{aligned} D^{(n_1,n_2)} &: \ \mathrm{SL}(2,\mathbb{C}) \to \mathrm{Aut}(\mathcal{D}_{(n_1,n_2)}) \\ D^{(n_1,n_2)}(g)F(z) &= F(g^{-1}z)\,, \quad z \in \mathbb{C}^2\,. \end{aligned}$$

The relevant representations for spin-foam models are those contained in the so-called principal series, characterized by the restriction $n_1 = -\bar{n}_2$. It is usual to redefine $n_1 = (-\nu + i\rho)/2$ and $n_2 = (\nu + i\rho)/2$ with $\nu \in \mathbb{Z}$, $\rho \in \mathbb{R}$, and to collect these variables in the label $\chi = (\nu, \rho)$. Such principal series representations are irreducible, and they are unitary under

the inner product

$$\langle F_1,\, F_2\rangle = \int_{\mathbb{C}\mathrm{P}} \omega\, \bar{F}_1(z), F_2(z)\,, \qquad \omega = \frac{i}{2}(z_1 \mathrm{d}z_2 - z_2 \mathrm{d}z_1) \wedge (\bar{z}_1 \mathrm{d}\bar{z}_2 - \bar{z}_2 \mathrm{d}\bar{z}_1)\,, \tag{B.1.1}$$

where $\mathbb{C}P$ denotes that the integral is to be computed over a section of the bundle $\mathbb{C}^{2*} \to \mathbb{C}P$, and the result is independent of the choice of section. There exists an intertwining isomorphism (defined up to normalization, possibly depending on $\chi$),

$$\mathcal{A}:\ \mathcal{D}_\chi \to \mathcal{D}_{-\chi} \qquad \mathcal{A}D^{-\chi}(g) = D^{\chi}(g)\mathcal{A}\,, \tag{B.1.2}$$

intertwining the $\chi$ and $-\chi$ representations, such that one may restrict themselves to $\nu, \rho \geq 0$. The homogeneous functions $F$ are uniquely determined by single-variable functions $\phi(z_2/z_1) := F(1, z_2/z_1)$, and in terms of this realization $\mathcal{A}$ reads [41]

$$\mathcal{A}\phi(z) = \frac{i}{2}\int \mathrm{d}y\, (y-z)^{-n_1-1}(\overline{y}-\overline{z})^{-n_2-1}\phi(y)\,.$$

It can be shown that complex-conjugation takes $\mathcal{D}_\chi \to \mathcal{D}_{-\chi}$, such that $\mathcal{A}$ can be used to construct a bilinear form $(\cdot,\cdot)$ on $\mathcal{D}_\chi$,

$$(F_2,\, F_2) := \langle \mathcal{J}F_1,\, F_2\rangle\ ,$$

having defined $\mathcal{J}F = \overline{\mathcal{A}F}$.

## B.2 Bases for the space of unitary representations

In order to make calculations more manageable, one might like to introduce orthonormal bases for the functions in $\mathcal{D}_\chi$. Two particularly useful realizations of these functions are provided by the so-called canonical- and pseudo-bases, obtained from representations of the compact and non-compact subgroups SU(2) and SU(1, 1), respectively. We shall go over their construction in the subsequent sections, following [42].

### B.2.1 The canonical basis

Due to the homogeneity property of $F \in \mathcal{D}_\chi$, such functions can be uniquely characterised by their values on $S^3 = \{z \in \mathbb{C}^2 \mid |z_1|^2 + |z_2|^2 = 1\}$ as

$$F(z_1, z_2) = (|z_1|^2 + |z_2|^2)^{i\rho/2-1} F\left(\frac{z_1}{\sqrt{|z_1|^2 + |z_2|^2}}, \frac{z_2}{\sqrt{|z_1|^2 + |z_2|^2}}\right),$$

and this allows us to realize $F$ on the unitary group. Indeed, using the well-known diffeomorphism between the sphere and SU(2),

$$u_1(z) = \frac{\overline{z}_1}{\sqrt{|z_1|^2 + |z_2|^2}}, u_2 = \frac{\overline{z}_2}{\sqrt{|z_1|^2 + |z_2|^2}},$$
$$u = \left(\begin{smallmatrix} u_1 & u_2 \\ -\bar{u_2} & \bar{u_1} \end{smallmatrix}\right) \in \mathrm{SU}(2),$$

$F$ may just as well be understood as a function $f \in \mathcal{C}^\infty(\mathrm{SU}(2))$ such that $f(u(z)) = F(z/|z|)$. Note that then

$$D^\chi(g) F(z/|z|) = f(u(z)g), \quad g \in \mathrm{SU}(2), \tag{B.2.1}$$

i.e. the representation of $\mathrm{SU}(2) \subset \mathrm{SL}(2, \mathbb{C})$ on such functions amounts to a right-translation. $F$ must still satisfy the homogeneity condition on the sphere,

$$F(e^{i\omega} z_1, e^{i\omega} z_2) = e^{-i\omega\nu} F(z_1, z_2)$$
$$\Rightarrow \; f(\gamma u) \stackrel{!}{=} e^{i\omega\nu} f(u), \quad \gamma = \begin{pmatrix} e^{i\omega} & 0 \\ 0 & e^{-i\omega} \end{pmatrix},$$

with $\omega \in \mathbb{R}$. To functions $f \in \mathcal{C}^\infty(\mathrm{SU}(2))$ satisfying this transformation property we will call *covariant*, as in [42]. It then turns out that one has a Plancherel-type theorem identifying the Hilbert space of square integrable covariant functions and $\mathcal{D}_\chi$,

$$L^2(\mathrm{SU}(2))_{\mathrm{cov}} \simeq \mathcal{D}_\chi,$$
$$\int_{\mathrm{SU}(2)} |f(u)|^2 \, \mathrm{d}u = \int_{\mathbb{C}P} |F(z_1, z_2)|^2 \, \omega.$$

Square-integrable functions on the unitary group can also be described in terms of its representations. According to the Peter-Weyl theorem for compact groups, there is an isomorphism

$$L^2(\mathrm{SU}(2)) \simeq \bigoplus_j \mathcal{H}^j \otimes \mathcal{H}^{j*}$$

between the supporting spaces $\mathcal{H}^j$ of unitary irreducible representations of SU(2) and the square-integrable functions. The set $\{\sqrt{\dim(D^j)}D^j_{mm'}\}$ of matrix coefficients in the $L^3$ eigenbasis furthermore constitutes an orthonormal basis for this space. Noting that $\gamma$ is an element of the one-parameter subgroup generated by $L^3$, one can then argue

$$\begin{aligned} D^j_{mm'}(\gamma u) &= \sum_l D^j_{ml}(e^{2i\omega L^3}) D^j_{lm'}(u) \\ &= e^{2im\omega} D^j_{mm'}(u)\,, \end{aligned} \tag{B.2.2}$$

showing that $D^j_{mm'}$ is covariant when $m = \nu/2$. Hence one has a final isomorphism

$$\begin{aligned} \mathcal{I}^\chi :\ \bigoplus_j \mathcal{H}^j &\to \mathcal{D}_\chi \\ |j, m\rangle &\mapsto |\chi; j, m\rangle\,, \end{aligned} \tag{B.2.3}$$

with the homogeneous function representation

$$F^\chi_{j,m}(z) = \sqrt{2j+1}\,(|z_1|^2 + |z_2|^2)^{i\rho/2-1} D^j_{\frac{\nu}{2}m}(u(z))\,. \tag{B.2.4}$$

This defines the canonical basis, together with a resolution of identity

$$\mathbb{1}_{(n,\rho)} = \sum_{j=\frac{n}{2}}^{\infty} \sum_{m=-j}^{m=j} |\chi; j\, m\rangle\langle\chi; j\, m|\,.$$

Observe that by equation (B.2.1) the map $\mathcal{I}^\chi$ commutes with the action of SU(2), i.e.

$$D^\chi(g) \circ \mathcal{I}^\chi |j, m\rangle = \mathcal{I}^\chi \circ D^j(g)|j, m\rangle\,, \quad g \in \mathrm{SU}(2). \tag{B.2.5}$$

### B.2.2 The pseudo-basis

In complete analogy to the previous SU(2) case, homogeneous functions in $\mathcal{D}_\chi$ are uniquely defined by their values on the hyperboloids (or pseudo-spheres) $H^3_\pm = \{\mathbf{z} \in \mathbb{C}^2 \mid |z_1|^2 - |z_2|^2 = \tau\,, \tau = \pm 1\}$ through

$$F(z_1, z_2) = \sum_\tau \Theta\left(\tau(|z_1|^2 - |z_2|^2)\right)\left(\tau(|z_1|^2 - |z_2|^2)\right)^{i\rho/2-1} \cdot \\ \cdot F\left(\frac{z_1}{\sqrt{\tau(|z_1|^2 - |z_2|^2)}}, \frac{z_2}{\sqrt{\tau(|z_1|^2 - |z_2|^2)}}\right)\,,$$

where $\Theta$ is the Heaviside function. Just as before, $F$ can be understood as a function $f_\tau \in \mathcal{C}^\infty(\mathrm{SU}(1,1))$ through an association of the hyperboloids with the non-compact group. This correspondence depends on $\tau$ as follows:

$$v = \begin{pmatrix} v_1 & v_2 \\ \bar{v_2} & \bar{v_1} \end{pmatrix} \in \mathrm{SU}(1,1)\,;$$

$$\tau = 1: \quad v_1 = \frac{\overline{z}_1}{\sqrt{|z_1|^2 - |z_2|^2}}\,, v_2 = \frac{-\overline{z}_2}{\sqrt{|z_1|^2 - |z_2|^2}}$$

$$\tau = -1: \quad v_1 = \frac{-z_2}{\sqrt{|z_2|^2 - |z_1|^2}}\,, v_2 = \frac{z_1}{\sqrt{|z_2|^2 - |z_1|^2}}\,,$$

and setting $F(z/|z|) = \sum_\tau \Theta\left(\tau(|z_1|^2 - |z_2^2|)\right) f_\tau(v(z))$ one has again the identity

$$D^\chi(g)F(z/|z|) = \sum_\tau \Theta\left(\tau(|z_1|^2 - |z_2|^2)\right) f_\tau(v(z)g)\,, \quad g \in \mathrm{SU}(1,1)\,. \tag{B.2.6}$$

There exists a Plancherel-type theorem [42] between covariant $\mathrm{SU}(1,1)$ functions and homogeneous functions,

$$L^2(\mathrm{SU}(1,1))_{\mathrm{cov}} \oplus L^2(\mathrm{SU}(1,1))_{\mathrm{cov}} \simeq \mathcal{D}_\chi\,,$$

$$\sum_\tau \int_{\mathrm{SU}(1,1)} |f_\tau(v)|^2\, \mathrm{d}v = \int_{\mathbb{C}P} |F(z_1, z_2)|^2\, \omega\,,$$

where the Haar measure is normalized as in appendix A. Unlike the case for SU(2), the space $\mathcal{D}_\chi$ is isomorphic to *two* copies of $L^2(\mathrm{SU}(1,1))_{\mathrm{cov}}$, labelled by $\tau$. Crucially, both $f_\tau$ will still need to satisfy the covariance condition $f_\tau(\gamma v) = e^{i\omega\nu\tau} f_\tau(v)$.

As in the case in the previous subsection, one would like to have a description of $\mathcal{D}_\chi$ in terms of unitary irreducible representations of $\mathrm{SU}(1,1)$. The Plancherel formula (A.4.1) of appendix A implies the isomorphism (recall $j = -1/2 + is$)

$$\bigoplus_k \mathcal{D}_k^+ \otimes \mathcal{D}_k^{+*} \bigoplus_k \mathcal{D}_k^- \otimes \mathcal{D}_k^{-*} \bigoplus_\delta \int^\oplus \mathrm{d}s\, \mathcal{C}_j^\delta \otimes \mathcal{C}_j^{\delta *} \simeq L^2(\mathrm{SU}(1,1))\,,$$

$$\begin{aligned}\sum_{mm'} \left[ \sum_\delta \int_{-\infty}^{\infty} \mathrm{d}s\, |\psi_{mm'}^{j,\delta}|^2 + \sum_k (|\psi_{mm'}^{k+}|^2 + |\psi_{mm'}^{k-}|^2) \right] \\ = \int_{\mathrm{SU}(1,1)} |f(v)|^2\, \mathrm{d}v\,,\end{aligned} \tag{B.2.7}$$

where the various $\psi$ are defined as follows:

$$\begin{aligned}\psi_{mm'}^{j,\epsilon} &= \left[ s \tanh^{1-4\delta}(\pi s) \right]^{1/2} \int_{\mathrm{SU}(1,1)} \mathrm{d}v\, f(v) D_{mm'}^{j(\delta)}(v)\,, \\ \psi_{mm'}^{k,\alpha} &= (-2k-1)^{1/2} \int_{\mathrm{SU}(1,1)} \mathrm{d}v\, f(v) D_{mm'}^{k(q)}(v)\,.\end{aligned}$$

According to the previous equations, the space of homogeneous functions should be isomorphic to two copies of the Hilbert space on the left-hand side of equation (B.2.7), constrained to satisfy covariance. Through a similar argument as the one used in (B.2.2), one may check that the representation functions of both the continuous and discrete series are constrained to $m = \tau\frac{\nu}{2}$ and that, among the discrete series representations, only those labelled by $q = \pm$ contribute to the expansion of $f_\pm$. We may thus unequivocally set $q = \tau$. The final isomorphism reads therefore

$$\mathcal{I}^\chi : \bigoplus_k \mathcal{D}_k^+ \bigoplus_k \mathcal{D}_k^- \bigoplus_\delta \int^\oplus \mathrm{d}s\, \mathcal{C}_j^\delta \bigoplus_\delta \int^\oplus \mathrm{d}s\, \mathcal{C}_j^\delta \simeq \mathcal{D}_\chi$$

$$|k_{(\tau)}, m\rangle \mapsto |\chi, \tau; k, m\rangle\,, \quad |j, \lambda, \sigma\rangle \mapsto |\chi, \tau; j, \lambda, \sigma\rangle\,,$$

having chosen the $K^2$ eigenbasis for the elements of $\mathcal{C}_j^\delta$. The vector $|\chi,\tau;k,m\rangle$ has the function representation

$$\begin{aligned} F^{\chi;\tau}_{k,m}(z) = &(-2k-1)^{1/2}\ \Theta\left(\tau(|z_1|^2-|z_2|^2)\right) \\ &\cdot\left(\tau(|z_1|^2-|z_2|^2)\right)^{i\rho/2-1} D^{k(\tau)}_{\frac{\tau\nu}{2},m}(v_\tau(z))\,, \end{aligned} \tag{B.2.8}$$

while $|\chi,\tau;j,\lambda,\sigma\rangle$ can be written in terms of mixed matrix coefficients as

$$\begin{aligned} F^{\chi;\tau}_{j,\delta,\lambda,\sigma}(z) = &\left[s\tanh^{1-4\delta}(\pi s)\right]^{1/2}\ \Theta\left(\tau(|z_1|^2-|z_2|^2)\right) \\ &\cdot\left(\tau(|z_1|^2-|z_2|^2)\right)^{i\rho/2-1} D^{j(\delta)}_{\frac{\tau\nu}{2},\lambda,\sigma}(v_\tau(z))\,. \end{aligned} \tag{B.2.9}$$

Both functions above are orthonormal relative to the inner product in $\mathcal{D}^\chi$, and we may write a resolution of identity as

$$\begin{aligned} \mathbb{1}_{(\nu,\rho)} = \sum_\tau\Bigg[&\int_{-\infty}^{\infty}\mathrm{d}s\sum_\sigma\int_{\mathbb{R}+i\alpha}\mathrm{d}\lambda\ |\chi,\tau;j,\delta,\lambda,\sigma\rangle\langle\chi,\tau;j,\delta,\overline{\lambda},\sigma| \\ &+\sum_{\substack{2k=-1\\ \text{s.t. } k+\frac{n}{2}\in\mathbb{N}^0}}^{-\infty}\ \sum_{m=-\tau k}^{\tau\infty}|\chi,\tau;k,m\rangle\langle\chi,\tau;k,m|\Bigg]\,. \end{aligned}$$

Observe once more the compatibility between $\mathcal{I}^\chi$ and $\mathrm{SU}(1,1)$ representations,

$$\begin{aligned} D^\chi(g)\circ\mathcal{I}^\chi|k_{(\tau)},m\rangle &= \mathcal{I}^\chi\circ D^{k(\tau)}(g)|k_{(\tau)},m\rangle\,, \\ D^\chi(g)\circ\mathcal{I}^\chi\,|j,\lambda,\sigma\rangle &= \mathcal{I}^\chi\circ D^{j(\delta)}(g)|j,\lambda,\sigma\rangle\,,\qquad g\in\mathrm{SU}(1,1)\,, \end{aligned}$$

following from equation (B.2.6).

## B.3 Harmonic Analysis

It is proven in [42] that there exists a Plancherel formula for smooth compactly-supported functions $f\in\mathrm{SL}(2,\mathbb{C})$. It reads

$$f(\mathbb{1}) = \frac{1}{2}\int_{-\infty}^{\infty}\sum_{\nu=-\infty}^{\infty}(\nu^2+\rho^2)\,\mathrm{Tr}_{\mathcal{D}_\chi}\left[\int\mathrm{d}g f(g)D^\chi(g)\right]\,,$$

with the Haar measure on the group normalized as

$$\mathrm{d}g = (2\pi)^{-4}|a_{22}|^{-2} Da_{12} Da_{21} Da_{22}\,, \tag{B.3.1}$$

for $a_{ij}$ the matrix coefficients of the defining representation and $Dz = \frac{i}{2}(\mathrm{d}z \wedge \mathrm{d}\overline{z})$.

# *Appendix C* : Homogeneous Spaces of the Lorentz Group

This small appendix lists the spaces of transitivity of $\mathrm{SO}(1,3)$ which figure frequently throughout the present work. Explicit parametrizations of the images of the spin homomorphism from section III.3.1 are also discussed, and the notation from that chapter is used.

*(a) The Euclidian sphere* $S^2$

As a subset of $\mathbb{R}^3$, it is defined as

$$S^2 = \{(x,y,z) \in \mathbb{R}^3 \mid x^2 + y^2 + z^2 = 1\} \quad \sim$$ 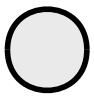 .

Using the SU(2) parametrization $g = e^{iL^3\phi} e^{iL^2\theta} e^{iL^3\psi}$, with $0 \le \phi, \psi < 2\pi$, $0 \le \theta < \pi$, it can be obtained through the maps

$$\begin{aligned} \pi_\pm : \mathrm{SU}(2)/\mathrm{U}(1) &\to S^2 \subset \mathbb{R}^3 \\ g &\mapsto \langle \pm \,|\, g^{-1}\sigma_i g \cdot \pm \rangle \, \hat{e}^i \,, \end{aligned}$$

such that $\langle \pm \,|\, g^{-1}\sigma_i g \cdot \pm \rangle \, \hat{e}^i = (-\sin\theta\cos\phi,\ \sin\theta\sin\phi,\ \cos\theta) \in S^2$.

*(b) The two-sheeted hyperboloid* $H^\pm$

As subsets of $\mathbb{R}^{1,2}$, the upper $(+)$ and lower $(-)$ sheet are defined as

$$H^\pm = \{(t,x,y) \in \mathbb{R}^{1,2} \mid t^2 - x^2 + y^2 = 1\,,\ \pm t > 0\} \quad \sim$$ 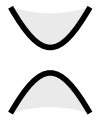 .

Using the $\mathrm{SU}(1,1)$ parametrization of equation (A.2.2)), $H^\pm$ may be obtained with

$$\begin{aligned} \pi_\pm : \mathrm{SU}(1,1)/\mathrm{U}(1) &\to H^\pm \subset \mathbb{R}^{2,1} \\ g &\mapsto \langle \pm \,|\, g^{-1}\, \varsigma_i g \ \cdot \pm \rangle \ , \end{aligned}$$

such that $\langle \pm \,|\, g^{-1}\, \varsigma_i g \cdot \pm \rangle \, \hat{e}^i = \pm(\cosh u, \cos\alpha \sinh u, -\sin\alpha\sinh u) \in H^\pm$.

*(c) The one-sheeted hyperboloid* $H^{\mathrm{sl}}$

Also a subset of $\mathbb{R}^{1,2}$, the one-sheeted space-like hyperboloid is defined as

$$H^{\mathrm{sl}} = \{(t,x,y) \in \mathbb{R}^{1,2} \mid t^2 - x^2 + y^2 = -1\} \quad \sim$$

.

With the $\mathrm{SU}(1,1)$ parametrization of equation (A.2.1)), it can be obtained via

$$\begin{aligned} \pi_\pm : \mathrm{SU}(1,1)/(\mathbb{Z}_2\, e^{iuK_2}) &\to H^{\mathrm{sl}} \subset \mathbb{R}^{2,1} \\ g &\mapsto -i\, \langle l^\pm \,|\, g^{-1}\, \varsigma_i g \;\cdot l^\pm \rangle\, \hat{e}^i\,, \end{aligned}$$

for which $\langle l^\pm \,|\, g^{-1}\, \varsigma_i g \;\cdot l^\pm \rangle\, \hat{e}^i = \pm(\sinh t, -\sin\alpha \cosh t, -\cos\alpha\cosh t) \in H^{\mathrm{sl}}$.

# *Appendix D* : Geometrical Aspects of 2+1 Minkowski Space-Time

In the following we make a couple of simple observations about convex geometry in $\mathbb{R}^{1,2}$. We will explicitely exclude light-like vectors from the analysis below[54]. The metric signature is taken to be sig $\eta_{(1,2)} = (+, -, -)$.

## D.1 Linear algebra for Minkowski space

*(a) Minkowski triangles and tetrahedra*

The convex hull of any three points not all colinear is a *Minkowski triangle* if:

1. The induced metric on the triangle - that is the metric obtained by restriction to vectors in the triangle - has either signature $(+, -)$ or $(-, -)$, and

2. Every edge of the triangle is non-null.

A triangle with signature $(+, -)$ is called *time-like*, while one with signature $(-, -)$ is termed *space-like*. The convex hull of any four points not all colinear is a *Minkowski tetrahedron* if:

1. Every edge of the tetrahedron is non-null, and

2. Every face is either space- or time-like.

*(b) Orthogonal projections*

The orthogonal projection of a vector $v$ onto $u$ is given by $\text{Proj}_u v = \|u\|^{-2} \langle v, u \rangle$. Let $\{u, u_i^\perp\}$ be a basis for $\mathbb{R}^{1,2}$ with $\langle u, u_i^\perp \rangle = 0$, and

[54]Light-like lines overlap the notions of orthogonality and parallelism, and are for this reason patological in our context.

define $v = au + b^i u_i^\perp$. Then the coefficient of $u$ in $v$ is given by the orthogonal projection of $v$ onto $u$,

$$\text{Proj}_u v = a \|u\|^{-2} \langle u, u \rangle = a.$$

*(c) Half-spaces*

Let $v$ be a unit vector, i.e. $\|v\|^2 = \pm 1$. The set $H_v = \left\{ x \mid \text{Proj}_v x \leq 0 \right\}$ defines a half-space through the origin such that $v \notin H$ and $\partial H_v \perp v$,

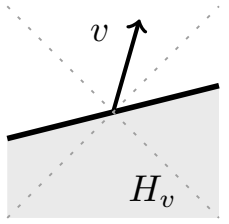

The set $H_v$ can be translated in the direction of $v$ by a positive amount $h$, defining the translated half-space

$$H_v^h = \left\{ x \mid \langle x, v \rangle \|v\|^2 \leq h \right\}.$$

Observe that the half-space so defined will always contain the origin.

*(d) Height of triangles and tetrahedra*

Consider the triangle $ABC$, and let $v$ be the outward unit normal vector to the opposite edge $AC$ to $B$. The height $h_B$ of the triangle $ABC$ from the vertex $B$ is defined to be the unique positive number such that $\vec{B} + h_B v$ lies on the line $AC$. In terms of the other edges, the triangle height is given by $h_B = |\text{Proj}_v \vec{AB}| = |\text{Proj}_v \vec{CB}|$,

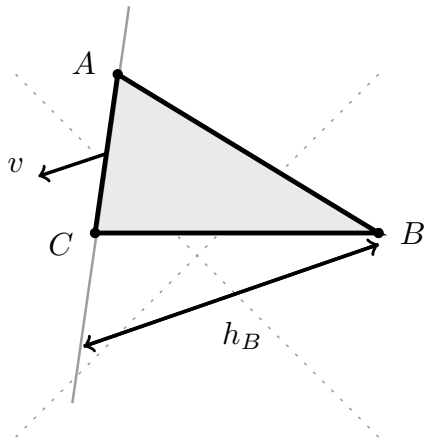

These definitions extend in the obvious manner to tetrahedra.

*(e) Squared areas and volumes of triangles and tetrahedra*

Consider a tetrahedron ABCD, and let $a, b$ and $c$ be edge vectors of the tetrahedron with the same base at vertex $D$ and end-points at A, B and C, respectively. We define the tetrahedron volume by $V_{abc}^2 = \frac{1}{(3!)^2} \left\|\star(a \wedge b \wedge c)\right\|^2$, and the area of the triangle with sides $a, b$ by $A_{ab}^2 = \frac{1}{(2!)^2} \left\|\star(a \wedge b)\right\|^2$. This is the same definition one may have for areas and volumes in Euclidean space, but now we allow for negative squared areas and volumes, depending on the signature of the metric of the space.

*(f) Signs of squared areas*

Given the above definition, the signs of the squared areas of triangles depend on their causal character. Indeed, let $a, b$ be edge vectors of a time-like triangle, both with base $A$. Since the face is time-like, the span of its edges must contain both space- and time-like vectors. Thus the quadratic $f(x) = ||ax + b||^2$ must change sign, implying its discriminant $\Delta$ must be positive. But we also have that $\Delta = -4||a \wedge b||^2$, and hence $A_{ab}^2 < 0$. If one considers a space-like triangle, the same argument implies that the polynomial must not change sign, and this shows that $A_{ab}^2 > 0$.

We may thus claim that time-like triangles are characterised by negative squared areas, while space-like triangles have positive squared areas.

*(g) Orthogonal vectors to triangles*

Consider again the triangle $ABC$ and the edges $a, b$, both with the same base point. Then the vector $v = \frac{\star(a \wedge b)}{2\sqrt{|A_{ab}^2|}}$ is a unit vector orthogonal to the triangle. Orthogonality follows from the properties of the Hodge star: let $\omega$ be the volume form induced by the metric. Then $\langle \star(a \wedge b), b \rangle \omega = - \langle a \wedge b, \star b \rangle \omega = a \wedge b \wedge b = 0$, and analogously $\langle \star(a \wedge b), a \rangle \omega = 0$. Since $\omega$ is non-degenerate by definition, orthogonality holds.

*(h) Squared volume formula for a tetrahedron in terms of boundary areas*

Let $a, b, c$ be edge vectors for a tetrahedron $ABCD$ as before, all having a common base point at D. Recall that $V_{abc}^2 = \frac{1}{(3!)^2} \left\|\star(a \wedge b \wedge c)\right\|^2$. Define the orthogonal unit vector to the face $ACD$ by $v = \frac{\star(a \wedge c)}{\sqrt{4|A^2_{ac}|}}$, pointing out of the tetrahedron, and the tetrahedron height from the vertex opposite to the same face by $h_B = |\mathrm{Proj}_v b|$. Then we may write $b = -h_B v + \alpha a + \beta c$, for some numbers $\alpha, \beta$. The following holds:

$$
\begin{aligned}
V^2_{abc} &= \frac{1}{(3!)^2} \left\| \star(a \wedge c \wedge (h_B v)) \right\|^2 \\
&= \frac{1}{(3!)^2} \det \begin{pmatrix} \langle a, a \rangle & \langle a, c \rangle & h_B \langle a, v \rangle \\ \langle c, a \rangle & \langle c, c \rangle & h_B \langle c, v \rangle \\ h_B \langle v, a \rangle & h_B \langle v, c \rangle & h_B^2 \langle v, v \rangle \end{pmatrix} \\
&= \frac{(2!)^2}{(3!)^2} h_b^2 \left\| v \right\|^2 A^2_{ac} = \frac{1}{3^2} h_b^2 |A^2_{ac}| \,,
\end{aligned}
$$

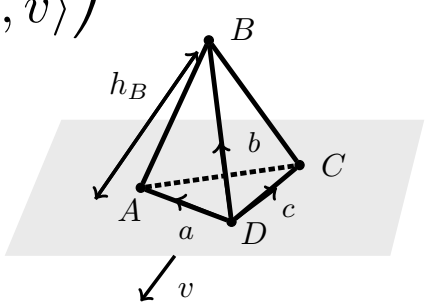

since $v \perp a, c$. We may conclude that Minkowskian tetrahedra, defined in this manner, have positive squared volumes independently of their causal character.

*(i) Areas and volumes of triangles and tetrahedra*

Because we have shown that squared volumes are always positive, we may define the volume of a tetrahedron $P$ by $V_P = \sqrt{V^2_P}$. For areas of triangles $Q$, we take $A_Q = \sqrt{|A^2_Q|}$.

*(j) Minkowski polygons and polyhedra*

Since we have a notion of orthogonal half-spaces, we define polyhedra as follows: a convex Minkowskian polyhedron is a finite and bounded intersection of half-spaces containing the origin,

$$
P = \bigcap_{\{v,h\}} H^h_v \,,
$$

with each half-space being characterised by an orthogonal non-null vector $v$ and a positive height $h$ from the origin along $v$. We further require that every edge of $P$ is non-null. Every face of such a polyhedron is a Minkowskian polygon.

*(k) Additivity of areas and volumes*

I have so far only defined areas of triangles and volumes of tetrahedra. Since our definitions have relied on the scalar product and the Hodge star - both linear maps - areas and volumes are naturally additive *in their own subspace of definition*. That is to say, although there is no sense in which one might add the area of a time-like triangle and a space-like triangle, our definitions naturally allow for summing areas of parallel triangles. Volumes, on the other hand, are defined on the whole three-dimensional space, are always positive, and may freely be added to each other. Note that then a tetrahedron has a total volume, but it may not have a total area.

*(l) Areas of polygons*

Consider a general convex polygon, and place a vertex in its interior. Now join every vertex on the boundary of the polygon to the interior vertex, obtaining a triangulation of the polygon. The polygonal area will be defined as the sum of the triangle areas.

*(m) Volumes of polyhedra*

Analogously, given a general convex polyhedron, consider its triangulation by tetrahedra: triangulate first every face as above, obtaining triangular faces $f$, and then join every vertex to a new vertex in the interior of the polyhedron[55]. The total volume of the polyhedron is well-defined as the sum of the volumes of the individual tetrahedra, and according to the previous discussion it is given by

$$V = \frac{1}{3} \sum_f h_f A_f \,, \tag{D.1.1}$$

[55] It is well-known that in more than two dimensions there exist non-convex polyhedra not admitting a triangulation by simplices.

where $A_f$ is the area of the face $f$ and $h_f$ is the tetrahedral height from the interior vertex to the plane defined by the face $f$.

*(n) Closure condition for polyhedra*

I show this for a tetrahedron, as the generalization to other polyhedra should be clear from the previous discussion. Let $a, b, c$ be edge vectors of the tetrahedron, all with the same base point, and consider the four vectors normal and outward-pointing to its faces: $\star(b\wedge a), \star(c\wedge b), \star(a\wedge c), \star[(c-b) \wedge (a-b)]$. Then it is immediate that

$$\star(b \wedge a) + \star(c \wedge b) + \star(a \wedge c) + \star[(c-b) \wedge (a-b)] = 0\,.$$

Thus, for any convex polyhedron,

$$\sum_f v_f A_f = 0\,,$$

where $v_f$ is the unit vector orthogonal to the face $f$ and outward-pointing.

## D.2 Angles in the Minkowski plane

In order to discuss some further properties of polytopes in Minkowski space-time we will need the notion of angle between two arbitrary non-null vectors. When the plane defined by the two vectors of interest is entirely space-like one may make use of Euclidean angles, which are defined in the usual way as

$$\cos\theta_{uv} = \frac{\langle u, v\rangle}{||u||\,||v||},$$

with the convention that $||u|| = \sqrt{||u||^2}$ and $||u|| = ix, x > 0$ whenever $||u||^2 < 0$. It might however be that the plane spanned by $u, v$ has the metric signature $(+,-)$, and this requires a more careful discussion. We will use the description of [136], since it allows for keeping the property of angle additivity and constructing an analogous Schläfli identity to the Euclidean case [137, 138].

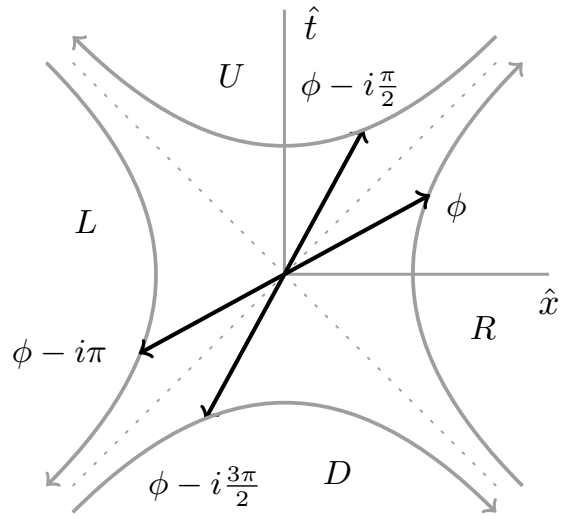

Figure D.1: Possible orientation for hyperbolas in $\mathbb{R}^{1,1}$. Just as Euclidean orthogonal vectors are separated by a $\frac{\pi}{2}$ angle, so too are Minkowski ones separated by $-i\frac{\pi}{2}$.

Choose an orientation for the Minkowski plane $\mathbb{R}^{1,1}$, e.g. the one of figure D.1, and consider a pair of non-null vectors $u, v$. The oriented angle $\phi_{uv}$ from $u$ to $v$ is defined to be a complex number of the form $\phi_{uv} = \theta_{uv} - i\frac{\pi}{2}k_{uv}$, $\theta_{uv} \in \mathbb{R}$, $k_{uv} \in \mathbb{Z}$ such that

1. $\theta_{uv} \geq 0$ whenever $(u, v)$ is positively ordered relative to the chosen orientation;

2. $\theta_{uv} = -\theta_{vu}$;

3. $\langle u, v\rangle = ||u||\,||v|| \cosh \phi_{uv}$
   $= ||u||\,||v|| (\cosh \theta_{uv} \cos \frac{\pi}{2}k_{uv} - i \sinh \theta_{uv} \sin \frac{\pi}{2}k_{uv})$;

4. if $(u, v)$ is positively ordered and $\langle u, v\rangle = 0$, then $\phi_{uv} = -i\frac{\pi}{2} + 2i\pi n$, $n \in \mathbb{Z}$.

5. if $t$ is a third non-null vector and $\phi_{uv} = \theta_{uv} - i\frac{\pi}{2}k_{uv}$, $\phi_{vt} = \theta_{vt} - i\frac{\pi}{2}k_{vt}$, then $\exists\, n$ s.t. $\phi_{ut} = (\theta_{uv} + \theta_{vt}) - i\frac{\pi}{2}(k_{uv} + k_{vt}) + 2i\pi n$.

Starting with a vector on the right-most hyperboloid $R$, property number 4 shows that each time $\phi$ jumps orthogonally between hyperboloids one gets an additional $-i\frac{\pi}{2}$ factor, modulo $2i\pi n$; in that sense the role of $-i\frac{\pi}{2}$ is analogous to the one of $\frac{\pi}{2}$ in Euclidian geometry. Observe furthermore that whenever two vectors have the same causal character the inner product formula involves a cosh, while otherwise it relates to a sinh function.

## D.3 Uniqueness and existence of Minkowski polyhedra

Having established the above definitions and properties, we now turn to formulating an analogous Minkowski theorem for Minkowskian polyhedra. We cite a famous result by Alexandrov [139, Theorem 1 of section 6.3]:

**Theorem D.1** (Alexandrov). *Let the term "face" stand for a vertex, edge or proper face of a polyhedron, and define two faces to be parallel if they are contained in parallel support hyperplanes. If for all pairs of parallel faces of two convex Euclidean polyhedra neither face can be placed strictly inside the other by parallel translation, then the polyhedra are translates of one another.*

Alexandrov proved this theorem in the context of Euclidean 3-space, but it clearly carries over to Minkowski tetrahedra.

**Corollary D.1.** *If for all pairs of parallel faces of two convex polyhedra in $\mathbb{R}^{1,2}$ neither face can be placed strictly inside the other by parallel translation, then the polyhedra are translates of one another.*

*Proof.* Note that there is an identity mapping $\mathbb{R}^{1,2} \to \mathbb{R}^3$, and that a set is convex in $\mathbb{R}^{1,2}$ if and only if it is convex in Euclidean space. Suppose that we are given polyhedra $P$, $P'$ in the conditions of the theorem, and consider their Euclidean image. These images satisfy the requirements of Theorem D.1, and thus they are translates of each other. But then, under the identity mapping, so too are the original polyhedra $P$, $P'$. □

We then have an immediate corollary on congruence of polyhedra depending on their face areas:

**Corollary D.2.** *Let $P$, $P'$ be two convex polyhedra in either $\mathbb{R}^{1,2}$ or $\mathbb{R}^3$, defined as above. Denote by $Q$ a face of $P$, and by $Q'$ a face of $P'$. Moreover, let* $\{v_Q, A_Q\}, \{v_{Q'}, A_{Q'}\}$ *be the sets of outward-pointing orthogonal vectors to their faces (none of them null), as well as their respective areas. If and only if both sets are the same, then* $P$ *and* $P'$ *are translates of each other. That is, convex polyhedra in Minkowski and Euclidian 3-space are uniquely characterised by their face areas and normals, up to translations.*

*Proof.* Since both sets of vectors are the same, the polyhedra share pairwise-parallel faces $(Q, Q')$. Consider the function $f(Q) = A_Q = \sqrt{|A_Q^2|}$, which is well-defined and monotonic on parallel polygons independently of their causal character. By virtue of the monotonicity of $f$, parallel faces of equal area cannot be placed strictly inside each other by a translation; Theorem D.1 then implies the result. □

On the other hand, we may also prove existence of polyhedra given some boundary data satisfying the closure condition, essentially repeating the proof due to Minkowski [139], thus establishing both uniqueness and existence:

**Theorem D.2** (Minkowski's theorem). *Let* $\{v_f, A_f\}$ *be a set consisting of unit vectors* $v_f$ *in either* $\mathbb{R}^3$ *or* $\mathbb{R}^{1,2}$ *and positive numbers* $A_f$*. Suppose such vectors are not all co-planar, no vector is light-like, and the following holds*

$$\sum_f v_f A_f = 0\,. \tag{D.3.1}$$

*Then there exists a unique convex polyhedron in* $\mathbb{R}^3$ *or* $\mathbb{R}^{1,2}$*, respectively, such that* $A_f$ *is the area of its face* $f$ *and* $v_f$ *points orthogonally outward to the face.*

*Proof.* Assume we are given $F$ such vectors and $F$ such numbers. Consider all sets $h$ containing $F$ non-negative numbers $h_f$, to be understood as distances from the origin. Equation (D.3.1) implies that the vectors $v_f$ cannot all point towards the same half-space, and thus, for every $h$, the intersection of half-spaces $H_{v_f}^{h_f}$ in the relevant space $\mathbb{R}^3$ or $\mathbb{R}^{1,2}$ defines convex polyhedra $P_h$. Let $\tilde{A}_f$ be the area associated to the hyperplane boundary of $H_{v_f}^{h_f}$, and set it to zero if that hyperplane does not define a (proper) face of $P_h$. Among all $P_h$, consider those satisfying the constraint

$$\sum_f h_f A_f = 1\,.$$

We now show that there exists a polyhedron which maximises the volume under the above condition. First, note that all $h_f$ are bounded from above by $h_f \leq 1/A_f$, so that the constraint is a closed condition and the set of admissible $h_f$ is compact. Hence the volume of $P_h$ attains a maximum in the domain of the constraint at some $P_h^*$. Then, resorting to Lagrange multipliers, the extrema are found at

$$\frac{\partial}{\partial h_f}\left(V(P_f) + \lambda\left(\sum_{f'} h_{f'} A_{f'} - 1\right)\right) = 0$$
$$\Rightarrow \frac{1}{3}\tilde{A}_f^* + \lambda A_f = 0\,,$$

where we used the volume formula (D.1.1). The maximum corresponds to a polyhedron $P_h$ with face normals $v_f$ and areas $\tilde{A}_f^* \propto A_f$. A suitable rescaling yields a polyhedron with areas $A_f$, and by Corollary D.2 this polyhedron is unique up to translations. □

## D.4 Rigidity of Minkowski polytopes

We now turn to the question of whether convex polytopes in space-time are rigid. We will call such a polytope rigid if every continuous displacement of its vertices leaving its edge lengths and internal face angles invariant and preserving its combinatorics amounts to an orthogonal transformation with respect to the space-time metric, i.e. a congruence, or rigid-body motion. That every convex polytope in Minkowski space of dimension $\geq 3$ is rigid, much like their Euclidean counterparts, can be shown straightforwardly by making use of a result from [136, Lemma 9].

**Lemma D.1.** *Let $P(t)$ be a smooth family of convex orientable polyhedra such that, for each t, each edge of $P(t)$ is non-null, its length is invariant, and each face carries a non-degenerate metric. Let $f_1, f_2$ denote the faces adjacent to an edge e. Denote by $n^e_{f_i}$ the unit vector which lies in $f_i$, is orthogonal to e, and points inside $f_i$, and by $m_{f_i}$ the outward-pointing normal unit vector to $f$. The dihedral angle (i.e. the angle between $m_{f_1}$ and $m_{f_2}$) at an edge $\theta_e(t)$ then satisfies the velocity equation*

$$\frac{d\theta_e}{dt} = \sum_i \langle \frac{dm_{f_i}}{dt}, n^e_{f_i} \rangle \ . \qquad \text{(D.4.1)}$$

With the goal of proving that polyhedral corners are rigid, we start with a lemma on the angular velocities of a deformation.

**Lemma D.2.** *Let $P$ be a polyhedral corner, and denote by e the unit edge vectors with base point at the vertex. Consider a deformation $P(t)$ as above, with the additional requirement that the internal face angles are held constant. Then $P(t)$ satisfies the closure condition*

$$\sum_e \epsilon(e) e \frac{d\theta_e}{dt} = 0 \, ,$$

*where $\epsilon(e)$ is the sign of the squared norm of e.*

*Proof.* Consider first the following identity

$$\begin{aligned}\sum_f \sum_{e\in\partial f} \epsilon(e) e \, \langle \frac{\mathrm{d}m_f}{\mathrm{d}t}, n^e_f \rangle &= \sum_f \sum_{e\in\partial f} \epsilon(e) \left[ \langle \frac{\mathrm{d}m_f}{\mathrm{d}t}, n^e_f \wedge e \rangle + n^e_f \, \langle \frac{\mathrm{d}m_f}{\mathrm{d}t}, e \rangle \right] \\ &= \sum_f \langle \frac{\mathrm{d}m_f}{\mathrm{d}t}, \sum_{e\in\partial f} \epsilon(e) n^e_f \wedge e \rangle \ ,\end{aligned}$$

where $f$ denotes a face of $P$ and the inner product between 1- and 2-vectors is short-hand for the interior product. In the second equality I used the fact that $\mathrm{d}m_f/\mathrm{d}t$ must be orthogonal to $m_f$ and tangent to the arc of rotation $\theta_e(t)$ (since the deformation is restricted to amount to hinge movements at the edges), and thus normal to $e$. Turning now to a single face, denote by $\overline{\star}$ and $\overline{\omega}$ the Hodge-star and volume form associated to the induced metric at $f$. Let also $e_1, e_2$ be the two edges incident to $f$. Then

$$\begin{aligned}\sum_{e\in\partial f} \epsilon(e) n_f^e \wedge e &= \epsilon(e_1) n_f^{e_1} \wedge e_1 + \epsilon(e_2) n_f^{e_2} \wedge e_2 \\ &= \pm\left(\epsilon(e_1)\overline{\star} e_1 \wedge e_1 - \epsilon(e_2)\overline{\star} e_2 \wedge e_2\right) \\ &= \pm\left(\epsilon(e_1)\langle e_1, e_1\rangle - \epsilon(e_2)\langle e_2, e_2\rangle\right)\overline{\omega} = 0\,,\end{aligned}$$

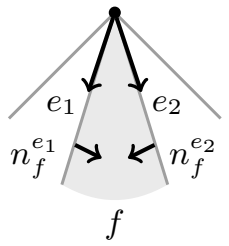

and the sign indeterminacy is due to the possible orientations of $\overline{\star}$. Interchanging summations it must be that

$$\sum_e \sum_f \epsilon(e) e \,\langle \frac{\mathrm{d} m_f}{\mathrm{d} t}, n_f^e \rangle = 0\,,$$

and the statement of the lemma follows from equation (D.4.1). □

Next we have a lemma on the combinatorics of corners.

**Lemma D.3.** *Consider a family of polyhedral corners $P(t)$, as in Lemma 2. Label the edges of $P(t)$ according to the sign of $\epsilon(e)\frac{d\theta_e}{dt}$, leaving an edge unlabelled if the associated angular velocity vanishes. Then either there are 4 or more sign changes as one goes around the vertex, or all edges are unlabelled.*

*Proof.* Clearly there must be an even number of sign changes, because in going around a vertex one always returns to the initial sign. We therefore need to show that 0 and 2 sign changes are impossible.

First, due to the closure condition, one cannot have 0 sign changes, as otherwise the linear combination $\sum_e \epsilon(e) e \frac{\mathrm{d}\theta_e}{\mathrm{d}t}$ would not vanish, contradicting the previous lemma.

Secondly, suppose there are exactly 2 sign changes. Assume without loss of generality that $e_1, ..., e_n$ are labelled with $+$, and $e_{n+1}, ..., e_m$ are labelled with $-$. Consider the half-space $H^+$ separating the two sets of edges and containing the ones labelled with a positive sign. Then clearly $\sum_{i=n+1}^{m} \epsilon(e_i) e_i \frac{\mathrm{d}\theta_{e_i}}{\mathrm{d}t}$ also lies in $H^+$, and thus the linear combination $\sum_e \epsilon(e) e \frac{\mathrm{d}\theta_e}{\mathrm{d}t}$ of all edges does too, contradicting the closure condition of the lemma above. □

From the previous two lemmas one finally has the following theorem.

**Theorem D.3.** *Let $P$ be a convex polytope in Minkowski space of dimension $d \geq 3$, defined as in section D.1. Then every smooth deformation of $P$ preserving the length of its edges and the internal face angles, and such that the metric at each face remains non-degenerate, is an isometry of the Minkowski metric, i.e. a congruence of $P$. That is, convex Minkowski polyhedra are rigid.*

*Proof.* We prove the statement first for 3 dimensions, since higher dimensional polytopes are characteristically more constrained as noted in [139].

Consider a smooth family $P(t)$ of convex 3 dimensional polyhedra containing $P$, and label the edges according to the rules of Lemma D.3. By the same lemma, at each vertex there can be either no labels at all or at least 4 sign changes. By Cauchy's combinatorial lemma [139, Section 2.1], in a planar graph where edges are either not labeled or labeled with a sign there must be less than 4 sign changes around a vertex. All convex polyhedra induce graphs through its vertices and edges which can be embedded on an Euclidean sphere [139], and thus those graphs are planar. It thus follows that no edge of $P$ can be labeled, and therefore all angular velocities vanish. This is enough to establish rigidity, since then every smooth deformation must preserve the dihedral angles.

For a given 4 dimensional polytope, consider its intersection with an Euclidean 3-sphere around one of its vertices. The resulting intersection is a 3 dimensional convex polyhedron, and it is rigid from the discussion above. Thus the polytope itself is rigid, and the argument can be extended to higher dimensions. □

# *Appendix E* : Critical Points in the 3d Model

This appendix contains the proof that the parameters $\alpha_{ab}$ appearing in the asymptotic analysis of the space-like 3-dimensional model (cf. section II.5 and equation (II.5.6)) are related to the dihedral angles of a tetrahedron. We shall take non-collinear boundary data, i.e. data for which Minkowski's theorem is applicable at each triple of boundary states, and we fix $g_4 = \mathbb{1}$.

Defining $\theta_{ab} := \ln \vartheta_{ab}$, the gluing condition (II.5.6) implies the system of equations

$$\begin{cases} n_{ab}^{-1} g_a^{-1} g_b n_{ba} |l^+\rangle = e^{\theta_{ab}} |l^+\rangle \\ n_{ab}^{-1} g_a^{-1} g_b n_{ba} |l^-\rangle = e^{-\theta_{ab}} |l^-\rangle \,, \quad n, g \in \mathrm{SU}(1,1)\,, \end{cases}$$

from where, since $|\pm\rangle$ spans $\mathbb{C}^2$, one sees that $n_{ab}^{-1} g_a^{-1} g_b n_{ba} = e^{\theta_{ab}\sigma_1}$. Then the following chain of equalities holds,

$$\begin{aligned} g_a^{-1} g_b n_{ba} n_{ab}^{-1} &= n_{ab} e^{\theta_{ab}\sigma_1} \sigma_3 n_{ab}^{\dagger} \sigma_3 \\ &= \cosh \theta_{ab} - i \sinh \theta_{ab}\, n_{ab} \sigma_2 n_{ab}^{\dagger}\, \sigma_3 \\ &= \cosh \theta_{ab} - i \sinh \theta_{ab}\, \sigma_3\, (v_{ab} \cdot \varsigma)\, \sigma_3 \\ &= e^{-i\theta_{ab}\sigma_3\, (v_{ab}\cdot\varsigma)\, \sigma_3}\,, \end{aligned}$$

with $v_{ab} = \pi(n_{ab})\hat{e}_2 \in H^{\mathrm{sl}}$ the geometrical vector associated to $n_{ab}$. The dot $(\cdot)$ stands for the scalar product with respect to $\eta_{(1,2)}$. We can construct the system of equations

$$\begin{cases} g_a^{-1} g_b n_{ba} n_{ab}^{-1} = e^{-i\theta_{ab}\sigma_3\, (v_{ab}\cdot\varsigma)\, \sigma_3} \\ g_c^{-1} g_b n_{bc} n_{cb}^{-1} = e^{-i\theta_{cb}\sigma_3\, (v_{cb}\cdot\varsigma)\, \sigma_3} \\ g_a^{-1} g_c n_{ca} n_{ac}^{-1} = e^{-i\theta_{ac}\sigma_3\, (v_{ac}\cdot\varsigma)\, \sigma_3}\,, \end{cases} \tag{E.1}$$

and by factoring out $g_a, g_b, g_c$ find

$$\begin{aligned} &e^{-i\theta_{ab}\sigma_3\,(v_{ab}\cdot\varsigma)\,\sigma_3} n_{ab} n_{ba}^{-1} \\ &\qquad = e^{-i\theta_{ac}\sigma_3\,(v_{ac}\cdot\varsigma)\,\sigma_3} n_{ac} n_{ca}^{-1}\, e^{-i\theta_{cb}\sigma_3\,(v_{cb}\cdot\varsigma)\,\sigma_3} n_{cb} n_{bc}^{-1}\,. \end{aligned} \tag{E.2}$$

To proceed we make the simplifying assumption that all matched boundary data are parallel, i.e. $n_{ab} = n_{ba}$; there is no loss of generality in doing so since, for a given solution to the gluing equations, each triple of boundary data can be gauge-rotated such that our assumption is satisfied. Equation (E.2) thus implies

$$\begin{aligned} &\cosh\theta_{ab} - i\sinh\theta_{ab}\sigma_3\,(v_{ab}\cdot\varsigma)\,\sigma_3 \\ &= [\cosh\theta_{ac} - i\sinh\theta_{ac}\sigma_3\,(v_{ac}\cdot\varsigma)\,\sigma_3]\,[\cosh\theta_{cb} - i\sinh\theta_{cb}\sigma_3\,(v_{cb}\cdot\varsigma)\,\sigma_3]\ . \end{aligned}$$

The Pauli matrices together with the identity are linearly independent, so that the previous equation splits into

$$\begin{cases} \cosh\theta_{ab} = \cosh\theta_{ac}\cosh\theta_{cb} - \sinh\theta_{ac}\sinh\theta_{cb}\, v_{ac}\cdot v_{cb} \\ \sinh\theta_{ab}\, v_{ab} = \cosh\theta_{ac}\sinh\theta_{cb}\, v_{cb} \\ \qquad\qquad + \cosh\theta_{cb}\sinh\theta_{ac}\, v_{ac} - \sinh\theta_{ac}\sinh\theta_{cb}\, v_{ac}\times v_{cb}\,, \end{cases}$$

where the completeness relation (II.2.4) for $\varsigma_i$ was used. Contracting the second equation with $v_{ab}\times v_{ac}$ yields

$$\theta_{cb} = 0 \quad \vee \quad \tanh\theta_{ac} = \frac{v_{cb}\cdot v_{ab}\times v_{ac}}{(v_{ac}\times v_{cb})\cdot(v_{ab}\times v_{ac})}\,,$$

or, employing the quadruple product identity $(a\times b)\times(c\times d) = a\cdot(b\times d)c - a\cdot(b\times c)d$,

$$\theta_{cb} = 0 \quad \vee \quad \tanh\theta_{ac} = \frac{v_{ac}\cdot[(v_{cb}\times v_{ac})\times(v_{ac}\times v_{ab})]}{(v_{ac}\times v_{cb})\cdot(v_{ab}\times v_{ac})}\,. \tag{E.3}$$

Equations (E.3) identify two *possible* solutions to the gluing equations. The first is given by $\theta_{cb} = 0$, or equivalently $g_b = g_c$. It then follows from

the system (E.1) that

$$e^{-i\theta_{ac}\sigma_3\,(v_{ac}\cdot\varsigma)\,\sigma_3} = e^{-i\theta_{ab}\sigma_3\,(v_{ab}\cdot\varsigma)\,\sigma_3}\,,$$

and - since $v_{ac}$ and $v_{ab}$ must not be colinear, as we assumed that the boundary data was not degenerate - it must be that also[56] $\theta_{ac} = \theta_{ab} = 0$. The same argument can be repeated with a fourth label $d$, from where it follows that $g_a = g_b = g_c = g_d$. By virtue of the gauge fixing $g_4 = \mathbb{1}$, one finally has that $g_a = \mathbb{1}$ for all $a = 1, .., 4$. Note that this solution is always present whenever the boundary data allows for a non-empty solution set.

Turning to the second equation in (E.3), observe first that the gluing equations (II.5.8) essentially describe a triangular development 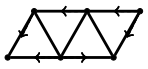 with identified edges and edge orientations. A moment of thought is enough to convince oneself that such a net can either correspond to 1) a degenerate tetrahedron (i.e. a triangle) or 2) to a proper tetrahedron and its reflection (depending on whether the "flaps" are closed above or below the bottom face)

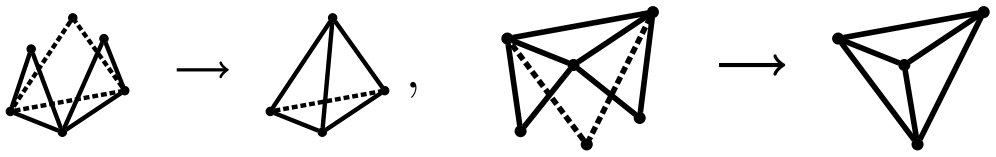

Constructive inspection shows that options 1) and 2) are mutually exclusive when the boundary data is not made up of four copies of an equilateral triangle: if the edge orientations are such that the net can be closed into a flat triangle, than it cannot correspond to a tetrahedron, and vice-versa. Moreover up to orientation signs (which depend on the particular boundary data) the vectors $v_{ab}$ can be identified with the sides of a (possibly

[56] Curiously, were we working with the time-like model the angles would be Euclidian, and there would be a second solution $\theta_{ab} = \pi$ showing that $g_a = \pm\mathbb{1}$. The sign is geometrically irrelevant, since the spin homomorphism is defined modulo $\mathbb{Z}_2$. For hyperbolic functions the solution $g_a = -\mathbb{1}$ is absent.

degenerate) tetrahedron,

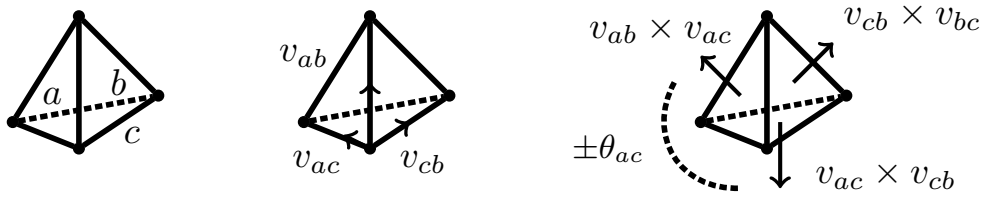

,

and the second equation of (E.3) clearly shows that $\theta_{ab}$ labels the dihedral angle between the triangular faces $a$ and $b$ modulo a sign. Thus the following two alternatives are possible: if the boundary data is that of a degenerate tetrahedron then all dihedral angles vanish, and the first and second solutions are simply identified; in that case there is a single solution to the gluing equations given by all $g_a = \mathbb{1}$. If however the boundary data is that of a proper tetrahedron then its reflected counterpart solves the second equation of (E.3) with

$$g_a = e^{-i\theta_{4a}\sigma_3\,(v_{4a}\cdot\varsigma)\,\sigma_3}\,, \quad \theta_{ac} = \text{arctanh}\,\frac{v_{cb}\cdot v_{ab}\times v_{ac}}{(v_{ac}\times v_{cb})\cdot(v_{ab}\times v_{ac})} \tag{E.4}$$

and this constitutes a different solution from $g_a = \mathbb{1}$ (which characterizes the original tetrahedron associated to the boundary data). Finally, in the marginal case where the boundary data is that of four copies of the same triangle, the net can be closed into a flat degenerate tetrahedron, into an equilateral tetrahedron or into its reflection. But since both proper tetrahedra have the exact same dihedral angles their associated critical points coalesce, and there is again a total of two critical configurations.

# Bibliography

[1] Event Horizon Telescope Collaboration, K. Akiyama *et al.*, "First M87 Event Horizon Telescope Results. I. The Shadow of the Supermassive Black Hole," *Astrophys. J. Lett.* **875** (2019) L1, `arXiv:1906.11238`.

[2] J. Butterfield and C. J. Isham, *Space-time and the philosophical challenge of quantum gravity*, pp. 33–89. Cambridge University Press, 1999. `arXiv:gr-qc/9903072`.

[3] S. Carlip, D.-W. Chiou, W.-T. Ni, and R. Woodard, "Quantum Gravity: A Brief History of Ideas and Some Prospects," *Int. J. Mod. Phys. D* **24** no. 11, (2015) 1530028, `arXiv:1507.08194`.

[4] S. W. Hawking, "Breakdown of Predictability in Gravitational Collapse," *Phys. Rev. D* **14** (1976) 2460–2473.

[5] J. Polchinski, "The Black Hole Information Problem," in *Theoretical Advanced Study Institute in Elementary Particle Physics: New Frontiers in Fields and Strings*, pp. 353–397. 2017. `arXiv:1609.04036`.

[6] K. Crowther and S. De Haro, "Four Attitudes Towards Singularities in the Search for a Theory of Quantum Gravity," `arXiv:2112.08531`.

[7] M. Bojowald, "Quantum cosmology: a review," *Rept. Prog. Phys.* **78** (2015) 023901, `arXiv:1501.04899`.

[8] D. Carney, Y. Chen, A. Geraci, H. Müller, C. D. Panda, P. C. E. Stamp, and J. M. Taylor, "Snowmass 2021 White Paper: Tabletop experiments for infrared quantum gravity," in *Snowmass 2021*. 3, 2022. `arXiv:2203.11846`.

[9] S. Carlip, "Is Quantum Gravity Necessary?," *Class. Quant. Grav.* **25** (2008) 154010, `arXiv:0803.3456`.

[10] C. Rovelli, "The Century of the incomplete revolution: Searching for general relativistic quantum field theory," *J. Math. Phys.* **41** (2000) 3776–3800, `arXiv:hep-th/9910131`.

[11] R. Penrose, "On the Gravitization of Quantum Mechanics 1: Quantum State Reduction," *Found. Phys.* **44** (2014) 557–575.

[12] A. Großardt, "Three little paradoxes: Making sense of semiclassical gravity," *AVS Quantum Sci.* **4** no. 1, (2022) 010502, `arXiv:2201.10452`.

[13] A. Einstein, "Die Grundlage der allgemeinen Relativitätstheorie," *Annalen der Physik* **354** no. 7, (1916) 769–822.

[14] M. Nakahara, *Geometry, Topology and Physics, Second Edition*. Graduate student series in physics. Taylor & Francis, 2003.

[15] M. Hamilton, J. D., *Mathematical Gauge Theory: With Applications to the Standard Model of Particle Physics*. Universitext. Springer International Publishing, Cham, 2017.

[16] K. Krasnov, *Formulations of General Relativity*. Cambridge Monographs on Mathematical Physics. Cambridge University Press, 11, 2020.

[17] J. Tambornino, "Relational Observables in Gravity: a Review," *SIGMA* **8** (2012) 017, `arXiv:1109.0740`.

[18] J. F. Donoghue, "The effective field theory treatment of quantum gravity," *AIP Conf. Proc.* **1483** no. 1, (2012) 73–94, `arXiv:1209.3511`.

[19] J. Polchinski, "What is string theory?," in *NATO Advanced Study Institute: Les Houches Summer School, Session 62: Fluctuating Geometries in Statistical Mechanics and Field Theory*. 11, 1994. `arXiv:hep-th/9411028`.

[20] C. Rovelli, "Loop quantum gravity," *Living Rev. Rel.* **1** (1998) 1, `arXiv:gr-qc/9710008`.

[21] D. Oriti, "The Group field theory approach to quantum gravity," `arXiv:gr-qc/0607032`.

[22] R. Loll, "Quantum Gravity from Causal Dynamical Triangulations: A Review," *Class. Quant. Grav.* **37** no. 1, (2020) 013002, `arXiv:1905.08669`.

[23] A. Eichhorn, "Asymptotically safe gravity," in *57th International School of Subnuclear Physics: In Search for the Unexpected*. 2, 2020. `arXiv:2003.00044`.

[24] C. Bambi, L. Modesto, and I. Shapiro, eds., *Handbook of Quantum Gravity*. Springer Singapore, 2023.

[25] R. P. Feynman, *The principle of least action in quantum mechanics*. PhD thesis, Princeton U., 1942.

[26] R. Oeckl, "A 'General boundary' formulation for quantum mechanics and quantum gravity," *Phys. Lett. B* **575** (2003) 318–324, `arXiv:hep-th/0306025`.

[27] R. Oeckl, "General boundary quantum field theory: Foundations and probability interpretation," *Adv. Theor. Math. Phys.* **12** no. 2, (2008) 319–352, `arXiv:hep-th/0509122`.

[28] R. Oeckl, "Probabilites in the general boundary formulation," *J. Phys. Conf. Ser.* **67** (2007) 012049, `arXiv:hep-th/0612076`.

[29] S. Holst, "Barbero's Hamiltonian derived from a generalized Hilbert-Palatini action," *Phys. Rev. D* **53** (1996) 5966–5969, `arXiv:gr-qc/9511026`.

[30] A. Ashtekar and R. Tate, *Lectures on Non-perturbative Canonical Gravity*. World Scientific, 1991.

[31] T. Thiemann, *Modern Canonical Quantum General Relativity*. Cambridge Monographs on Mathematical Physics. Cambridge University Press, 2007.

[32] G. Immirzi, "Real and complex connections for canonical gravity," *Class. Quant. Grav.* **14** (1997) L177–L181, `arXiv:gr-qc/9612030`.

[33] A. Ashtekar, "New Variables for Classical and Quantum Gravity," *Phys. Rev. Lett.* **57** (1986) 2244–2247.

[34] J. F. Barbero G., "Real Ashtekar variables for Lorentzian signature space times," *Phys. Rev. D* **51** (1995) 5507–5510, `arXiv:gr-qc/9410014`.

[35] C. Rovelli and L. Smolin, "Discreteness of area and volume in quantum gravity," *Nucl. Phys. B* **442** (1995) 593–622, `arXiv:gr-qc/9411005`. [Erratum: Nucl.Phys.B 456, 753–754 (1995)].

[36] K. Krasnov, "On the constant that fixes the area spectrum in canonical quantum gravity," *Class. Quant. Grav.* **15** (1998) L1–L4, `arXiv:gr-qc/9709058`.

[37] A. Ashtekar, J. Baez, A. Corichi, and K. Krasnov, "Quantum geometry and black hole entropy," *Phys. Rev. Lett.* **80** (1998) 904–907, `arXiv:gr-qc/9710007`.

[38] N. Barros e Sa, "Hamiltonian analysis of general relativity with the Immirzi parameter," *Int. J. Mod. Phys. D* **10** (2001) 261–272, `arXiv:gr-qc/0006013`.

[39] J. C. Baez, "An Introduction to Spin Foam Models of $BF$ Theory and Quantum Gravity," *Lect. Notes Phys.* **543** (2000) 25–93, `arXiv:gr-qc/9905087`.

[40] A. Perez, "Spin foam models for quantum gravity," *Class. Quant. Grav.* **20** (2003) R43, `arXiv:gr-qc/0301113`.

[41] I. Gelfand, M. Graev, and N. Vilenkin, *Generalized Functions: Integral geometry and representation theory*, vol. 5. Acad. Press, 1966.

[42] W. Ruhl and W. Rühl, *The Lorentz Group and Harmonic Analysis*. Mathematical physics monograph series. W. A. Benjamin, 1970.

[43] L. Freidel and K. Krasnov, "A New Spin Foam Model for 4d Gravity," *Class. Quant. Grav.* **25** (2008) 125018, `arXiv:0708.1595`.

[44] J. C. Baez and J. W. Barrett, "The Quantum tetrahedron in three-dimensions and four-dimensions," *Adv. Theor. Math. Phys.* **3** (1999) 815–850, `arXiv:gr-qc/9903060`.

[45] J. W. Barrett and L. Crane, "Relativistic spin networks and quantum gravity," *J. Math. Phys.* **39** (1998) 3296–3302, `arXiv:gr-qc/9709028`.

[46] J. W. Barrett and L. Crane, "A Lorentzian signature model for quantum general relativity," *Class. Quant. Grav.* **17** (2000) 3101–3118, arXiv:gr-qc/9904025.

[47] J. Engle, E. Livine, R. Pereira, and C. Rovelli, "LQG vertex with finite Immirzi parameter," *Nucl. Phys. B* **799** (2008) 136–149, arXiv:0711.0146.

[48] J. Engle, R. Pereira, and C. Rovelli, "Flipped spinfoam vertex and loop gravity," *Nucl. Phys. B* **798** (2008) 251–290, arXiv:0708.1236.

[49] M. Montesinos, "Alternative symplectic structures for SO(3,1) and SO(4) four-dimensional BF theories," *Class. Quant. Grav.* **23** (2006) 2267–2278, arXiv:gr-qc/0603076.

[50] B. Dittrich and T. Thiemann, "Testing the master constraint programme for loop quantum gravity. I. General framework," *Class. Quant. Grav.* **23** (2006) 1025–1066, arXiv:gr-qc/0411138.

[51] E. R. Livine and S. Speziale, "A New spinfoam vertex for quantum gravity," *Phys. Rev. D* **76** (2007) 084028, arXiv:0705.0674.

[52] A. Perelomov, *Generalized Coherent States and Their Applications*. Theoretical and Mathematical Physics. Springer, 1986.

[53] J. W. Barrett, R. J. Dowdall, W. J. Fairbairn, F. Hellmann, and R. Pereira, "Lorentzian spin foam amplitudes: Graphical calculus and asymptotics," *Class. Quant. Grav.* **27** (2010) 165009, arXiv:0907.2440.

[54] F. Conrady and J. Hnybida, "A spin foam model for general Lorentzian 4-geometries," *Class. Quant. Grav.* **27** (2010) 185011, arXiv:1002.1959.

[55] A. F. Jercher, D. Oriti, and A. G. A. Pithis, "Complete Barrett-Crane model and its causal structure," *Phys. Rev. D* **106** no. 6, (2022) 066019, arXiv:2206.15442.

[56] E. Alesci and C. Rovelli, "The Complete LQG propagator. I. Difficulties with the Barrett-Crane vertex," *Phys. Rev. D* **76** (2007) 104012, arXiv:0708.0883.

[57] R. E. Livine, "Immirzi parameter in the Barrett-Crane model?," arXiv:gr-qc/0103081.

[58] M. Geiller and K. Noui, "A note on the Holst action, the time gauge, and the Barbero-Immirzi parameter," *Gen. Rel. Grav.* **45** (2013) 1733–1760, arXiv:1212.5064.

[59] J. C. Baez, J. D. Christensen, T. R. Halford, and D. C. Tsang, "Spin foam models of Riemannian quantum gravity," *Class. Quant. Grav.* **19** (2002) 4627–4648, arXiv:gr-qc/0202017.

[60] E. Bianchi, L. Modesto, C. Rovelli, and S. Speziale, "Graviton propagator in loop quantum gravity," *Class. Quant. Grav.* **23** (2006) 6989–7028, arXiv:gr-qc/0604044.

[61] M. P. Reisenberger, "On relativistic spin network vertices," *J. Math. Phys.* **40** (1999) 2046–2054, arXiv:gr-qc/9809067.

[62] A. Baratin and D. Oriti, "Quantum simplicial geometry in the group field theory formalism: reconsidering the Barrett-Crane model," *New J. Phys.* **13** (2011) 125011, arXiv:1108.1178.

[63] W. Kaminski, M. Kisielowski, and H. Sahlmann, "Asymptotic analysis of the EPRL model with timelike tetrahedra," *Class. Quant. Grav.* **35** no. 13, (2018) 135012, arXiv:1705.02862.

[64] H. Liu and M. Han, "Asymptotic analysis of spin foam amplitude with timelike triangles," *Phys. Rev. D* **99** no. 8, (2019) 084040, arXiv:1810.09042.

[65] J. D. Simão and S. Steinhaus, "Asymptotic analysis of spin-foams with timelike faces in a new parametrization," *Phys. Rev. D* **104** no. 12, (2021) 126001, arXiv:2106.15635.

[66] S. K. Asante, B. Dittrich, and H. M. Haggard, "The Degrees of Freedom of Area Regge Calculus: Dynamics, Non-metricity, and Broken Diffeomorphisms," *Class. Quant. Grav.* **35** no. 13, (2018) 135009, arXiv:1802.09551.

[67] P. Dona and S. Speziale, "Asymptotics of lowest unitary SL(2,C) invariants on graphs," *Phys. Rev. D* **102** no. 8, (2020) 086016, arXiv:2007.09089.

[68] P. Donà, F. Gozzini, and G. Sarno, "Numerical analysis of spin foam dynamics and the flatness problem," *Phys. Rev. D* **102** no. 10, (2020) 106003, arXiv:2004.12911.

[69] F. Conrady and L. Freidel, "On the semiclassical limit of 4d spin foam models," *Phys. Rev. D* **78** (2008) 104023, arXiv:0809.2280.

[70] P. Dona, M. Han, and H. Liu, "Spinfoams and high performance computing," arXiv:2212.14396.

[71] B. Bahr and S. Steinhaus, "Investigation of the Spinfoam Path integral with Quantum Cuboid Intertwiners," *Phys. Rev. D* **93** no. 10, (2016) 104029, arXiv:1508.07961.

[72] M. Han, Z. Huang, H. Liu, and D. Qu, "Complex critical points and curved geometries in four-dimensional Lorentzian spinfoam quantum gravity," *Phys. Rev. D* **106** no. 4, (2022) 044005, arXiv:2110.10670.

[73] M. Han, Z. Huang, H. Liu, D. Qu, and Y. Wan, "Spinfoam on a Lefschetz thimble: Markov chain Monte Carlo computation of a Lorentzian spinfoam propagator," *Phys. Rev. D* **103** no. 8, (2021) 084026, arXiv:2012.11515.

[74] F. Gozzini, "A high-performance code for EPRL spin foam amplitudes," *Class. Quant. Grav.* **38** no. 22, (2021) 225010, `arXiv:2107.13952`.

[75] S. K. Asante, J. D. Simão, and S. Steinhaus, "Spin-foams as semiclassical vertices: Gluing constraints and a hybrid algorithm," *Phys. Rev. D* **107** no. 4, (2023) 046002, `arXiv:2206.13540`.

[76] S. K. Asante, B. Dittrich, and H. M. Haggard, "Effective Spin Foam Models for Four-Dimensional Quantum Gravity," *Phys. Rev. Lett.* **125** no. 23, (2020) 231301, `arXiv:2004.07013`.

[77] S. K. Asante, B. Dittrich, and H. M. Haggard, "Discrete gravity dynamics from effective spin foams," *Class. Quant. Grav.* **38** no. 14, (2021) 145023, `arXiv:2011.14468`.

[78] S. K. Asante, B. Dittrich, and J. Padua-Arguelles, "Effective spin foam models for Lorentzian quantum gravity," *Class. Quant. Grav.* **38** no. 19, (2021) 195002, `arXiv:2104.00485`.

[79] R. P. Geroch and J. B. Hartle, "Computability and physical theories," *Found. Phys.* **16** (1986) 533–550, `arXiv:1806.09237`.

[80] S. K. Asante, B. Dittrich, and S. Steinhaus, "Spin foams, Refinement limit and Renormalization," `arXiv:2211.09578`.

[81] D. Oriti, "Group Field Theory and Loop Quantum Gravity," `arXiv:1408.7112`.

[82] J. Engle and C. Rovelli, "The accidental flatness constraint does not mean a wrong classical limit," *Class. Quant. Grav.* **39** no. 11, (2022) 117001, `arXiv:2111.03166`.

[83] S. Steinhaus, "Coarse Graining Spin Foam Quantum Gravity—A Review," *Front. in Phys.* **8** (2020) 295, `arXiv:2007.01315`.

[84] B. Dittrich, E. Schnetter, C. J. Seth, and S. Steinhaus, "Coarse graining flow of spin foam intertwiners," *Phys. Rev. D* **94** no. 12, (2016) 124050, `arXiv:1609.02429`.

[85] B. Dittrich, F. C. Eckert, and M. Martin-Benito, "Coarse graining methods for spin net and spin foam models," *New J. Phys.* **14** (2012) 035008, `arXiv:1109.4927`.

[86] B. Bahr and S. Steinhaus, "Hypercuboidal renormalization in spin foam quantum gravity," *Phys. Rev. D* **95** no. 12, (2017) 126006, `arXiv:1701.02311`.

[87] C. Delcamp, B. Dittrich, and A. Riello, "Fusion basis for lattice gauge theory and loop quantum gravity," *JHEP* **02** (2017) 061, `arXiv:1607.08881`.

[88] P. Donà, A. Eichhorn, and R. Percacci, "Matter matters in asymptotically safe quantum gravity," *Phys. Rev. D* **89** no. 8, (2014) 084035, `arXiv:1311.2898`.

[89] M. Ali and S. Steinhaus, "Toward matter dynamics in spin foam quantum gravity," *Phys. Rev. D* **106** no. 10, (2022) 106016, arXiv:2206.04076.

[90] E. Bianchi, M. Han, C. Rovelli, W. Wieland, E. Magliaro, and C. Perini, "Spinfoam fermions," *Class. Quant. Grav.* **30** (2013) 235023, arXiv:1012.4719.

[91] M. Han and C. Rovelli, "Spin-foam Fermions: PCT Symmetry, Dirac Determinant, and Correlation Functions," *Class. Quant. Grav.* **30** (2013) 075007, arXiv:1101.3264.

[92] W. J. Fairbairn, "Fermions in three-dimensional spinfoam quantum gravity," *Gen. Rel. Grav.* **39** (2007) 427–476, arXiv:gr-qc/0609040.

[93] L. Freidel and S. Speziale, "Twisted geometries: A geometric parametrisation of SU(2) phase space," *Phys. Rev. D* **82** (2010) 084040, arXiv:1001.2748.

[94] L. Freidel and S. Speziale, "From twistors to twisted geometries," *Phys. Rev. D* **82** (2010) 084041, arXiv:1006.0199.

[95] E. R. Livine, "Towards a Covariant Loop Quantum Gravity," arXiv:gr-qc/0608135.

[96] S. Speziale and W. M. Wieland, "The twistorial structure of loop-gravity transition amplitudes," *Phys. Rev. D* **86** (2012) 124023, arXiv:1207.6348.

[97] W. M. Wieland, "Twistorial phase space for complex Ashtekar variables," *Class. Quant. Grav.* **29** (2012) 045007, arXiv:1107.5002.

[98] J. Rennert, "Timelike twisted geometries," *Phys. Rev. D* **95** no. 2, (2017) 026002, arXiv:1611.00441.

[99] J. D. Simão, "Biquaternions, Majorana spinors and time-like spin-foams," arXiv:2401.10324 [gr-qc].

[100] J. D. Simão, "A new 2+1 coherent spin-foam vertex for quantum gravity," arXiv:2402.05993 [gr-qc].

[101] E. R. Livine and J. Tambornino, "Spinor Representation for Loop Quantum Gravity," *J. Math. Phys.* **53** (2012) 012503, arXiv:1105.3385.

[102] S. Speziale and M. Zhang, "Null twisted geometries," *Phys. Rev. D* **89** no. 8, (2014) 084070, arXiv:1311.3279.

[103] A. Kirillov, *Lectures on the Orbit Method*. Graduate studies in mathematics. American Mathematical Society, 2004.

[104] E. R. Livine, S. Speziale, and J. Tambornino, "Twistor Networks and Covariant Twisted Geometries," *Phys. Rev. D* **85** (2012) 064002, arXiv:1108.0369.

[105] P. Woit, *Quantum Theory, Groups and Representations*. Springer, 2017.

[106] F. Conrady and J. Hnybida, "Unitary irreducible representations of SL(2,C) in discrete and continuous SU(1,1) bases," *J. Math. Phys.* **52** (2011) 012501, `arXiv:1007.0937`.

[107] V. Bargmann, "Irreducible unitary representations of the Lorentz group," *Annals Math.* **48** (1947) 568–640.

[108] G. Lindblad and B. Nagel, "Continuous bases for unitary irreducible representations of $\mathrm{SU}(1,1)$," *Ann. De L'I.H.P., Sec. A.* **13** (1970) 27–56.

[109] M. Combescure and D. Robert, *Coherent States and Applications in Mathematical Physics*. Theoretical and Mathematical Physics. Springer Netherlands, 2012.

[110] G. Ponzano and T. E. Regge, "Semiclassical limit of Racah coefficients,". `http://cds.cern.ch/record/461451`.

[111] L. C. Biedenharn, J. Nuyts, and N. Straumann, "On the unitary representations of $SU(1,1)$ and $SU(2,1)$," *Ann. De L'I.H.P., Sec. A.* **3** (1965) 13–39.

[112] G. Lindblad, "Eigenfunction expansions associated with unitary irreducible representations of su(1,1)," *Phys. Scripta* **1** (1970) 201–207.

[113] L. Freidel, "A Ponzano-Regge model of Lorentzian 3-dimensional gravity," *Nucl. Phys. B Proc. Suppl.* **88** (2000) 237–240, `arXiv:gr-qc/0102098`.

[114] L. Freidel, E. R. Livine, and C. Rovelli, "Spectra of length and area in (2+1) Lorentzian loop quantum gravity," *Class. Quant. Grav.* **20** (2003) 1463–1478, `arXiv:gr-qc/0212077`.

[115] S. K. Asante, B. Dittrich, and J. Padua-Argüelles, "Complex actions and causality violations: applications to Lorentzian quantum cosmology," *Class. Quant. Grav.* **40** no. 10, (2023) 105005, `arXiv:2112.15387`.

[116] S. Davids, "Semiclassical limits of extended Racah coefficients," *J. Math. Phys.* **41** (2000) 924–943, `arXiv:gr-qc/9807061`.

[117] S. Davids, "A State sum model for (2+1) Lorentzian quantum gravity," doctoral thesis, 2000.

[118] J. M. Garcia-Islas, "(2+1)-dimensional quantum gravity, spin networks and asymptotics," *Class. Quant. Grav.* **21** (2004) 445–464, `arXiv:gr-qc/0307054`.

[119] L. Hörmander, *The Fourier Transformation*, pp. 158–250. Springer Berlin Heidelberg, Berlin, Heidelberg, 2003.

[120] P. Donà, M. Fanizza, G. Sarno, and S. Speziale, "Numerical study of the Lorentzian Engle-Pereira-Rovelli-Livine spin foam amplitude," *Phys. Rev. D* **100** no. 10, (2019) 106003, `arXiv:1903.12624`.

[121] R. U. Sexl, *Relativity, groups, particles: Special relativity and relativistic symmetry in field and particle physics*. Springer Vienna, 2001.

[122] T. Bommer, "Dominant contributions in the semi-classical limit of Lorentzian spin foams with time-like faces: the role of lightlike vectors." *Master's thesis*.

[123] P. Miller and A. M. Society, *Applied Asymptotic Analysis*. Graduate studies in mathematics. American Mathematical Society, 2006.

[124] A. Ciarkowski, "Uniform asymptotic expansion of an integral with a saddle point, a pole and a branch point," *Proceedings of the Royal Society of London. A. Mathematical and Physical Sciences* **426** no. 1871, (1989) 273–286.

[125] P. Donà, M. Fanizza, G. Sarno, and S. Speziale, "SU(2) graph invariants, Regge actions and polytopes," *Class. Quant. Grav.* **35** no. 4, (2018) 045011, `arXiv:1708.01727`.

[126] E. R. Livine and D. Oriti, "Implementing causality in the spin foam quantum geometry," *Nucl. Phys. B* **663** (2003) 231–279, `arXiv:gr-qc/0210064`.

[127] B. Dittrich and A. Kogios, "From spin foams to area metric dynamics to gravitons," *Class. Quant. Grav.* **40** no. 9, (2023) 095011, `arXiv:2203.02409`.

[128] T. Hahn, "CUBA: A Library for multidimensional numerical integration," *Comput. Phys. Commun.* **168** (2005) 78–95, `arXiv:hep-ph/0404043`.

[129] S. Speziale, "Boosting Wigner's nj-symbols," *J. Math. Phys.* **58** no. 3, (2017) 032501, `arXiv:1609.01632`.

[130] M. Han, H. Liu, and D. Qu, "Complex critical points in Lorentzian spinfoam quantum gravity: Four-simplex amplitude and effective dynamics on a double-$\Delta 3$ complex," *Phys. Rev. D* **108** no. 2, (2023) 026010, `arXiv:2301.02930`.

[131] P. Ramacher, "Singular equivariant asymptotics and the moment map I," 2009.

[132] M. Han, "Cosmological Constant in LQG Vertex Amplitude," *Phys. Rev. D* **84** (2011) 064010, `arXiv:1105.2212`.

[133] L. Pukanszky, "On the Plancherel theorem of the $2 \times 2$ real unimodular group," *Bull. Amer. Math. Soc.* **69** no. 504, (1963) .

[134] R. Takahashi, "Sur les Fonctions Spheriques et la Formule de Plancherel dans le Groupe Hyperbolique," *Japan. J. Math.* **31** (1961) 55–90.

[135] M. Naimark and H. Farahat, *Linear Representations of the Lorentz Group*. ISSN. Elsevier Science, 2014.

[136] V. Alexandrov, "Flexible polyhedra in Minkowski 3-space," *Manuscripta Math.* no. 111, (2003) 341–356.

[137] R. Souam, "The Schläfli formula for polyhedra and piecewise smooth hypersurfaces," *Differential Geometry and its Applications* **20** no. 1, (2004) 31–45.

[138] E. Suárez-Peiró, "A Schläfli differential formula for simplices in semi-riemannian hyperquadrics, Gauss-Bonnet formulas for simplices in the de Sitter sphere and the dual volume of a hyperbolic simplex," *Pacific Journal of Mathematics* **194** (2000) 229–255.

[139] A. Alexandrov, *Convex Polyhedra.* Springer Monographs in Mathematics. Springer Berlin Heidelberg, 2005.